\pdfoutput=1
\documentclass[11pt,twoside,a4paper,cmspaper,final,collab]{cms-tdr}
\def\svnVersion{3e5052c}\def\svnDate{2025/04/11}\def\cmsCernNoTag{CERN-EP-2024-068}\def\cmsCernDate{\today}\def\cmsMessage{Published in Physics Reports as \href{http://dx.doi.org/10.1016/j.physrep.2024.09.006}{\doi{10.1016/j.physrep.2024.09.006}.}}
\begin{document}\cmsNoteHeader{EXO-23-007}

\newcommand{\keVns}{\ensuremath{\text{ke\hspace{-.08em}V}}\xspace}

\newcommand{\mMM}{\ensuremath{m_{\PGm\PGm}}\xspace}
\newcommand{\mEE}{\ensuremath{m_{\Pe\Pe}}\xspace}
\newcommand{\mMMMM}{\ensuremath{m_{4\PGm}}\xspace}
\newcommand{\ptMM}{\ensuremath{\pt^{\PGm\PGm}}\xspace}
\newcommand{\drMM}{\ensuremath{\deltar_{\PGm\PGm}}\xspace}

\DeclareRobustCommand{\PZD}{{\HepParticle{Z}{D}{}}\Xspace}

\newcommand{\ns}{\unit{ns}}

\newlength\cmsTabSkip\setlength{\cmsTabSkip}{0.3ex}
\newlength\cmsBigSkip\setlength{\cmsBigSkip}{0.6ex}
\providecommand{\cmsTable}[1]{\resizebox{\textwidth}{!}{#1}}

\providecommand{\cmsLeft}{left\xspace}
\providecommand{\cmsRight}{right\xspace}

\newcommand{\nsubjet}{\ensuremath{\tau_{32}}\xspace}
\newcommand{\mjj}{\ensuremath{m_{\mathrm{jj}}}\xspace}
\newcommand{\mjjj}{\ensuremath{m_{\mathrm{jjj}}}\xspace}
\newcommand{\ptscout}{\ensuremath{\pt^\text{scout}}\xspace}
\newcommand{\ptoff}{\ensuremath{\pt^\text{off}}\xspace}

\newcommand{\pp}{\ensuremath{{\Pp}{\Pp}}\xspace} 
\newcommand{\bparking}{\ensuremath{\PB\,\text{parking}}\xspace} 
\newcommand{\egamma}{\ensuremath{\Pe/\PGg}\xspace}

\newcommand{\sci}[2]{\ensuremath{#1\times10^{#2}}\xspace}
\newcommand{\Lint}{\ensuremath{\mathcal{L}_{\text{int}}}\xspace}
\newcommand{\Linst}{\ensuremath{\mathcal{L}_{\text{inst}}}\xspace}

\newcommand{\invcms}{\ensuremath{\text{cm}^{-2}\text{s}^{-1}}\xspace}
\newcommand{\gbs}{\unit{GB/s}}

\newcommand{\ptgen}{\ensuremath{\pt^{\text{gen}}}\xspace}
\newcommand{\etagen}{\ensuremath{\eta^{\text{gen}}}\xspace}
\newcommand{\Et}{\ensuremath{E_{\text{T}}}\xspace}

\newcommand{\ipsig}{\ensuremath{\text{IP}_{\text{sig}}}\xspace}
\newcommand{\dxy}{\ensuremath{d_{\text{xy}}}\xspace}
\newcommand{\dxyerr}{\ensuremath{\sigma_{d_{\text{xy}}}}\xspace}
\newcommand{\dxysig}{\ensuremath{\dxy/\dxyerr}\xspace}
\newcommand{\lxy}{\ensuremath{l_{\text{xy}}}\xspace}
\newcommand{\lxyerr}{\ensuremath{\sigma_{\lxy}}\xspace}
\newcommand{\deltar}{\ensuremath{{\Delta}R}\xspace}

\newcommand{\ate}{\ensuremath{\mathcal{A}\epsilon}\xspace}
\newcommand{\eff}{\ensuremath{\varepsilon}\xspace}
\newcommand{\pvtx}{\ensuremath{P_{\text{vtx}}}\xspace}
\newcommand{\pmis}{\ensuremath{P_{\text{mis}}}\xspace}

\newcommand{\BF}{\ensuremath{\mathcal{B}}\xspace}
\newcommand{\qsq}{\ensuremath{q^2}\xspace}
\newcommand\qsqmin{\ensuremath{q^2_{\text{min}}}\xspace}
\newcommand\qsqmax{\ensuremath{q^2_{\text{max}}}\xspace}
\newcommand{\PfivePrime}{\ensuremath{P_{5}^{\prime}}\xspace}

\newcommand{\ELLELL}{\ensuremath{\Pell^+\Pell^-}\xspace}

\newcommand{\bto}{\ensuremath{\PQb\to}\xspace}
\newcommand{\cto}{\ensuremath{\PQc\to}\xspace}
\newcommand{\btosll}{\ensuremath{\bto\PQs\Pell\Pell}\xspace}
\newcommand{\btosee}{\ensuremath{\bto\PQs\Pep\Pem}\xspace}
\newcommand{\btosmm}{\ensuremath{\bto\PQs\PGm\PGm}\xspace}
\newcommand{\btoclnu}{\ensuremath{\bto\PQc\Pell\PGn}\xspace}
\newcommand{\btocmunu}{\ensuremath{\bto\PQc\PGm\PGn}\xspace}
\newcommand{\btomux}{\ensuremath{\bto\PGm\PX}\xspace}
\newcommand{\bctomux}{\ensuremath{\bto\cto\PGm\PX}\xspace}
\newcommand{\btoctomux}{\ensuremath{\PQb(\to\PQc)\to\PGm\PX}\xspace}

\newcommand{\bstomm}{\ensuremath{\PBzs\to\MM}\xspace}

\newcommand{\RX}{\ensuremath{R_{\text X}}\xspace}
\newcommand{\RK}{\ensuremath{R_\PK}\xspace}
\newcommand{\RKst}{\ensuremath{R_{\PKst}}\xspace}
\newcommand{\RKstar}{\ensuremath{R_{\PK^{(\ast)}}}\xspace}
\newcommand{\RD}{\ensuremath{R_{\PD}}\xspace}
\newcommand{\RDst}{\ensuremath{R_{\PD^{\ast}}}\xspace}
\newcommand{\RDstar}{\ensuremath{R_{\PD^{(\ast)}}}\xspace}

\newcommand{\BKll}{\ensuremath{\PBp\to\PKp\ELLELL}\xspace}
\newcommand{\BKmm}{\ensuremath{\PBp\to\PKp\MM}\xspace}
\newcommand{\BKee}{\ensuremath{\PBp\to\PKp\EE}\xspace}
\newcommand{\BKJp}{\ensuremath{\PBp\to\JPsi\PKp}\xspace}
\newcommand{\BKJpll}{\ensuremath{\PBp\to\JPsi(\to\ELLELL)\PKp}\xspace}
\newcommand{\BKJpmm}{\ensuremath{\PBp\to\JPsi(\to\MM)\PKp}\xspace}
\newcommand{\BKJpee}{\ensuremath{\PBp\to\JPsi(\to\EE)\PKp}\xspace}

\renewcommand{\PKstz}{\ensuremath{\PK^{\ast}(892)^{0}}\xspace}
\newcommand{\BKst}{\ensuremath{\PBz\to\PKstz}\xspace}
\newcommand{\BKstll}{\ensuremath{\PBz\to\PKstz\ELLELL}\xspace}
\newcommand{\BKstmm}{\ensuremath{\PBz\to\PKstz\MM}\xspace}
\newcommand{\BKstee}{\ensuremath{\PBz\to\PKstz\EE}\xspace}
\newcommand{\BKstJpll}{\ensuremath{\PBz\to\JPsi(\to\ELLELL)\PKstz}\xspace}
\newcommand{\BKstJpmm}{\ensuremath{\PBz\to\JPsi(\to\MM)\PKstz\MM}\xspace}
\newcommand{\BKstJpee}{\ensuremath{\PBz\to\JPsi(\to\EE)\PKstz}\xspace}

\newcommand{\PKstp}{\ensuremath{\PK^{\ast}(892)^{+}}\xspace}
\newcommand{\BpKst}{\ensuremath{\PBp\to\PKstp}\xspace}
\newcommand{\BpKstll}{\ensuremath{\PBp\to\PKstp\ELLELL}\xspace}
\newcommand{\BpKstmm}{\ensuremath{\PBp\to\PKstp\MM}\xspace}
\newcommand{\BpKstee}{\ensuremath{\PBp\to\PKstp\EE}\xspace}
\newcommand{\BpKstJpll}{\ensuremath{\PBp\to\JPsi(\to\ELLELL)\PKstp}\xspace}
\newcommand{\BpKstJpmm}{\ensuremath{\PBp\to\JPsi(\to\MM)\PKstp\MM}\xspace}
\newcommand{\BpKstJpee}{\ensuremath{\PBp\to\JPsi(\to\EE)\PKstp}\xspace}

\newcommand{\PBpz}{\ensuremath{\PB^{+(0)}}\xspace}
\newcommand{\PKpst}{\ensuremath{\PK^{+(\ast)}}\xspace}
\newcommand{\BpzKst}{\ensuremath{\PBpz\to\PKpst}\xspace}
\newcommand{\BpzKstll}{\ensuremath{\PBpz\to\PKpst\ELLELL}\xspace}
\newcommand{\BpzKstmm}{\ensuremath{\PBpz\to\PKpst\MM}\xspace}
\newcommand{\BpzKstee}{\ensuremath{\PBpz\to\PKpst\EE}\xspace}
\newcommand{\BpzKstJpll}{\ensuremath{\PBpz\to\PJGy(\to\ELLELL)\PKpst}\xspace}
\newcommand{\BpzKstJpmm}{\ensuremath{\PBpz\to\PJGy(\to\MM)\PKpst\MM}\xspace}
\newcommand{\BpzKstJpee}{\ensuremath{\PBpz\to\PJGy(\to\EE)\PKpst}\xspace}

\newcommand{\bdtokshortjpsi}{\ensuremath{\PBz\to\PJGy\PKzS}\xspace}
\newcommand{\bstojpsiphi}{\ensuremath{\PBzs\to\PJGy\PGf}\xspace}
\newcommand{\bctojpsilnu}{\ensuremath{\PBpc\to\PJGy\Pell\PGn}\xspace}
\newcommand{\bctojpsitaunu}{\ensuremath{\PBpc\to\PJGy\PGt\PAGn_{\PGt}}\xspace}
\newcommand{\bctojpsik}{\ensuremath{\PBpc\to\PJGy\PKp}\xspace}

\newcommand{\btodtaunu}{\ensuremath{\PB^{-}\to\PD^{0}\PGt^-\PAGn_{\PGt}}\xspace}
\newcommand{\btodtautomunu}{\ensuremath{\PB^{-}\to\PD^{0}\PGt^-(\to\PGm^-\PAGn_{\PGm}\PGn_{\PGm})\PAGn_{\PGt}}\xspace}
\newcommand{\btodlnu}{\ensuremath{\PB^{-}\to\PD^{0}\ell^-\PAGn_{\Pell}}\xspace}

\newcommand{\rdstar}{\ensuremath{R_{\PD^{(\ast)}}}\xspace}
\newcommand{\btodstlnu}{\ensuremath{\PBz\to\PD^{\ast{-}}\Pell^+\PGn_{\Pell}}\xspace}
\newcommand{\btodstmunu}{\ensuremath{\PBz\to\PD^{\ast{-}}\PGm^+\PGn_{\PGm}}\xspace}
\newcommand{\btodsttaunu}{\ensuremath{\PBz\to\PD^{\ast{-}}\PGt^+\PGn_{\PGt}}\xspace}
\newcommand{\btodsttautomunu}{\ensuremath{\PBz\to\PD^{\ast{-}}\PGt^+(\to\PGm^+\PGn_{\PGm}\PAGn_{\PGm})\PGn_{\PGt}}\xspace}

\newcommand{\taue}{\ensuremath{\PGt^-\to\Pe^-\PAGn_{\Pe}\PGn_{\PGt}}\xspace}
\newcommand{\taumu}{\ensuremath{\PGt^-\to\PGm^-\PAGn_{\PGm}\PGn_{\PGt}}\xspace}
\newcommand{\tauhad}{\ensuremath{\PGt_{\text{h}}}\xspace}
\newcommand{\tauthreeprong}{\ensuremath{\PGt^-_{\text{h}}\to\PGp^-\PGp^+\PGp^-(\PGp^0) \PGn_\PGt}\xspace}

\newcommand{\Ve}{\ensuremath{V_{\Pe\PN}}\xspace}
\newcommand{\Vu}{\ensuremath{V_{\PGm\PN}}\xspace}
\newcommand{\Vt}{\ensuremath{V_{\PGt\PN}}\xspace}
\newcommand{\VV}{\ensuremath{\abs{V_{\PN}}^2}\xspace}
\newcommand{\VVe}{\ensuremath{\abs{V_{\Pe\PN}}^2}\xspace}
\newcommand{\VVu}{\ensuremath{\abs{V_{\PGm\PN}}^2}\xspace}
\newcommand{\VVt}{\ensuremath{\abs{V_{\PGt\PN}}^2}\xspace}
\newcommand{\tauN}{\ensuremath{\tau_\PN}\xspace}
\newcommand{\ctauN}{\ensuremath{c\tau_\PN}\xspace}
\newcommand{\mN}{\ensuremath{m_\PN}\xspace}
\newcommand{\rehnl}{\ensuremath{r_\Pe}\xspace}
\newcommand{\ruhnl}{\ensuremath{r_\PGm}\xspace}
\newcommand{\rthnl}{\ensuremath{r_\PGt}\xspace}

\cmsNoteHeader{EXO-23-007}
 
\title{Enriching the physics program of the CMS experiment via data scouting and data parking}

\author*[cern]{The CMS Collaboration}

\date{\today}

\abstract{
Specialized data-taking and data-processing techniques were introduced by the CMS experiment in Run 1 of the CERN LHC to enhance the sensitivity of searches for new physics and the precision of standard model measurements. These techniques, termed data scouting and data parking, extend the data-taking capabilities of CMS beyond the original design specifications. The novel data-scouting strategy trades complete event information for higher event rates, while keeping the data bandwidth within limits. Data parking involves storing a large amount of raw detector data collected by algorithms with low trigger thresholds to be processed when sufficient computational power is available to handle such data. The research program of the CMS Collaboration is greatly expanded with these techniques. The implementation, performance, and physics results obtained with data scouting and data parking in CMS over the last decade are discussed in this Report, along with new developments aimed at further improving low-mass physics sensitivity over the next years of data taking.
}

\hypersetup{%
pdfauthor={CMS Collaboration},%
pdftitle={Enriching the physics program of CMS through data scouting and data parking},%
pdfsubject={CMS},%
pdfkeywords={CMS, data scouting, data parking, trigger, BSM physics}}

\maketitle

\tableofcontents

\newpage

\section{Introduction to data scouting and data parking} \label{sec:TheIntro}

The Compact Muon Solenoid (CMS) experiment at CERN's Large Hadron Collider (LHC) has achieved remarkable success in its mission to probe the fundamental structure of the universe. 

Results from CMS and other experiments have considerably constrained the available parameter space for physics beyond the standard model (BSM), excluding the possibility of new states with masses up to several \TeVns predicted by a wide range of models of new physics. 
It has also scrutinized the realm of strong and weak interactions with great precision, including the discovery of the Higgs boson (\PH) and the measurements of its couplings~\cite{CMS-PAPERS-HIG-19-004, CMS-PAPERS-HIG-19-001, CMS-PAPERS-HIG-19-002,CMS:2022dwd}. As we delve deeper into the extensive data set afforded by the LHC, the absence of clear signals for new BSM physics prompts us to explore further avenues of investigation.

In this report, we describe the data-scouting and data-parking techniques, which involve the nonstandard use of the trigger, data acquisition (DAQ), and offline computing and software environments of CMS. Data scouting and data parking can overcome the limits of the conventional data processing strategies employed within CMS, by leveraging the capability and flexibility of the DAQ and offline computing systems. These techniques also exploit the advanced capabilities of the smart algorithms embedded within the level-1 (L1) trigger firmware and the sophisticated software-based event reconstruction algorithms used by the high-level trigger (HLT). Data scouting and data parking were introduced during the early running period of proton-proton (\pp) collisions at the LHC, have been employed ever since, and equip CMS with the ability to substantially extend its sensitivity to low-mass and rare phenomena.

\subsection{Report structure}

This report is organized as follows. Section~\ref{sec:TheIntro} introduces the physics motivations, the common data processing challenges and the solutions adopted to mitigate them, and the evolution of data scouting and data parking since they were initially introduced in CMS. Section~\ref{sec:CMS} describes the CMS detector and trigger system, and details the typical event reconstruction workflow used in the experiment. In Section~\ref{ch:run1run2scouting}, the scouting strategy adopted in 2010--2012 (Run~1 period) and 2015--2018 (Run~2 period) is discussed, along with the main physics results obtained with this technique. Section~\ref{ch:run3scouting} describes new scouting developments for the ongoing Run~3 (started in 2022, and planned to continue through 2025). Section~\ref{ch:parking_run2} introduces the original data-parking implementation in 2012 and then focuses on the \PB parking strategy developed in Run~2 to increase the CMS sensitivity to flavor physics processes. In Section~\ref{ch:parking_run3}, new parking improvements designed for Run~3 are discussed, which are meant to complement the existing standard triggers with a large variety of physics goals in mind. Finally, Section~\ref{sec:summary} summarizes the main features and achievements of the data-scouting and data-parking strategies in CMS.

\subsection{Physics motivations}\label{sec:Motivation}

The search for new physics often leads to scenarios where hypothetical particles have low masses and feeble couplings. Processes involving such particles are difficult to detect, given the large rate of standard model (SM) backgrounds at the LHC~\cite{Agrawal:2021dbo,Antel:2023hkf}.
In order to maintain a manageable overall trigger rate, traditional data acquisition protocols frequently necessitate relatively high thresholds to mitigate SM backgrounds. Consequently, intriguing signal events characterized by lower energy and momenta may inadvertently be discarded.
 Low-mass BSM particles that decay into final states involving low-energy jets or lepton pairs therefore present considerable challenges at the LHC. These challenges stem from the huge cross sections associated with jet production. Analogously, the large quantum chromodynamics (QCD) cross section and the subsequent (semi)leptonic decays of hadrons pose similar problems to searches for new physics in light dilepton final states.

In addition to direct searches for new physics, we pursue indirect strategies where new physics may manifest as significant deviations between precise SM predictions and experimental measurements. One example is the study of rare \PB meson decays, involving particles with momenta in the few \GeVns range. However, online selection of these events presents formidable challenges, which are compounded by the need to collect a substantial amount of data to achieve sufficient statistical precision.

\subsection{Challenges and solutions}
\label{sec:Challenges}

The LHC facility features two adjacent parallel circular beamlines, each containing a bunched beam of protons traveling in opposite directions around the 27\unit{km} ring~\cite{Bruning:782076}. Each proton bunch orbits the ring at close to the speed of light 11,245 times per second. The proton beams are directed by superconducting magnets, and made to intersect at various points around the ring, where the \pp collisions take place. In 2011 and 2012, the protons were accelerated to energies of 3.5 and 4\TeV, respectively. Starting from 2015, the energy was raised to 6.5\TeV and then, from 2022, further increased to 6.8\TeV.

The LHC orbit is divided into a total of 3564 time windows, each 25\ns in duration (bunch crossing slots) and potentially containing a colliding proton bunch. The actual collision rate depends on the number of colliding bunches and the structure of the filling scheme, which varies with time. Bunches are grouped into ``trains'' with 25\ns spacing (50\ns before 2015), and larger gaps between trains. The largest number of colliding bunches, 2544, was reached in 2017 and 2018, corresponding to an average collision rate of almost 30\unit{MHz}. 
Moreover, the number of multiple \pp interactions within the same or adjacent bunch crossings, termed ``pileup'', has also varied in time, ranging between ${\approx}20$ on average in Run~1, ${\approx}40$ in Run~2, and finally ${\approx}50$ in Run~3. 

Protons are delivered in ``fills'' and, in 2023, the peak instantaneous luminosity (\Linst) at the beginning of each fill was above $\sci{2}{34}\invcms$. This level was typically maintained (``luminosity leveling'') for six hours, after which the \Linst slowly decayed to lower values for the remainder of the fill (usually lasting several hours).
The process of luminosity leveling entails deliberately diminishing the instantaneous luminosity from its maximal potential by slightly defocusing and/or separating the beams. This adjustment is crucial to prevent excessive pileup in experiments
Starting in 2024, the LHC aims to further increase the integrated luminosity ($\Lint$) delivered by extending the duration of the luminosity-leveling period.
This continuous push for improvements in the performance of the LHC operations requires the experiments to develop innovative trigger and DAQ strategies in order to continue recording data sets rich with physics potential.

The traditional paradigm for data analysis at the LHC is that \pp collision events are selected online by a trigger system, stored to disk in raw data format, and finally reconstructed and analyzed. The offline reconstruction aims to provide the best physics objects for analysis and, 
since it is not bound to be executed at the same pace of data acquisition and with the same low latency, as opposed to the trigger-level reconstruction, it achieves this goal at the cost of being computationally expensive.

The CMS experiment uses a two-tiered trigger system to filter the interesting collision events. The first level, L1, composed of custom hardware processors, relies on information from the calorimeters and muon detectors to select events up to a rate of around 100\unit{kHz} within a fixed latency of about 4\mus~\cite{CMS:2020cmk}. The second level, HLT, consists of a farm of processors running a version of the event reconstruction software optimized for fast processing~\cite{CMS:2016ngn}. The HLT reduces the event rate to several \unit{kHz} before data storage.

There are various constraints imposed on the trigger system and on the data processing framework that limit the number of events that can be selected, recorded and analyzed in this way: 

\begin{itemize}
    \item \textbf{L1 acquisition rate}. The rate of events that are accepted by the L1 system is limited to ${{\approx}100\unit{kHz}}$, determined from the finite bandwidth of the detector readout systems and the amount of raw detector information transmitted per event~\cite{CMS:2006myw,CMS:2000mvk,Sphicas:2002gg}. This is a hard constraint dictated by the detector design. Operating the system at rates beyond this threshold would result in dead time (the recording time lost because the readout system is not ready to transmit data for a new event)~\cite{CMS:2020cmk} and, effectively, no additional DAQ capability.
    \item \textbf{Event-processing time at the HLT}. The processing capacity of the HLT farm is proportional to the number of computing cores available and to the speed of such cores. The maximum processing time per event is therefore determined by the rate of L1-accepted events passed on to the HLT and by the total capacity of the farm. In 2018, it corresponded to a limit of about 600\unit{ms} per event assuming 100\unit{kHz} of L1 throughput. This rate is somewhat less of a hard constraint, as the HLT computing farm can be and is continuously being expanded via new acquisitions or via the replacement of older machines.
    \item \textbf{DAQ output bandwidth}. The DAQ throughput, increased from a few \gbs in Run~2 to about 20\gbs in Run~3, is not considered to be a limiting factor. More relevant are the restrictions on the output bandwidth from the DAQ system, imposed by the size of the temporary raw data storage buffer at the site hosting the CMS experiment and by the bandwidth of the link transferring the raw data from the temporary to the permanent storage at the main CERN site. These limit the product of the HLT output rate and the event size, which in turn opens the possibility of collecting data at higher rates in exchange for reduced event sizes. 
    \item \textbf{Prompt reconstruction of recorded data}. Normally, the full offline reconstruction of freshly recorded data, called ``prompt reconstruction'', starts with only a short delay of about 48\unit{hours} once various detector calibration and alignment data are available~\cite{Cerminara:2015hov}. 
    Routine performance measurements of high-level physics objects and simulation-to-data corrections are often essential requirements for analyses and therefore time critical. The available computing resources allow for the prompt reconstruction of data with an approximately constant 48\unit{hours} turnaround time for HLT rates up to a few \unit{kHz} on average. 
    \item \textbf{Finite permanent data storage}. Ultimately, data storage is the remaining potential bottleneck to consider in the DAQ chain. Data can be stored on disks as well as on tapes. Disks are faster to access, but offer reduced storage relative to tapes. Very large data sets may stay on disk only for short periods of time until they are processed and stored in higher-level, smaller-sized data formats. After that, they must be moved to tape, where their retrieval is not immediate. However, this is also a soft constraint, as the purchase of additional disk storage is less costly compared to the purchase of computing cores.
\end{itemize}

The trigger system selects interesting events for physics analysis at a rate that is four orders of magnitude smaller than the bunch crossing rate. As the LHC performance improves over time, the higher \Linst values delivered impact the operations of the trigger, DAQ, and computing systems. Higher \Linst values imply higher pileup, which can degrade the performance of the trigger algorithms and increase both the event size and the computational load from the event reconstruction. 

\begin{figure*}[!htb]
    \centering
    \includegraphics[width=0.95\textwidth]{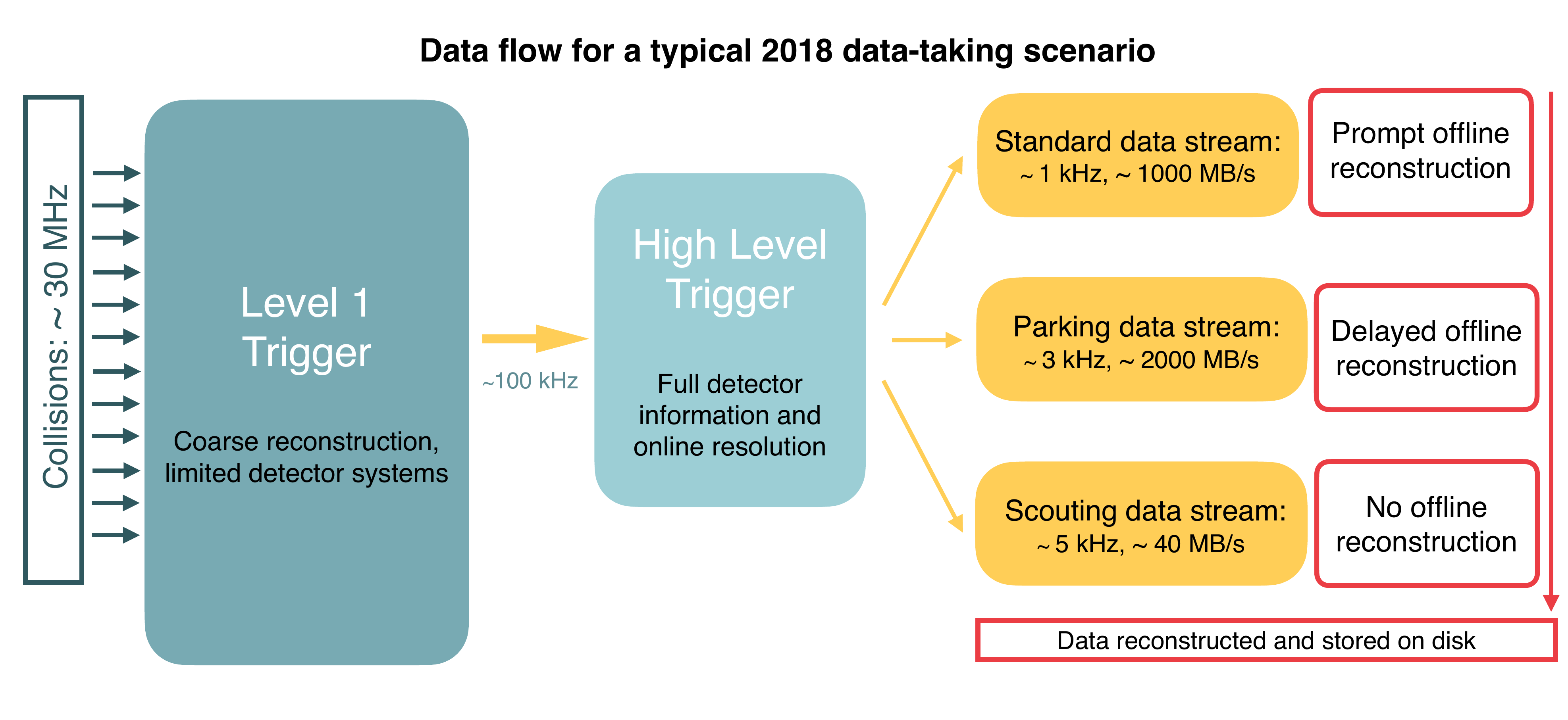}
    \caption{A schematic view of the typical Run~2 data flow during 2018 showing the data acquisition strategy with scouting and parking data streams, along with the standard data stream. A value of ${\Linst = \sci{1.2}{34}\invcms}$ over a typical 2018 fill, corresponding to an average pileup of 38, is considered.}
    \label{fig:Run2DataFlow}
\end{figure*}

The data-scouting and data-parking strategies~\cite{CMS:2012ScoutingParking} overcome two of the main limitations in the CMS data acquisition chain, namely the finite bandwidth available to write data to permanent storage and the finite ability to promptly process (\ie, reconstruct) the data as they are recorded. These two techniques are illustrated schematically in Fig.~\ref{fig:Run2DataFlow} for a representative year of data taking. Data scouting is a novel concept that CMS first prototyped in 2011 at the end of Run~1, used throughout Run~2~\cite{EXO-14-005}, and developed substantially for Run~3. Data parking is novel at the LHC, borrowing from a frequently used strategy by fixed-target experiments in which raw data are recorded and subsequently processed for analysis much later in time. 

The data-scouting strategy enhances sensitivity to low-energy physics processes by significantly lowering the HLT thresholds and storing a reduced event content on disk. Events reconstructed at the HLT are selected based on the kinematic quantities of their reconstructed objects using looser trigger thresholds than those applied in the standard trigger paths. For each event passing these looser selections, only high-level physics objects (such as jets or leptons) reconstructed at the HLT are stored on disk. No raw data from  detector channels are stored for later offline analysis. These dedicated data samples are then used offline to perform physics analysis. The excellent performance of the HLT online reconstruction, which closely approximates the performance of the standard offline reconstruction, is the basis of the success of this strategy.

The data-parking strategy also lowers the thresholds used by the trigger algorithms, thereby increasing the experimental acceptance to low-mass physics processes. The event collection rate is thus substantially increased, potentially beyond the capacity of the computational resources available to promptly reconstruct the events as they are acquired. In this case, the data parking stream is transferred, unprocessed, to tape storage and is kept in a raw format until sufficient computational resources are available for the events to be reconstructed, such as between data-taking periods. 

\begin{figure*}[!htb]
    \centering
    \includegraphics[width=0.85\textwidth]{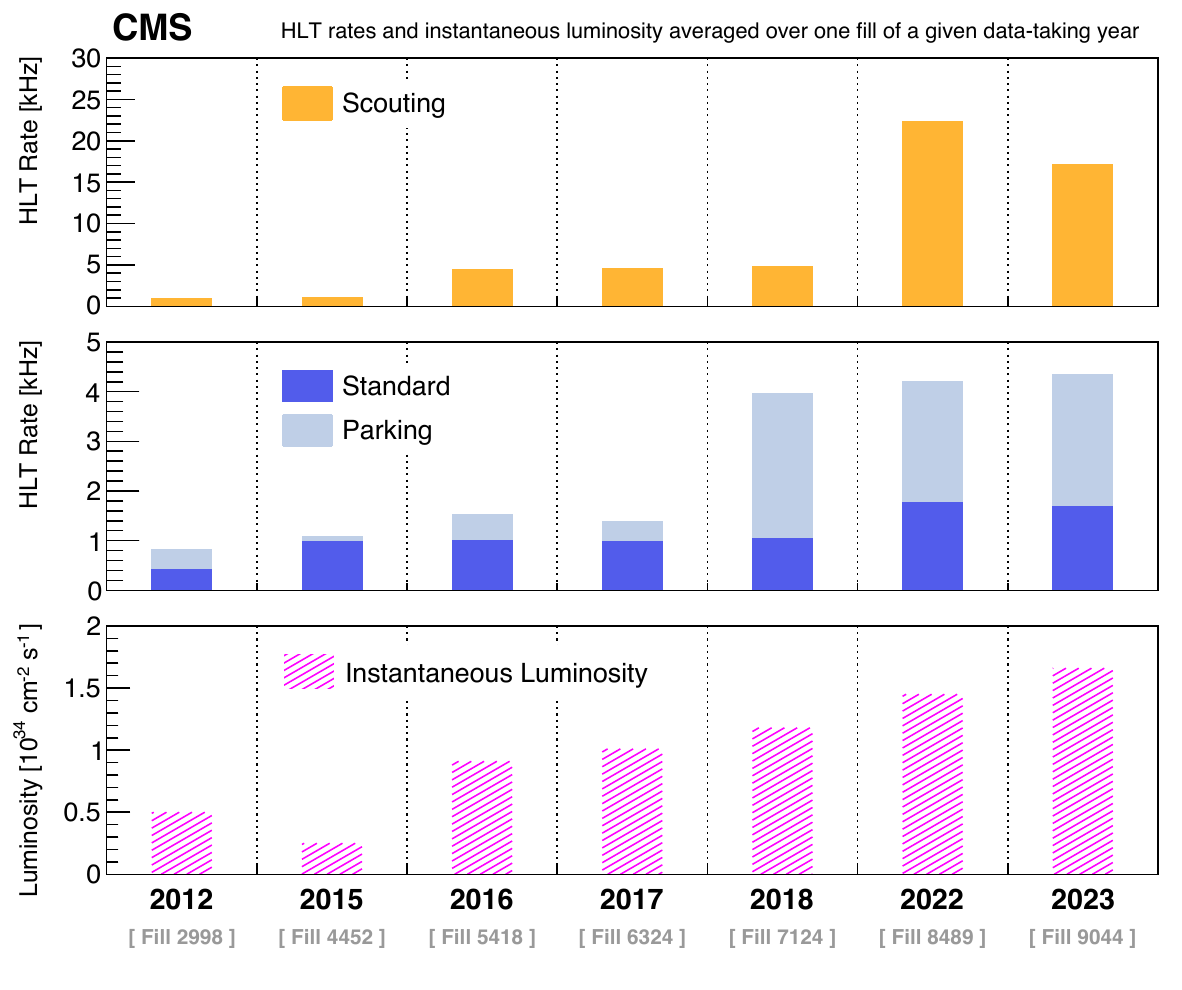}
    \caption{Comparison of the typical HLT rates of the standard, parking, and scouting data streams from Run~1 to Run~3. The \Linst averaged over one typical fill of a given data-taking year is shown in pink.}
    \label{fig:hlt_rates_evolution}        
\end{figure*}

Figure~\ref{fig:hlt_rates_evolution} shows the time evolution of the output HLT rates for the standard, data-scouting, and data-parking streams, averaged over one typical fill of a given data-taking year.

The CMS efforts have helped set a now-established trend in our field. Similar to the CMS data scouting, the LHCb and ATLAS Collaborations have ``turbo''~\cite{Benson:2015yzo} and ``trigger-level analysis''~\cite{ATLAS:2018qto} streams, respectively, which were implemented during Run~2. 
Concurrently with the CMS data-parking developments in 2012, the ATLAS Collaboration developed a comparable ``delayed stream''~\cite{Aad:1735492_AtlasDelayedStream} approach. 
Finally, one additional DAQ technique at the LHC that circumvents limitations in the ``standard'' infrastructures is the ALICE ``triggerless readout system''~\cite{Antonioli:2013ppp,Kvapil:2021tuj}.

\subsection{Origin and evolution of the data-scouting model}
\label{sec:ScoutingEvolution}

The history of data scouting starts in the last weeks of the 2011 data-taking period. The CMS Collaboration aimed to preserve the physics sensitivity of searches for resonances with sub-\TeVns masses decaying to a pair of jets (dijet). The low-mass regime had become inaccessible because of the more stringent trigger selections applied to multijet event topologies in response to increases in the peak \Linst values delivered by the LHC. These additional selections were required because the cross section for the production of two jets mediated by the strong interaction grows substantially (roughly according to a power law) as the dijet mass decreases. During Run~1, the particle-flow (PF) algorithm~\cite{Sirunyan:2017ulk} was introduced in the HLT online reconstruction. The PF algorithm, described in more detail in Section~\ref{subsubsec:event reconstruction PF}, aims to reconstruct and identify each individual particle (called a PF candidate) in an event with an optimized combination of information from the various CMS subdetectors. 
In 2011, the HLT jet algorithms adopted the PF reconstruction providing a better jet momentum and spatial resolutions. Lowering the thresholds of the jet-based HLT triggers to select interesting low-mass events led to considerably higher trigger rates and to a higher data volume. The standard trigger and DAQ pipelines were not designed to handle such large amounts of data, which would have exceeded the resources allocated for the entire physics program of the experiment.

The proposed solution was to maintain low thresholds in the HLT PF jet algorithms and mitigate the high trigger bandwidth issue by permanently recording only a reduced data format (about a hundred times smaller than the standard one) in order to satisfy the design constraints of the DAQ system at the time. The data consisted essentially of the four-momenta of the jets reconstructed at the HLT and little additional information. This new data-scouting approach, which aimed to explore previously inaccessible regions of the mass-coupling model parameter space, was successfully tested in the last days of the data-taking period in 2011 and employed to produce a first preliminary result in a search for dijet resonances. It was the first attempt of its kind at the LHC. 

Since its inception, data scouting has evolved, becoming a well-established approach in CMS, as described in Section~\ref{ch:run1run2scouting}. In 2012, the final year of Run~1, the data-scouting stream was used to search for dijet resonances with jets reconstructed from the calorimeter energy deposits alone~\cite{EXO-14-005}. In 2015, after the first long shutdown (LS1) of the LHC in 2013--2014, the data-scouting approach was consolidated by introducing a comprehensive event record based on the PF algorithm. The jet, muon, and electron candidates provided by the PF algorithm were all added to the scouting event record, which in turn allowed complex analyses to be performed, similar to what is possible with standard CMS data. An example is the study of jet substructure originating from the hadronic decay of a Lorentz-boosted resonance, as described in Section~\ref{sec:ScoutingRun2Jets}. 
New muon-based trigger algorithms were introduced to select events containing a pair of muons (dimuon) with transverse momenta of only a few \GeVns. These algorithms allowed extended searches for new dimuon resonances below 40\GeV, almost down to the kinematic threshold of twice the muon mass. The excellent performance of scouting muons in Run~2 is presented in Section~\ref{sec:ScoutingRun2Muons}.
The scouting strategy in Run~2 enabled CMS to embark on pioneering searches for low-mass resonances, including pairs of jets or muons, promptly produced or displaced with respect to the primary \pp interaction vertex, and complex decay chains involving multiple jets in the final state. An overview of these results is presented in Section~\ref{sec:Run2ScoutingPhysicsResults}.

The primary constraint in implementing the scouting strategy was found to be the HLT event processing time for the PF reconstruction algorithm.
By the end of the second long shutdown of the LHC (LS2), in 2019--2021, the computing capabilities of the HLT system were greatly improved, thanks to the new computing farm equipped with graphics processing units (GPUs). Within this new GPU-based model, events are reconstructed at the HLT with a novel scouting PF algorithm that exploits charged-particle tracks built solely from information provided by the innermost silicon-pixel tracker. The substantial reduction in the average HLT event processing time (by over 40\%) contributes to the increase in the maximum event rate that can be processed by the scouting stream. 

At the beginning of Run~3, a single, unifying data-scouting stream has been available, comprising a complete PF-based event record for all events that satisfied the requirements imposed by at least one of several L1 algorithms based on jets, muons, and electrons or photons. The development of a suitably compact, yet complete, event record for scouting relied on the substantial experience developed within CMS~\cite{Petrucciani_2015, Peruzzi_2020}. The total rate of events accepted by the combination of L1 algorithms fed to the scouting stream has increased to ${{\approx}30\unit{kHz}}$ in 2022. A minimal event selection is then applied based on the PF candidates reconstructed at the HLT before scouting events are recorded permanently for analysis. The event output rate of the data-scouting stream reached a maximum value of ${{\approx}30\unit{kHz}}$ in 2022 and ${{\approx}26\unit{kHz}}$ in 2023, roughly a factor of 10 higher than the standard data stream. In addition to jets, muons, and electrons, photons and individual PF candidates (such as hadron candidates) are now stored in the reduced scouting data format. More complex objects, such as hadronically decaying tau leptons or jets coming from the decay of heavy quarks, can in principle be reconstructed later from the constituent PF candidates. More information on the data-scouting strategy and event content in Run~3 is provided in Sections~\ref{subsec:run3scouting} and~~\ref{sec:Run3EventContent}, respectively.

Data scouting made it possible to significantly reduce trigger thresholds with respect to the standard data stream, thereby increasing the sensitivity to previously unexplored new physics domains. As an example, scouting analyses with jet-based final states can select events with \HT greater than about 300\GeV, where \HT is defined as the scalar sum of the jet \pt momenta reconstructed in the event. This can be compared to analyses using standard triggers, which require ${\HT \gtrsim 1000\GeV}$. Similarly, the scouting dimuon triggers require two muons with ${\pt > 3 \GeV}$, compared to ${\pt > 17}$ and 8\GeV for the leading and subleading \pt muons, respectively, for the standard inclusive triggers. One of the key aspects of the success of data scouting is the outstanding quality of the HLT online reconstruction. The muon reconstruction at the HLT is very similar to the offline one, which guarantees excellent performance in terms of identification efficiency and momentum resolution. Jets reconstructed from PF candidates and electron and photon objects also show comparable online and offline performance, in terms of energy scale and resolution, and particle tagging capabilities. The Run~3 performance of jets and muons, and initial studies on electron and photon objects, are presented in Sections~\ref{subsec:run3jets},~\ref{subsec:run3muons},~and~\ref{subsec:run3electronsAndPhotons}, respectively.

\subsection{Evolution of the data-parking program}
\label{sec:ParkingEvolution}

The LHC delivered \pp collisions at center-of-mass energies of 7 and 8\TeV during Run~1. Towards the end of that data-taking period, CMS was accumulating data for the core \pp physics program with an HLT rate of {300--350\unit{Hz}}. In 2012, the data-parking technique was first deployed in CMS. 
New trigger algorithms, or existing ones with relaxed kinematical thresholds, were introduced to accumulate an additional 350\unit{Hz} of data, which were subsequently parked in raw format and later reconstructed during 2013. The triggers targeted a range of SM and BSM physics scenarios, including vector boson fusion (VBF) topologies and Higgs boson measurements, \PB physics measurements, and searches for models of compressed supersymmetry (SUSY) and dark matter (DM). Section~\ref{sec:run1altParking} summarizes the Run~1 activities and the physics analyses served by these data-parking streams.

Early in Run~2, a data parking stream was enabled as a monitoring tool for the PF-based data scouting stream. As the LHC approached the end of Run~2, CMS initiated a powerful data-parking program to enable measurements of observables connected to the ``flavor anomalies''. This collective term refers to several measurements of rare \PQb hadron decays that exhibit some level of discrepancy with respect to the SM predictions~\cite{doi:10.1146/annurev-nucl-102020-090209}. These measurements have been the subject of substantial interest in the field since 2015.

In early 2018, a new trigger strategy was designed and implemented to identify muons originating from \PQb hadron decays and thus accumulate a high-purity sample of \PQb quark-antiquark (\bbbar) pairs. Kinematical requirements on the muon were progressively relaxed in the L1 and HLT algorithms during an LHC fill: the L1 algorithms were adjusted such that the system operated at or near its design limit throughout the entire LHC fill; the higher trigger rates from relaxing thresholds in the HLT algorithms were mitigated by the parking strategy.  This approach minimized the impact on the core physics program while maximizing the sample size of \bbbar events: Around $10^{10}$ \bbbar events were recorded in 2018, which enabled a new program of measurements involving \PQb hadron decays. The ``tag-side'' \PQb hadron decays to a muon (responsible for the positive trigger decision) and other particles allowed for precision measurements of rare and low-mass signatures. 
Furthermore, the sample also crucially provided an unprecedented sample of $10^{10}$ unbiased decays from the other \PQb hadron in each event, which allowed the studies of final states involving low-\pt leptons and hadrons that previously could not be probed with existing triggers. In comparison, other data sets highly enriched in \PQb hadrons collected during the same period comprise at most 5\ten{8} unbiased \PQb decays. The data sample also has rich potential for BSM searches involving, \eg, low-mass states and very rare decays, which is complementary to the data sets that serve the high-\pt searches typical at the LHC, and thus substantially extends the reach of the CMS physics program. This data sample was parked and subsequently reconstructed in 2019 during LS2. Section~\ref{sec:bparking_run2} motivates and describes the strategy in detail, and summarizes some key physics results based on the analysis of the \bparking data set.

The evolution of the \bparking physics program during the LHC Run~3 was facilitated by 
the L1 trigger and DAQ and HLT systems operating at capacities beyond their original design specifications. For instance, the L1 system routinely operated at ${{\approx}110\unit{kHz}}$ in 2023, which is exploited by the data-parking programs. Perhaps most crucially, additional computing resources are opportunistically available that allow CMS to reconstruct more data, by accommodating higher trigger rates from the HLT system. These operational developments directly and significantly enhance the scope of the CMS physics program. The improvements in LHC performance in Run~3 provide exciting new opportunities as well as challenges for data-parking strategies. The luminosity leveling periods impose constraints on available resources while enabling improved sensitivity to a wide variety of new physics searches and precision SM measurements. Section~\ref{sec:bparking_run3} describes the \bparking strategy for Run~3.

The aims of the \bparking strategy in Run~3 are twofold: collecting events with dimuon final states inclusively, and collecting events with dielectron final states inclusively. The dimuon approach simplifies the array of exclusive dimuon-based triggers that served much of the \PB physics program in Run~2. The new dimuon trigger consolidates the \PB physics program in many ways: a more efficient use of allocated trigger, DAQ, and computing resources; a common trigger strategy for the \PB physics group as a whole; and substantial gains in yields for \PQb hadron decay modes (\eg, by more than a factor of 10 for \bdtokshortjpsi) that were poorly served during Run~2. The dimuon trigger logic and physics performance are described in Section~\ref{sec:dimuon-2022}. The dielectron trigger primarily targets a measurement of the \RK observable~\cite{Hiller:2003js, Bordone:2016gaq, Isidori:2020acz, Isidori:2022bzw} with a precision that is substantially improved with respect to that achieved using the single-muon trigger strategy of Run~2. However, the dielectron trigger logic is sufficiently inclusive to provide a data set that is also rich in possibility with regards to low-mass BSM searches. The dielectron trigger adopts the same approach as the single-muon trigger algorithms in 2018, by progressively lowering kinematical thresholds at L1 and HLT during the LHC fill; transverse energy (\Et) thresholds for each electron candidate as low as 5\GeV are deployed in the L1 system towards the end of an LHC fill. Section~\ref{sec:di-ele-2022} describes the trigger logic and characterizes the data set recorded in 2022.

Improvements to the dielectron trigger strategy in 2023 opened up possibilities to further diversify the data-parking program to cover a wider range of physics topics, as it was originally conceived in 2012. Several triggers were added to the data-parking streams, with changes to their algorithms in both the L1 and HLT systems, providing improved sensitivity to a range of interesting physics processes beyond the scope of the \PB physics program. The VBF production mode for the Higgs boson is covered by a suite of triggers that identify pairs of jets in the forward regions of the CMS detector. The Higgs boson self-coupling is a key parameter of the Higgs potential that remains unmeasured. Thus, optimized triggers that provide sensitivity to the pair production of Higgs bosons via final states containing pairs of \PGt leptons and jets from \PQb quark decays were developed. Finally, new triggers were added to provide sensitivity to the distinctive experimental signatures of long-lived particles (LLPs), predicted by many BSM models. Examples include triggers that identify displaced dijets, or make use of timing information from the electromagnetic calorimeter (ECAL) subdetector. These additional data-parking strategies are discussed in Section~\ref{sec:run3altParking}.

\section{The CMS detector, trigger, and event reconstruction}
\label{sec:CMS}

This section introduces the CMS detector and describes in more detail the standard CMS trigger and event reconstruction workflow. These topics are relevant for the subsequent discussion of the data-scouting and data-parking techniques developed in the remainder of the Report.

\subsection{The CMS detector}

The central feature of the CMS apparatus is a superconducting solenoid of 6\unit{m} internal diameter, providing a magnetic field of 3.8\unit{T}. Within the solenoid volume are a silicon pixel and strip tracker, a lead tungstate crystal ECAL, and a brass and scintillator hadron calorimeter (HCAL), each composed of a barrel and two endcap sections. Forward calorimeters extend the pseudorapidity coverage provided by the barrel and endcap detectors. Muons are measured in gas-ionization detectors embedded in the steel flux-return yoke outside the solenoid. At the start of 2017, a new pixel detector was installed~\cite{Phase1Pixel} to provide four-hit pixel coverage in the pseudorapidity range ${\abs{\eta} < 2.5}$. A more detailed description of the CMS detector, together with a definition of the coordinate system used and the relevant kinematic variables, can be found in Refs.~\cite{CMS:2008xjf,CMS:2023gfb}.

\subsection{Trigger and data acquisition}

All CMS analyses rely heavily on an efficient trigger system that is able to separate the interesting processes from the huge number of background events produced in \pp collisions at the LHC. The trigger system in CMS is split into two levels, the first one relying on custom-design hardware boards that use a minimal amount of information from the subdetectors with the fastest response, and the second one exploiting the complete event information for trigger decisions.

\textbf{The L1 trigger}~\cite{CMS:2020cmk} utilizes high-bandwidth optical links and large field programmable gate arrays (FPGAs) to process the information from the calorimeters and the muon trigger system to build trigger primitives (TPs). These trigger objects along with their kinematic features are used in a large set of algorithms for the final decision of the global trigger (GT). 
The products of these algorithms are referred to as L1 seeds.
The L1 calorimeter trigger operates in two stages (layers). The calorimeter TPs (local energy deposits) from the electromagnetic, hadronic, and hadronic forward calorimeters are received by the first layer and calibrated. Then the ECAL and HCAL TPs are combined into single trigger towers and transmitted to the second layer for further processing. Jet, electron, photon, and tau lepton candidates are reconstructed and calibrated by the second layer and then fed to the GT together with the computed energy sums. 
The L1 muon trigger receives TPs from the overlapping muon subdetectors and feeds the reconstructed muon tracks into the GT. Finally, the inputs received by the GT are evaluated with a suite of algorithms and selection criteria, collectively called the trigger menu. The flexibility of the GT hardware allows regular updates of the L1 menu in response to physics program choices and to changes in the LHC beam conditions. 

\textbf{The HLT}~\cite{CMS:2016ngn} operates on fully assembled events that contain the entire event information, for example reconstructing tracks of particle trajectories and providing precise energy measurements from various subdetectors, making use of the full detector granularity and resolution. The HLT uses software algorithms running asynchronously on commercial computing hardware. 
Compared to the L1, the HLT menu defines a set of more complex algorithms and selection criteria, reconstructing online physics objects and filtering events. 
The HLT enables a more refined event selection than the L1 for storage and posterior analysis. It is also more computationally intensive, requiring longer processing times compared to the L1 trigger.

The data processing in the HLT is based on the concept of a trigger ``path'', which is a set of algorithmic processing steps run in a predefined order that both reconstruct physics objects and apply selections on these objects based on physics requirements. 
The HLT paths targeting similar physics processes are grouped into common primary data sets (PDs). The PDs are defined to keep the total event rate balanced and within the limits imposed by the available data offline processing resources. While events can end up in more than one PD because of different trigger selections, significant effort is made to keep the event overlap to a minimum. Collections of PDs are organized into data ``streams'', for efficient data handling. A data stream consists of a set of HLT paths and a well-defined event content.

In addition to the algorithms used to record events for physics analyses, the HLT also contains specific paths and data streams to gather information for detector calibrations and to conduct online data quality monitoring during data taking. Over the years, the HLT processing requirements increased notably in response to the evolving LHC and detector conditions, and the HLT computing capacity has been gradually scaled up to reflect the needs of the experiment. The Run~3 system was largely renewed by including general-purpose GPUs to provide cost-effective computing acceleration. Further details about the main Run~3 changes relevant to this Report are described in Section~\ref{ch:run3scouting}. 

\textbf{The DAQ system} provides the data pathway and time decoupling between the synchronous detector readout and data reduction, the asynchronous selection of interesting events in the HLT, their intermediate or temporary local storage at the experiment site, and the transfer to Tier-0 for offline permanent storage and analysis. The DAQ system includes software to perform data handling, a hierarchical system to control the electronics components, monitoring systems to collect relevant metrics, and several monitoring clients to interpret those metrics. More details on the CMS DAQ and offline computing systems can be found in Refs.~\cite{CMS:2008xjf,PhaseTwoUpgradeDAQ2021,CMS-TDR-022}.

\subsection{Online and offline event reconstruction}

This section describes the physics event reconstruction workflow of CMS, both online (at the HLT) and offline. We focus on the physics objects that are also used in scouting analyses. A more complete description of the event reconstruction in CMS can be found in the references provided in the next sections.

\subsubsection{Tracks and primary vertices} \label{subsec:tracks-vertices}

The tracking and vertex reconstruction algorithms employed by CMS~\cite{CMS:2014pgm} aim to precisely reconstruct the trajectories of charged particles and pinpoint the locations of \pp interaction vertices within collision events.

In the initial stages of offline tracking, raw detector signals are converted into ``hits'' representing particle interactions with the various layers of the CMS detector, including the silicon pixel tracker and the silicon strip tracker. These hits are then utilized in a multi-step track reconstruction process: track seed generation, track finding, and track fitting. During track seed generation, potential track candidates are identified using subsets of hits, and various algorithms evaluate their compatibility with charged-particle trajectories. The subsequent track-finding stage propagates these candidates through the detector layers iteratively, refining their parameters to best fit the observed hit positions. Finally, track fitting optimizes the track parameters, such as position, direction, and momentum, by minimizing the discrepancies between the predicted and measured hit positions. The HLT uses track reconstruction software that is identical to that used for the offline reconstruction summarized above but configured to meet the constraints of the available central processing unit (CPU) resources at the HLT. This primarily involves reducing the number of iterations in the iterative tracking and/or only running iterations around objects of interest. Additionally, for some purposes at the HLT, the performance of pixel-only tracks is sufficiently robust to omit the time-consuming pixel-plus-strip tracking step, thus increasing the tracking rate for the same CPU budget. In general, the simplified HLT tracking reduces the efficiency and resolution of track parameters in some regions of phase space compared to the standard offline reconstruction.

The vertex reconstruction aims to measure the location and associated uncertainty of all \pp interaction vertices in each event, including the vertex from the hard parton scattering and any additional pileup vertices, using the available reconstructed tracks. It consists of three steps: (i) selection of the tracks, (ii) clustering of the tracks that appear to originate from the same interaction vertex, and (iii) fitting for the position of each vertex using its associated tracks.

The first stage is to select high-quality tracks that are likely to be associated with the primary interaction. This involves applying track quality criteria to filter out noise and low-quality tracks. Track clustering is then performed using a ``deterministic annealing'' algorithm~\cite{CMS:2014pgm} converging towards a set of vertex candidates.
Once the initial vertex seeds are found, a vertex fitting algorithm is employed. This algorithm iteratively refines the vertex positions and uncertainties by considering the selected tracks associated with each vertex candidate. 
The primary vertex (PV) is taken to be the vertex corresponding to the hardest scattering in the event, evaluated using tracking information alone, as described in Section 9.4.1 of Ref.~\cite{CMS-TDR-15-02}.

At the HLT, pixel tracks can be used in the reconstruction of the vertex position. A ``gap'' algorithm~\cite{CMS:2014pgm} and a ``density-based'' algorithm~\cite{Bocci:2020pmi} are used in Run~2 and Run~3, respectively. The pixel vertex reconstruction improves the overall speed while sacrificing some of the efficiency and resolution.

\subsubsection{Calorimeters}

Calorimeters are used to measure the energies of the various particles produced in each collision.
The ECAL measures the energy of electrons and photons by absorbing them completely. Hadrons typically pass through the ECAL and are absorbed and measured by the HCAL. The local reconstruction of energy deposits in ECAL and HCAL is described in the following paragraphs.

\paragraph{The ECAL}

The ECAL consists of lead tungstate ($\mathrm{PbWO}_{\mathrm{4}}$) crystals emitting scintillation light when particles interact within their volumes. The 75,848 crystals are arranged in a central, cylindrical barrel section (EB), with pseudorapidity coverage up to ${\abs{\eta}=1.48}$, closed by two flat endcap sections (EE), extending the coverage to ${\abs{\eta}=3.0}$. The scintillation light produced inside the crystals is collected by photodetectors, creating an electrical signal amplified and shaped using a multigain preamplifier, which provides analog outputs that are converted into digital signals by analog-to-digital converters. Because of the increased \Linst provided by the LHC and thus the higher number of overlapping signals from neighboring bunch crossings in Run~2 compared with Run~1 (resulting from the LHC bunch spacing changing from 50 to 25\ns), a novel ECAL amplitude reconstruction algorithm was developed in Run~2. The algorithm is based on a template fit called ``multifit'', introduced since 2017, which attempts to resolve the many overlapping signals coming from pulses emitted in different bunch crossings, and has replaced the Run~1 method that was based on a digital filtering technique~\cite{ECAL:CMS:2020xlg}. This ``multifit'' algorithm is robust and fast enough to be used both in the offline CMS reconstruction and at the HLT.

The energy response of the ECAL changes with time due to ageing of the crystals and of the photodetectors, caused by the high radiation levels at the LHC~\cite{CMS-EGM-18-002}. A dedicated monitoring system, using lasers that inject light during the LHC orbit gap, which contains no proton collisions, is used to measure the transparency of each crystal and the photodetector response. For energy measurements at the HLT level, correction factors for the change in transparency are derived using measurements from the light monitoring system recorded in the preceding hours or days. In Run~3, these corrections are updated once per LHC fill, which is deemed sufficient given the existing running conditions.
The finer time granularity of these offline corrections, which enables an accurate monitoring of the evolution of the detector response during an LHC fill, introduces some differences between online and offline reconstructed ECAL energy deposits.

A clustering algorithm is required to sum the energy deposits of adjacent channels that are associated with a single electromagnetic shower~\cite{CMS:2020uim}. Corrections are applied to rectify the cluster partial containment effects.
The ECAL clusters are dynamically combined into larger clusters to capture the full energy deposit from an electron or photon that might have undergone bremsstrahlung emission or conversions in the inner pixel tracker.

\paragraph{The HCAL}

The HCAL system includes several sections: the barrel (HB), endcaps (HE), outer (HO), and forward (HF). The HB and HE are sampling calorimeters made of interleaved brass and scintillating material, stationed outside the ECAL and inside the solenoid magnet. The HO is a plastic scintillator placed outside the solenoid and designed to catch highly-energetic hadrons. Finally, the HF is a quartz fiber Cherenkov calorimeter with steel absorbers also located outside the solenoid. Scintillation light produced inside the HB and HE are collected with wavelength-shifting fibers, optically summed, and sent to photodetectors to form analog electric signals. These signals are digitized by a charge integrator over a 25\ns interval, the latter known as a time sample (TS). Each recorded pulse shape consists of 10\unit{TSs} (8 since 2018 to reduce the data volume). 

In the HB and HE, approximately 85--90\% of the integrated energy occurs in a 50\ns window (2\unit{TSs}), while the LHC has delivered proton bunches every 25\ns since Run~2. The overlapping signals from nearby bunch crossings required the development of dedicated algorithms to estimate energy deposition in the HCAL. Used prior to 2015, a method based on the simple corrected sum of charges deposited in 2\unit{TSs} became unsuitable with the 25\ns bunch spacing. Consequently, several algorithms \cite{HCAL:CMS:2023lqq} were developed, based on fitting pulse-shape templates similar to the ECAL local reconstruction. 
Since 2018, the ``minimization at HCAL, iteratively'' (MAHI) algorithm, based on a fast chi-square minimization, has been used. 
This algorithm, deployed both offline and online, leads to a smaller difference between the offline and online reconstruction performance compared to the previous methods developed in 2016--2017. 

\subsubsection{Muon detectors}

The CMS detector was designed with subdetectors dedicated to muon identification and to muon triggering, as well as to the measurement of the muon momentum and charge over a broad range of kinematic parameters~\cite{CMS:2018rym}. The drift tubes and cathode strip chambers are located in the regions ${\abs{\eta}<1.2}$ and ${0.9 < \abs{\eta} < 2.4}$, respectively, and are complemented by resistive plate chambers in the range ${\abs{\eta}<1.9}$. Three regions are distinguished, naturally defined by the cylindrical geometry of CMS, referred to as the barrel (${\abs{\eta}<0.9}$), overlap (${0.9 < \abs{\eta} < 1.2}$), and endcap (${1.2 < \abs{\eta} < 2.4}$) regions. The chambers are arranged to maximize the coverage and to provide some overlap where possible.

Muons and other charged particles that traverse a muon subdetector ionize the gas in the chambers, which eventually causes electric signals to be produced on the wires and strips. These signals are read out by electronics and are associated with well-defined locations, generically called ``hits", in the detector. The precise location of each hit is reconstructed from the electronic signals using different algorithms depending on the detector technology.

\subsubsection{Particle flow}
\label{subsubsec:event reconstruction PF}

The PF algorithm~\cite{Sirunyan:2017ulk} aims to reconstruct and identify each individual particle (called a PF candidate) in an event, with an optimized combination of information from the various elements of the CMS detector. The energy of photons is obtained from the ECAL measurement. The energy of electrons is determined from a combination of the electron momentum at the primary interaction vertex as measured by the tracker, the energy of the corresponding ECAL cluster, and the energy sum of all bremsstrahlung photons spatially compatible with originating from the electron track. The energy of muons is obtained from the curvature of the corresponding track. The energy of charged hadrons is determined from a combination of their momentum measured in the tracker and the matching ECAL and HCAL energy deposits, corrected for the response function of the calorimeters to hadronic showers. Finally, the energy of neutral hadrons is obtained from the corresponding corrected ECAL and HCAL energies.

Offline PF reconstruction is used in the vast majority of physics analyses in CMS, and has also been deployed at the HLT. To cope with the stringent timing constraints, the HLT relies on a simplified PF algorithm. 
Offline, most of the processing time is spent reconstructing the inner tracks for the PF algorithm. 
The online version of the PF algorithm runs with two minor differences compared to its offline counterpart: the electron and isolated photon identification and reconstruction tasks are not included, and the reconstruction of tracks arising from nuclear interactions in the tracker material is not performed.

\subsubsection{Jets}
\label{subsec:jets}

One important aspect of event reconstruction in CMS is the identification and reconstruction of jets. Jets are collimated streams of particles that arise from the fragmentation and hadronization processes of quarks and gluons produced in high-energy collisions. Reconstructing jets is crucial for understanding the properties of the particles involved in the collision and for identifying potential new physics phenomena.

The offline jets considered in this Report are reconstructed using the infrared- and collinear-safe anti-\kt (AK) algorithm~\cite{Cacciari:2008gp, Cacciari:2011ma}. 
The default distance parameters used by the algorithm are 0.4 or 0.8, to reconstruct AK4 jets from single quarks/gluons or AK8 jets from the decay of Lorentz-boosted hadronic resonances, respectively. 
The inputs to the clustering algorithm are the four-momentum vectors of calorimeter energy deposits or PF reconstructed particles, which result in a calorimeter (Calo) jet or a PF jet, respectively.

\paragraph{Calo and PF jets}~

Calo jets are reconstructed from energy deposits in the calorimeter towers. A calorimeter tower consists of one or more HCAL cells and the geometrically corresponding ECAL crystals. In this process, the contribution from each calorimeter tower is assigned a momentum, the absolute value and direction of which are given by the energy measured in the tower and by the coordinates of the tower. The jet energy is obtained from the sum of the tower energies, and the jet momentum by the vector sum of the tower momenta. The jet energies are then corrected to establish a relative uniform response of the calorimeter in $\eta$ and a calibrated absolute response in \pt.

In contrast, PF jets are reconstructed by clustering the four-momentum vectors of PF candidates. The jet momentum is determined as the vector sum of all the particle momenta in the jet. Pileup interactions can contribute extra tracks and calorimetric energy depositions to the jet momentum. To mitigate this effect in offline analysis, the jets are subject to the charged-hadron subtraction (CHS) or the pileup-per-particle identification  (PUPPI)~\cite{PUPPI:Sirunyan_2020foa,PUPPI:Bertolini_2014bba} algorithms.
In CHS, charged particles identified as originating from pileup vertices are discarded and an offset is applied to correct for remaining contributions. In PUPPI, the effect of pileup is mitigated at the reconstructed particle level, making use of local shape information, event pileup properties, and tracking information. While pileup charged particles are discarded, the momenta of neutral particles are rescaled according to their probability to originate from the PV, which is deduced from a local shape variable.

Calo jets result from a relatively simple yet robust approach and were widely used in early CMS publications. However, as the performance of the PF reconstruction has proven reliable and more powerful, PF jets have become the norm in CMS analyses. The advantages of using PF jets over Calo jets include a more complete event description as well as improved jet momenta and spatial resolutions, stemming from the combined use in PF of tracking detectors and of the high granularity of the ECAL.

\paragraph{Jet calibration}~

Jet energy corrections are derived from simulated samples to bring the measured response of jets to that of particle-level jets on average. In situ measurements of the momentum balance variable in dijet, ${\PGg + \text{jet}}$, ${\PZ + \text{jet}}$, and multijet events are used to account for any residual differences in the jet energy scale (JES) between data and simulation~\cite{JEC:CMS_2016lmd}. The PF jet energy resolution (JER) typically amounts to 15--20\% at 30\GeV, 10\% at 100\GeV, and 5\% at 1\TeV~\cite{JEC:CMS_2016lmd}. Additional selection criteria are applied to each jet to remove jets potentially dominated by anomalous contributions from various subdetector components or reconstruction failures~\cite{qgl_old}.

The HLT reconstruction of jets uses the same clustering algorithm as its offline counterpart but differs in the calorimeter energy deposits or the PF candidates provided as input, as discussed in previous sections. Similarly to offline jets, jet energy corrections are derived from simulation to correct the response of HLT reconstructed jets. Dedicated studies to quantify and account for residual differences in jet energy scale and resolution between online (with scouting) and offline reconstructed jets are presented in Sections~\ref{sec:jet-reco-performance}~and~\ref{subsec:run3jets}.

\paragraph{Jet substructure}~
\label{par:jetsubstructure}

The collisions at the LHC can produce heavy particles with large transverse momenta. In events that contain \PW and \PZ gauge bosons, Higgs bosons, top quarks, or even new resonances predicted in new physics scenarios, it is possible to achieve a high selection efficiency through the use of their hadronic decay channels. At sufficiently large Lorentz boosts (typically with \pt of a few hundreds of \GeVns), the final-state hadrons from decays of such resonances merge into a single large-radius jet. In these cases, the analysis of jet substructure can be used to distinguish between those jets arising from a resonance decay and those arising from the numerous SM events composed uniquely of jets produced through the strong interaction, referred to as QCD multijet events~\cite{CMS:2014rsx}.

The jet mass is one of the most powerful observables to discriminate resonance jets from background jets (\ie, jets stemming from the hadronization of light-flavor quarks or gluons). Contributions from initial-state radiation, the underlying event, and pileup can strongly impact the jet mass. Jet ``grooming'' techniques (such as the jet trimming~\cite{Krohn:2009th} employed at the HLT) are applied to remove low-energy or uncorrelated radiation contributions from jets, thus improving the jet mass scale and resolution.
Powerful machine learning (ML) techniques based on particle-level information have been recently used in offline analyses to identify and classify hadronic decays of highly Lorentz-boosted resonances~\cite{CMS:2020poo}. In the analysis of the scouting data described in Sections~\ref{sec:jet-reco-performance} and~\ref{sec:MultiSearches}, techniques without ML were used to identify the three-prong substructure from boosted top quarks or trijet resonances decays (the $N$-subjettiness ratio \nsubjet~\cite{Thaler:2010tr}), and the two-prong substructure of boosted \PW and \PZ bosons or dijet resonance decays (the $N^1_2$ variable based on energy correlation functions~\cite{Moult:2016cvt}).

\paragraph{Tagging of \texorpdfstring{\PQb}{b} jets}~

Jets from the hadronization and subsequent decay of bottom quarks (or \PQb quarks) are called \PQb jets. The hadronization of a \PQb quark produces a \Pb hadron that traverses the detector before decaying within the tracker volume. This phenomenon results in distinctive attributes within the emerging \PQb jet, exemplified by the presence of a displaced secondary vertex (SV) that exhibits a displacement from the PV exceeding the CMS tracker resolution. The tracks stemming from this secondary vertex have a large impact parameter. Occasionally, the \PQb jet is accompanied by a tertiary vertex (an outcome of the decay of the \Pb hadron into a charm hadron), or by a lepton via the semileptonic decay of the \Pb hadron or the charm hadron from a \Pb cascade decay.

Physics analyses with \PQb jets in the final state rely greatly on the identification, or tagging, of \PQb jets. The \PQb jets can be discriminated from jets produced by the hadronization of light quarks based on characteristic attributes of \Pb hadrons, such as those described above. The CMS experiment employs a variety of \PQb tagging algorithms. During Run~1, the principal tools employed for \PQb jet identification consisted of likelihood-based discriminators~\cite{BTagCSV:Chatrchyan:2012jua}. Subsequently, in Run~2 and Run~3, the evolution of \PQb tagging algorithms led to the adoption of multilayer perceptrons~\cite{BTagCSVandDeepCSV:Sirunyan_2018}, deep neural network multiclassifiers~\cite{BTagDeepJet:Bols_2020, BTagDeepJet:CMS-DP-2018-058}, and graph convolutional neural networks~\cite{PNet:Qu_2020}. Each successive algorithm yielded notable enhancements in the efficiency of \PQb jet identification. Similar algorithms, trained with the online reconstructed objects as input, were employed at the HLT in Run~2 and Run~3 to increase the online selection of events containing \PQb jets.  
Tagging of \PQb jets has not been employed so far in scouting-based analyses, but information that would allow such an analysis has been stored in the scouting data set since the beginning of Run~3.

\subsubsection{Muons}

Muons are crucial objects for the physics program of CMS since the original design of the detector. Thanks to their very clean experimental signature as they pass through the detector, muons are excellent probes to study known SM processes and to search for the production of new particles at colliders.

Following the hardware-driven reconstruction steps within the L1 trigger system, the standard reconstruction of muon objects and their trajectories takes place via a two-step process at the software level.
First, muons are reconstructed within the muon system only, which produces level-2 (L2) muons. Then, tracks produced in the pixel tracker are combined with the information from the muon spectrometer to reconstruct the full trajectory of the muon through the detector, which are termed level-3 (L3) muons. 

The L2 muon reconstruction can refine the initial estimate of the muon trajectory by applying more accurate algorithms that are not feasible at the L1 trigger. Standalone muon tracks are constructed by combining information from all muon subdetectors along a muon trajectory with a Kalman filter technique~\cite{Fruhwirth:178627}. This iterative algorithm executes pattern recognition on a detector layer basis while concurrently refining the trajectory parameters. 

The L3 muon reconstruction uses all available information about the muon trajectory from both the muon detectors and the tracker. Different L3 algorithms were used over the data-taking years. Generally, global muon tracks are built via an outside-in (OI) matching between a standalone muon track and a tracker track. The information from both tracks is used to perform a combined fit with the Kalman filter. Tracker muon tracks are instead constructed with an inside-out (IO) extrapolation by looking for a loose match between the tracker tracks and at least one muon detector segment. With the installation of a new pixel detector before the beginning of the 2017 data taking, a new iterative algorithm was adopted. It works in three steps. The first two, the OI and the IO steps, are both seeded by L2 muons. The second step considers only muons that were not already reconstructed by the previous step. Then, an additional IO step seeded by L1 muons is performed to recover candidates that could not be matched to an already reconstructed L3 muon. This IO step recovers some of the efficiency loss observed in previous steps, ensuring excellent performance for high-\pt muons and for muons in high pileup scenarios.

At the HLT~\cite{CMS:2021yvr}, the procedure to build L2 muons as seeds for the track reconstruction in the inner tracker is identical to the one used for offline standalone muons. However, the HLT computing time constraints preclude conducting the full track reconstruction based on the multi-iteration approach across the complete volume of the inner tracker. The L3 reconstruction algorithms are performed only in smaller regions of the detector based on the presence of L1 or L2 muons. As a result, high reconstruction efficiency is achieved while minimizing computing resources. Differences between online and offline muons are typically small in terms of muon momentum scale and resolution, as described in Section~\ref{subsec:run3muons}.

\subsubsection{Electrons and photons}
\label{subsubsec:eg_reco_desc}

Electrons and photons in the CMS detector are reconstructed with high purity and efficiency, and excellent resolution, making them ideal to use both in SM precision measurements and in BSM searches. Electrons and photons deposit almost all of their energy in the ECAL. In addition, electrons produce hits in the tracker layers. As electrons and photons propagate through the material in front of the ECAL, they may interact with the medium, with electrons emitting bremsstrahlung photons and photons converting into electron-positron pairs. Thus, by the time they reach the ECAL, they could consist of a shower of multiple electrons and photons. Their resulting clusters are combined into a single supercluster (SC) object to recover the energy of the primary electron or photon. Additionally, for an electron that loses momentum by emitting bremsstrahlung, the curvature of its trajectory changes in the tracker. A tracking algorithm based on a Gaussian sum filter (GSF)~\cite{Adam_2005} is used to estimate the track parameters of electrons even in the presence of such emissions.

In CMS, there are three main ways to reconstruct an electron: seeded by the ECAL, seeded by the tracker, and with a special low-\pt electron reconstruction. The ECAL-driven approach starts by combining ECAL clusters into a SC. For each SC found, compatible pixel hits in the inner tracker are sought, and any matches are used to seed the GSF tracking algorithm that builds the electron candidate. The tracker-driven approach takes the standard track collection and looks for one track that is compatible with ECAL energy clusters after applying some preselection. It then uses that track to seed the GSF tracking step. All ECAL clusters compatible with the track are associated with a single SC. Finally, the low-\pt electron reconstruction is a variant of the tracker-driven one and optimized for very low track momenta. Because of the high CPU cost to reconstruct all tracks in the event, only the ECAL-driven algorithm is available at the HLT and thus all electrons at the HLT require at least two hits in the inner tracker.

The differences between the ECAL-driven HLT and offline reconstruction algorithms are minimal and primarily driven by the limited CPU time available at the HLT and by the lack of final calibrations, which are not promptly computed during the data-taking period. The main distinction is in the GSF tracking algorithm, which is applied with fewer iterations compared to the offline reconstruction. Additionally, the formation of SCs is purely calorimeter-based, and not refined with tracking information, which would more accurately account for energy deposits that may be compatible with bremsstrahlung interactions. 

More details on electron and photon reconstruction in CMS can be found in Ref.~\cite{CMS:2020uim}. The dedicated offline reconstruction and identification algorithm optimized for electrons with $\pt < 10\GeV$ is described in Section~\ref{sec:lowpt-electrons}.

\subsubsection{Missing transverse momentum}

The presence of particles that do not interact with the detector material is indirectly measured by the missing transverse momentum (\ptmiss). The measurement captures the momentum carried away by undetected or invisible particles, such as neutrinos or other weakly interacting particles. The \ptvecmiss vector is computed as the negative vector sum of the transverse momenta of the input objects in an event. The inputs can be calorimeter towers, PF candidates or jets (in the latter case denoted by the symbol \mht). Similar to jets, the offline and online missing transverse momentum reconstruction algorithms mainly differ in the inputs fed to the algorithm.

More details on the reconstruction and calibration of these objects are provided in Ref.~\cite{MET:2019ctu}.

\subsubsection{Tau leptons}

The tau lepton (\PGt), with a mass of about 1.78\GeV, is the only lepton sufficiently massive to decay into hadrons. About one third of the time, tau leptons decay into an electron or a muon, plus two neutrinos. The neutrinos escape undetected, but the electron and muon are reconstructed and identified through the usual techniques available for such leptons, as described in previous sections. Almost all of the remaining decay final states of tau leptons contain hadrons, typically with a combination of charged and neutral mesons, and a tau neutrino.

Hadronic \PGt lepton decays (\tauh) are reconstructed from jets, using the hadrons-plus-strips (HPS) algorithm~\cite{Sirunyan:2018pgf}, which combines one or three tracks with energy deposits in the calorimeters to identify the tau lepton decay modes. Neutral pions are reconstructed from  electrons and photons as strips with dynamic size in the $\eta$-$\phi$ plane, where the strip size varies as a function of the \pt of the electron or photon candidate.

To distinguish genuine \tauh decays from jets originating from the hadronization of quarks or gluons, and from electrons and muons, the \textsc{DeepTau} algorithm is used~\cite{CMS:2022prd}. Information from all individual reconstructed particles near the \tauh axis is combined with properties of the \tauh candidate and of the event. 

The HLT system runs a version of the \tauh reconstruction that is slightly different from the one used offline. This is achieved via specialized, fast, and regional versions of the reconstruction algorithms, and via the implementation of a multistep selection logic, designed to reduce the number of events processed by the more complex, and therefore more time-consuming, subsequent steps. 
Reconstructed tau leptons have not been employed so far in scouting-based analyses, but information that would allow such an analysis is stored in the Run~3 scouting data set.

\section{Data scouting in Run 1 and Run 2}
\label{ch:run1run2scouting}

This section details the development and application of the scouting technique by the CMS Collaboration during the first two periods of LHC operation. Two scouting data streams were defined, one based on jets and the other on muons. First, we describe in Section~\ref{sec:ScoutingRun2Strategy} the general trigger and reconstruction strategy for the scouting streams throughout the Run~1 and Run~2 data-taking periods. In Section~\ref{sec:ScoutingRun2Details} we focus on the definition of the triggers used to select interesting collision events and describe the corresponding event content of data stored with those triggers. 
In Sections~\ref{sec:ScoutingRun2Jets}~and~\ref{sec:ScoutingRun2Muons}, we report efficiency measurements of the scouting triggers and of the reconstruction performance for jet and muon objects, respectively. Finally, Section~\ref{sec:Run2ScoutingPhysicsResults} showcases the physics results obtained with scouting-based analyses.

\subsection{General strategy of data scouting}\label{sec:ScoutingRun2Strategy}

The scouting strategy at the HLT was originally designed and tested in 2011 to improve access to the enormous amount of data collected by the CMS detector, totaling over a hundred million individual readout channels. Scouting events are selected with a dedicated set of L1 algorithms and at a higher HLT rate with respect to the standard streams to provide additional sensitivity to specific parts of the CMS physics program. These events are then processed in real time by the HLT computer farm and written on disk with reduced content. The majority of scouting events are reconstructed as part of the standard HLT event selection workflow. The CPU count dedicated to scouting thus constituted less than 5\% of the total HLT farm resources, which in 2018 featured approximately 30,000 CPU cores. In Run~2, the scouting event rate accepted by the HLT was approximately 5\unit{kHz} on average and 6\unit{kHz} at the highest value of \Linst, while the total allocated rate for the standard CMS physics program was approximately 1\unit{kHz}. 

A comparison of the typical rates for each data stream during Run~1 and Run~2 operation is reported in Table~\ref{tab:Run1Run2DataStreams}, ranging from the initial tests performed in 2011 to the final configuration reached in 2018. 

\begin{table*}[!htb]
    \centering
    \topcaption{Comparison of the typical HLT trigger rates of the standard, parking, and scouting data streams during Run~1 and Run~2. The average \Linst over one typical fill of a given data-taking year and the average pileup (PU) are also reported, consistent with the scenarios reported in Fig.~\ref{fig:hlt_rates_evolution}.}
    \renewcommand{\arraystretch}{1.3}
    \begin{tabular}{lccccc}
    Year      & \Linst [$\invcms$] & PU & Standard rate [Hz] & Parking rate [Hz] & Scouting rate [Hz]  \\
    \hline
    2012 & \sci{0.5}{34} &28 & 420 & 400 & 1000 \\ [\cmsTabSkip]
    2016 & \sci{0.9}{34} &35 & 1000 & 500 & 4500 \\
    2017 & \sci{1.0}{34} &43 & 1000 & 400 & 4500 \\
    2018 & \sci{1.2}{34} &38 & 1000 & 3000 & 5000 \\
    \end{tabular}
    \label{tab:Run1Run2DataStreams}
\end{table*}

\subsection{Trigger definitions and event content}\label{sec:ScoutingRun2Details}

The CMS trigger system is a dynamical entity, with operational parameters that are adjusted frequently to adapt to changing data-taking conditions in the short term, and less frequently to adjust to different physics goals in the long term. This section describes the specific event content and the algorithms designed for each scouting stream, as well as the dedicated rate budget available for data scouting. Most of the information is reported for the 2018 data-taking scenario, because it represents the final configuration achieved after various developments in Run~1 and Run~2, thus serving as a useful benchmark reference. 

The initial scouting development in Run~1 focused on dijet triggers to search for low-mass hadronic resonances. Dedicated trigger paths based on calorimeter jets and on PF jets were successfully commissioned in the final months of 2011, leading to the first preliminary results from dijet resonance searches. In Run~2, a new set of dimuon algorithms was employed to feed the scouting reconstruction in addition to the existing hadronic algorithms. Two versions of the hadronic trigger path were still in place: one using the calorimeter information and the other an optimized version of the PF reconstruction, which relied on additional tracking algorithms needed to improve the momentum resolution. As a result, two scouting data sets were produced and stored on disk: one from the ``Calo'' scouting stream, including both the muon and the hadronic triggers, and one from the ``PF'' scouting stream. This notation will be used in the following sections to identify the various groups of triggers. The event content of the PF scouting stream includes all PF candidates, resulting in a significant event size increase relative to the Calo scouting stream. Finally, a complementary data set that includes both the scouting event content and the complete CMS raw detector output was also defined, and used to collect events at a much lower rate. This data set is used to fully reconstruct a subset of scouting events offline, providing a useful way to validate the scouting reconstruction performance.

Table~\ref{tab:L1_HLT_Thr} lists the most important L1 and HLT triggers deployed in 2018 to collect scouting events. The dimuon scouting triggers were fully commissioned during 2017 with the aim of substantially lowering the muon \pt thresholds compared to the standard triggers. The L1 requirements on the dimuon invariant mass \mMM and angular separation \drMM help reduce the trigger rates. Lower-mass resonances are typically produced with considerable Lorentz boosts at the LHC, leading to final-state muons with significant momentum vector collimation (or low values of \drMM). The hadronic triggers are based on the \HT content of the event. In the Calo and PF scouting streams, only jets with ${\pt > 40\GeV}$ are considered in the \HT sum. In the hadronic PF trigger, the L1 threshold was below 300\GeV in 2016 but subsequently raised to 360\GeV because of the increased pileup in 2017 and 2018. In parallel, new single-jet and double-jet L1 algorithms were added to better serve low-mass dijet analyses.  

To maintain the event rate, data set size, and processing time within the allocated resources, minimal additional selection criteria are implemented in the scouting paths at the HLT. The dimuon and triple-muon L1 algorithms require each muon to have ${\pt > 3\GeV}$, without imposing the need for muon tracks to point back to the nominal interaction point. The Calo scouting stream affords an \HT threshold at the HLT as low as 250\GeV, while maintaining a reasonable rate and good energy scale and resolution. The PF scouting stream, in contrast, requires a higher threshold of ${\HT > 410\GeV}$ because of the larger event content compared to the Calo stream. A summary of the typical trigger rates achieved for each stream is reported in Table~\ref{tab:run2_rates}, for a scenario corresponding to the end of the 2018 data taking.

\begin{table*}[!htb]
    \centering
    \topcaption{List of L1 and HLT thresholds for the most relevant scouting triggers in Run~2. The list corresponds to the 2018 thresholds that were valid for the overall Run~2 data-taking period. Differences with respect to the 2016 or 2017 scenario are reported in parentheses. Muons and photons are annotated as \PGm and \PGg, respectively, while OS stands for opposite-sign muon pairs. In cases where the same threshold is applied to all selected objects in an event, a single number is shown, while if different thresholds are applied to the objects, they are shown separated by slashes from the highest to the lowest.}
    \renewcommand{\arraystretch}{1.3}
    \centering\begin{tabular}{lll}
    Stream & L1 thresholds & HLT thresholds  \\ \hline
    \multirow{7}{*}{Calo} &  $1\PGm$, $\pt > 22\GeV$  (not in 2017) &
    \multirow{7}{*}{
            $\left\}\begin{array}{l} \\  \\ \\ \pt > 3 \GeV \\ \\ \\ \\ \end{array}\right.$} \\
    & $2\PGm$, $\pt > 15/7\GeV$ \\
    & $2\PGm$, $\pt > 4.5\GeV$, $\abs{\eta}<2.0$, $\mathrm{OS}$, $7<\mMM<18$\GeV\\	
    & $2\PGm$, $\pt > 4.5\GeV$, $\abs{\eta}<2.0$, $\mathrm{OS}$, $\mMM>7\GeV$ (not in 2017)\\	
    & $2\PGm$, $\pt > 0\GeV$, $\abs{\eta}<1.5$, $\mathrm{OS}$, $\Delta R<1.4$ \\
    & $2\PGm$, $\pt > 4\GeV$, $\abs{\eta}<2.5$, $\mathrm{OS}$, $\Delta R<1.2$ \\
    & $3\PGm$, $\pt > 5/3/3\GeV$   \\ \cline{2-3}
    &  $\HT>360\GeV$ (200\GeV in 2016) & $\HT > 250\GeV$ \\ \hline
    \multirow{4}{*}{PF} &
    $\HT>360\GeV$ & $\HT > 410\GeV$  \\ 
    & 1 jet, $\pt>180\GeV$ & -  \\
    & 2 jets, $\pt >30\GeV$, $\abs{\eta}<2.5$, $\Delta \eta<1.5$, $\mjj >300\GeV$ & - \\
    & $1\PGg$, $\pt>60\GeV$ & $\pt > 200\GeV$\\
    \end{tabular}
    \label{tab:L1_HLT_Thr}
\end{table*}

\begin{table*}[!htb]
    \centering
    \topcaption{Comparisons of the event rate, event size, and total bandwidth between the standard and scouting trigger strategies, for an LHC fill corresponding to data collected in 2018 with  $\Linst \approx \sci{1.8}{34}\invcms$ at the start of the fill, one of the highest at the LHC in Run~2, and pileup around 50.}
    \renewcommand{\arraystretch}{1.3}
    \begin{tabular}{lccc}
    Data stream & Event rate [Hz] & Event size & Total bandwidth [MB/s] \\
    \hline
    Standard muons & 600 & 0.86\unit{MB} & 485 \\
    Standard jets/\HT & 400 & 0.87\unit{MB} & 385 \\ [\cmsTabSkip]
    Scouting Calo muons and Calo \HT & 5970 & 8.9\unit{KB} & 45 \\
    Scouting PF jets and PF \HT & 1766 & 14.8\unit{KB} & 25 \\
    \end{tabular}
    \label{tab:run2_rates}
\end{table*} 

Table~\ref{tab:run2scout_vars} summarizes the event content of the Run~2 scouting streams. Since there is no offline reconstruction in the scouting streams, the scouting event content comprises physics objects reconstructed online by the HLT. In the Calo stream, the jet information includes the kinematic observables of jets reconstructed with the calorimeter, which are stored if they satisfy ${\pt > 20\GeV}$ and ${\abs{\eta} < 3}$. 
In addition, the \ptmiss and the average energy density per unit area in the event ($\rho$)~\cite{Cacciari:2007fd} are also stored. This stream also includes muon objects in events with at least two reconstructed muons accepted by the muon scouting triggers. Muon information includes kinematic and identification observables, such as the muon track momentum and the number of hits in the tracker and muon detectors, and information about the dimuon vertices such as the three-dimensional (3D) vertex position and corresponding uncertainty. These objects add up to about 10\unit{KB} per event, compared to roughly 1\unit{MB} in a typical standard event. In the PF scouting stream, the information stored per event consists of all PF candidates with ${\pt > 0.6\GeV}$, as well as PF jets, leptons, and photons as reconstructed at the HLT. In addition, the \ptmiss object reconstructed with all PF candidates and the collection of primary vertices along with $\rho$ are also stored.

\begin{table*}[!htb]
    \centering
    \topcaption{List of observables saved in the scouting output during Run~2. The upper part of the table lists the observables present in the Calo stream and the lower part lists the contents of the PF stream. The PF candidates are sorted into charged and neutral hadrons, muons, electron and photons, hadronic and electromagnetic deposits in HF.}
    \renewcommand{\arraystretch}{1.3}
    \begin{tabular}{ll}
    Observable & Definition \\ \hline
    \multicolumn{2}{c}{\textit{Calo scouting stream}} \\ 
    \hline
    ($m^{\text{jet}}$, $\pt^{\text{jet}}$, $\eta^{\text{jet}}$, $\phi^{\text{jet}}$) & Calo jet four-momentum \\
    $A^{\text{jet}}$ & Jet area \\
    $E^{\text{EM}}_{\text{max.}}$ & Maximum energy in electromagnetic towers \\
    $E^{\text{had.}}_{\text{max.}}$ & Maximum energy in hadronic towers \\
    $E^{\text{EM}}_{\text{HB,HF,HE}}$ & Electromagnetic energy in the HB, HE, and HF \\
    $E^{\text{had.}}_{\text{HB,HF,HE}}$ & Hadronic energy in the HB, HE, and HF \\
    $A^{\text{towers}}$ & Area of the EM and hadronic towers \\
    \hline
    \ptmiss, $\phi^\text{miss}$, $\rho$ & Missing transverse momentum, angle, energy density \\
    \hline
    ($E^{\PGm}$, $\pt^{\PGm}$, $\eta^{\PGm}$, $\phi^{\PGm}$) & Muon four-momentum \\
    $d_{0} \pm \sigma_{d_{0}}$, $d_{z} \pm \sigma_{d_{z}}$ & Muon impact parameters and uncertainties \\
    $I_{\rm E}$, $I_{\rm H}$, $I_{\rm T}$ & ECAL, HCAL, and tracker isolation \\
    $N_{\rm P}$, $N_{\rm S}$, $N_{\rm M}$ & Number of pixel, strip, and muon detector hits \\
    $N_{\rm L}^{\rm S}$, $N_{\rm L}^{\rm T}$ & Number of muon stations and tracker layers with hits \\
    ($\pt^{\text{track}}$, $\eta^{\text{track}}$, $\phi^{\text{track}}$) & Track three-momentum \\
    $\chi^2$, dof & Track $\chi^2$ and number of degrees of freedom \\
    ($q/p \pm \sigma_{q/p}$, $\lambda \pm \sigma_\lambda$, $\phi \pm \sigma_\phi$, $d_{sz} \pm \sigma_{d_{sz}}$) & Fitted track parameters and uncertainties \\
    $i_{\text{vertex}}$ & Reference to the corresponding dimuon vertex \\ 
    ($x \pm \sigma_x$, $y \pm \sigma_y$, $z \pm \sigma_z$) & List of 3D positions and uncertainties of dimuon vertices \\ \hline
    \multicolumn{2}{c}{\textit{PF scouting stream}}  \\ \hline
    ($m^{\text{jet}}$, $\pt^{\text{jet}}$, $\eta^{\text{jet}}$, $\phi^{\text{jet}}$) & PF jet four-momentum \\
    $A^{\text{jet}}$ & Jet area \\
    $E_i$, $N_i$ & Energy fractions and multiplicity for $i^{th}$ particle type in jet \\
    \hline
    \ptmiss, $\phi^\text{miss}$, $\rho$ & Missing transverse momentum, angle, energy density \\
    \hline
    ($m$, $\pt$, $\eta$, $\phi$), id, $i_{\text{vertex}}$ & PF candidate four-momentum, type, vertex index \\
    \hline
    ($x \pm \sigma_x$, $y \pm \sigma_y$, $z \pm \sigma_z$) & List of 3D positions and uncertainties of primary vertices \\
    \end{tabular}
    \label{tab:run2scout_vars}
\end{table*}

The next sections demonstrate the feasibility of using scouting jet and muon objects with a reduced event content, and without applying the offline reconstruction algorithms, making scouting a valuable technique for several physics analyses.

\subsection{Jets}\label{sec:ScoutingRun2Jets}

Jets are the experimental signature of quarks and gluons produced in high-energy collisions such as the \pp interactions at the LHC. The understanding of jet properties is a key ingredient of several physics measurements and searches for BSM physics. Jets have been extensively employed in past CMS searches for new hadronic resonances with the data-scouting technique. This section presents the performance of the scouting jet triggers, showing the large increase in trigger efficiency for low-energy signals compared to the standard data stream. The reconstruction performance of jets in data scouting is also analyzed, demonstrating the feasibility of constructing and applying jet substructure variables with data scouting.

\subsubsection{Jet trigger performance} \label{sec:JetTriggerRun2}

\begin{figure*}[!htb]
    \centering
    \includegraphics[width=0.48\textwidth]{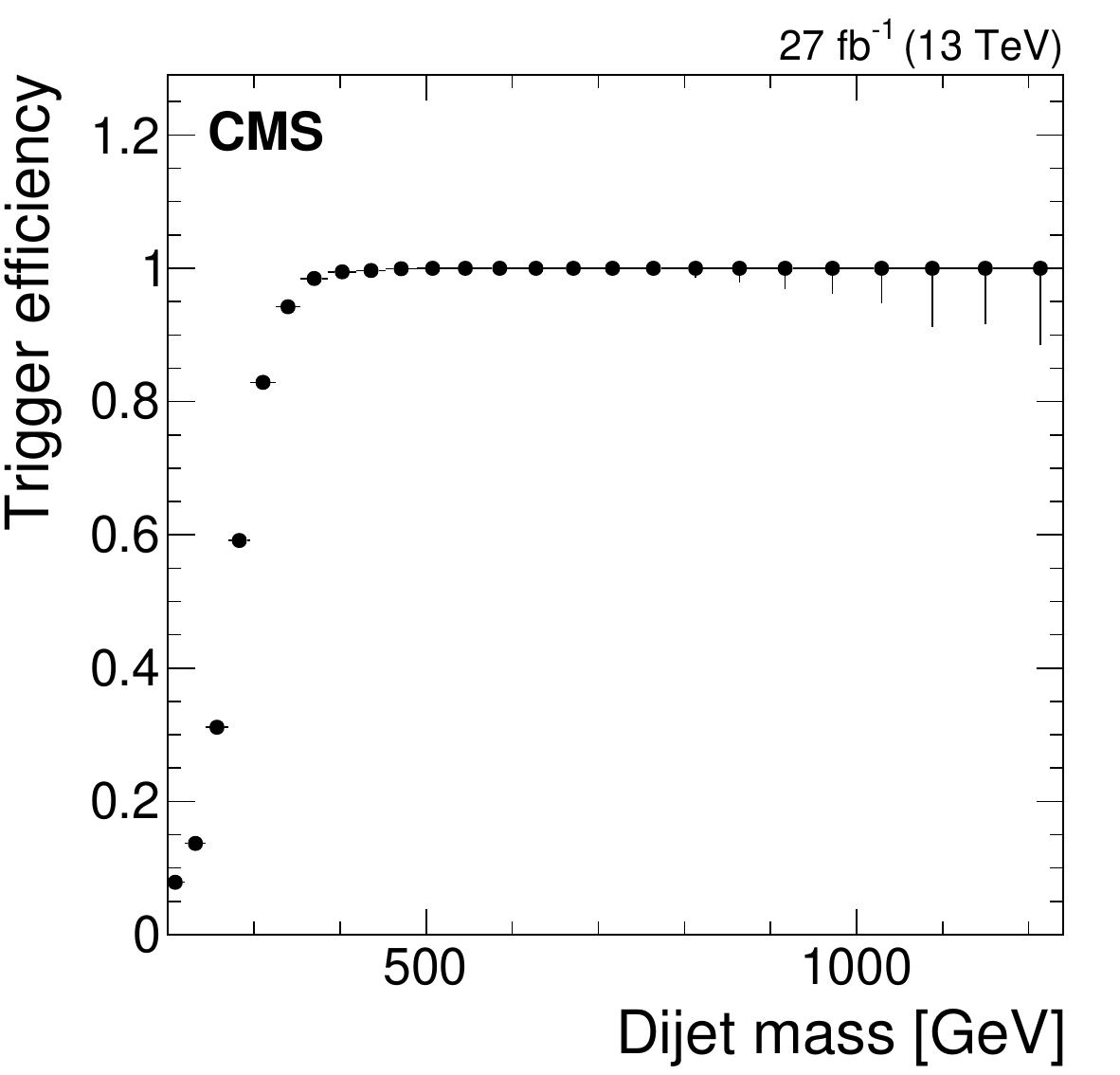}
    \includegraphics[width=0.48\textwidth]{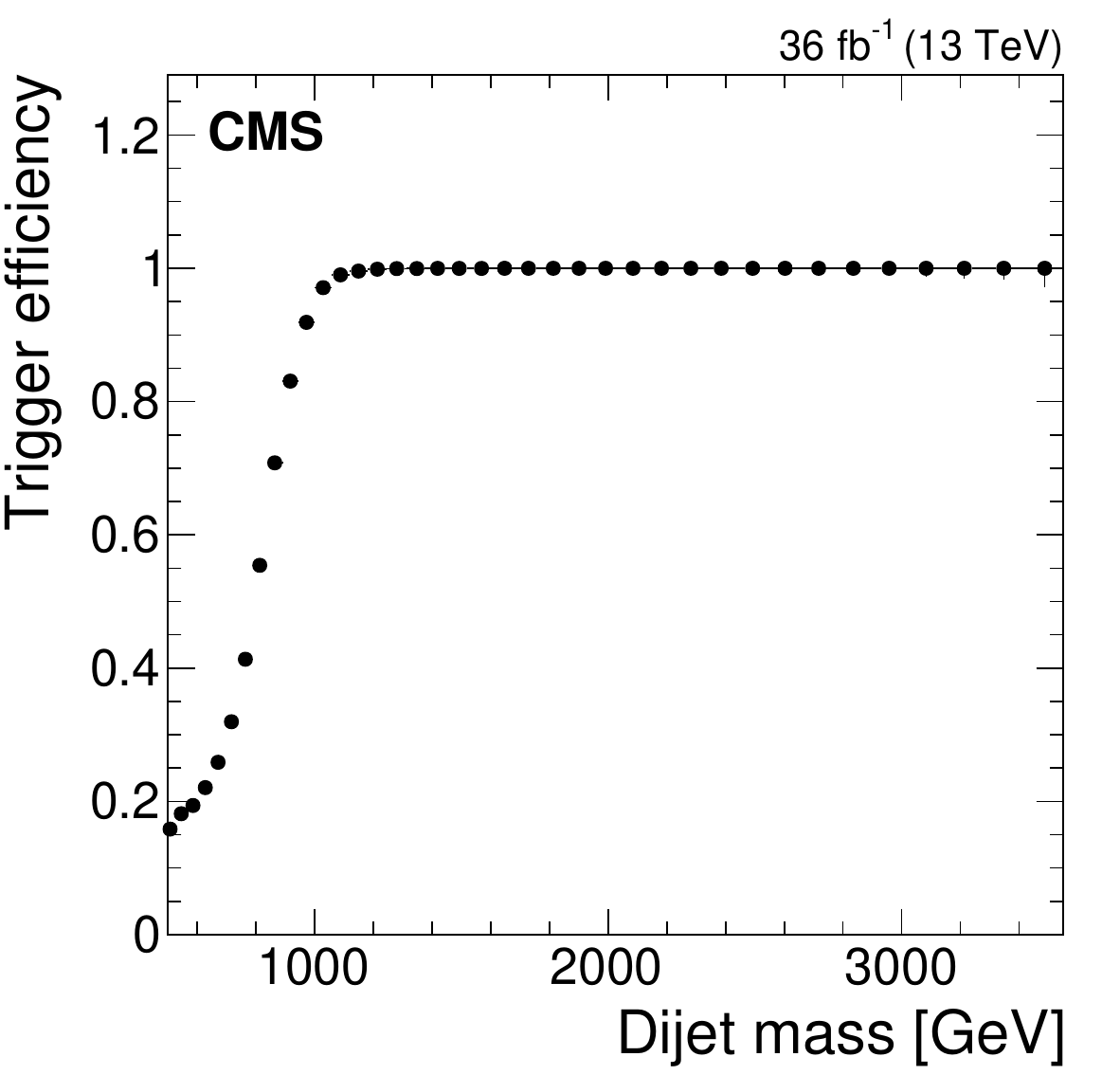}
    \caption{The efficiency of the Run~2 Calo scouting (left) and standard (right) jet triggers as a function of the reconstructed mass of the dijet system. Figures taken from Ref.~\cite{EXO-16-056}.}
    \label{fig:EXO-16-056_TriggerEff}
\end{figure*}

The Calo scouting stream was active during both Run~1 and Run~2, and included jets from energy deposits in the ECAL and HCAL. The main trigger selection requires \HT larger than 250\GeV at the HLT, compared to {$\HT > 800$--900\GeV} for the triggers in the standard data stream. Although designed to select generic collision events that include jets in the final state, this trigger was primarily used to perform searches for new resonances decaying to pairs of jets, as described in Section~\ref{sec:DijetSearches}. This analysis searches for a resonance peak in the invariant mass distribution of the two leading jets (the dijet mass \mjj) and it provides a benchmark for testing the performance of the data-scouting approach. Figure~\ref{fig:EXO-16-056_TriggerEff} shows the total trigger efficiency as a function of \mjj for the scouting (left) and the standard (right) triggers. While the standard trigger becomes fully efficient only for ${\mjj > 1.25\TeV}$, the scouting trigger efficiency reaches the 100\% plateau at around 500\GeV, thus significantly extending the sensitivity of searches for low-mass resonances. Given the generic design of the \HT trigger, a similar improvement from scouting compared to the standard triggers is also expected for other new-physics signatures with final states dominated by the presence of high-\pt jets.

\begin{figure*}[!htb]
    \centering
    \includegraphics[width=0.48\textwidth]{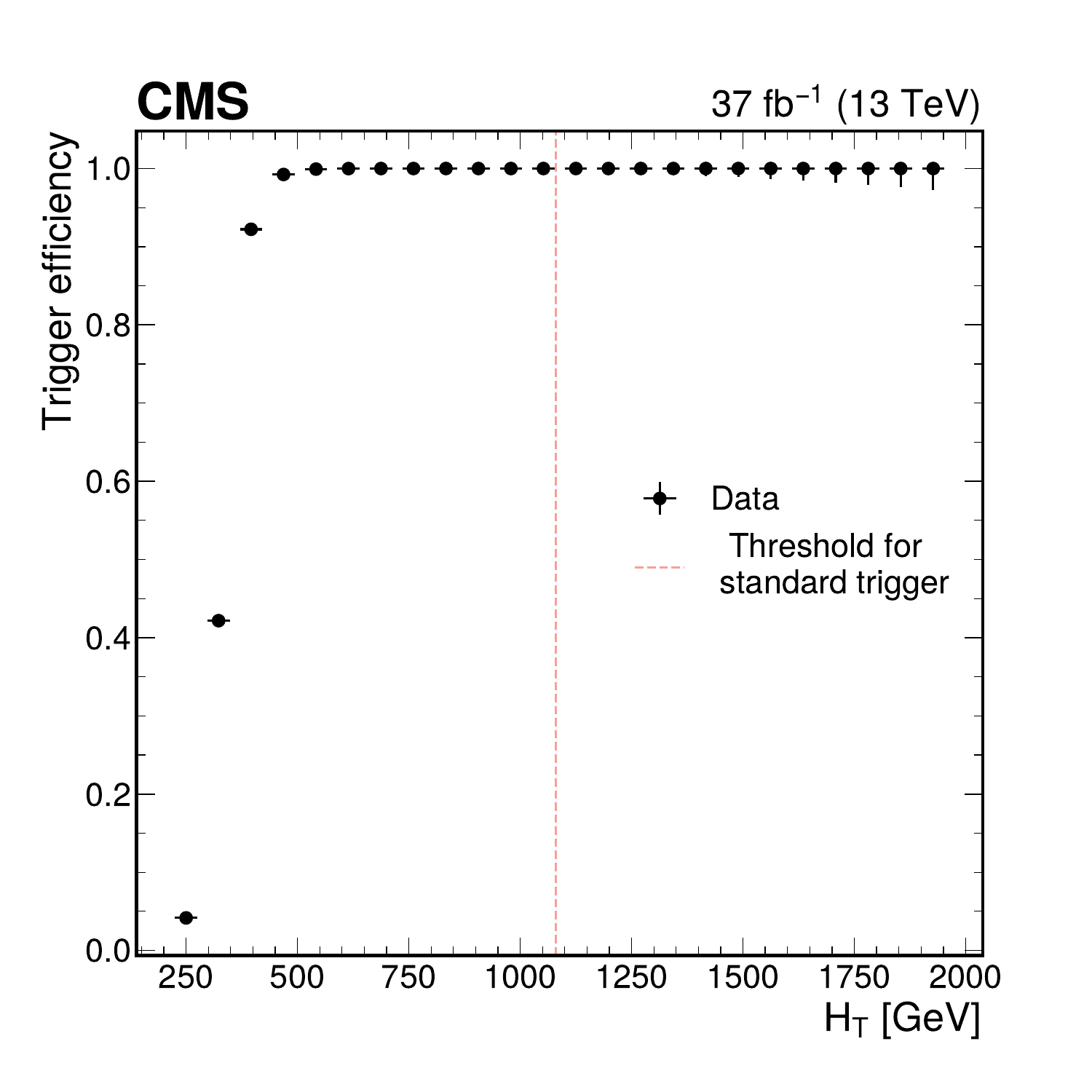}
    \includegraphics[width=0.48\textwidth]{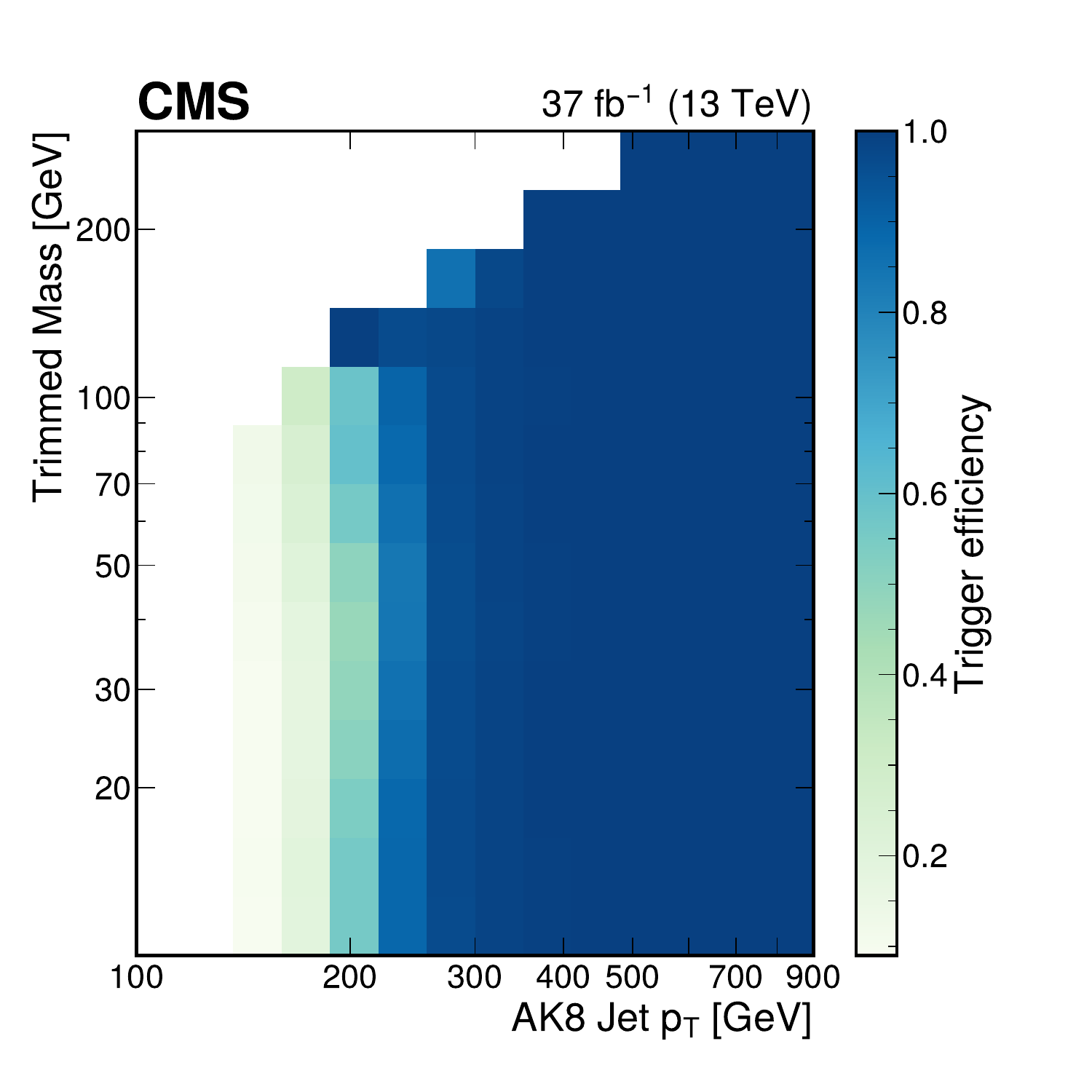}
    \caption{The efficiency of the Run~2 PF scouting jet triggers as a function of \HT (left) and as a function of the leading large-radius jet \pt and trimmed jet mass (right).}
    \label{fig:PF_trigger_eff2}
\end{figure*}

The PF scouting data stream was introduced in Run~2 and was primarily used to perform searches for new resonances decaying to multijet final states. The data collected in Run~2 and used for physics analysis correspond to ${\Lint = 128\fbinv}$. The main trigger selection requires ${\HT > 410\GeV}$, where here \HT is calculated with jet ${\pt > 40\GeV}$. Analyses using this trigger typically estimate the trigger efficiency as a function of \HT. The standard jet triggers require a threshold at the HLT of ${\HT > 1080\GeV}$. Figure~\ref{fig:PF_trigger_eff2} (left) shows that the scouting PF \HT trigger is fully efficient at around ${\HT > 500\GeV}$, offering a significant improvement in signal efficiency for low-energy multijet signals compared to standard triggers. The PF scouting data stream also contains information about the individual particles as reconstructed by the PF algorithm. Their availability enables the reconstruction of jets with different cluster radii, for example large-radius jets with distance parameter of 0.8, which is useful for identifying resonances with high Lorentz boost that decay to jets. In the case of signals featuring merged decays of individual quarks, the trigger efficiency is measured as a function of the \pt of the leading large-radius jet and the jet mass, the latter being related to the resonance mass. Figure~\ref{fig:PF_trigger_eff2} (right) indicates that the PF scouting \HT trigger is fully efficient when ${\pt > 300\GeV}$, for any trimmed jet-mass (described in Section~\ref{par:jetsubstructure}), while the standard triggers are fully efficient for jet momenta that are twice as high. These properties make the trigger suitable for new-resonance searches with a wide range of mass hypotheses.

\subsubsection{Jet reconstruction performance} \label{sec:jet-reco-performance}

Jets in events collected by scouting triggers are formed from input calorimeter energy deposits or from PF candidates reconstructed at the HLT. To meet the stringent HLT time constraints, the online algorithms used to construct these inputs are in general simplified versions of those applied in the standard offline reconstruction. This can cause differences in the JES and JER between the online and offline jet objects. These effects are studied in this section, focusing on the performance of both Calo and PF jet reconstruction in Run~2 data scouting.

\begin{figure*}[!htb]
    \centering
    \includegraphics[width=0.48\textwidth]{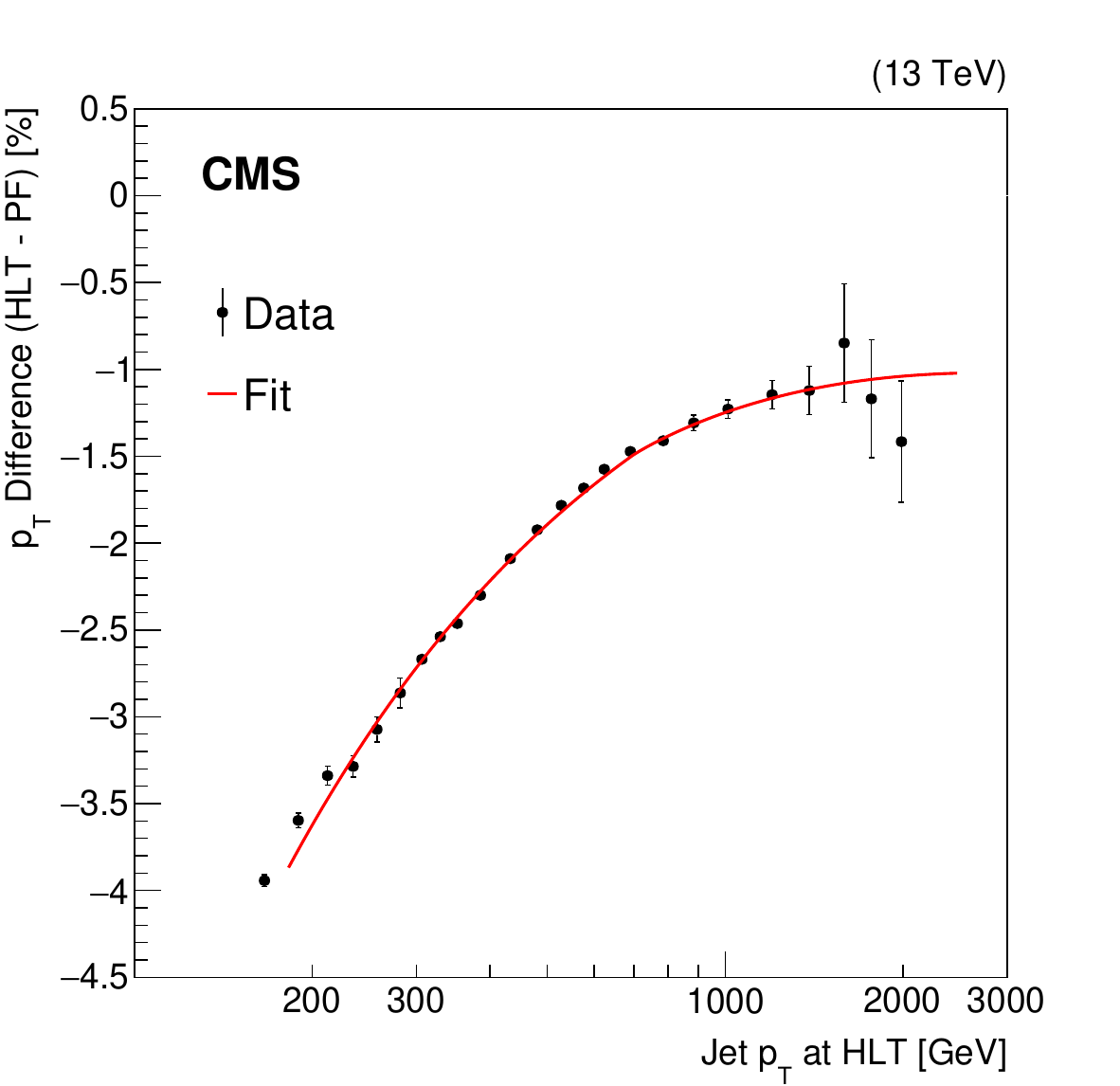}
    \caption{The observed percent difference between the \pt of Calo jets at the HLT and the \pt of PF jets reconstructed offline (points), fitted to a smooth function (curve), \vs the Calo jet \pt. Both Calo and PF jets are calibrated with corrections derived from simulation. Figure taken from Ref.~\cite{EXO-16-056}.}
    \label{fig:EXO-16-056_CaloJetCalibration}
\end{figure*}

The JES of scouting Calo jets that are reconstructed online is calibrated to the one obtained with PF jets reconstructed offline. A monitoring data set has been defined, including both Calo jets at the HLT and the offline reconstructed PF jets, to measure the \pt difference between the two types of jets. A tag-and-probe method~\cite{JEC:CMS_2016lmd} is used to obtain these measurements. Figure~\ref{fig:EXO-16-056_CaloJetCalibration} shows the observed \pt difference between the two collections as a function of jet \pt. The measured points are fitted with a smooth function and the resulting curve is used to calibrate the Calo jets collected by scouting triggers. The JES of Calo jets at the HLT is slightly smaller, by around 4\% at low \pt and 1\% at high \pt, compared to the PF jets reconstructed offline. With the dijet asymmetry method~\cite{JEC:CMS_2016lmd}, we estimate that the scouting Calo JER is only about 10\% worse compared to offline PF jets. These results confirm the good performance of Calo jet reconstruction in data scouting in the high jet \pt range considered.

\begin{figure*}[!htb]
    \centering
    \includegraphics[width=0.426\textwidth]{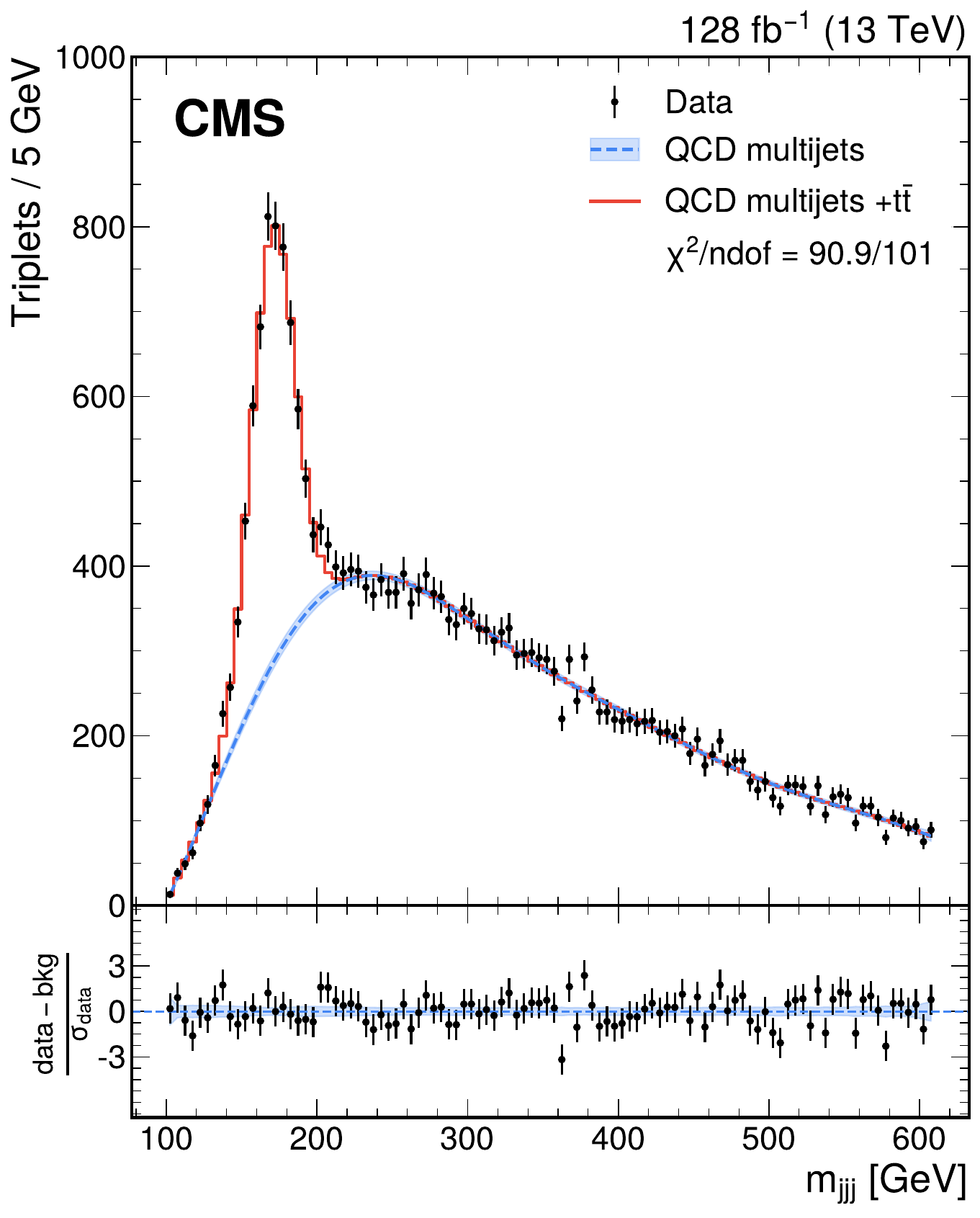}
    \includegraphics[width=0.445\textwidth]{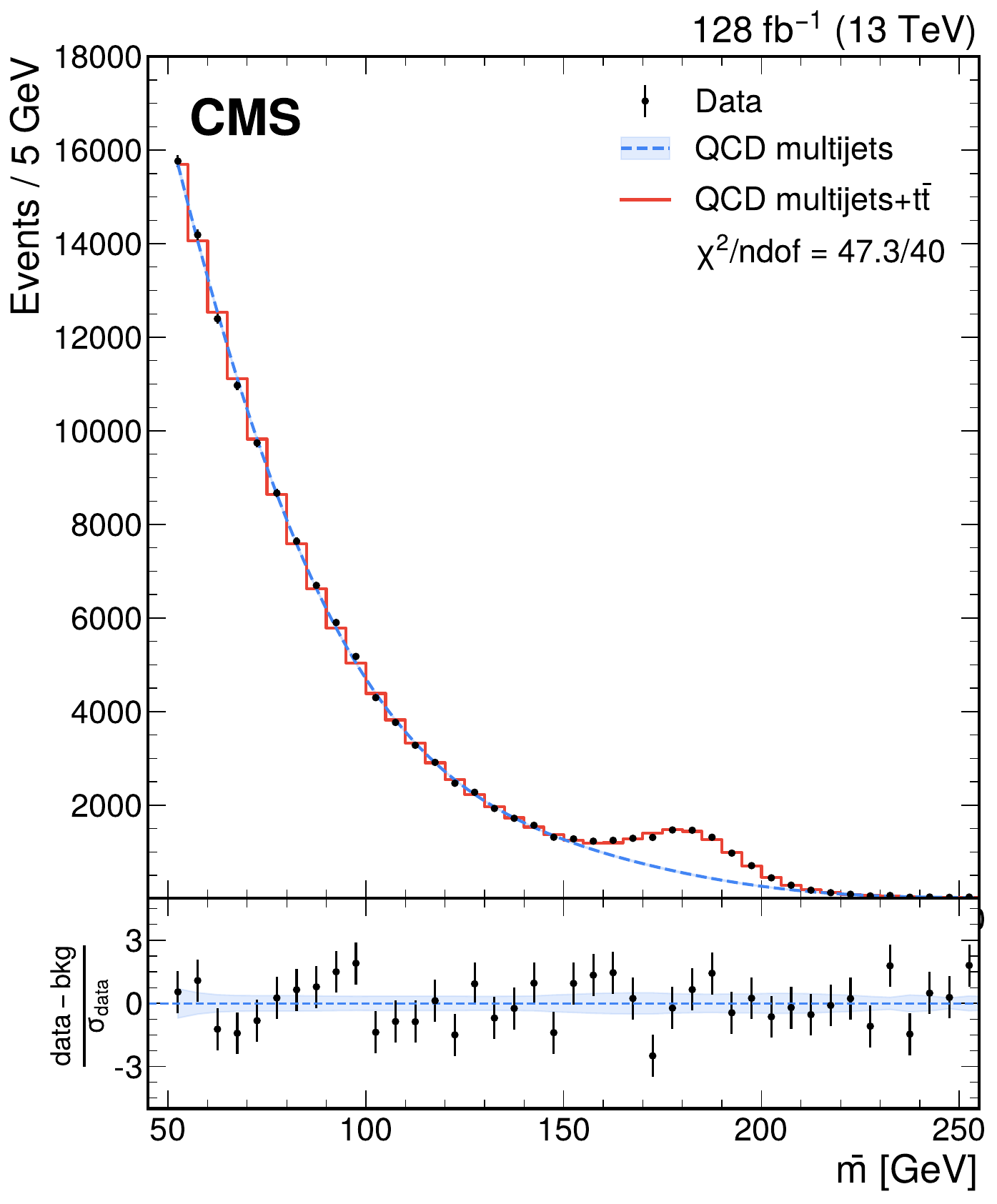}    
    \caption{The distribution of \mjjj for the resolved three-jet search (left), and average jet mass {($\tilde{m} = (m_1 + m_2)/ 2$)} for the merged three-parton search (right), adapted from Ref.~\cite{EXO-21-004}. Both analyses use PF jets. The peak around 170\GeV in both distributions corresponds to the all-hadronic decay of the top quark. The data (points) are compared to the background-only prediction (blue) and the full background fit including simulations of the top quark resonance (red).}
\label{fig:top_quark}
\end{figure*}

Studies similar to those presented above, which concern the comparison between PF jets reconstructed at HLT and offline, are described in detail in Section~\ref{subsec:run3jets} based on Run~3 data. In this section we focus on studying the calibration level of PF jets at the HLT by reconstructing SM resonances. The HLT PF jet energies are corrected with the standard methods described in Section~\ref{subsec:jets} using both simulation and data. The corrections are derived from data sets that have undergone full offline reconstruction, rather than from the scouting data set.
The top quark, which can be reconstructed as a resonance in three-quark final states, is clearly visible in Fig.~\ref{fig:top_quark}. The figure shows the invariant mass distribution of three PF jets and the mass distribution of single PF jets selected using jet substructure techniques that indicate they are likely to contain decays of three separate partons. The observed top quark mass peak positions agree with the expected ones from \ttbar simulation within less than 2\%, while the resolution in data is only 5\% worse.

To increase the sensitivity of multijet searches in scouting, new techniques are employed. One innovation is the use of the information in PF candidates within a jet to construct a quark-gluon discriminator (QGD) that enhances signals featuring decays to quarks while suppressing QCD backgrounds consisting largely of gluon jets.
The QGD is constructed from observables sensitive to fundamental differences in the fragmentation properties of quarks and gluons, such as the number of constituents and the jet radius. It uses a neural network (NN) architecture based on the \textsc{DeepSets} technique~\cite{NIPS2017_f22e4747}. The NN inputs are the normalized four-momenta information along with the particle type of each jet constituent (PF candidates).

\begin{figure*}[!htb]
    \centering
    \includegraphics[width=0.463\textwidth]{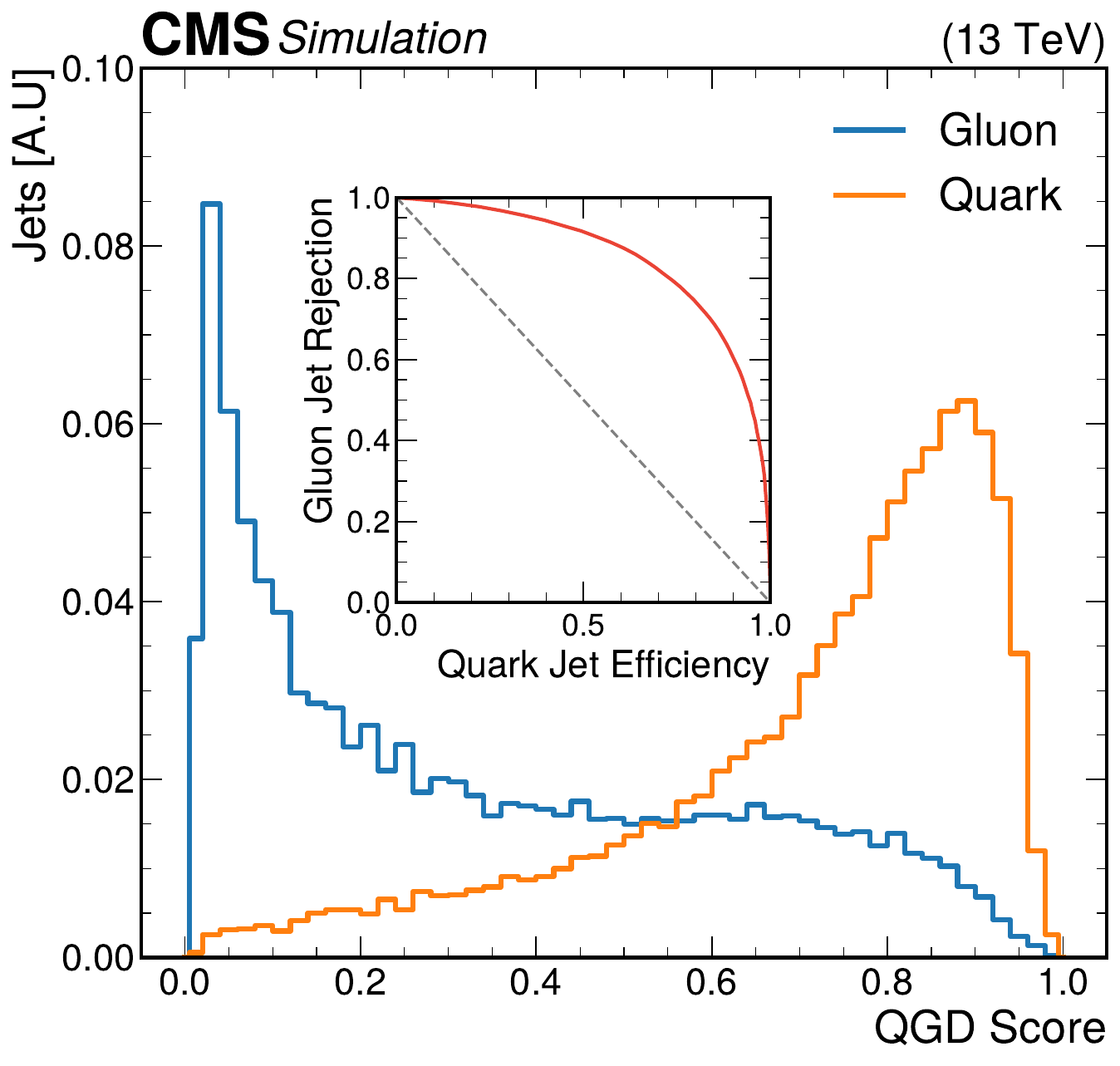}
    \includegraphics[width=0.427\textwidth]{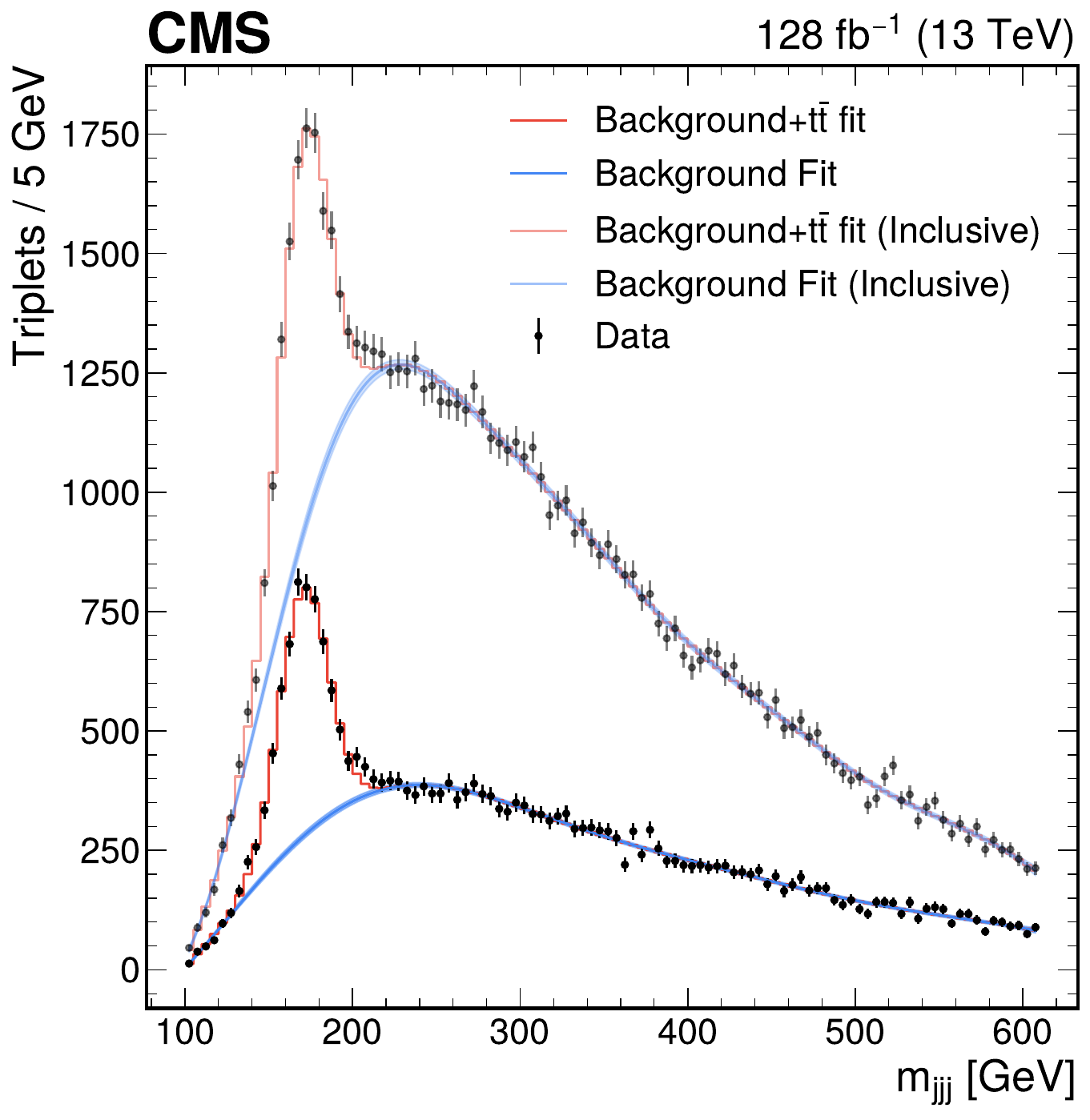}
    \caption{Left: output of the QGD for quark (orange) and gluon (blue) jets. The corresponding receiver operating characteristic (ROC) curve is also shown. Right: observation of fully hadronic top quark decays in the invariant mass of three jets with a QCD multijet background, for an inclusive selection (no QGD), and for a selection including the QGD score. Figure adapted from Ref.~\cite{EXO-21-004}.}
    \label{fig:qgl_opt}
\end{figure*}

The QGD NN selects quarks and rejects gluons with better performance than traditional methods that rely on jet multiplicity and jet mass~\cite{qgl_old}. Figure~\ref{fig:qgl_opt} (left) shows the QGD score distributions obtained for quarks and gluons, indicating a clear separation. Different working points are considered in analyses, corresponding to quark signal efficiency (gluon background rejection rates) of 98\% ( 31\%) , 83\% (70\%) , and 61\% (87\%), respectively, for loose, medium, and tight selections on the QGD score. In Fig.~\ref{fig:qgl_opt} (right), the tight selection on the QGD score is applied to the invariant mass of jet triplets in the search for $R$-parity violating (RPV) gluinos, described in Section~\ref{sec:MultiSearches}. The top quark peak is clearly seen above the QCD background. The jets in the top quark peak are mostly quarks, while the continuum background contains a large component of jets originating from gluons. The QGD significantly reduces the continuum background, in comparison to the inclusive selection (without QGD), while preserving the top quark signal. This demonstrates the power of the scouting technique in advancing the jet-based physics program of CMS, despite the limited event content stored on disk.

\subsection{Muons}
\label{sec:ScoutingRun2Muons}

Muons are often indicators of interesting physics because they can be produced via the electroweak interaction, where \PZ or \PW bosons are involved, or via new hypothetical interactions featuring unknown gauge bosons. The CMS detector is particularly well suited to the task of reconstructing and identifying muons, as explained in Section~\ref{sec:CMS}. The clean signature of a pair of muons in the final state is exploited in a large number of analyses, from SM precision measurements to searches for new physics up to the \TeVns scale. Thus, increasing the number of collected dimuon events was a key motivation in the development of the scouting strategy.

\subsubsection{Muon trigger performance}
\label{subsec:ScoutingRun2MuonsTriggers}

Dedicated trigger algorithms targeting dimuon events with significantly lower muon \pt thresholds than those of the standard trigger paths were implemented and fully commissioned in 2017. The trigger definitions are discussed in Section~\ref{sec:ScoutingRun2Details}. The number of selected events at low dimuon masses ($\mMM < 40\GeV$) is substantially increased by reducing the size of the event content at the HLT. The data collected during the last two years of Run~2 (2017 and 2018) correspond to a total \Lint value of 101.1\fbinv, 96.6\fbinv of which are used for analysis. The scouting dimuon triggers provided an overall rate of approximately 2\unit{kHz} for ${\Linst \approx \sci{1.5}{34}\invcms}$, about a hundred times higher than the standard dimuon triggers.

\begin{figure*}[!htb]
    \centering
    \includegraphics[width=0.74\textwidth]{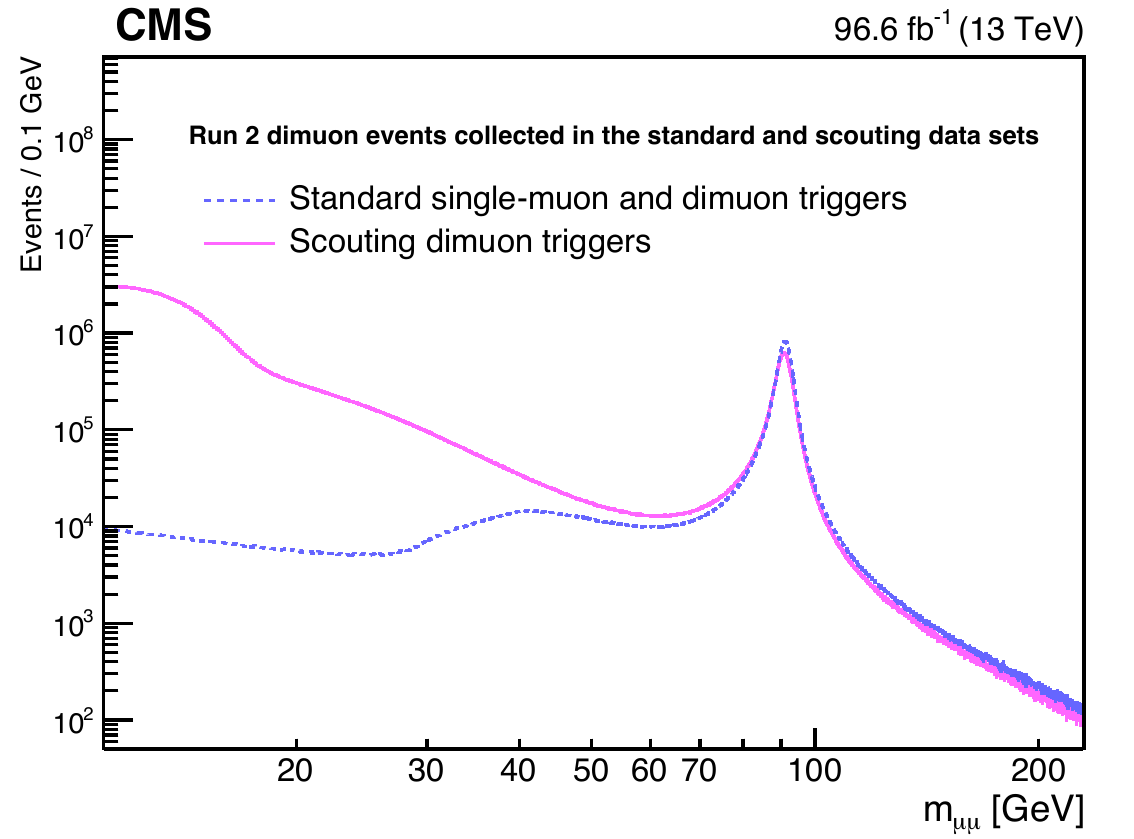}
    \caption{Dimuon invariant mass distribution of events selected with the standard muon triggers (blue, dashed) and scouting muon triggers (pink, solid) in the mass range 11--240\GeV, normalized to ${\Lint = 96.6\fbinv}$, corresponding to the scouting data collected in 2017 and 2018. The selection applied to obtain each distribution is described in Ref.~\cite{EXO-19-018}.}
    \label{fig:muons_scoutingVSstandard}
\end{figure*}

Dimuon invariant mass spectra obtained using data collected with the standard and scouting triggers are compared in Fig.~\ref{fig:muons_scoutingVSstandard}. The two curves are normalized to the amount of data collected with the scouting triggers in 2017 and 2018. The standard triggers show significant acceptance losses below about 40\GeV because of the higher \pt thresholds on the leading and subleading \pt muons. For the standard trigger strategy, these thresholds are 17 and 8\GeV, respectively. The acceptance is considerably recovered by the scouting triggers thanks to the looser HLT selections. These selections include the reconstruction of at least two muons at the HLT, each with ${\pt > 3\GeV}$. These requirements are minimal relative to the L1 selections, which are described in Table~\ref{tab:L1_HLT_Thr} for 2017 and 2018. The distributions in Fig.~\ref{fig:muons_scoutingVSstandard} are obtained by selecting events with offline muon \pt requirements of 20 and 10\GeV in the standard dimuon triggers, and 4\GeV for both muons in the scouting dimuon trigger. These selections ensure operation on the plateau of the trigger efficiency curves. We note that the comparison between the two data sets has some limitations. The standard thresholds of the dimuon selections in 2016 were lower than the ones adopted in 2017. In addition, the single-muon L1 path was only added to the dimuon scouting stream in 2018. This explains the  discrepancy in event yields near the \PZ boson mass peak and at higher masses, observed when normalizing both curves to the same \Lint value.

\begin{figure*}[!htb]
    \centering
    \includegraphics[width=0.80\textwidth]{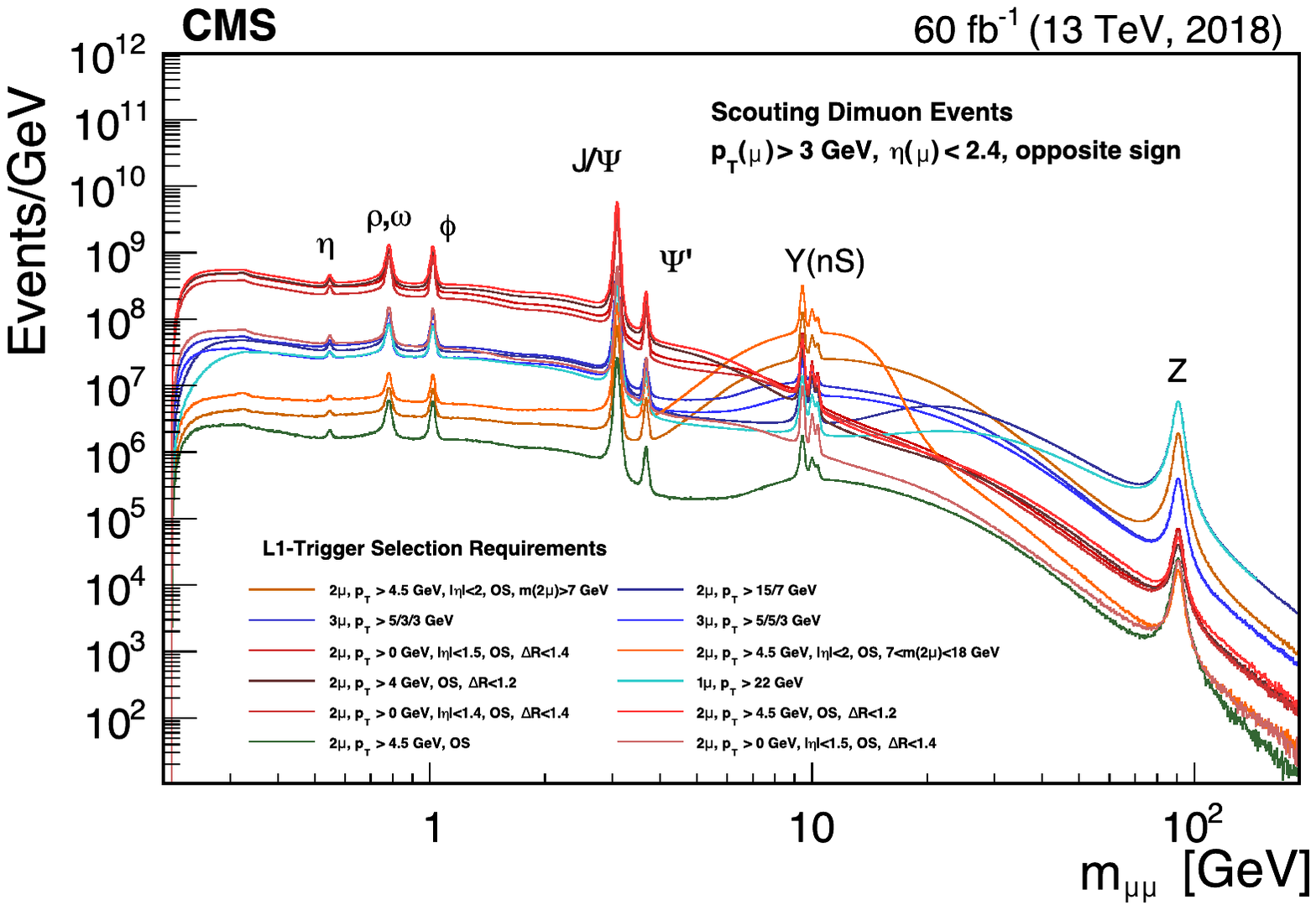}
    \caption{Dimuon invariant mass spectrum and event rate of each L1 seed (legend) obtained with the scouting stream reconstructed at the HLT, using data collected in 2018 corresponding to ${\Lint = 60\fbinv}$. Well-known dimuon resonances from various meson decays or from \PZ boson decays are indicated above each peak.}
    \label{fig:muons_dimuonSpectrum_l1t}
\end{figure*}

The high-rate dimuon scouting stream with lower transverse momentum selections at the L1 (as low as the ones at the HLT) enables the exploration of an otherwise inaccessible phase space at low dimuon masses, down to about twice the muon mass at ${\approx210\MeV}$, which is the dimuon kinematic threshold. Figure~\ref{fig:muons_dimuonSpectrum_l1t} shows the dimuon invariant mass distribution obtained with the various L1 algorithms in 2018 and reconstructed at the HLT.
    
\begin{figure*}[!htb]
    \centering
    \includegraphics[width=0.78\textwidth]{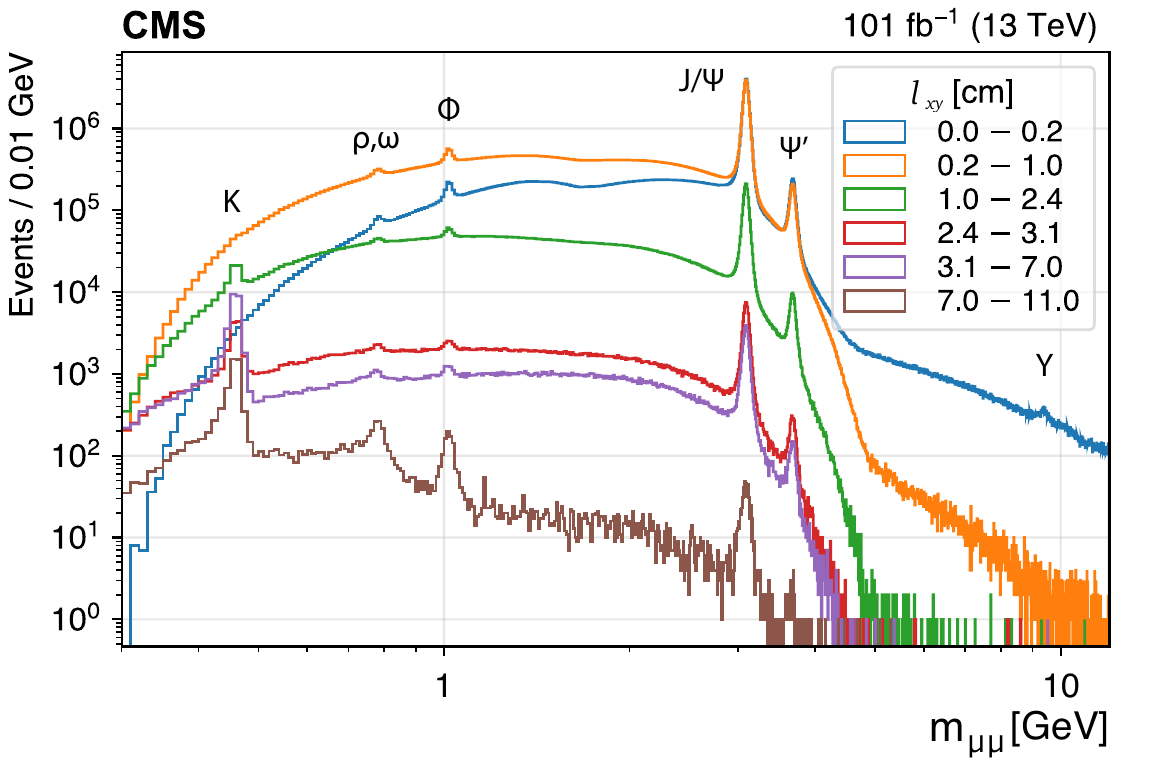}
    \caption{Dimuon invariant mass distributions in bins of transverse displacement from the PV (\lxy). Figure adapted from Ref.~\cite{EXO-20-014}.}
    \label{fig:muons_dimuonSpectrum_displacement}
\end{figure*}

To extend the physics use case of the scouting stream, the scouting triggers utilize HLT reconstruction algorithms that lack any association between muons and the PV. This enables scouting searches for resonances that have nonzero displacement from the PV. The dimuon invariant mass distributions for different values of the dimuon transverse displacement from the interaction point (referred to as \lxy) are shown in Fig.~\ref{fig:muons_dimuonSpectrum_displacement}. The maximum transverse displacement of about 11\cm is determined by the requirement that muon tracks deposit energy in at least two layers of the CMS pixel tracker. The definition of the muon scouting triggers in Run~3 have been updated to remove this requirement, providing sensitivity to resonances with even higher transverse displacements, as described in Section~\ref{sec:trigger_strategy_run3}. 

\begin{figure*}[!htb]
    \centering
    \includegraphics[width=0.49\textwidth]{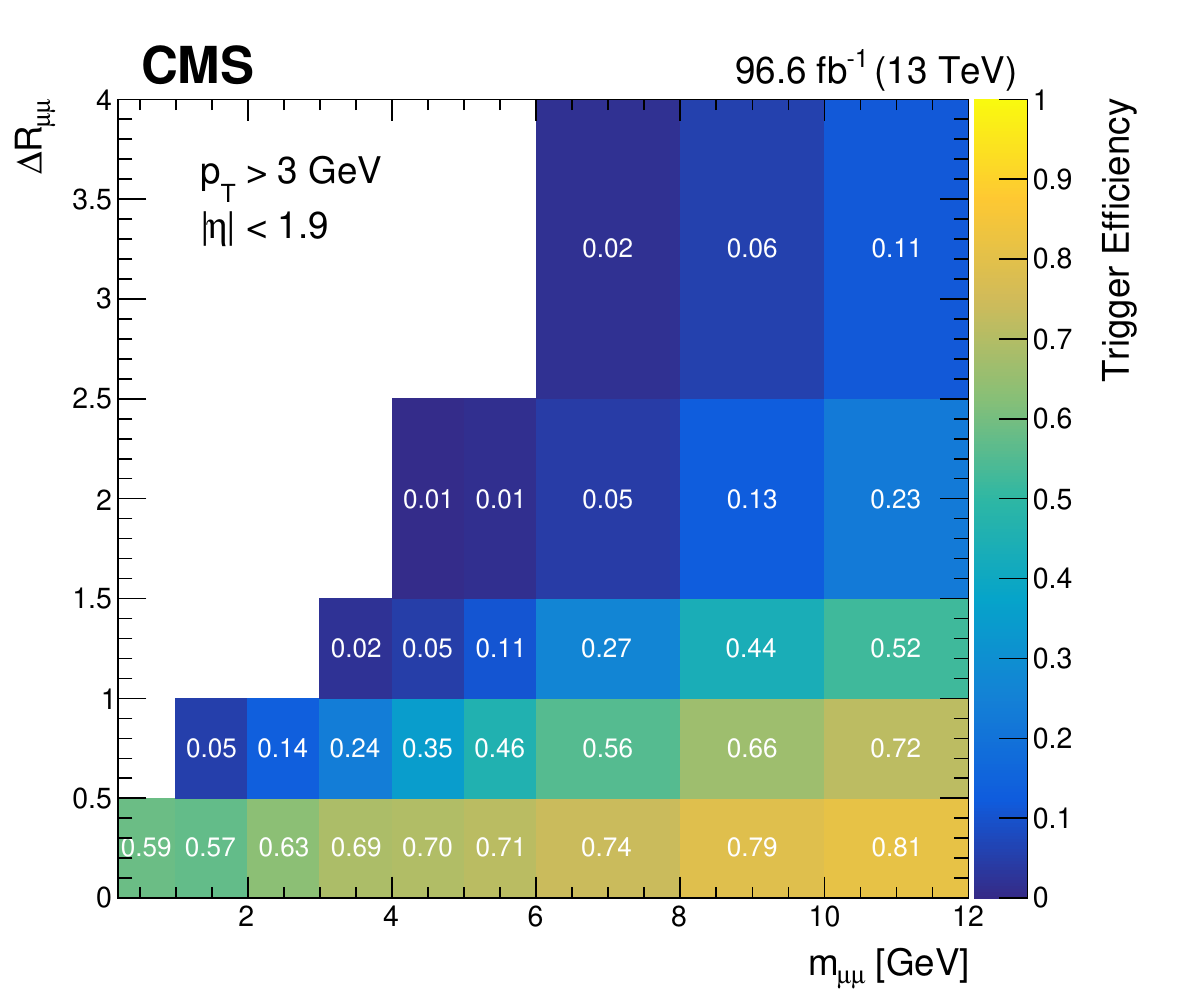}
    \includegraphics[width=0.49\textwidth]{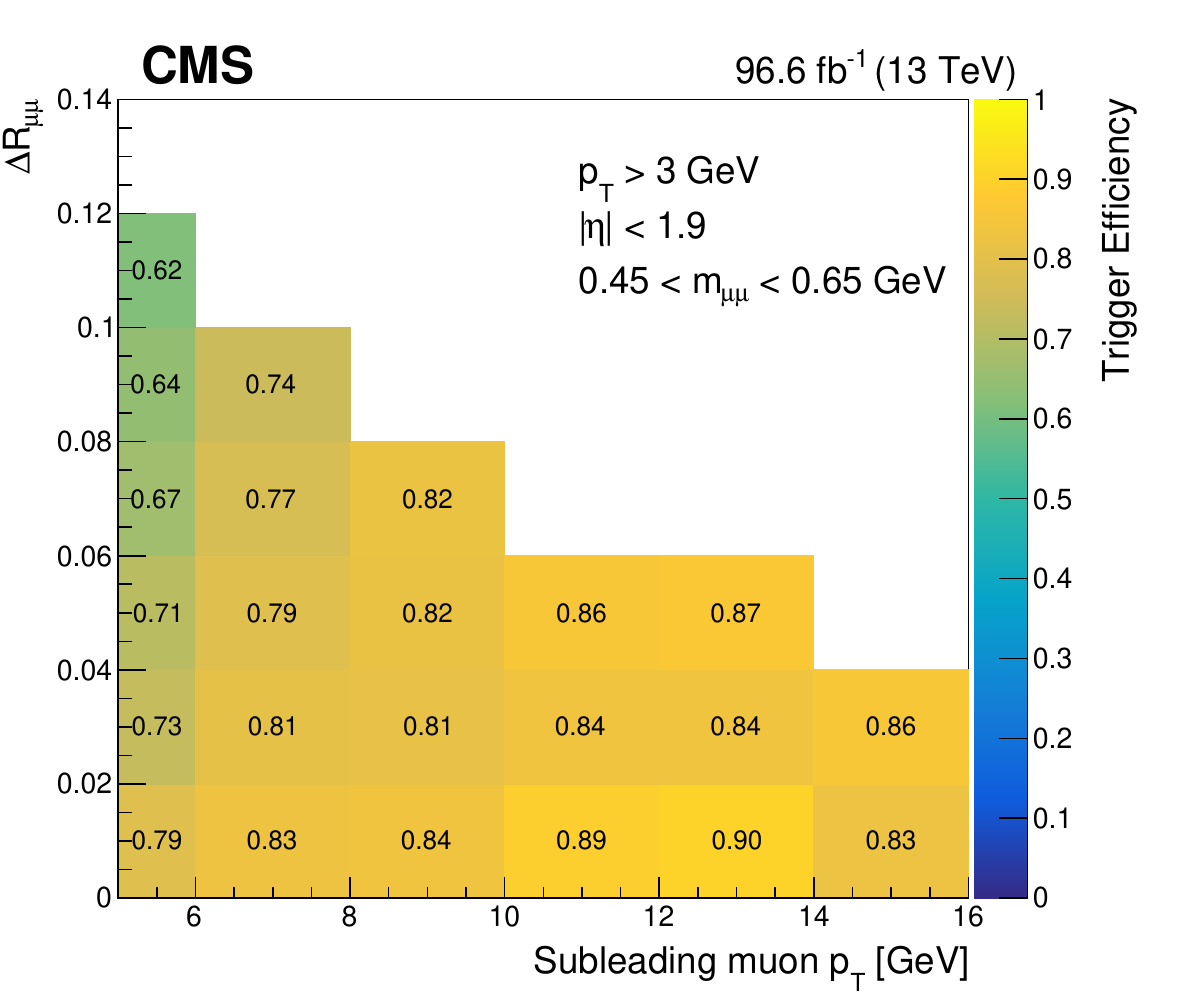}
    \caption{Efficiency of the dimuon scouting trigger and logical ``OR'' of all L1 triggers measured with 2017 and 2018 data. The efficiency is shown as a function of the angular separation between the two muons \drMM and the dimuon mass \mMM (\cmsLeft), and as a function of \drMM and the subleading muon \pt (\cmsRight). The selection in the \cmsRight plot also requires ${0.45 < \mMM < 0.65\GeV}$ to focus on the \PGh meson resonance region. The statistical uncertainty in the measured values is generally less than 3\% per bin on the left plot and less than 15\% per bin on the right plot.}
    \label{fig:muons_eff}
\end{figure*}

In Fig.~\ref{fig:muons_eff}, the efficiency of the dimuon scouting triggers, including the HLT selection and the main L1 paths, is presented in two-dimensional (2D) maps as a function of the angular separation $\deltar_{\mu\mu}$ between the two muons and the dimuon invariant mass \mMM (\cmsLeft plot), and as a function of \drMM and the subleading muon \pt (\cmsRight plot). These maps are needed to account for correlations between muons in the L1 algorithms. For very low mass resonances, the muon momenta are typically collinear, with low values of \drMM. Therefore, L1 requirements on \drMM are applied as the \pt threshold is lowered to focus on the relevant physics expected in this kinematic region, as reported in Table~\ref{tab:L1_HLT_Thr}. The efficiency measurements are performed with orthogonal data sets that are independent of the presence of muons. All muons with ${\pt > 3\GeV}$ and ${\abs{\eta} < 1.9}$ are considered. The additional selection ${0.45 < \mMM < 0.65\GeV}$ is applied to the \cmsRight plot to focus on events likely to contain $\PGh\to\PGmp\PGmm$ decays. The \PGh meson, with a mass around 0.55\GeV, is one of the lightest resonances decaying to pairs of muons for which CMS has sensitivity, and thus serves as a useful proxy to study the performance of low-mass dimuon reconstruction.

The \cmsLeft plot in Fig.~\ref{fig:muons_eff} shows that the efficiency to trigger on muon pairs with small angular separation (${\drMM < 1.0)}$ ranges from 60 to 80\%. The efficiency gradually decreases for higher \drMM and lower \mMM. Near the \PGh meson mass, it ranges from 50 to 90\% \vs subleading muon \pt, within the region ${\drMM < 0.06}$, which contains most of the $\PGh\to\PGmp\PGmm$ decays as determined from simulation. For a fixed value of \mMM, higher values of \drMM imply lower muon \pt. This explains the lack of events in the upper left region of the \cmsLeft plot. In the \cmsRight plot, for fixed muon \pt, higher \drMM implies higher \mMM. Since the selection in this plot includes ${0.45 < \mMM < 0.65\GeV}$, no events are found with \drMM values above 0.12. 

\subsubsection{Muon reconstruction performance}
\label{subsubsec:mu_recoPerformance}

The process of reconstructing muon objects within the scouting stream mirrors that of the standard stream, differing only in the removal of the vertex constraint. However, recorded events contain only a limited amount of information compared to the offline muons. Therefore, dedicated identification (ID) criteria were developed to select muons from the scouting data set.

A customized selection based on standard physics variables (cut-based selection)  was initially designed in the context of a search for dimuon prompt production in the mass range 11--45\GeV, as described in Section~\ref{par:promptDimuonResonances}. It relies on some requirements applied to the muon track, such as the number of tracker pixel hits, the total number of tracker layers containing energy deposits, and the quality and relative isolation of the muon track. This selection is not ideal for lower mass resonances, however. The angular separation between the muons is small when ${\mMM < 10\GeV}$. The muon isolation is less efficient for the boosted system because the isolation cone of the two muons may partially overlap. An optimized selection based on a multivariate analysis (MVA) technique was therefore developed to improve sensitivity to lower mass signals, by increasing the signal muon efficiency and suppressing the rate of background muons, mainly coming from decays in flight of hadrons. 

The set of input variables for training the MVA classifier contains a combination of muon and vertex variables, such as the number of pixel hits and tracker layers, the muon track and vertex $\chi^2$, the track isolation, and the vertex transverse displacement from the interaction point. These parameters are combined into a single discriminator using a boosted decision tree (BDT)~\cite{BDT_ROE2005577}.

\begin{figure*}[!htb]
    \centering
    \includegraphics[width=0.6\textwidth]{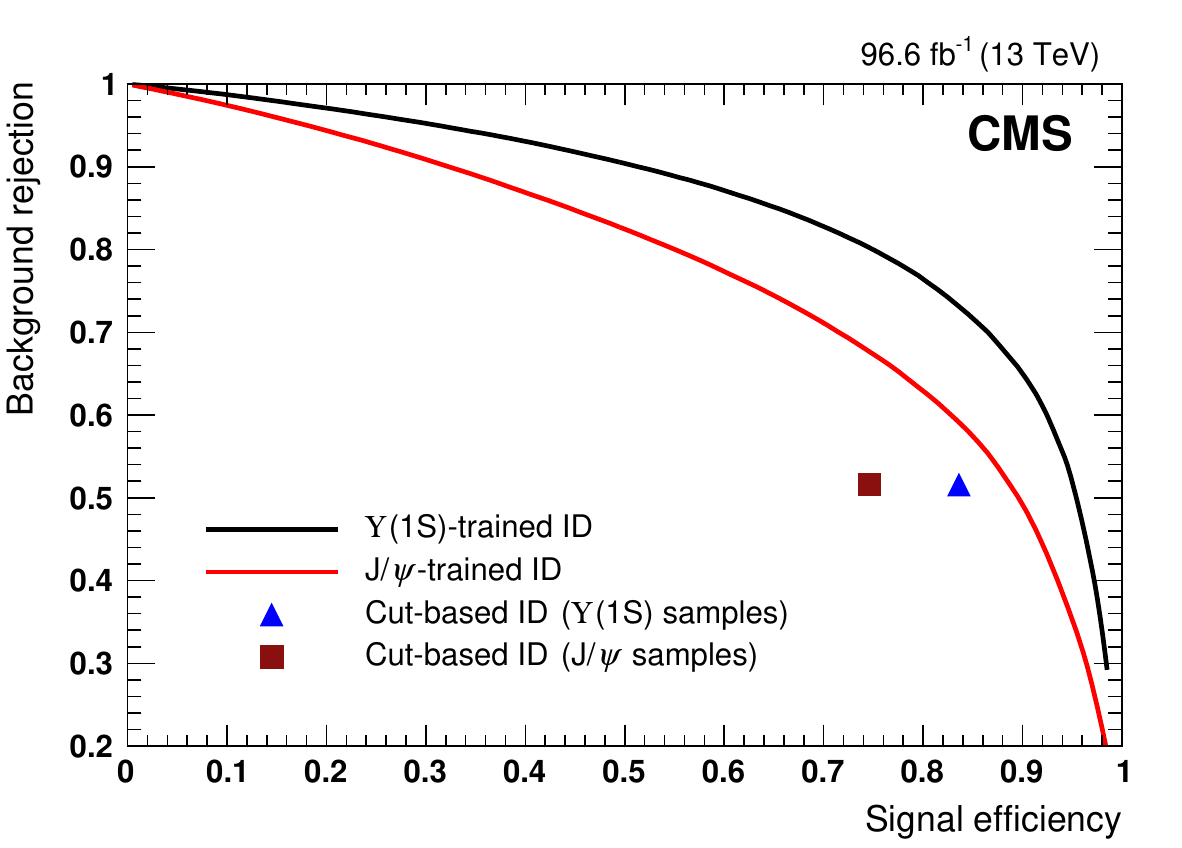}
    \caption{Background rejection \vs signal efficiency of the new MVA-based muon ID strategies evaluated on scouting data: \PgUa-trained MVA (black line), \PJGy-trained MVA (red line). A comparison with the performance of the previous cut-based selection, which was optimized for signals with masses higher than 11\GeV, is also shown for the \PgUa (blue triangle) and \PJGy (brown square) signals. Figure adapted from Ref.~\cite{EXO-21-005}.}
\label{fig:muons_IDs}
\end{figure*}

The MVA is optimized separately for the higher mass ({4--10\GeV}) and lower mass ({$<4\GeV$}) regions. The signal samples used for the MVA training and validation are extracted from events in data containing \PgUa and \PJGy meson decays, for the high-mass and low-mass regions respectively, while same-charge muon events are used as background samples. The background rejection \vs signal efficiency of all IDs are summarized in Fig.~\ref{fig:muons_IDs}, demonstrating the improved sensitivity of the MVA-based selections relative to the cut-based ones in comparable mass regions. Considering a similar signal efficiency for \PgUa and \PJGy signals, the new MVA ID achieves significantly higher background rejection.

\begin{figure*}[!hbt]
    \centering
    \includegraphics[width=0.65\textwidth]{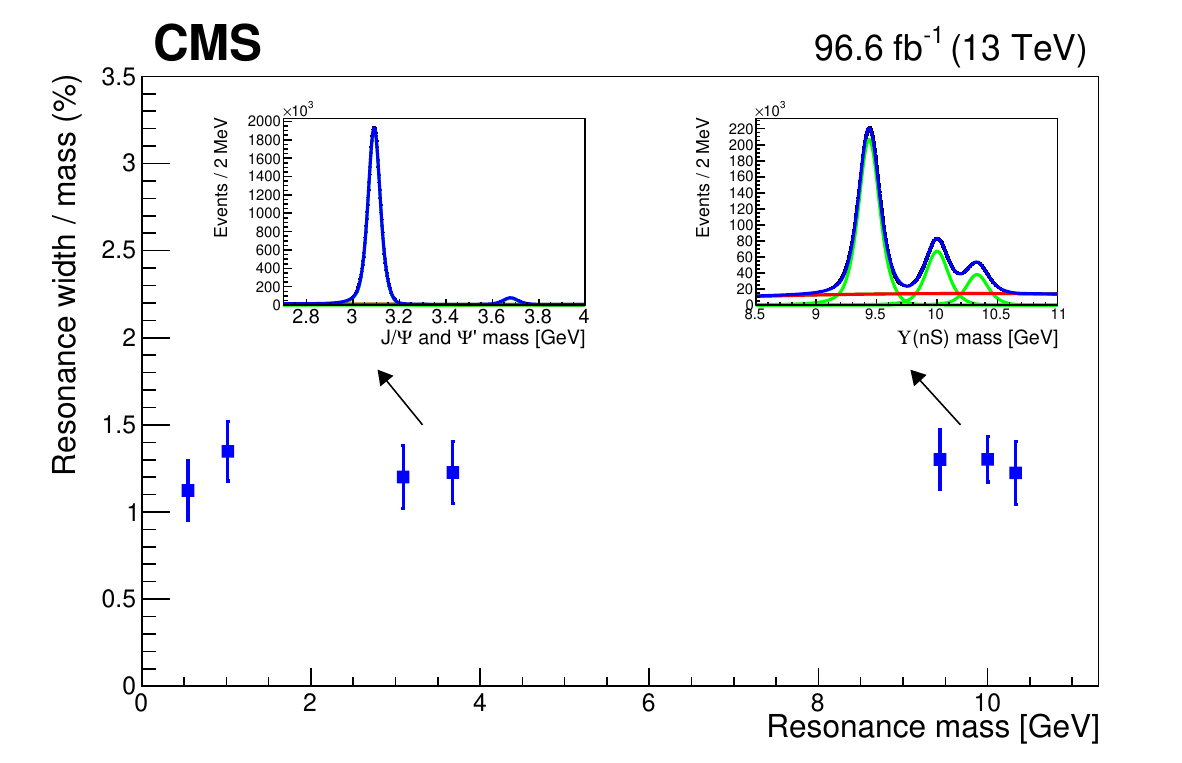}
    \caption{Relative width of dimuon resonances as a function of mass, measured in 2017 and 2018 scouting data. The fits are performed separately for the 2017 and 2018 data sets. The values shown are the average width of each fit, weighted by the \Lint value corresponding to the data accumulated in each year. From left to right, the \PGh, \PGf, \PJGy, \Pgy, \PgUa, \PgUb, and \PgUc resonances are shown. The inserts display fits of the \PJGy and \PGU peaks obtained with scouting data (black markers) separately for the signal (green) and background (red) components, and for their sum (blue).}   
    \label{fig:muons_resWidth}  
\end{figure*}

The performance of the muon scouting stream was also measured by means of the dimuon mass resolution. Figure~\ref{fig:muons_resWidth} shows mass resolutions measured for various SM resonances. The resolution was determined by fitting a signal-plus-background model to the dimuon mass spectrum in 2017 and 2018 data around known resonances, such as the \PGh, \PGf, \PJGy, and \PGU, and then extracting the relative width of the peaks. The signal is modeled with the sum of a double-sided Crystal Ball (CB) function~\cite{crystalball-1,crystalball-2} and a Gaussian function in a mass window of ${\pm}5\%$ around the mean peak value, while the background is described by a third-order Bernstein polynomial. The signal resolution, estimated with the $\sigma_{\mathrm{CB}}$ parameter of the Gaussian core of the CB function, is found to be ${\approx}1.3\%$ and roughly independent of year, mass hypothesis, or detector region. The uncertainty in the resolution is evaluated as the variation introduced by alternative signal models, such as a double-Gaussian function, and measured to be 13--18\% depending on the resonance. 
In the low-\pt regime (the region of interest) the absolute difference between the mass resolution of scouting muons and that of offline muons is less than 1\%. This demonstrates the remarkable capabilities of data scouting in boosting the muon-based physics program of CMS.

\subsection{Physics results}\label{sec:Run2ScoutingPhysicsResults}

We now describe some of the CMS physics results obtained with the scouting data sets collected during Run~2. The results are organized in three sections: searches for new physics in hadronic final states, searches for new physics in leptonic final states, and observations of rare SM decays.

\subsubsection{Searches in hadronic final states}

Searches for direct production of hadronic resonances are particularly important at the LHC, as any hypothetical particle produced via the strong interaction in \pp collisions can also decay to quarks and gluons, which hadronize to form jets. The main source of backgrounds consists of SM multijet QCD backgrounds. The rates for these processes are typically very large compared to those of the potential signals of new physics, increasing substantially for lower resonance masses. Consequently, the event rate and the amount of data that must be recorded in order to carry out physics analyses in this low-energy regime rapidly increases. In this context, the data-scouting approach plays a crucial role in probing the low-mass region in final states with jets.

The CMS hadronic physics program covers a wide range of experimental signatures, such as resonances decaying to a pair of jets (dijet), dijet resonances in association with initial-state radiation, resonances decaying to three jets, and pair-produced resonances resulting in final states with four or more jets. In this section, we describe CMS searches that apply the data scouting technique to extend sensitivity to new physics in the mass region below the \TeVns scale.

\paragraph{Dijet resonances}
\label{sec:DijetSearches}~

Proton-proton collisions can produce two or more energetic jets when the constituent partons scatter with large momentum transfer. The invariant mass distribution of the two jets with the largest \pt is predicted to fall steeply and smoothly with increasing mass, based on known multijet processes. Many proposed extensions of the SM predict the existence of new states coupling to quarks and gluons, which would appear as resonances on top of this smooth background in the dijet mass spectrum. 

A review of searches for dijet resonances at hadron colliders can be found in Ref.~\cite{Harris:2011bh}. The first searches for dijet resonances were presented by the UA1 and UA2 experiments after collecting data at the Sp$\overline{\mathrm{p}}$S accelerator with ${\sqrt{s} = 630\GeV}$. These results were later extended to higher resonance masses by the CDF and D0 experiments, using the Fermilab proton-antiproton Tevatron collider, which operated with center-of-mass energies of 1.8 and 1.96\TeV. Finally, the mass reach was increased further by ATLAS and CMS, relying on \pp collisions at the LHC with ${\sqrt{s}= 7}$, 8, and 13\TeV. 

Results obtained with different collider energies were compared by translating the upper limits on the quark-quark resonance cross sections into upper bounds on the coupling constant $g$ between the new resonance and a pair of partons as reported in Ref.~\cite{Dobrescu:2013cmh}. That study demonstrated that the existing searches, up until early Run~1, were not sensitive to the presence of low-mass (${<1\TeV}$) resonances with small couplings to quarks (${g_B<1}$). In particular, LHC experiments were affected by the aforementioned limitations in the conventional DAQ approaches for triggering, processing, and storing data, resulting in decreased sensitivity to lower-mass resonances.

To address this issue, in the last days of the 2011 data-taking period, the CMS Collaboration tested the new data scouting approach for the first time~\cite{CMS:2012ScoutingParking}. 
A preliminary search for dijet resonances using this special data set was performed with \pp collision data corresponding to 0.13\fbinv at ${\sqrt{s}=7\TeV}$. This search improved the limits on the production cross section of new dijet resonances in the 0.6--0.9\TeV range, a region otherwise inaccessible with standard triggers. Since 2012, data scouting has become a well-established approach in CMS, leading to the publication of several physics results on searches for dijet resonances, which are summarized in the following paragraphs.

The basic strategy of these searches consists in reconstructing the two jets that correspond to the pairs of quarks or gluons arising from the decay of a new particle. We therefore look for a peak in the invariant mass distribution of the reconstructed dijet system (\mjj), with characteristic shape compatible with the one expected from a resonance decay. The main background from QCD multijet production is predicted by fitting the \mjj distribution with an empirical functional form that describes well the QCD simulation and the data in absence of a new physics signal. The main trigger for dijet resonance searches requires \HT to exceed a predefined threshold. As discussed in Section~\ref{sec:JetTriggerRun2}, the scouting trigger has a lower threshold than standard triggers and becomes fully efficient for ${\mjj > 500\GeV}$, compared to ${\mjj>1.25\TeV}$ required by the standard triggers. The data scouting approach is thus able to extend the search for resonances down to the sub-\TeVns mass range. 

The analyses combine calorimeter jets originally reconstructed with the standard anti-\kt algorithm at the HLT with distance parameter ${R < 0.4}$ or 0.5 (AK jets) into ``wide jets'', which are then used to measure the mass spectrum and to search for new physics. The partons from the decay of heavy objects can radiate additional partons, which are often produced at large angles with respect to the original parton direction and thus clustered into a separate AK jet. To reduce this effect, the two \pt-leading AK jets are used as seeds and the four-momenta of all other jets, if within ${\deltar < 1.1}$ of the seed jet, are added to it to obtain two wide jets, which then form the dijet system. Wide jets collect more of the final-state radiation compared to AK jets and therefore improve the mass resolution of dijet resonances.

\begin{figure*}[!htb]
    \centering
    \includegraphics[width=0.475\textwidth]{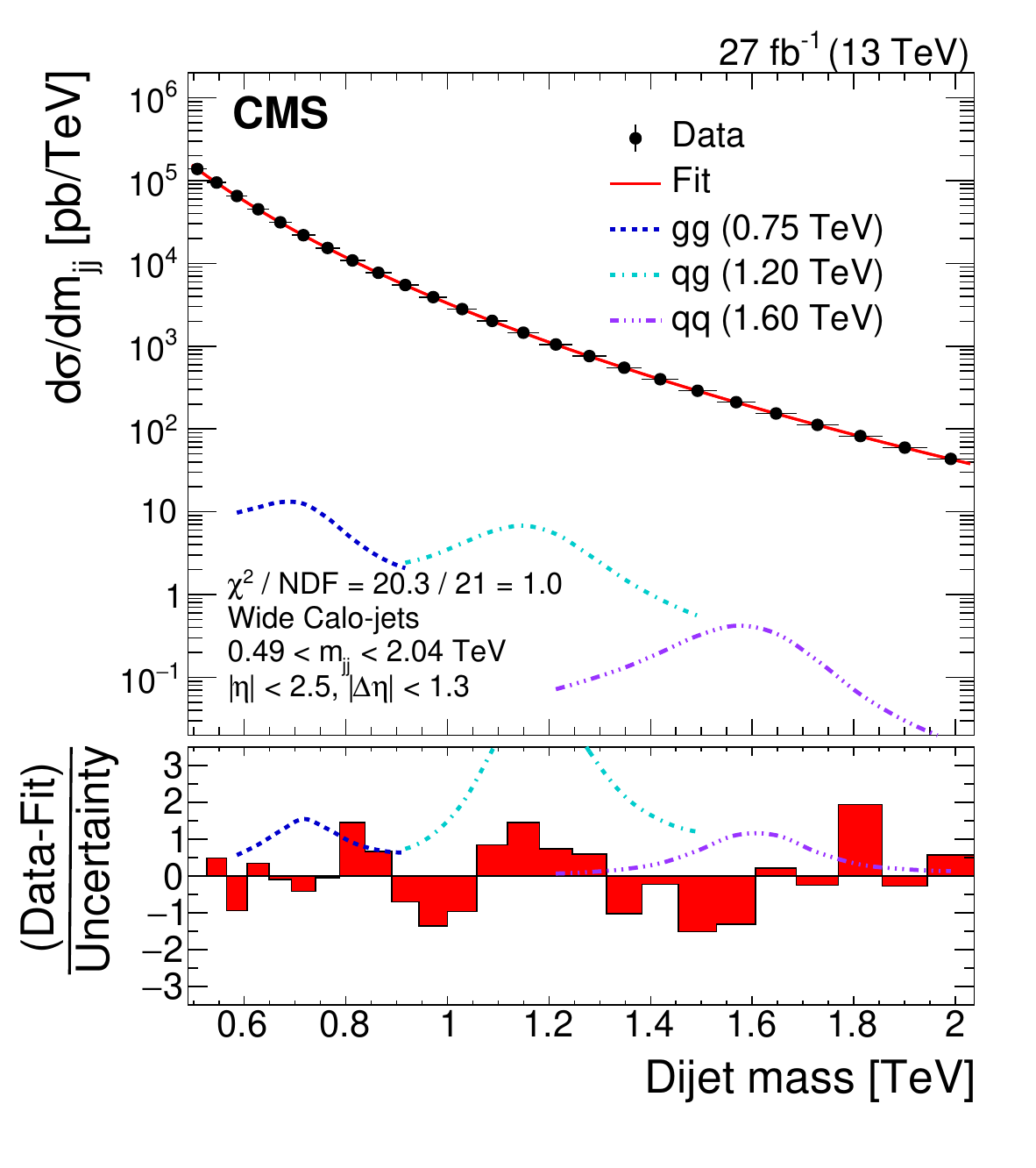}
    \includegraphics[width=0.475\textwidth]{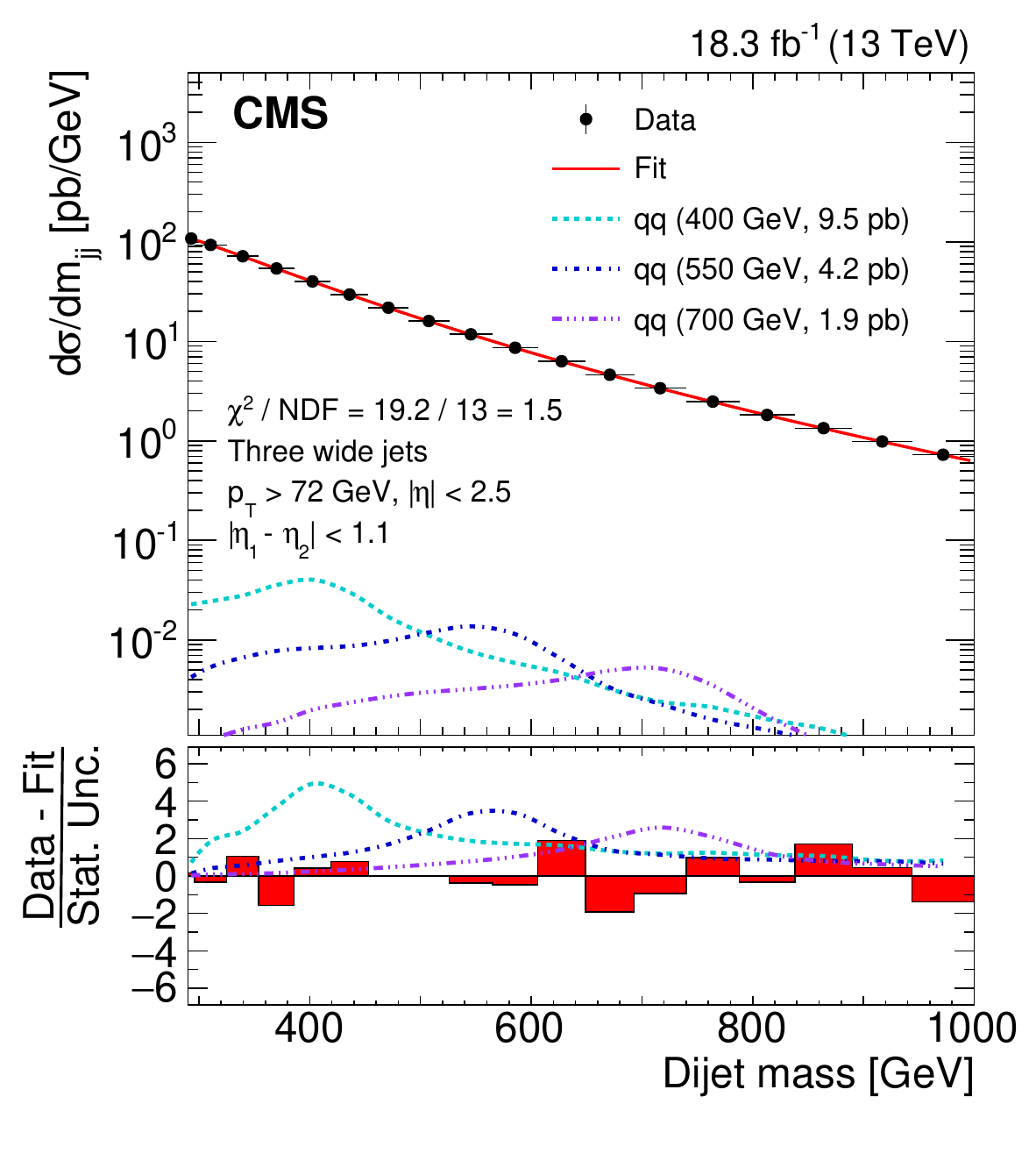}
    \caption{Left: dijet mass spectra (points) compared to a fitted parametrization of the background (solid curve) for the inclusive search performed in Ref.~\cite{EXO-16-056}. Right: dijet mass spectrum (points) compared to a fitted parametrization of the background (solid curve) for the three-jet analysis performed in Ref.~\cite{EXO-19-004}. The lower panel shows the difference between the data and the fitted parametrization, divided by the statistical uncertainty of the data. Examples of predicted signals from narrow gluon-gluon, quark-gluon, and quark-quark resonances are shown with cross sections equal to the observed upper limits at 95\%~\CL. Figures taken from Refs.~\cite{EXO-16-056} (left) and \cite{EXO-19-004} (right).}
    \label{fig:dijetMassSpectra}
\end{figure*}

Inclusive searches for dijet resonances have been published using data scouting at both $\sqrt{s} = 8\TeV$~\cite{EXO-14-005} and 13\TeV~\cite{EXO-16-032, EXO-16-056}. The dijet mass spectrum for the most recent analysis at 13\TeV, shown in Fig.~\ref{fig:dijetMassSpectra} (\cmsLeft), is well described by a smooth background parametrization, and no evidence for the production of new particles is observed. The spectrum is only shown up to about 2\TeV, as standard offline reconstructed data is used for higher dijet masses. Upper limits at 95\%~confidence level (\CL) are reported on the production cross section for narrow resonances, with masses between 0.6 and 1.6\TeV. The limits range between about 0.1 and 50\unit{pb} depending on the final state considered for the signal model and the resonance mass hypothesis.

To probe resonance masses below 600\GeV, we focus on events where at least one additional high-\pt jet is produced in association with a dijet resonance, resulting in a three-jet final state. The requirement of an additional jet (for example coming from initial state radiation) provides enough energy in the event to satisfy the trigger selection. A search for a dijet resonance decaying to a pair of jets with mass between 350 and 700\GeV is performed using events containing at least three jets ~\cite{EXO-19-004}. The dijet invariant mass spectrum, calculated from the two jets with the largest transverse momenta in the event, is used to search for a resonance. No significant excess over a smoothly falling background is found, as shown in Fig.~\ref{fig:dijetMassSpectra} (\cmsRight). Limits at 95\%~\CL are set on the production cross section of a narrow resonance in the range between 1 and 20\unit{pb}, depending on the resonance mass. The three-jet final state provides sensitivity to even lower resonance masses than in previous searches with the data-scouting technique.

\begin{figure*}[!htb]
    \centering
    \includegraphics[width=0.99\textwidth]{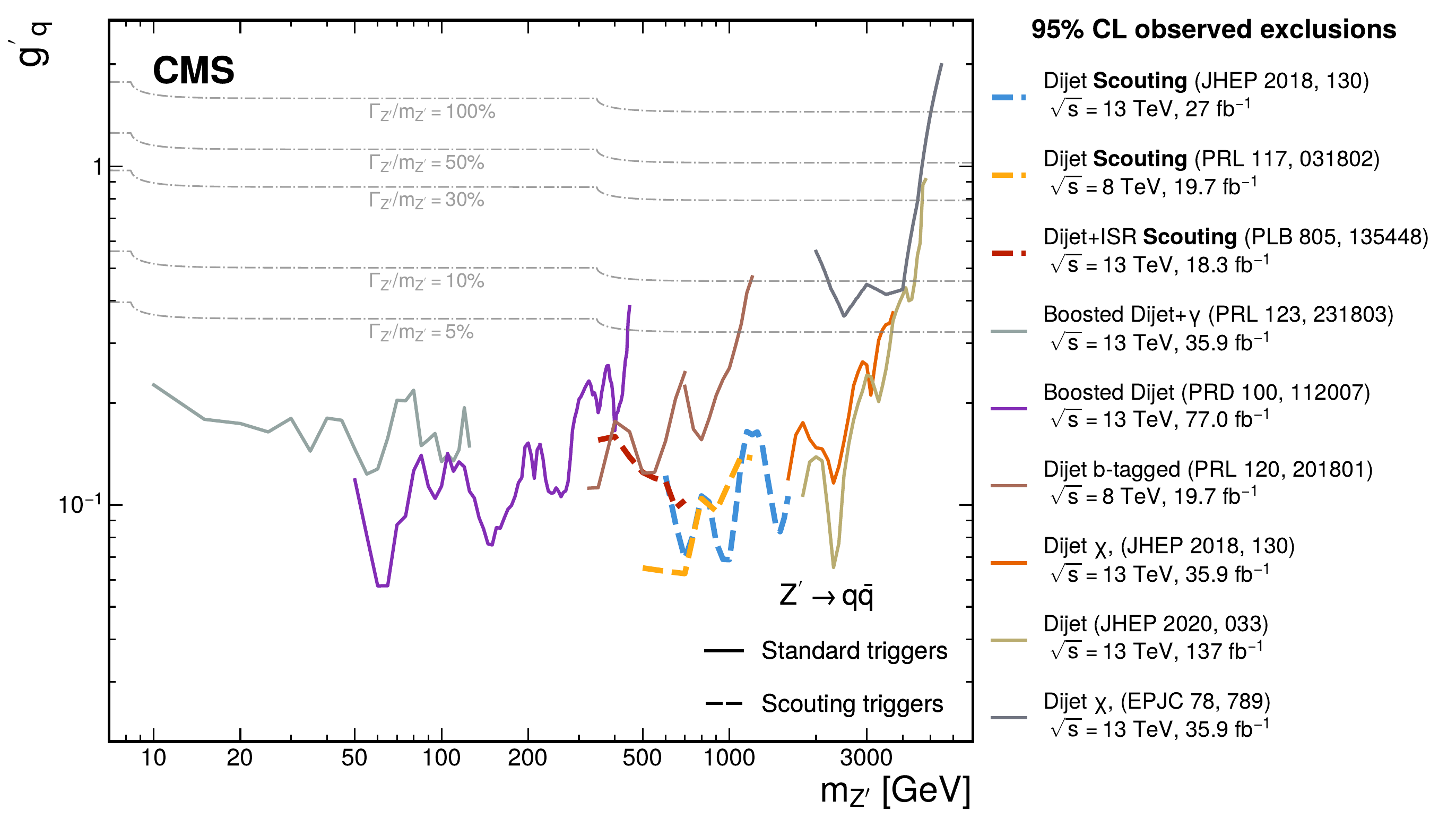}
    \caption{Observed limits on the universal coupling $g^\prime_\mathrm{q}$ between a leptophobic \PZpr boson and quarks~\cite{EXO-16-032} from various CMS dijet analyses. Regions above the lines are excluded at 95\%~\CL. The grey dashed lines show the $g^\prime_\mathrm{q}$ values at fixed values of ${\Gamma_{\PZpr}/m_{\PZpr}}$. Limits from scouting-based analyses are indicated with bold-dashed lines.} 
    \label{fig:coupling_vs_mass_dijet_limits}
\end{figure*}

Following the method presented in Ref.~\cite{EXO-16-056}, the model-independent upper limits on the cross section of dijet resonances are translated into 95\%~\CL upper limits on the coupling $g'_q$ of a hypothetical leptophobic resonance ${\PZpr \to \PQq \PAQq}$ as a function of its mass. Figure~\ref{fig:coupling_vs_mass_dijet_limits} shows the upper limits obtained by various CMS searches for dijet resonances. These results improve upon those obtained from previous experiments at the Sp$\overline{\mathrm{p}}$S and Tevatron colliders at lower center-of-mass energies. The aforementioned analyses with the data scouting technique provide the best limits in the mass region from 400 to 1600\GeV.

The mass region below 350\GeV is probed by analyses that rely on standard triggers and study events where the hypothetical resonance is produced with sufficiently high transverse momentum such that its decay products are merged into a single jet, with a two-prong substructure (boosted dijet)~\cite{EXO-17-001,EXO-17-027}. A signal would be identified as a peak over a smoothly falling background in the distribution of the invariant mass of the jet, using jet substructure techniques. These analyses study resonances produced in association with a high-\pt photon or jet and probe the resonance mass range 10--125\GeV and 50--500\GeV, respectively. For full efficiency with respect to the standard trigger requirements, events are selected by demanding the presence of a photon with ${\pt > 200\GeV}$ in the first case or a jet with ${\pt > 500\GeV}$ in the second. Future developments for these analyses include exploiting the scouting triggers, which would significantly reduce the photon and jet \pt trigger thresholds and hence improve the signal efficiency and sensitivity of these analyses. A similar approach has been investigated for the study of boosted ${\PH \to \PQb \PAQb}$ decays, as reported in Section~\ref{subsec:run3jets}. As demonstrated in Section~\ref{sec:jet-reco-performance}, the use of jet substructure techniques is now established in data scouting and could be also applied to the case of boosted dijet resonance searches to reconstruct the jet mass and identify the two-prong jet substructure.

\paragraph{Multijet resonances}
\label{sec:MultiSearches}~

Scouting data is ideal for new physics searches in regions of parameter space dominated by large backgrounds, such as resonances decaying to multiple jets. New physics signatures with multijet final states can be produced in several ways, including from the decay of new colored particles. Relevant SUSY models include RPV squarks and gluinos, which can produce paired dijets and paired jet triplets, respectively~\cite{Barbier:2004ez}. For low masses, the partons arising from the decays of these squarks and gluinos can merge into a single jet. To probe such low masses, scouting analyses utilize jet substructure techniques~\cite{Thaler:2010tr,Moult:2016cvt} . Here we highlight the results of searches for pairs of two- and three-parton resonances~\cite{EXO-21-004}, interpreted as RPV squarks and RPV gluinos, respectively. The analysis relies on a data set collected with the PF scouting triggers, which require thresholds as low as ${\HT > 410\GeV}$ at the HLT. The data set also stores relevant jet substructure variables, allowing us to probe low-mass resonances in which the partons merge into single jets, as discussed in Section~\ref{subsec:jets}.

The investigation of the multijet phenomena follows three different paths within the analysis:
pairs of large-radius jets with substructure consistent with three underlying quarks ({\em merged trijets}), pairs of large-radius jets with two-quark substructure ({\em merged dijets}), and pairs of well-resolved triplets of jets ({\em resolved trijets}).
Given the characteristic decay of RPV gluinos into three quarks in the final state, high-mass gluino pair production is studied with resolved triplets of jets while gluinos of lower mass are studied with dijet events wherein each jet exhibits substructure indicative of the 
merging of three partons into a single large jet. The RPV squarks undergo decay into pairs of quarks. These decays manifest as events with two jets, where the substructure of each jet aligns with the fusion of two partons. The merged dijet study centers on squark pair production scenarios characterized by masses below 200\GeV.  The resolved trijet analysis exhibits sensitivity to RPV gluinos across the mass spectrum of 200--2000\GeV. In contrast, the merged trijet analysis leverages the \nsubjet jet substructure variable introduced in Section~\ref{subsec:jets}, formulated utilizing the designed decorrelated tagger (DDT) technique~\cite{DDT}. With this analysis, the search sensitivity is extended to resonance masses as low as 70\GeV.

\begin{figure*}[!htb]
    \centering
    \includegraphics[width=0.99\textwidth]{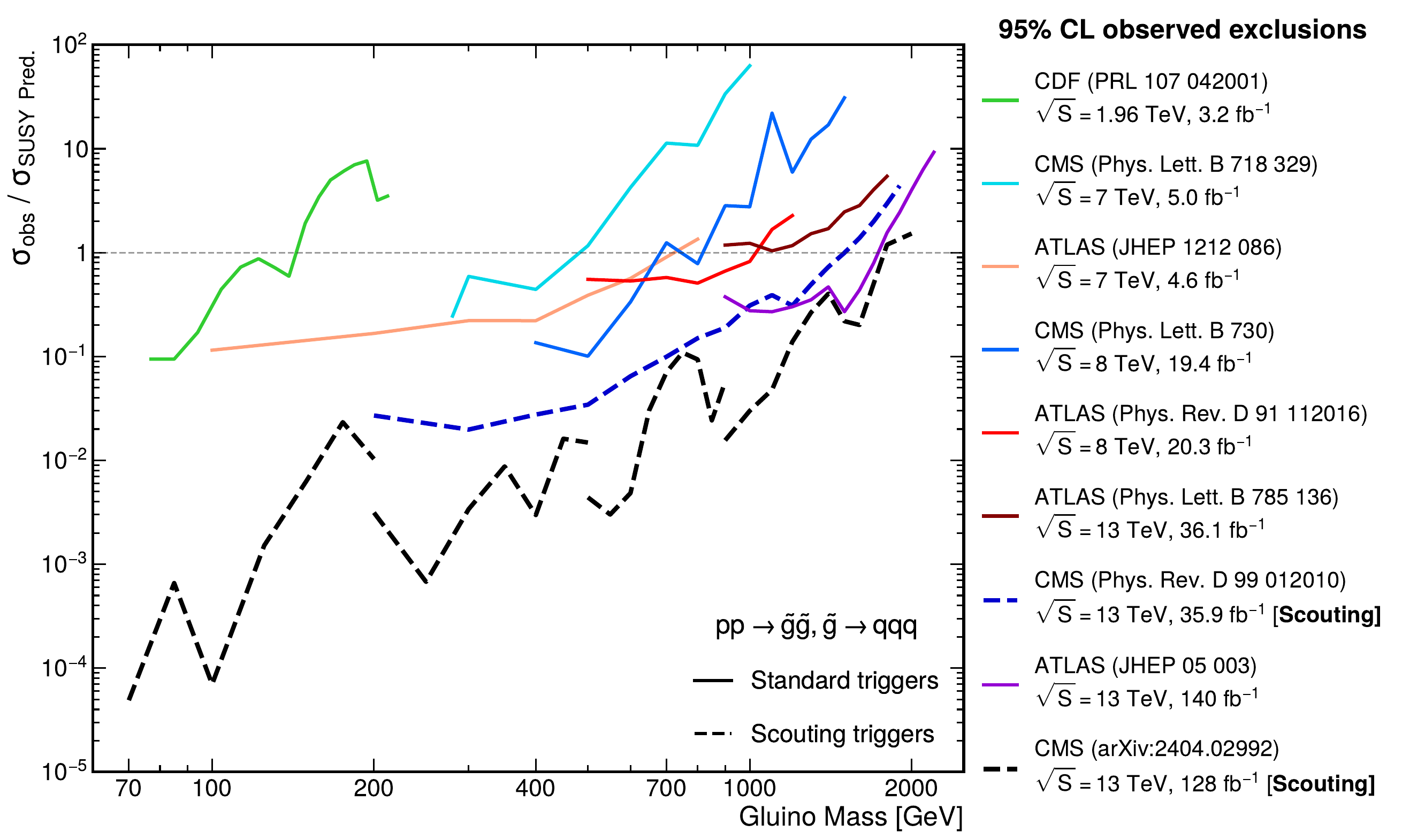}
    \caption{Comparison of limits from searches for RPV gluinos decaying to three partons. Regions above the lines are excluded at 95\%~\CL. The two CMS analyses that use data scouting are also indicated with bold-dashed lines.} 
    \label{fig:multijet_limit_comparison}
\end{figure*}

For the resolved trijet analysis, all pairs of jet triplets are analyzed using kinematic variables that differentiate between multijet backgrounds and signal triplet pairs. The jets within the triplets are subjected to QGD methods, as described in Section~\ref{sec:jet-reco-performance}.  Figure~\ref{fig:top_quark} shows the invariant three-jet mass for triplets that pass all selection criteria. In that figure, we show the lowest mass range used in the analysis, as well as the jet mass distribution for the low mass gluino search where the three partons merge into a single jet. The top quark mass is clearly discernible in both distributions. The sensitivity to gluinos in searches using scouting data is better than searches performed by the CDF Experiment at the Tevatron~\cite{cdfmultijets} and other searches by LHC experiments~\cite{atlas8, atlas13,gluino2011,gluino2012,gluino2017}, achieving both a lower mass reach and lower cross section limits as shown in Fig.~\ref{fig:multijet_limit_comparison}.

\begin{figure*}[!htb]
    \centering
    \includegraphics[width=0.55\textwidth]{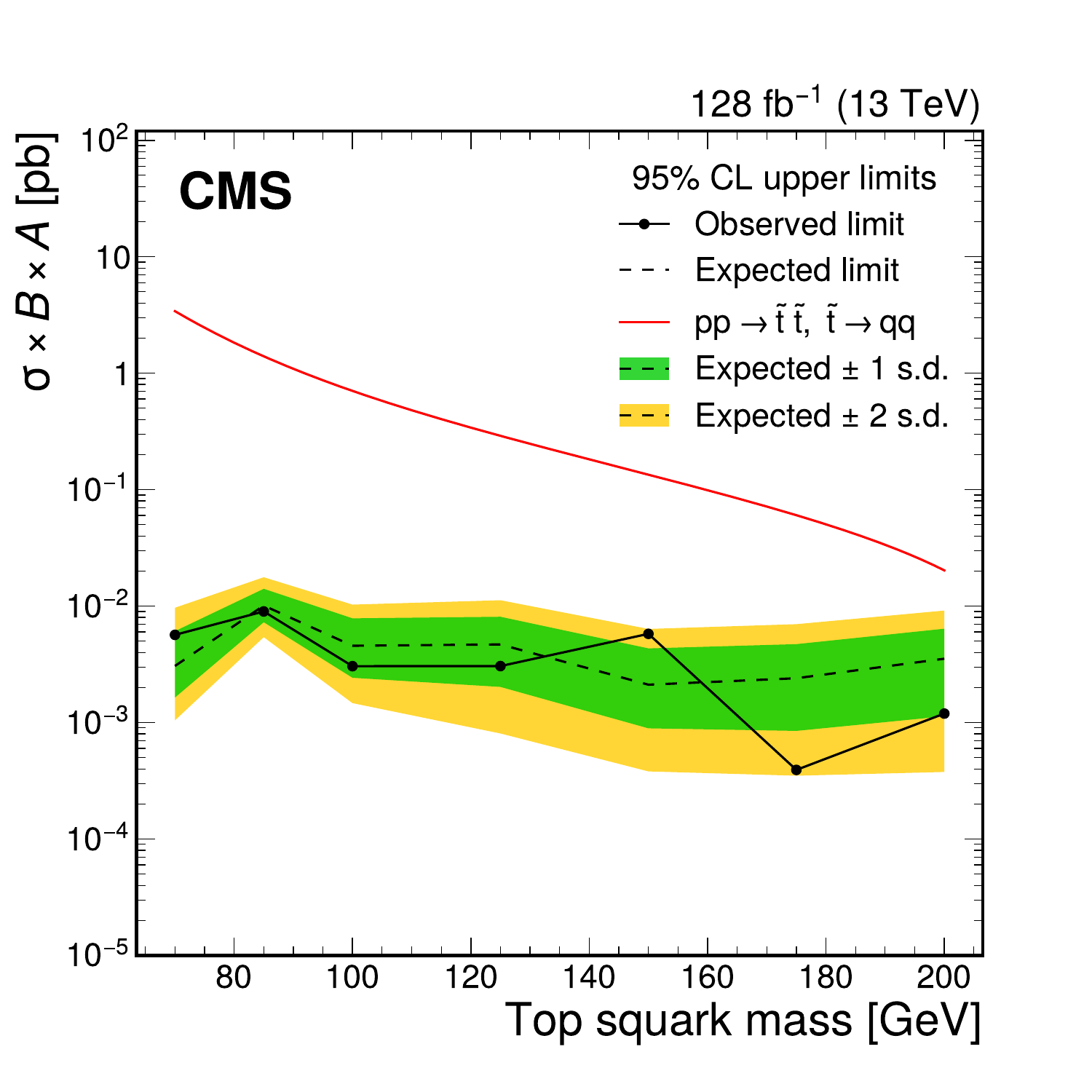}
    \caption{Observed (points) and expected (dashes) limits on the product of production cross section, branching fraction, and acceptance for pair-produced merged two-quark resonances. The variations at the one and two standard deviation levels in the expected limits are displayed with shaded bands. A comparison with the theoretical predictions for top squark production (red) is also shown. Figure taken from Ref.~\cite{EXO-21-004}.}
    \label{fig:rpv_squarks_limit}
\end{figure*}

Searches for final states consisting of pairs of two merged partons were previously performed using standard CMS triggers~\cite{cmsstop13,CMS:2022usq}, with limits of roughly 500\GeV set on RPV squarks. Limits on the production cross section of RPV squarks with the scouting data are shown in Fig.~\ref{fig:rpv_squarks_limit}. Using data scouting and $N_{2,\mathrm{DDT}}$ jet substructure techniques, we have extended the sensitivity to RPV squark masses down to 70\GeV.

In summary, with the scouting technique CMS has achieved unprecedented sensitivity to hadronic resonances with low masses. In the case of new particles decaying to three partons, we are sensitive to weak production (Higgsino) cross sections. 

\subsubsection{Searches in muon final states}

Searches for resonant pair production in dilepton final states played a crucial role in the development of the SM, leading to, \eg, the discovery of the charm and bottom quarks via ${\PJGy \to \PGm\PGm}$ and ${\PGU \to \PGm\PGm}$ decays, respectively, and of the \PZ boson via ${\PZ \to \Pe\Pe/\PGm\PGm}$ decays. Today the same approach is used to investigate still unexplored regions of phase space and to look for new particles in dilepton mass spectra. Here we present searches for the production of dimuon resonances in the mass range below 200\GeV. Both scenarios where the resonance decays promptly or displaced from the interaction point are considered.

\paragraph{Prompt dimuon resonances}
\label{par:promptDimuonResonances}~

The availability of the scouting data set makes it possible to focus on searches for prompt dimuon resonances in the mass range below 40\GeV. Several sources of cosmological evidence point to the existence of a hidden sector, an idea that is imperative to investigate. The hypercharge portal is one of three fully renormalizable portals between the SM and a hidden sector. It features a spontaneously broken dark gauge symmetry $U(1)_D$, which is mediated by a new vector boson called the dark photon, \PZD. The dark photon interacts with the SM via its kinetic mixing with the hypercharge gauge boson, and that mixing is controlled by the kinetic mixing coefficient $\epsilon$. If there are no other hidden-sector states below the \PZD mass, this mixing causes the dark photon to decay exclusively to SM particles, with a sizable branching fraction to leptons. Given the extraordinary capabilities of CMS to reconstruct and identify muons down to a \pt of just a few \GeVns, searching for dark photons in the dimuon channel is a clear natural target.

Two searches were performed using the full muon scouting data set collected during Run~2, corresponding to ${\Lint = 96.6\fbinv}$. The mass spectrum up to 200\GeV was scanned to search for a narrow resonance with subsequent prompt decay to a pair of oppositely charged muons. Different strategies were adopted according to the mass window. The regions around the known resonances, namely the \PJGy,  \Pgy, \PgUa, and \PZ, were excluded because of the difficulty in looking for a new particle in the vicinity of existing resonances with the same final-state signature.

The first search~\cite{EXO-19-018} investigated the resonance mass ranges of 45--75 and 110--200\GeV by exploiting conventional trigger paths and event reconstruction techniques. The coverage was extended down to 11.5\GeV by exploiting the scouting triggers. The dimuon mass resolution depends strongly on the pseudorapidity of the muons. The \pt resolution of muons with ${\pt < 50\GeV}$ is around 1\% in the barrel region of the detector and 3\% in the endcaps. Therefore, events are divided in two categories based on the pseudorapidity of the muons. In the search performed with the scouting triggers, events are required to contain two muons of opposite charge  that are consistent with same-vertex production. The muons are required to be well isolated and to pass selection requirements based on the track quality information available in the scouting event content. To suppress sources of background involving muons originating from heavy ﬂavor decays that typically have low \pt, the muons with the largest and second largest \pt are required to have ${\pt > \mMM/3}$ and ${\pt > \mMM/4}$, respectively.

\begin{figure*}[!hbt]
    \centering
    \includegraphics[width=0.65\textwidth]{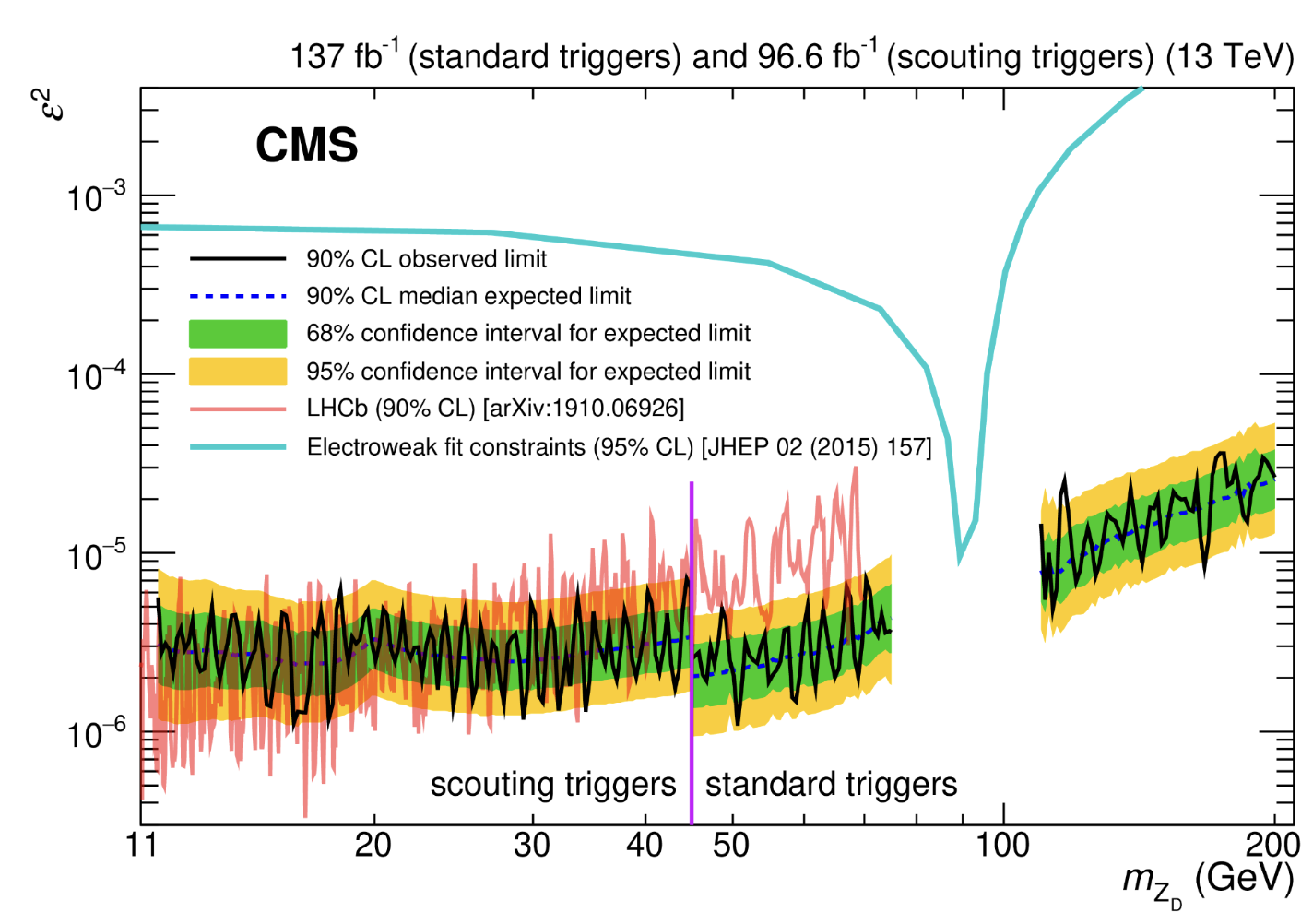}
    \caption{Expected and observed upper limits at 90\%~\CL on the square of the kinetic mixing coefficient ($\epsilon^2$) as a function of dark photon mass. Results obtained with scouting triggers are displayed to the left of the vertical purple line, while those obtained with standard triggers are shown to the right. Limits at 90\%~\CL obtained from the search performed by the LHCb Collaboration~\cite{EXO-19-018_LHCb_PhysRevLett.124.041801} are shown in red, and constraints at 95\%~\CL from the measurements of electroweak observables are shown in light blue~\cite{EXO-19-018_EW_Curtin:2014cca}. Figure taken from Ref.~\cite{EXO-19-018}.}
    \label{fig:exo-19-018_limit}
\end{figure*}

The data are found to be consistent with the background prediction. The results of this search are interpreted in the context of the dark photon model introduced earlier. Upper limits are provided at 95\%~\CL on the product of the signal cross section, branching fraction to a pair of muons, and kinematic and geometrical CMS acceptance of a narrow resonance. Moreover, expected and observed upper limits at 90\%~\CL on $\epsilon^2$ as a function of \PZD mass are obtained and compared with the existing results by the LHCb Collaboration~\cite{EXO-19-018_LHCb_PhysRevLett.124.041801}, as shown in Fig.~\ref{fig:exo-19-018_limit}. The search using scouting data sets stringent constraints on dark photon production in the 11--45\GeV mass range.

\begin{figure*}[!hbt]
    \centering
    \includegraphics[width=0.7\textwidth]{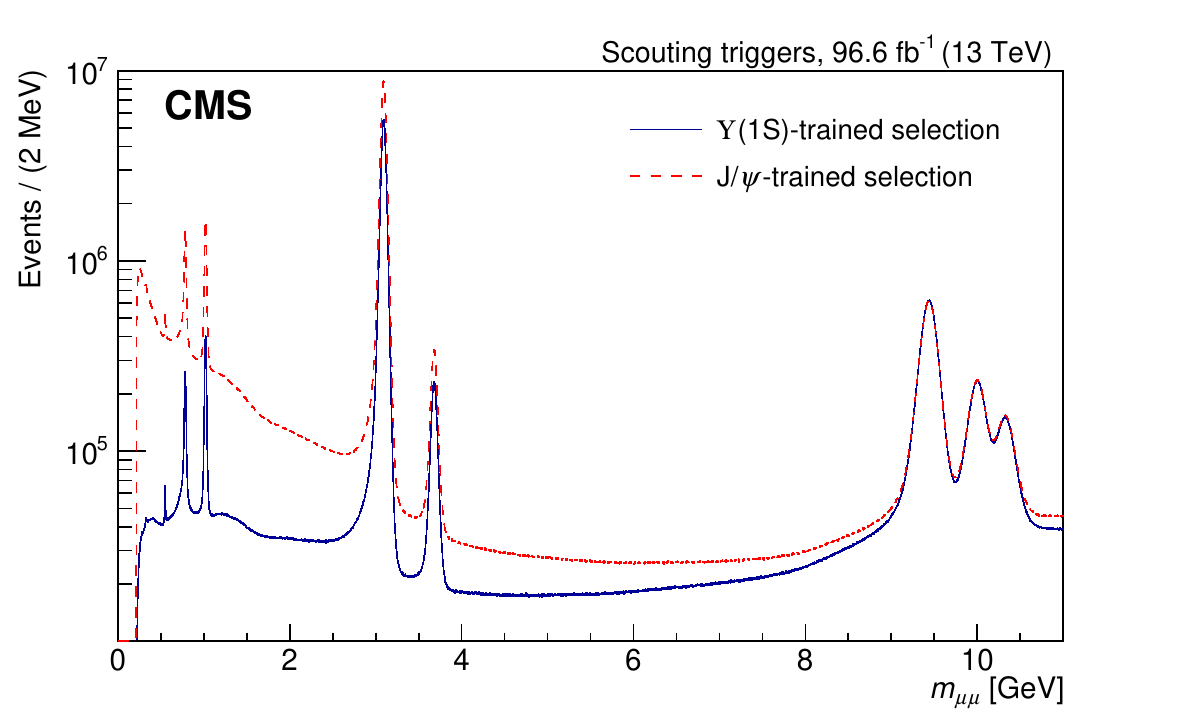}
    \caption{The \mMM distribution obtained with the scouting data collected during 2017 and 2018 with two sets of selections: the \PJGy-trained (red) and the \PgUa-trained (blue) MVA-based muon identification algorithms. Figure taken from Ref.~\cite{EXO-21-005}.}
    \label{fig:exo-21-005_ID}
\end{figure*}

The second search~\cite{EXO-21-005} is an extension of the first, focused on the mass window below the \PGU resonance peak, in the 1.1--2.6 and 4.2--7.9\GeV mass ranges. The region around the \PJGy peak was excluded. A dedicated MVA muon identification technique trained on control samples in data was used to enhance the sensitivity to this very low mass region. This strategy allowed the optimization of the selection of a promptly produced dimuon resonance while minimizing the rate of muon misidentification. Details are provided in Section~\ref{subsubsec:mu_recoPerformance}. The training of the algorithm was performed on \PJGy and \PgUa events for the lower and higher mass windows, respectively. The algorithm trained on \PJGy events recovered the selection efficiency in the very low mass region, compared to the one trained on \PGU events, as demonstrated by the dimuon spectra obtained with the two different selections in Fig.~\ref{fig:exo-21-005_ID}. 

Event candidates are selected by requiring at least one interaction vertex, as reconstructed by the HLT system, and a pair of oppositely charged muons originating from this vertex. A dedicated mass-dependent vertex displacement criterion is applied to focus on promptly produced dimuon resonances. Two selections are defined: an inclusive one and a high-\pt one, focusing on Drell--Yan and gluon fusion production, respectively. The two signal categories, referred to as inclusive and boosted, are used to obtain an interpretation of the results in the context of two specific models: a minimal dark photon model (as in the previous search), and a model with two Higgs doublets as well as an extra complex scalar singlet (2HDM+S). The inclusive selection requires muon ${\pt > 4\GeV}$ in the pseudorapidity region ${\abs{\eta}<1.9}$, while the boosted selection requires muon ${\pt > 5\GeV}$ and a dimuon \pt larger than 35 (20)\GeV in the mass region below (above) 4\GeV.

\begin{figure*}[!hbt]
    \centering
    \includegraphics[width=0.99\textwidth]{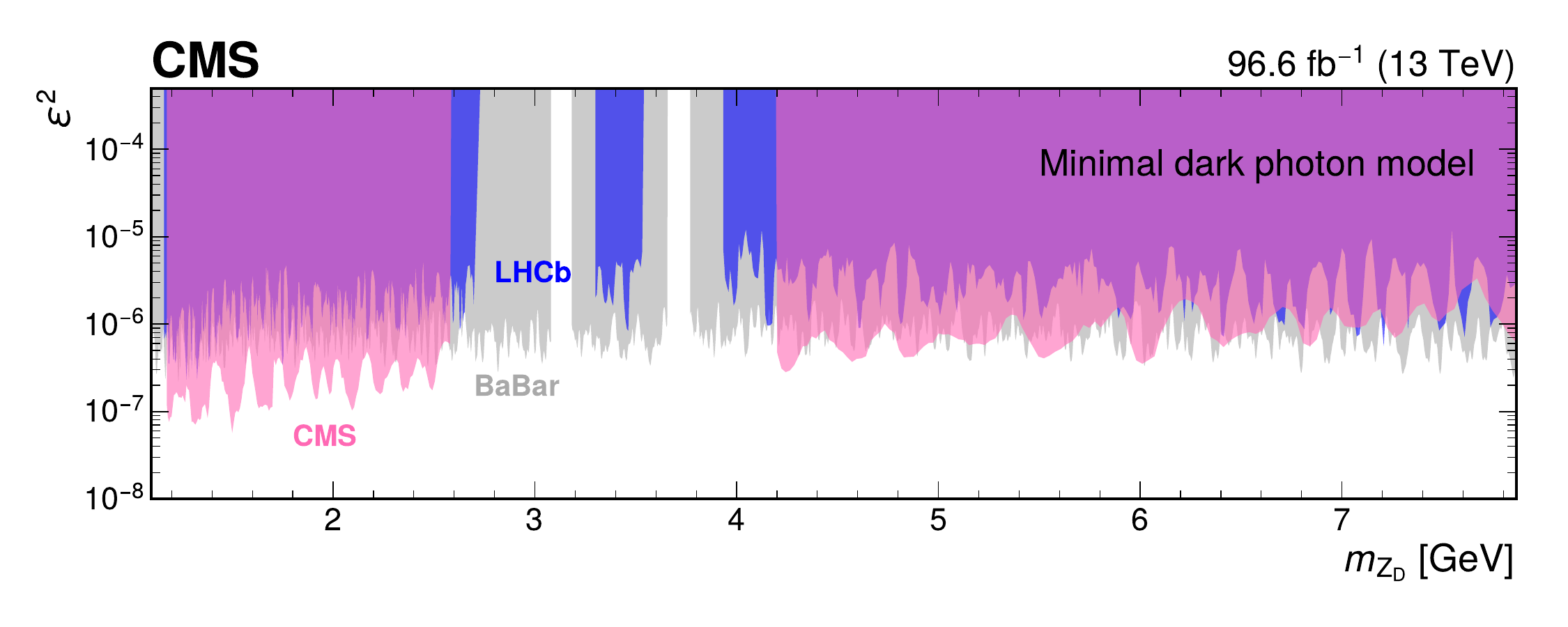}
    \caption{Upper limits at 90\%~\CL on the square of the kinetic mixing coefficient ($\epsilon^2$) in the minimal dark photon model obtained as a recast of model-independent limits on the production rates of dimuon resonances for the inclusive category. The CMS limits (pink) are compared with the existing limits at 90\%~CL provided by LHCb~\cite{EXO-21-005_LHCb:2020ysn} (blue) and BaBar~\cite{BaBar:2012wey} (gray). 
    In the CMS analysis, the background-model fit of the mass distribution becomes unreliable when the tails of \PJGy and \Pgy resonances enter the fit mass window, so the mass range 2.6–4.2 \GeV is excluded from the search.
    Figure taken from Ref.~\cite{EXO-21-005}.}
    \label{fig:exo-21-005_interpretation}
\end{figure*}

The signal is extracted from maximum likelihood fits to the \mMM distribution in data corresponding to selected events. The fit relies on a signal-plus-background model under the assumption that the natural width of the new resonance is much smaller than the detector dimuon mass resolution. Various empirical functions are investigated and used to model the background shape and to estimate the associated systematic uncertainties. For each mass hypothesis, the fit is performed over a mass window spanning $\pm5$ or $\pm8$ times the mass resolution around the hypothesized resonance mass for the inclusive and boosted scenario, respectively.
Model-independent limits on the product of the resonance production cross section, branching fraction to muons, and geometrical acceptance are computed. 
The results are also interpreted as constraints on the parameters of the two models introduced earlier. The model-specific limits rely on the theoretical calculation of cross sections and branching fractions, and on the experimental acceptance derived with simulation.
The sensitivity to the kinetic mixing coefficient $\epsilon$ is significantly improved or comparable to the one obtained by the LHCb~\cite{EXO-21-005_LHCb:2020ysn} and BaBar~\cite{BaBar:2012wey} Collaborations, as shown in Fig.~\ref{fig:exo-21-005_interpretation}. 

\paragraph{Displaced dimuon resonances}
\label{paragraph:exo-20-014}~

The scouting triggers do not require muons to be associated with the reconstructed PV, as described in Section~\ref{subsec:ScoutingRun2MuonsTriggers}. Therefore, the muon scouting data stream can also be used to search for displaced muon signatures. A search for narrow, long-lived dimuon resonances~\cite{EXO-20-014} was performed based on dimuon data collected with the CMS experiment during the LHC Run~2 in 2017 and 2018 using the dimuon scouting trigger stream, with muon ${\pt>3\GeV}$ and ${\abs{\eta}<2.4}$, and probing resonance masses down to ${\approx}2 m_{\PGm}$. The selected data correspond to ${\Lint = 101\fbinv}$.

\begin{figure*}[!htb]
\centering
\includegraphics[width=0.4\textwidth]{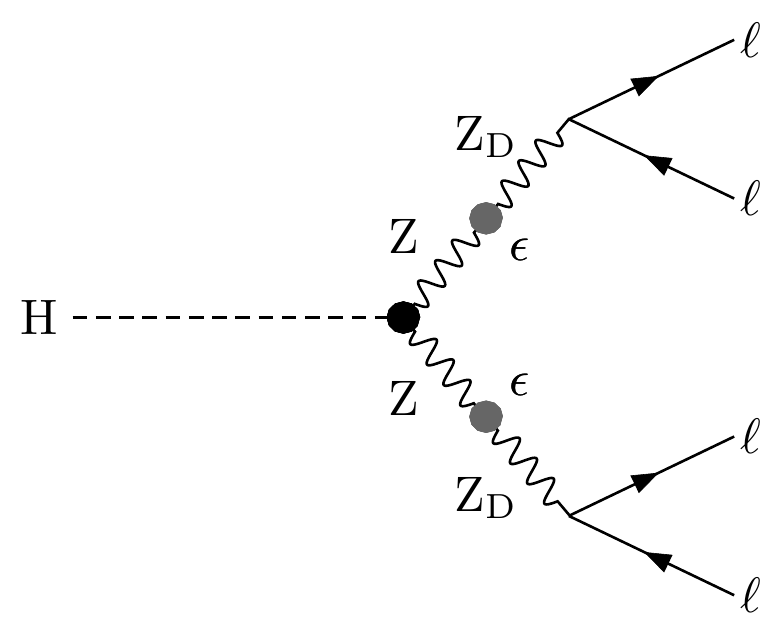}
\hspace{0.1\textwidth}
\includegraphics[width=0.3\textwidth]{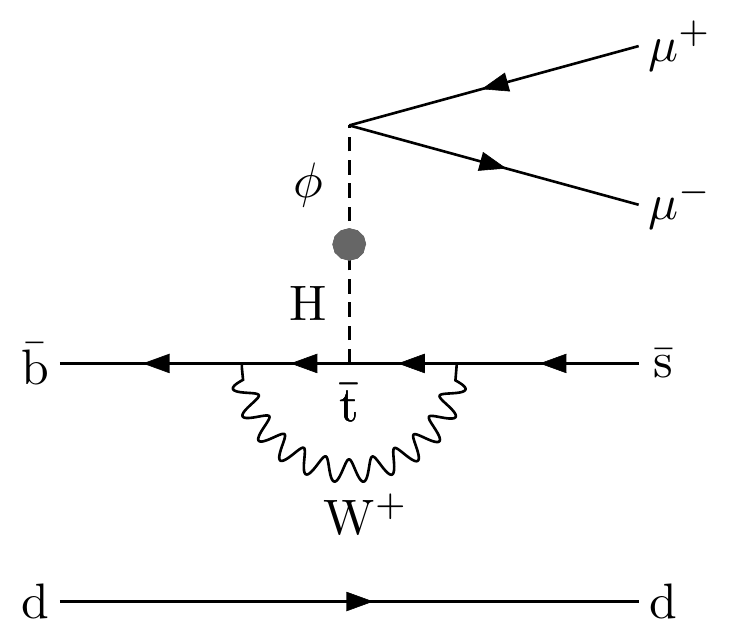}
\caption{Left: diagram illustrating an SM-like Higgs boson (\PH) decay to four leptons ($\ell$) via two intermediate dark photons (\PZD). Right: diagram illustrating the production of a scalar resonance \PGf in a \PQb hadron decay, through mixing with an SM-like Higgs boson. Figure taken from Ref.~\cite{EXO-20-014}.
}
\label{fig:EXO-20-014_feyn}
\end{figure*}

The search targets narrow, low-mass, long-lived resonances decaying into a pair of oppositely charged muons, where the lifetime of the long-lived particle is such that the transverse displacement \lxy of its decay vertex is within 11\cm of the PV. As mentioned in Section~\ref{subsec:ScoutingRun2MuonsTriggers}, this constraint arises from the requirement for muon tracks to leave a hit in at least two layers or disks of the pixel tracker. The selected muons are used in pairs to form dimuon vertices, considering all possible pairs. These vertices are hereafter referred to as SVs, and they may or may not be displaced from the PV. The signal is expected to appear as a narrow peak on top of the dimuon mass continuum, with an intrinsic resonance width smaller than the experimental mass resolution. Such a signal is predicted in BSM frameworks with the Higgs boson decaying into a pair of long-lived dark photons, as shown in Fig.~\ref{fig:EXO-20-014_feyn} (left), or with a long-lived scalar resonance arising from a decay of a \PQb hadron, as shown in Fig.~\ref{fig:EXO-20-014_feyn} (right).

Events are required to contain at least one pair of opposite-sign muons associated with a selected SV. The events that contain a single muon pair are then categorized according to the decay length \lxy, the \pt of the muon pair (${\ptMM < 25}$ and ${\ptMM > 25\GeV}$), and isolation (to distinguish isolated, partially isolated, and nonisolated topologies).
The \lxy categorization is intended to maximize the search sensitivity to a range of potential BSM signal lifetimes, and is based on the CMS pixel tracker geometry. 
The categorization in \ptMM, on the other hand, improves the sensitivity to signal models with different production modes and boost distributions.

In each (\lxy, \ptMM, isolation) bin, we define mass windows sliding along the dimuon invariant mass spectrum, and perform a search for a resonant peak in each mass window. The size of the sliding windows is set to ${\pm5\,\sigma_{\PGm\PGm}^{\mathrm{mass}}}$ around the signal mass hypothesis, where ${\sigma_{\PGm\PGm}^{\mathrm{mass}}}$ is determined from simulation and equals about 1\% of the mass.
Mass regions corresponding to known resonances decaying either to a pair of muons or to a pair of charged hadrons are not considered, \ie, they are ``masked'' for the purpose of this search.
In events with two muon pairs, each associated with an SV, we further require the difference between the two dimuon masses to be within 5\% of their mean, and the four-muon mass to be consistent with the mass of the SM Higgs boson 
(${115 < \mMMMM < 135\GeV}$). The selected four-muon events are treated as an exclusive independent category, with a single four-muon mass window centered around the known Higgs boson mass (\ie, ${115 < \mMMMM < 135\GeV}$). This additional region is aimed at improving the search sensitivity to models of BSM physics where an SM Higgs boson decays to a pair of \PZD bosons, each decaying to two muons.

In each mass window, the signal is parametrized using the sum of a Gaussian function and a double-sided CB function. The SM background is modeled by means of different functional forms, that include Bernstein polynomials, exponential
functions, and combinations of the two. 
Then, binned maximum likelihood fits to the data are performed simultaneously in all search bins, under either background-only or background-plus-signal hypotheses.

\begin{figure*}[!htb]
\centering
\includegraphics[width=0.9\textwidth]{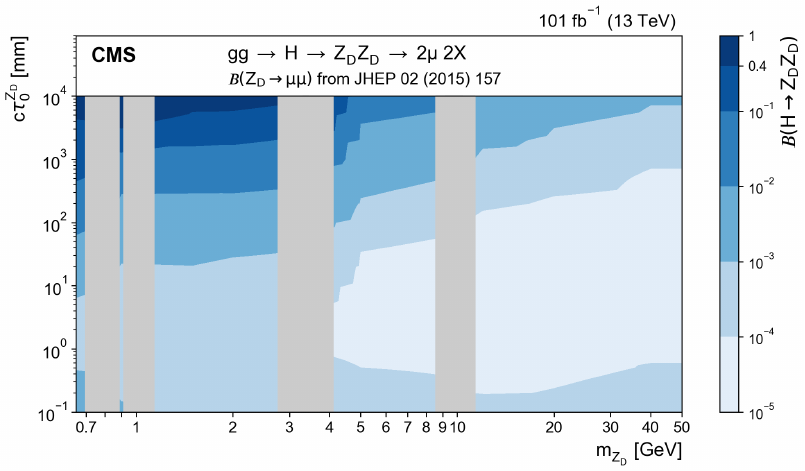}
\includegraphics[width=0.9\textwidth]{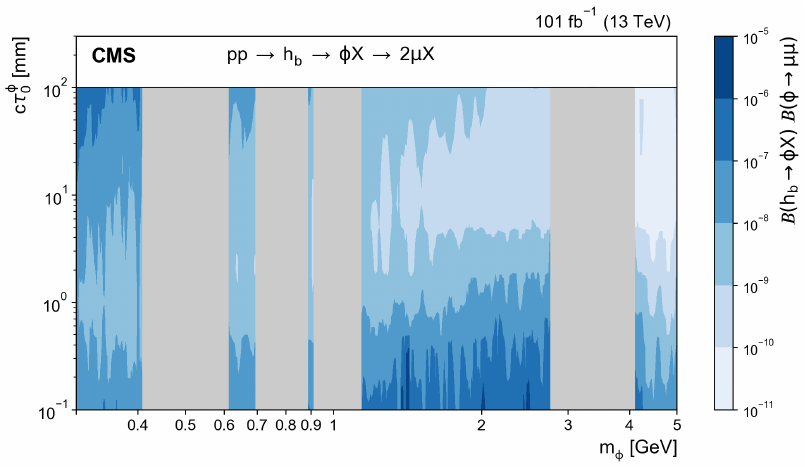}
\caption{
Observed limits at 95\%~\CL on (upper) the branching fraction $\mathcal{B}(\PH \to \PZD\PZD)$ and (lower) the branching fraction product ${\mathcal{B}(\mathrm{h}_{\PQb}\to\PGf \mathrm{X})\cdot\mathcal{B}(\PGf\to\PGm\PGm)}$ as contours in the parameter space containing the signal mass ($m_{\PZD}$ or $m_{\PGf}$, respectively) and the signal lifetime $c\tau_0$. The vertical gray bands indicate mass ranges containing known SM resonances, which are masked for this search. The limits are obtained using the combination of all dimuon and four-muon event categories. Figures taken from Ref.~\cite{EXO-20-014}.
}
\label{fig:EXO-20-014_limits_massVSctau}
\end{figure*}

No significant peak-like structures are observed in data. The background-plus-signal fits are used to set upper limits at 95\%~\CL on a wide range of mass and lifetime hypotheses for models of BSM physics where a Higgs boson decays to a pair of long-lived dark photons, or where a long-lived scalar resonance arises from the decay of a \PQb hadron. These constraints are the most stringent to date for substantial regions of the explored parameter space. For illustrative purposes, the exclusion limits obtained for the ${\PH \to \PZD\PZD}$ model of Ref.~\cite{EXO-19-018_EW_Curtin:2014cca} and for the ${\mathrm{h}_{\PQb}\to\PGf \mathrm{X}}$ inflaton model of Ref.~\cite{Bezrukov:2013fca} are shown in Fig.~\ref{fig:EXO-20-014_limits_massVSctau}. The limits on the inflaton model are more stringent or comparable to the ones obtained by the LHCb Collaboration~\cite{LHCb:2015nkv,LHCb:2016awg} for $m_{\PGf}$ greater than approximately $1\GeV$. The search is mostly sensitive to signatures with a dimuon resonance produced at nonzero displacement from the PV. At large displacement values, the sensitivity is degraded because the transverse displacement must be within the first three layers of the pixel detector, namely ${\lxy < 11\cm}$.

\subsubsection{Observation of the rare \texorpdfstring{$\PGh \to 4\PGm$}{eta4mu} decay channel}
\label{sec:rare_decays}

The scouting data set also enables the study of rare light-meson decays to be considered. The lower muon momentum thresholds considerably expand the mass range of particles that can be probed. The power of the data set to measure rare SM decays was demonstrated by the first observation of the four-muon decay of the \PGh meson. The production rate of the \PGh meson falls quickly \vs \pt, so lowering the muon momentum thresholds is essential to enhance the collection of events involving the \PGh meson.

\begin{figure*}[!hbt]
    \centering
    \includegraphics[width=0.7\textwidth]{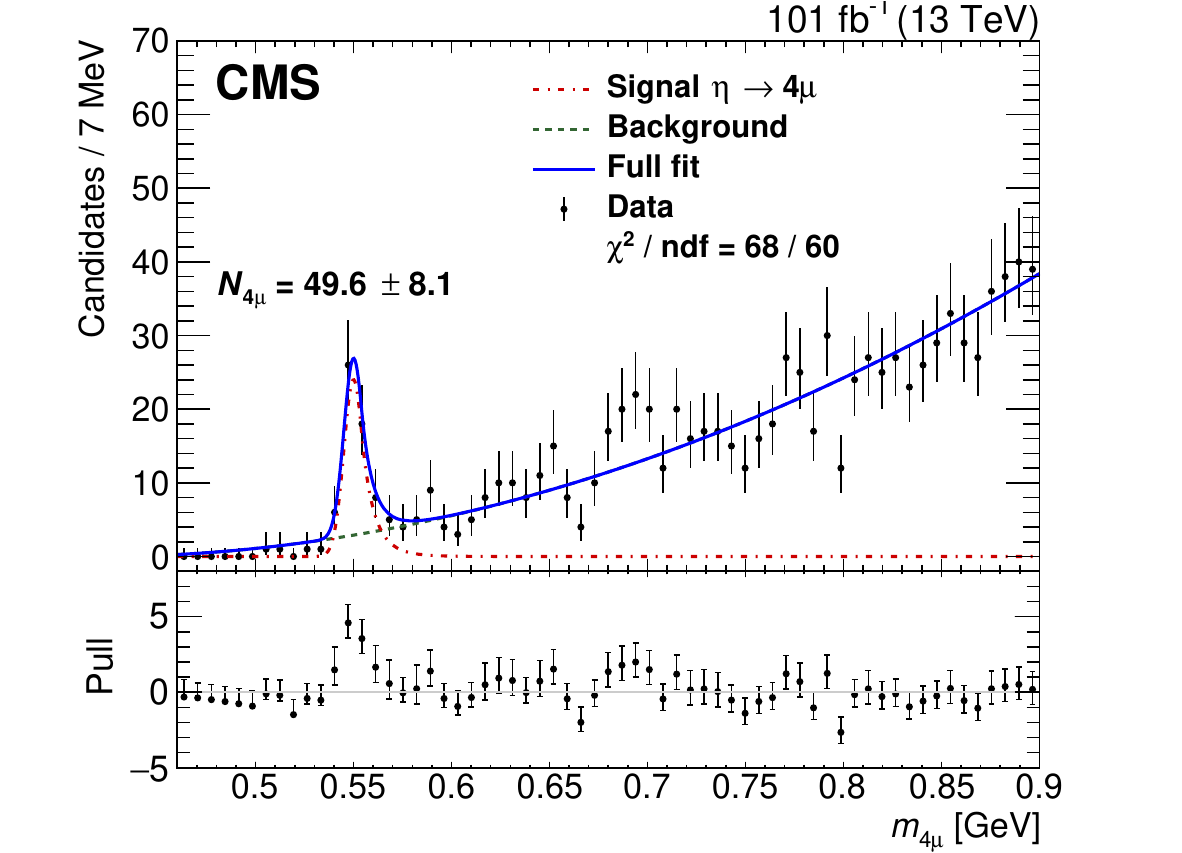}
    \caption{The four-muon invariant mass (\mMMMM) distribution in the range 0.46--0.90\GeV, obtained with pp collision data collected during 2017--2018. The observed distribution (points) is compared to the background-only prediction (green dashed) and to the full background fit including simulations of the signal (solid blue). The peak observed in the mass window 0.53--0.57\GeV corresponds to the \PGh meson. The pull distribution in the lower panel is shown relative to the background component of the fit model and defined as ${(\mathrm{Data} - \mathrm{Fit})/ \mathrm{Uncertainty}}$, where the uncertainty is statistical only. Figure taken from Ref.~\cite{BPH-22-003}.}
    \label{fig:threshold_fit}
\end{figure*}

The ${\PGh \to \PGmp \PGmm \PGmp \PGmm}$ decay was observed by the CMS Collaboration using a four-muon selection in the scouting data set. Figure~\ref{fig:threshold_fit} shows the measured distribution of the four-muon invariant mass, after requiring four muons that are compatible with same-vertex production. About 50 signal events ($N_{4\mu}$) are observed on top of a background of roughly 17 events, corresponding to a statistical significance much greater than 5 standard deviations. 

The branching fraction of the newly observed decay channel was measured relative to the reference channel ${\PGh \to 2\PGm}$, which is known with a precision of about 14\%. The additional ingredients needed for this measurement are the yields of the two-muon reference channel, $N_{2\mu}$, and the products $A$ of the CMS detector geometric acceptance and reconstruction efficiency for both channels, which were determined with simulation. The relative branching fraction is computed via
\begin{equation*}
    \frac{\mathcal{B}_{4\mu}}{\mathcal{B}_{2\mu}} = \frac{N_{4\mu}}{ \sum\limits_{i,j} N^{i,j}_{2\mu} \frac{A^{i,j}_{4\mu}}{A^{i,j}_{2\mu}}},
    \label{eq:BR}
\end{equation*}
where $N_{4\mu}$ is the total four-muon signal yield, and $A_{4\mu}^{i,j}$, $A_{2\mu}^{i,j}$, and $N_{2\mu}^{i,j}$ are the four-muon $A$, two-muon $A$, and two-muon yields in bin $i$ of the candidate \PGh meson \pt and bin $j$ of the \PGh meson rapidity, respectively. We define 32 bins in \pt, in the range 7--70\GeV, and 2 bins in $\abs{y}$.

\begin{figure*}[!hbt]
    \centering
    \includegraphics[width=0.7\textwidth]{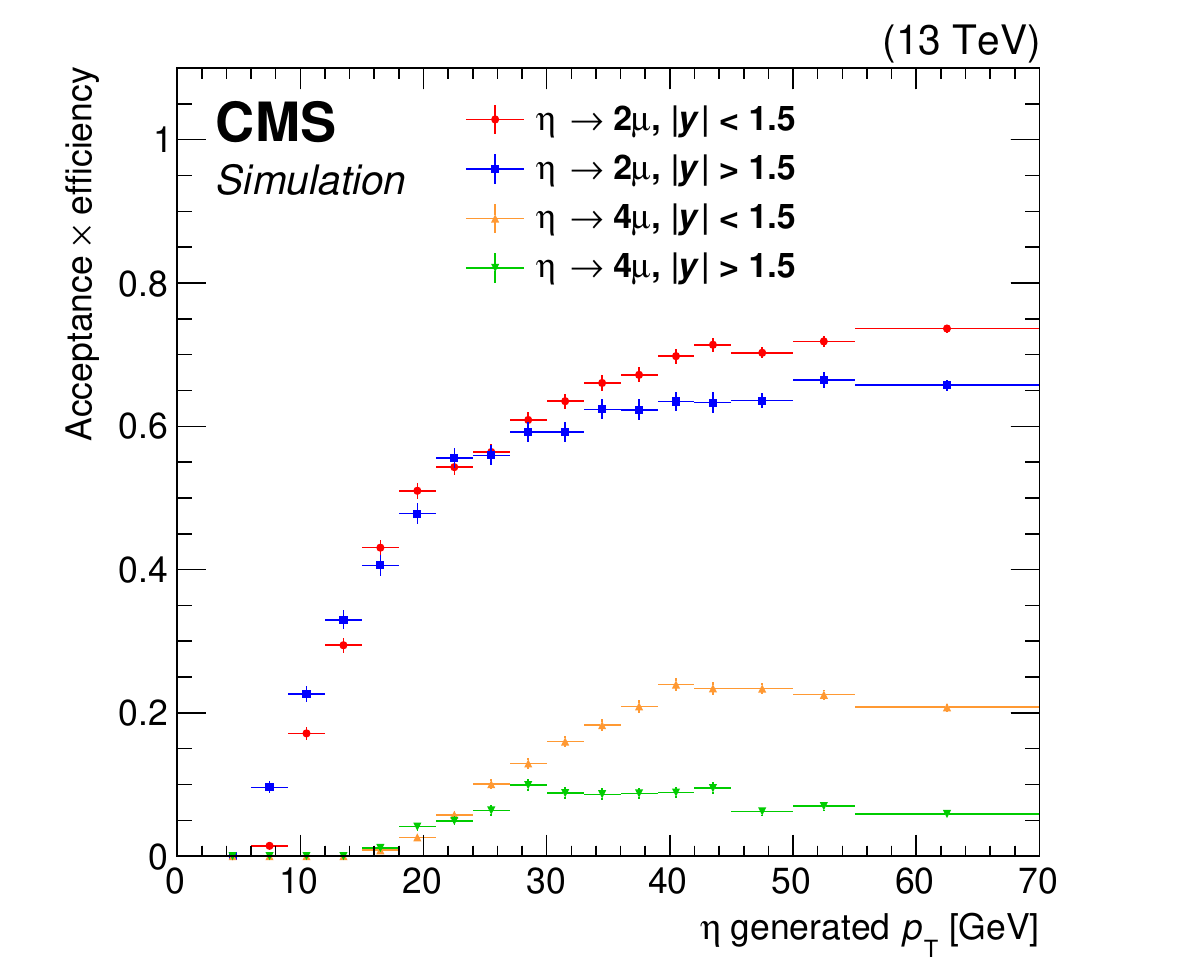}
    \caption{The product of acceptance ($A$) and efficiency as a function of the generated \PGh meson \pt for the two-muon (red circles and blue squares) and four-muon (orange up triangles and green down triangles) \PGh meson decays. The product is evaluated using simulated samples. Figure taken from Ref.~\cite{BPH-22-003}.}
    \label{fig:acceptance_2mu_4mu}
\end{figure*}

Figure~\ref{fig:acceptance_2mu_4mu} shows $A$ for both channels as determined by simulation. The efficiency of the two-muon channel is limited by the trigger efficiency, while that of the four-muon channel is constrained by the efficiency to reconstruct all four signal muons. This is more challenging for higher \PGh meson \pt, since the muons are more collimated, leading to overlapping tracks and decreased reconstruction efficiency.

The measured relative branching fraction is 
\begin{equation*}
\frac{B_{4\mu}}{B_{2\mu}} = [ 0.86 \pm 0.14 \stat \pm 0.12 \syst ] \times 10^{-3},
\end{equation*}
where the systematic and statistical uncertainties are roughly balanced. Using the world average value for the reference branching fraction ${\mathcal{B}(\PGh \to 2\PGm)}$, the absolute branching fraction of the four-muon decay is measured as 
\begin{equation*}
	\mathcal{B}(\PGh \to 4\PGm) = [ 5.0 \pm 0.8 \stat \pm 0.7 \syst \ \pm 0.7 \, (B_{2\mu}) ] \times 10^{-9},
\end{equation*}
where the last term is the uncertainty in the branching fraction of the reference channel. This result is in agreement with theoretical predictions, \eg, ${\mathcal{B}(\PGh \to 4\PGm) = (3.98 \pm 0.15) \times 10^{-9}}$ from Ref.~\cite{escribanoDatadrivenApproachPi2018}, and improves on the precision of previous upper-limit measurements~\cite{Workman:2022ynf} by more than 5 orders of magnitude.

\section{Data scouting in Run 3}
\label{ch:run3scouting}

The CMS data scouting technique was explored, developed, and brought to maturity during Run~1 and Run~2, proving to be an innovative trigger strategy and a successful paradigm for data analysis. During this period, the primary constraint in implementing the scouting strategy was found to be the HLT event processing time. In Run~3, the computing capabilities of the HLT were greatly improved, as described in Ref.~\cite{CMS:2023gfb}. The availability of a new GPU-equipped HLT farm, provided as a way to gain expertise for the next phase of operation of the CMS detector during HL-LHC, and the subsequent improvement of the HLT reconstruction (as detailed in the next section) facilitated a significant broadening of the scouting scope. It is interesting to note that comparable performance based on an HLT farm equipped only with CPUs would require a higher overall cost (by approximately 15\%) and higher power consumption (by about 30\%). This expansion of data scouting is illustrated in Fig.~\ref{fig:hlt_rates_evolution}, where the average rates allocated to the standard, parking, and scouting streams are reported from Run~1 to Run~3.

This chapter outlines the improvements to the data scouting strategy in Run~3, providing insights into the relevant updates with respect to the Run~2 approach described previously in Chapter~\ref{ch:run1run2scouting}. First, the trigger rates and event content of the 2022 and 2023 data-taking periods are discussed in Sections~\ref{subsec:run3scouting}~and~\ref{sec:Run3EventContent}. Then, we investigate the performance of data scouting in the context of jets (Sections~\ref{subsec:run3jets}), muons (Section~\ref{subsec:run3muons}) and electrons and photons (Section~\ref{subsec:run3electronsAndPhotons}). 

\subsection{The new Run~3 scouting strategy at the HLT}
\label{subsec:run3scouting}

Following the hardware upgrade and increased usage of GPUs during Run~3, the HLT algorithms were redesigned to harness the capabilities of parallel architectures. The primary focus was to offload HLT reconstruction steps to GPUs, particularly for complex tasks. As a result, during 2021 and 2022 a new GPU-based approach was developed and fully commissioned for the calorimeter reconstruction, the pixel local reconstruction, and the pixel-based tracking.

Within this new GPU-based paradigm, events are reconstructed with a novel scouting PF algorithm that exploits tracker tracks built solely with pixel hits, using the \textsc{Patatrack} algorithm~\cite{Bocci:2020pmi}. These pixel-only tracks replace the tracks reconstructed with the combined information from the pixel and strip trackers, as done with the standard PF algorithm described earlier in Section~\ref{subsubsec:event reconstruction PF}. The main advantage of the \textsc{Patatrack} pixel-only tracking is the possibility of offloading the track reconstruction to GPUs, thereby notably accelerating event processing at the cost of a slightly worse track resolution compared to standard tracks. The degradation is more significant in high-\pt tracks~\cite{CMS-DP-2021-005}. As low-\pt tracks are most relevant to scouting, this acceleration particularly benefits the scouting strategy. Quality criteria based on the momentum resolution and on the distance in the longitudinal plane from the two leading vertices are applied to the pixel-only tracks before offering them as input to a modified PF algorithm. Vertices are reconstructed using pixel-only tracks and the measured transverse coordinates are computed relative to the online measurement of the interaction point. 

The processing time required for the full reconstruction of scouting events and the application of the selection criteria are shown in Table~\ref{tab:run3_scouting_timing_per_path} for the scouting paths active during 2023. The difference between scenarios with and without outsourcing of certain steps to GPUs is also presented. The significant speed-up provided by the GPUs is clearly seen.  

\begin{table*}[!htb]
    \centering
    \topcaption{Time required for the object reconstruction and selection criteria in the scouting paths seeded by different L1 algorithms, using only a CPU or accelerating certain steps with a GPU. For the comparison, we pick a representative run recorded in 2023. The time needed to run the full HLT menu reconstruction including non-scouting paths is also shown for reference.}
    \renewcommand{\arraystretch}{1.3}
    \begin{tabular}{lcc}
        Scouting path & CPU-only [ms] & CPU+GPU [ms]  \\
        \hline
        1 electron/photon  & 76.0 & 49.5  \\
        ${\geq}2$ electrons/photons  & 9.3 & 6.8  \\
        ${\geq}2$ muons  & 69.0 & 41.6  \\
        Jets or MET  & 83.3 & 52.1  \\
        \hline
        Full HLT menu  & 578.4 & 377.7  \\
    \end{tabular}
    \label{tab:run3_scouting_timing_per_path}
\end{table*}

In the Run~3 scouting strategy, unlike in Run~2, only jets clustered from PF inputs are reconstructed and stored. The reconstruction of PF jets follows the same method employed in Run~2 (see Section~\ref{subsec:jets}), but uses PF candidates reconstructed from pixel-only tracks. All PF candidates are used to calculate the missing transverse momentum. More details on the performance of jet objects and algorithms is presented in Section~\ref{subsec:run3jets}. 

Muon candidates are reconstructed based on information from the silicon tracker and the muon system, as described in Section~\ref{sec:ScoutingRun2Muons}. The reconstruction of scouting muons benefited from the integration of ML-based outside-in and inside-out seeding improvements in the standard muon reconstruction at HLT for Run~3~\cite{CMS:2023gfb}. A relevant difference between the online and offline reconstruction is represented by the removal of the requirement on the minimum number of hits in the pixel layers, which was introduced at the beginning of Run~3. This adjustment enhances the sensitivity to signals with displaced muons in the final state, since, depending on their displacement from the collision region, they may not create many hits in the pixel layers. More details are available in Section~\ref{subsec:run3muons} where we discuss the Run~3 muon performance. 

The electron and photon reconstruction is a novelty in Run~3 scouting, made possible by the newly available resources arising from the offloading of pixel-only tracking to GPUs. The reconstruction of electrons and photons themselves is identical to the standard online reconstruction, as discussed in Chapter~\ref{sec:CMS}. The scouting path runs reconstruction over the full ECAL volume. Since 2023, an HLT preselection is applied, as shown in the third column of Table~\ref{tab:run3_L1_Thr}, in the scouting trigger paths seeded by the L1 \egamma triggers. The preselection reduces the number of events on which the scouting reconstruction is run and frees up resources to enable reconstruction over the full ECAL volume irrespective of the L1 seed. Most other HLT paths reconstruct SCs in a geometrical region matching an L1 ECAL trigger tower. The full ECAL reconstruction results in larger efficiency for low-energy \egamma in the scouting events. The performance of electron and photon reconstruction is further discussed in Section~\ref{subsec:run3electronsAndPhotons}.

\begin{figure*}[!hbt]
  \centering
  \includegraphics[width=0.6\textwidth]{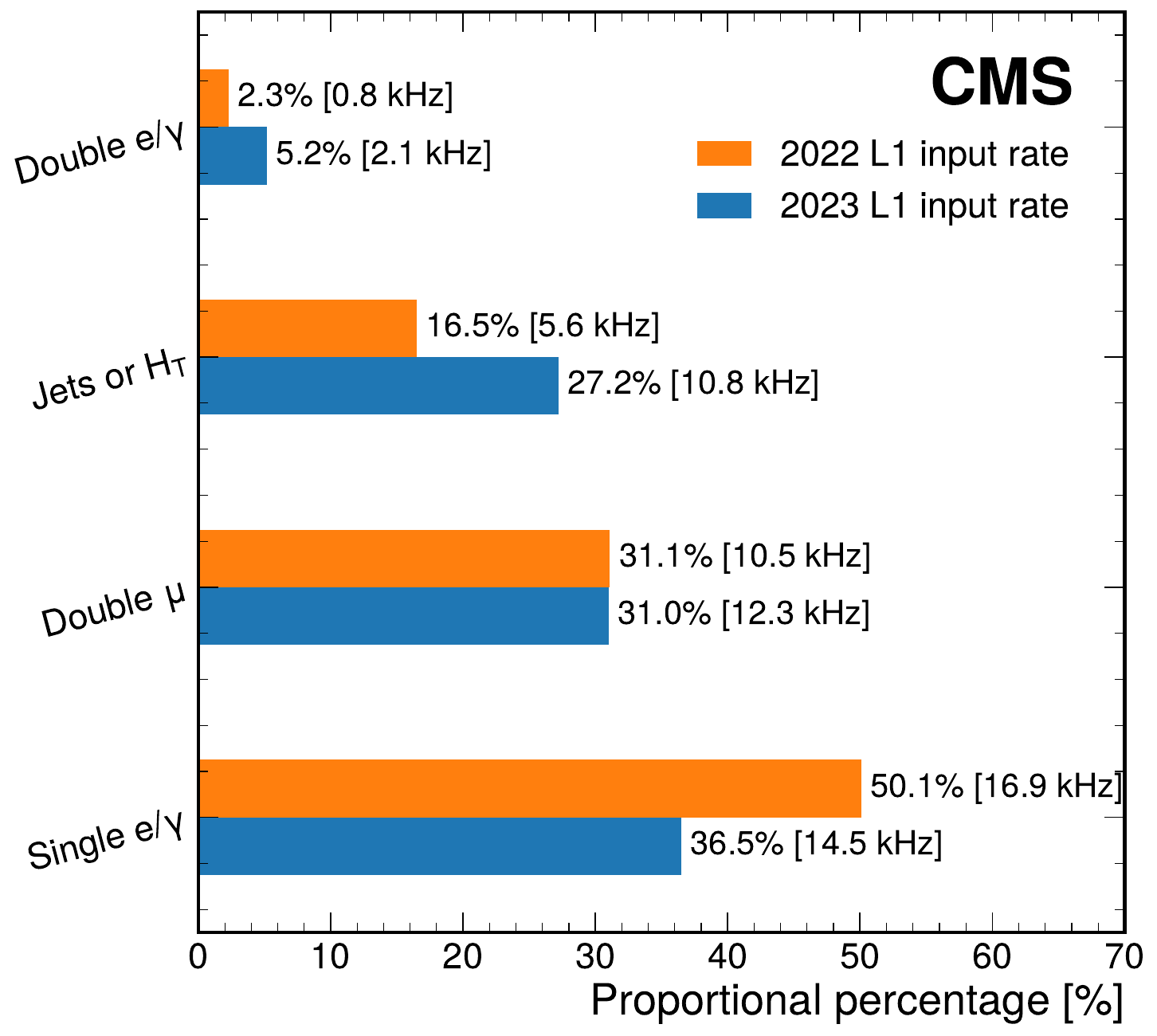}
  \caption{Relative rate of each L1 algorithm category, shown as the fraction of the total rate based on the 2022 (orange) and 2023 (blue) configurations. The proportional rate of each category with respect to the total is shown on the right. The values are computed using reference runs with average pileup of 60.}
  \label{fig:l1_rates}
\end{figure*}

The substantial improvement in reconstruction speed enables an expansion of the maximum L1 event rate that can be processed by the scouting stream and thereby a deeper exploration of physics processes at lower masses and weaker couplings. For events to be reconstructed in the scouting data stream, they must be selected by one of several L1 seeds targeting signatures such as one or two photons or electrons, muons with low \pt, one or two jets, or a moderate amount of \HT. 
The scouting streams in Run~3 sustained an L1 input rate of approximately 30\unit{kHz} for high pileup data-taking scenarios. Figure~\ref{fig:l1_rates} illustrates the fractional rate relative to the scouting L1 input rate of each L1 category for two reference runs in 2022 and 2023, with average pileup of 60. The list of L1 seeds used in Run~3 is summarized in Table~\ref{tab:run3_L1_Thr}. The most notable change with respect to the list of L1 seeds previously reported for Run~2 in Section~\ref{sec:ScoutingRun2Details} is the inclusion of dedicated algorithms targeting events with one or two electrons or photons. The single \egamma trigger has been included to target single-photon signatures that exploit the notable lowering of the photon \pt threshold in scouting. The list of algorithms for muons and jets/\HT remained largely unchanged, except for the temporary removal of the lowest threshold dimuon seed in 2022 (which was added back in 2023) and the lowering of the dijet invariant mass and \HT requirements in 2023. The latter is evident in the increased proportional rate of the hadronic category. The rate variation of the \egamma categories is due to changes in the online data-taking conditions at L1 and not due to updates of the algorithms.

As discussed in Section~\ref{ch:run1run2scouting}, due to the limited computational resources in Run~2, the scouting data was split into two separate streams: PF scouting, which involved the computationally intensive PF reconstruction, and Calo scouting, which used fewer resources. But with the increased computing power available in Run~3, a transition was made to a single scouting data stream where the PF reconstruction is run for all events passing one of the input L1 algorithms. After reconstruction, a minimal event selection is applied before storing events on disk. Therefore, the HLT scouting data stream achieved a data output rate of approximately 26\unit{kHz} for high pileup data-taking scenarios during the beginning of Run~3.

In 2023, optimized selections were added to improve the purity of certain final-state topologies. These selections vary based on the L1 algorithm seed used to collect a given event. The selection criteria are listed in Table~\ref{tab:run3_L1_Thr}. No further selection was applied to scouting events seeded by the jet and \HT algorithms. In the dimuon path, events were only stored if they contained at least two scouting muons, each with $\pt>3\GeV$. Events seeded by the single \egamma path required an \egamma candidate with ${\pt>30\GeV}$, and the L1 double \egamma path required at least two \egamma candidates, each with ${\pt>12\GeV}$.

\begin{table*}[!htb]
    \centering
    \topcaption{List of L1 and HLT thresholds for the lowest unprescaled scouting triggers during Run~3. The list corresponds to the 2022 and 2023 data-taking periods. Conditions that changed between the two years are annotated in bold face. The four separate HLT paths and corresponding thresholds were only present in 2023. No thresholds were applied at the HLT in 2022. When the same threshold is applied to all selected objects in an event, a single number is shown. If different thresholds are applied for each object, they are shown separately, from highest to lowest threshold. The notation ``OS'' stands for opposite-sign muon pairs and ``SC'' for calorimeter superclusters. }
    \renewcommand{\arraystretch}{1.3}
    \begin{tabular}{lll}
            Type & L1 threshold  & HLT threshold (\textbf{2023})\\
            \hline
            \multirow{2}{*}{\egamma}  & 1 \egamma, $\pt>30\GeV$, $\abs{\eta}<2.1$ & 1 SC (loose), $\pt > 30\GeV$ \\ 
            &  2 \egamma, $\pt>18/12\GeV$, $\abs{\eta}<1.5$  & 2 SC (loose), $\pt > 12\GeV$ \\
            \hline
            \multirow{5}{*}{$\mu$} &  $2\mu$, $\pt>15/7\GeV$  &  \multirow{5}{*}{
            $\left\}\begin{array}{l} \\  \\ 2\mu, \pt > 3\GeV \\ \\ \\  \end{array}\right.$} \\
            & $2\mu$, OS, $\pt > 4.5\GeV$, $\abs{\eta}<2$, $\mMM > 7\GeV$ &  \\	
            & $2\mu$, OS, $\pt > 4\GeV$, $\abs{\eta}<2.5$, $\Delta R<1.2$ &  \\
            & $2\mu$, OS, $\pt > 0\GeV$, $\abs{\eta}<1.5$, $\Delta R<1.4$ (\textbf{2023}) &  \\
            & $3\mu$, $\pt > 5/3/3\GeV$ &  \\
            \hline
            \multirow{3}{*}{Jets/\HT} &  $\HT> 280~(\textbf{2023}),~360~(\textbf{2022})\GeV$  \\ 
            & 1 jet, $\pt>180\GeV$   \\
            & 2 jets, $\pt >30\GeV$, $\abs{\eta}<2.5$, $\Delta \eta<1.5$, & \\ 
            & \hspace{9.5mm} $\mjj>250 ~(\textbf{2023}),~300~(\textbf{2022})\GeV$ & \\
    \end{tabular}
    \label{tab:run3_L1_Thr}
\end{table*}

A comparison of the average HLT output rates for the standard, parking, and scouting streams during Run~2 and Run~3 is reported in Table~\ref{tab:Run3DataStreams} and also shown in Fig.~\ref{fig:hlt_rates_evolution}. The average scouting rate decreased from 2022 to 2023 as a result of the additional object and event selections listed earlier. Table~\ref{tab:run3_scouting_rates_per_path} shows the peak rates of the scouting paths, corresponding to two reference runs with an average pileup of 60 recorded during 2022 and 2023. The rates are consistent with the scenarios shown for the L1 input rates in Fig.~\ref{fig:l1_rates}.

\begin{table*}[!htb]
\centering
\topcaption{Comparison of the typical HLT trigger rates of the standard, parking, and scouting data streams during 2018 (Run~2), 2022, and 2023 (Run~3). The average \Linst value over one typical fill of a given data-taking year and the average pileup (PU) are also reported, coherently with the scenarios reported in Fig.~\ref{fig:hlt_rates_evolution}.
}
\renewcommand{\arraystretch}{1.3}
\centering\begin{tabular}{lccccc}
Year      & \Linst [$\invcms$] &PU & Standard rate [Hz] & Parking rate [Hz] & Scouting rate [Hz]  \\
\hline
2018 & \sci{1.2}{34} & 38 &1000 & 3000 & 5000 \\
2022 & \sci{1.5}{34} & 46 &1800 & 2440 & 22000 \\
2023 & \sci{1.7}{34} & 48 &1700 & 2660 & 17000 \\
\end{tabular}
\label{tab:Run3DataStreams}
\end{table*}

\begin{table*}[!htb]
    \centering
    \topcaption{HLT rates for the scouting paths seeded by different L1 algorithms during two reference runs with an average pileup of 60 recorded in 2022 and 2023.}
    \renewcommand{\arraystretch}{1.3}
    \begin{tabular}{lcc}
        Configuration & 2022 & 2023\\  
        \hline
        Scouting path & \multicolumn{2}{c}{Rate per path [kHz]}\\ \hline
        1 \egamma  & \NA & 9.1  \\
        ${\geq}2$ \egamma  & \NA & 0.3  \\
        ${\geq}2$ muons  & \NA & 3.4  \\
        Jets or \HT  & \NA & 11.0  \\
        \egamma, ${\geq}2$ muons, jets or \HT  & 31.3 & \NA  \\
    \end{tabular}
\label{tab:run3_scouting_rates_per_path}
\end{table*}

\subsection{Event size and content}
\label{sec:Run3EventContent}

The event size reduction needed for the scouting strategy is achieved by applying selection criteria on the reconstructed physics objects and by storing high-multiplicity quantities with reduced numerical precision on the mantissa (10 bits). The amount of information in the data scouting output increased in Run~3 compared to Run~2 due to the added reconstruction of electrons and photons, and the storage of additional quantities. Nevertheless, the overall event size for the Run~3 scouting was kept smaller than 10\unit{KB} thanks to the selection and precision optimization.

Reconstructed PF candidates are stored together with their kinematic variables and their particle type as identified by the PF algorithm, if they have ${\pt>0.6\GeV}$ and ${\abs{\eta}<3}$ (similar to the Run~2 approach). If a PF candidate is a charged particle with an associated pixel-only track, basic track parameters and a reference to the vertex associated with that track are also stored. Track-related quantities of PF candidates were added in Run~3 that, for example, are used as input when training neural networks such as  \textsc{ParticleNet}~\cite{PNet:Qu_2020} for tasks, such as jet flavor identification. Additionally, for reconstructed tracks with ${\pt > 3\GeV}$, the kinematic information, hit pattern, track parameters, and track fit quality, including the covariance matrix for the refitting of vertices, are stored. Finally, for vertices reconstructed from the pixel-only or muon tracks, their position and associated uncertainties as well as the $\chi^2$ and number of degrees of freedom in the vertex fit, are stored.

\begin{table*}[!htb]
    \centering
        \topcaption{List of observables related to the newly introduced scouting electrons (e) and photons ($\gamma$) stored in the scouting data set in Run~3. The calorimeter observables shared by \egamma objects are listed in the upper part of the table, while the track features specific to electrons are reported in the lower part. Tracker-based isolation for photons is to be computed offline from the stored PF candidates.}
        \renewcommand{\arraystretch}{1.3}
        \begin{tabular}{ll}
        Observable & Definition \\ \hline
        \multicolumn{2}{c}{\textit{Electron/photon common quantities (calorimeter-based)}} \\ \hline
        (E, \pt, $\eta$, $\phi$) & ECAL SC four-momentum \\
        $\sigma_{i\eta i\eta}$ & Spread of the ECAL shower from the central crystal\\
        $H/E$ & Ratio of energy deposit in HCAL to ECAL \\
        $I_{\rm E}$ & ECAL isolation  \\
        $I_{\rm H}$ & HCAL isolation \\
        $r_{9}$ & Relative energy deposit in 3x3 $\eta$-$\phi$ matrix\\
        seed ID & Crystal number of the central crystal  \\
        energy matrix & Energy deposit in each crystal of the SC  \\
        detector ID & Crystal number of each crystal of the SC  \\
        time matrix & Time stamp of each crystal of the SC  \\
        $s_{\rm minor}$ and $s_{\rm major}$ & Second moments of the SC energy matrix  \\
        rechitZeroSuppression & Flag indicating events with nonzero reconstructed hits  \\ 
         \hline
        \multicolumn{2}{c}{\textit{Electron quantities only (tracker-based)}}  \\ \hline
        \text{track} (E, \pt, $\eta$, $\phi$) & GSF track four-momentum \\
        \text{track} $d_{0}$ & Track $d_{0}$ \\
        \text{track} $d_{z}$ & Track $d_{z}$  \\
        \text{track} $\chi^{2}/$\text{dof} & Reduced-$\chi ^{2}$ of the track fit \\
        \text{track} missing hits & Missing hits in the tracker inner pixel region  \\
        \text{track} $q$ & Track charge 
        \\
        $\Delta \eta^{\rm seed}_{\rm in}$ & Difference in $\eta$ between central ECAL crystal and inner track \\
        $\Delta \phi_{\rm in}$ & Difference in $\phi$ between SC and inner track \\
        $1/E - 1/p$ & Difference between the inverse of SC $E$ and  track $p$ \\
        $I_{\rm track}$ & Track isolation\\
        \end{tabular}
       \label{tab:run3scout_egvars}
\end{table*}

Jet selection criteria require that jets have ${\pt>20\GeV}$ and ${\abs{\eta}<3}$ to be stored in the scouting stream. The same variables as in Run~2 are retained for each jet, which is detailed in Section~\ref{sec:ScoutingRun2Details}. While the Run~3 event content stores only AK4 jets, the inclusion of PF candidates allows for subsequent offline clustering of PF jets with any distance parameter. The missing transverse momentum reconstructed from the PF candidates is stored on an event basis.

Muons with ${\abs{\eta}<2.4}$ are stored with their kinematic and isolation quantities, as previously discussed in Section~\ref{sec:ScoutingRun2Details} for the Run~2 scenario. Detailed track information is also available, enabling the refitting of dimuon vertices. All reconstructed dimuon vertices are stored separately with their positions, associated uncertainties, and fit qualities. Finally, electrons and photons are stored if they satisfy ${\pt>2\GeV}$ and ${\abs{\eta}<2.5}$. To reduce the processing workload, the time-consuming track reconstruction for electrons is initiated only when the energy in the HCAL directly behind the ECAL supercluster (within a cone ${\deltar < 0.15}$) is less than 20\% of the supercluster energy, and at least two hits are observed in the pixel tracker. The list of quantities stored in the scouting stream that relate to electrons and photons is reported in Table~\ref{tab:run3scout_egvars}.

The reconstruction and identification performance of the jets, muons, electrons, and photons stored in the Run~3 scouting stream is discussed in the next sections. A special scouting monitoring data set that collects a randomly chosen reduced number of events (about 1\% of the total) is used for most of the studies described in the following sections. Both offline and scouting objects are available in this data set, allowing for an easier comparison between the two strategies over the same set of events.

\subsection{Jets}
\label{subsec:run3jets}

\begin{figure*}[!htb]
  \centering
  \includegraphics[width=0.32\textwidth]{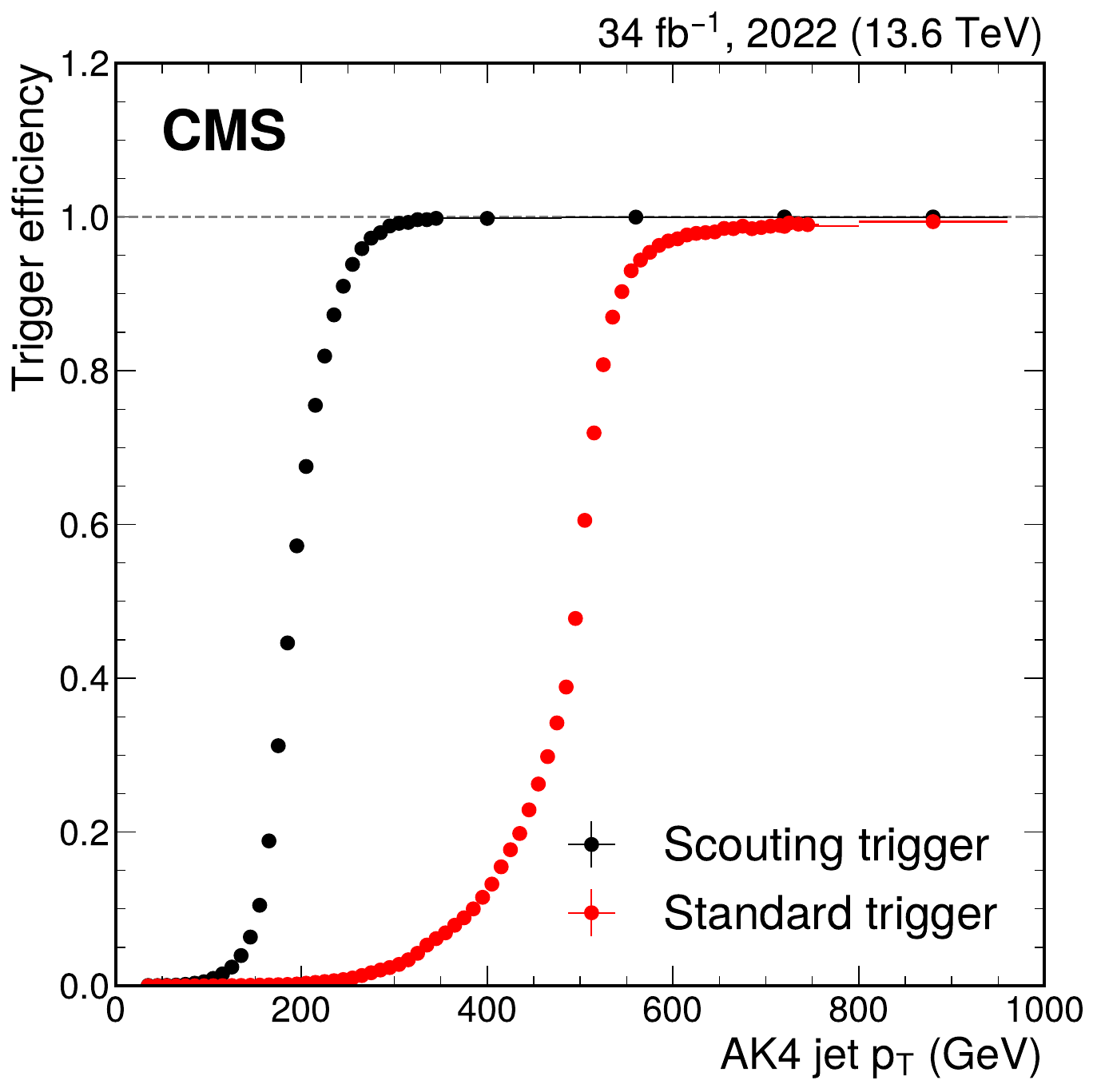}
  \includegraphics[width=0.32\textwidth]{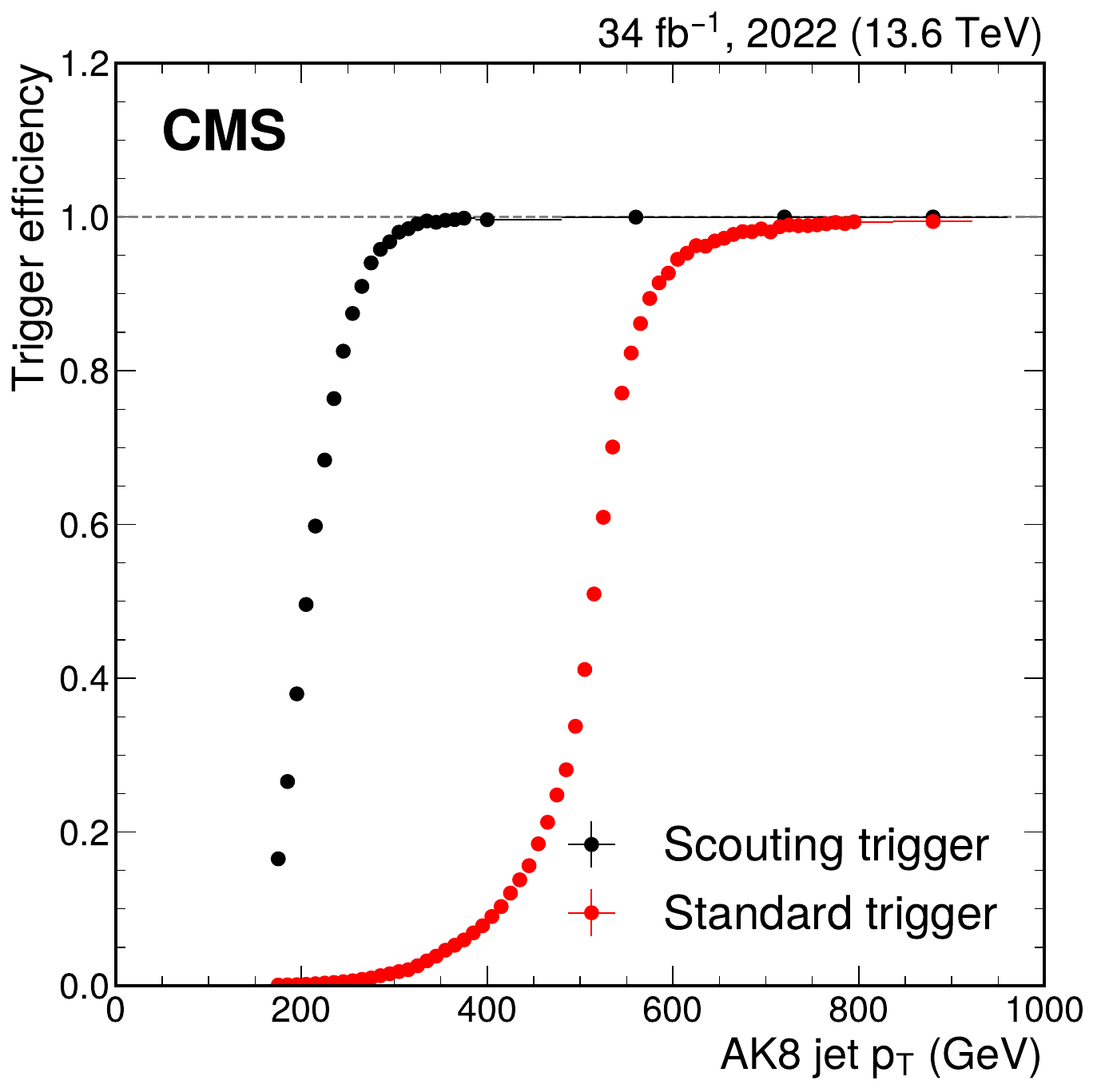}
  \includegraphics[width=0.32\textwidth]{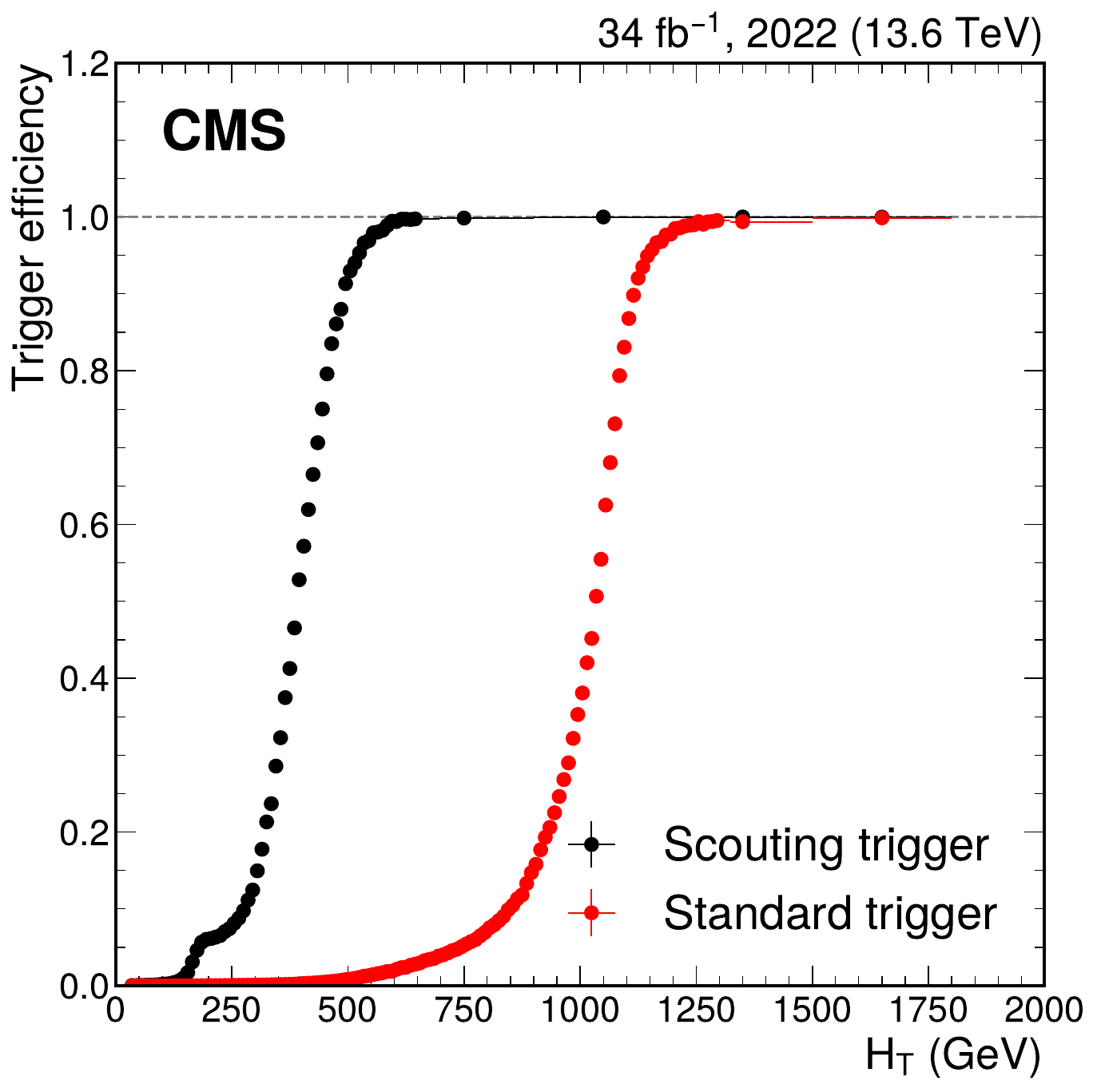}
  \caption{Trigger efficiency as a function of AK4 jet \pt (left), AK8 jet \pt (center), and \HT (right). The efficiency is computed from collision data recorded in 2022 by the scouting (black points) and standard (red points) streams.}
  \label{plot:jets trigger efficiency}
\end{figure*}

The scouting strategy enables lower hadronic trigger thresholds than the standard strategy relying on offline reconstructed data. As listed in Table~\ref{tab:run3_L1_Thr}, the data scouting trigger in 2022 included unprescaled L1 seeds targeting events based on the presence of at least one jet with \pt exceeding 180\GeV or \HT exceeding 360\GeV. In comparison, the lowest unprescaled triggers in the standard trigger strategy required jet \pt or \HT to exceed 500\GeV or 1050\GeV, respectively. The benefits of the lower trigger thresholds in scouting are quantified in Fig.~\ref{plot:jets trigger efficiency}, where the trigger efficiencies of the scouting and standard trigger selections are compared using collision data recorded in 2022. The trigger selection is a logical ``OR'' expression of all L1 seeds and HLT triggers. Events are selected based on the presence of at least one energetic jet or sufficiently energetic \HT. The efficiency is displayed as a function of the offline reconstructed AK4 PF jet \pt (left), AK8 PF jet \pt (center), and PF \HT (right).

The efficiency is measured using an unbiased sample of events, collected with a single-muon trigger and containing only one well-identified and isolated muon outside of the jet cone. Events with additional muons are excluded. At least one well-reconstructed PF jet is required in the event, and jets must further pass identification criteria that reject poorly reconstructed jets or jets arising from detector noise. The AK4 PF jets are required to have ${\abs{\eta} < 2.5}$ and ${\pt>30\GeV}$, whereas the AK8 PF jets require ${\abs{\eta} < 2.5}$ and ${\pt>170\GeV}$. The efficiency is defined as the ratio of the number of events where an offline reconstructed PF jet is selected by the data scouting or standard triggers, relative to the total number of events with an offline reconstructed PF jet.

The low thresholds of the data scouting triggers are visible in the plot of each jet observable. The efficiency to select AK4 and AK8 scouting jets is about 100\% for ${\pt>300\GeV}$. In comparison, the standard trigger is fully efficient only from around 700--800\GeV. Similarly, data scouting is fully efficient for ${\HT>600\GeV}$, compared to roughly 1300\GeV for the standard trigger. Therefore jet-based analyses relying on data scouting are able to probe regions of phase space inaccessible with the standard trigger strategy. By lowering the \HT threshold from 360 to 280\GeV in 2023, as discussed in Section~\ref{subsec:run3scouting}, the scouting trigger improves even further the CMS acceptance to hadronic resonances. 

The quality of the Run 3 scouting jets is evaluated by computing the JES and JER with data recorded in 2022. The scouting jets in this study are reconstructed offline, using as input the scouting PF candidates that were reconstructed online during the data-taking. The offline reconstructed jets of the standard triggers are used as a reference. Before clustering, pileup is mitigated via the PUPPI technique for offline jets and via the CHS technique for scouting jets, as described in Section~\ref{subsec:jets}. Jets are corrected by applying detector response corrections computed from simulated samples. The corrections applied to the scouting jets are derived specifically for HLT jets, and differ from those applied to the offline jets. The same corrections are applied to simulated and observed jets. No in situ corrections are applied to observed jets either in the scouting or offline data sets.

\begin{figure*}[!htb]
  \centering
  \includegraphics[width=0.46\textwidth]{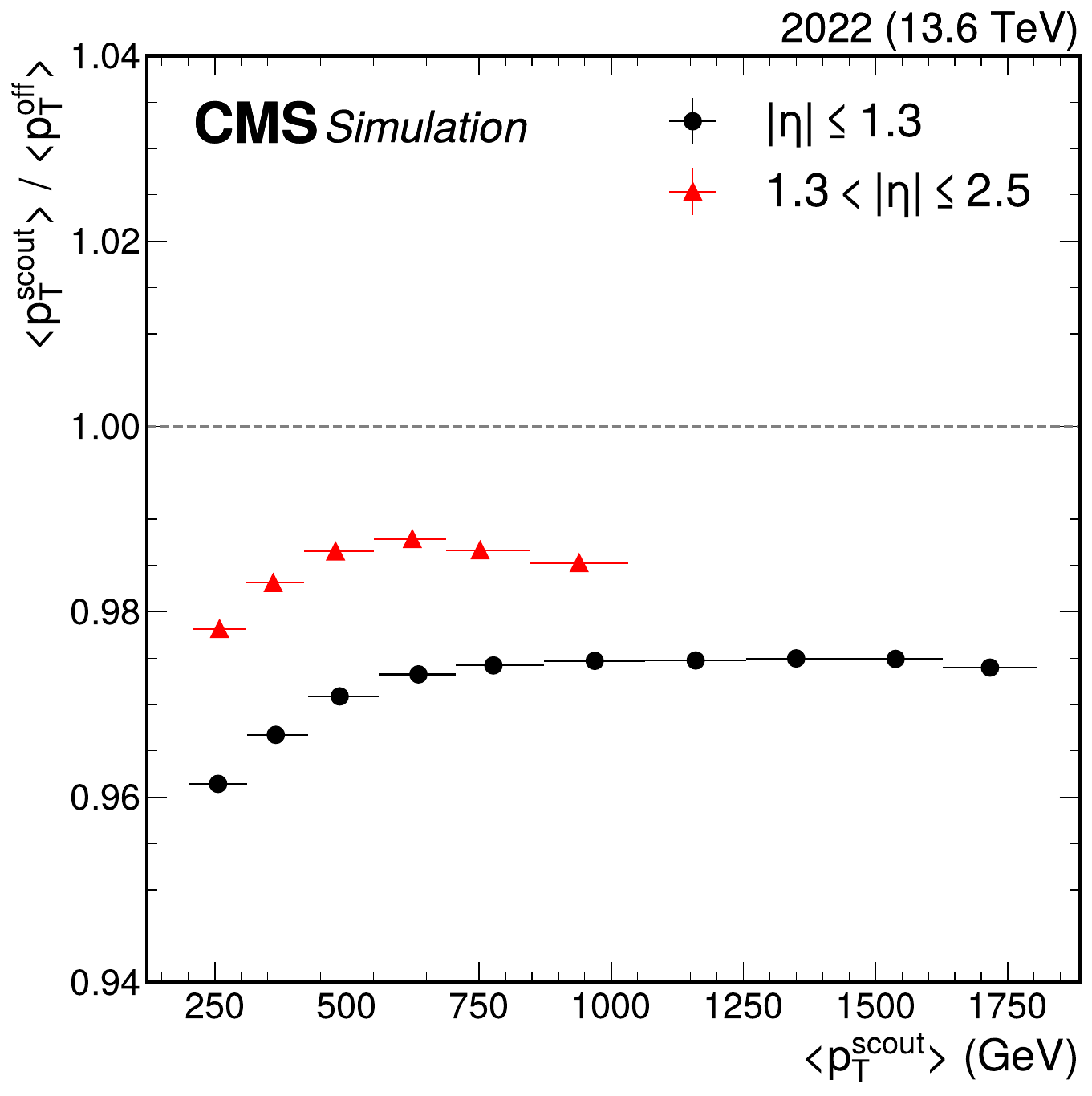}
  \includegraphics[width=0.46\textwidth]{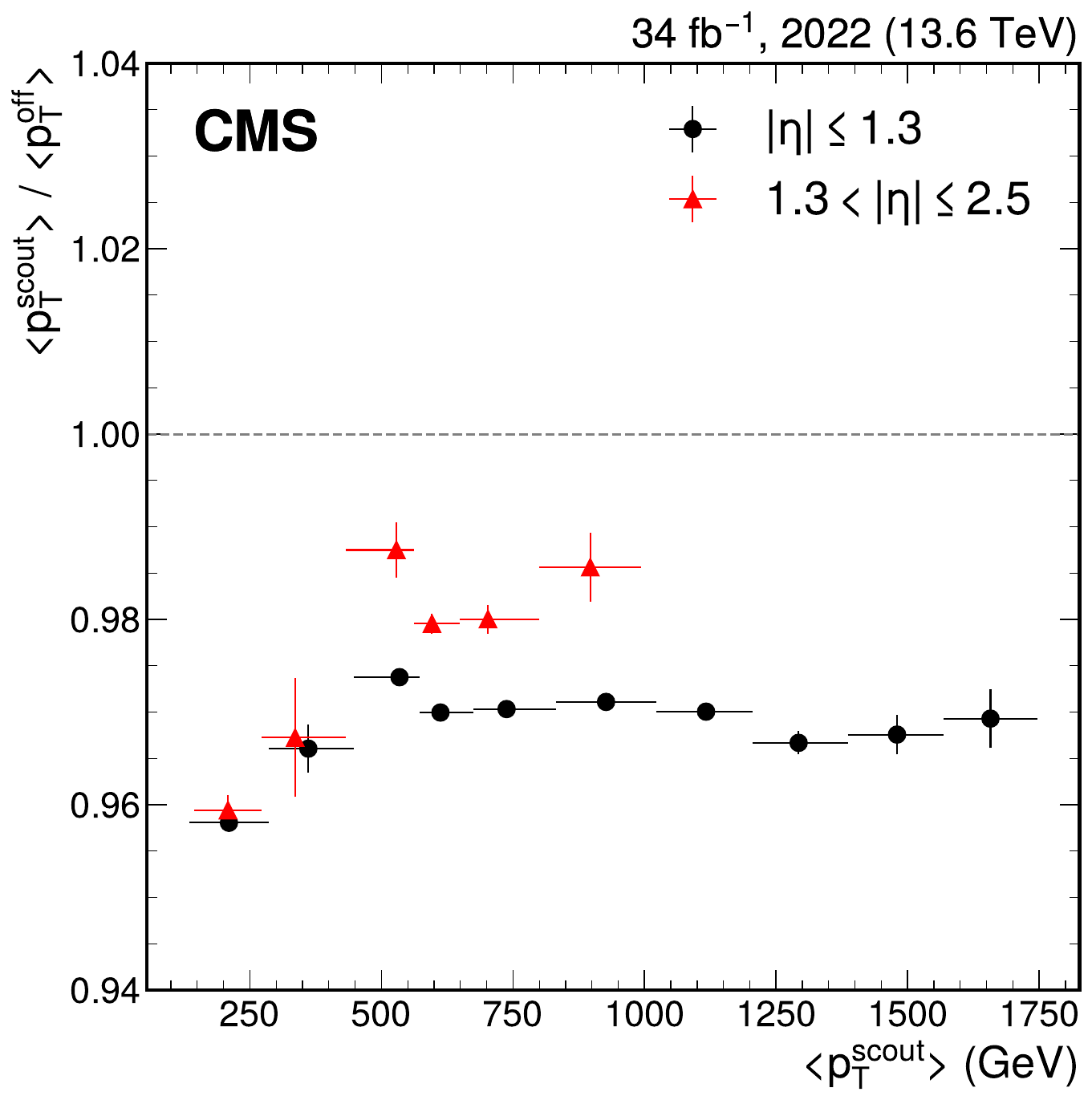}
  \caption{The JES as a function of mean scouting jet \pt  derived from simulated (left) and recorded (right) events. The red and black points correspond to events where the two leading jets have ${\abs{\eta} <1.3}$ and ${1.3 < \abs{\eta} < 2.5}$, respectively.}
  \label{plot:run 3 jes}
\end{figure*}

The JES and JER derivations are performed with the same methods as described in Section~\ref{sec:jet-reco-performance}. The measurement is performed in bins of jet $\eta$, where both leading jets are required to have ${\abs{\eta} < 1.3}$ or ${1.3 < \abs{\eta} < 2.5}$. Events where they are in different $\eta$ regions are discarded. The final result of the JES measurement is presented in Fig.~\ref{plot:run 3 jes}, where \ptscout is the \pt of the scouting object, and \ptoff is the \pt of the offline object. The creation of mean scouting jet \pt ($\langle \ptscout \rangle$) involves a mapping from \ptoff, resulting in varying bin widths across different $\eta$ regions, as well as discrepancies between bin widths for simulation and recorded data. The uncertainties vary as a result of the trigger selection. The JES is similar between simulated and recorded events, at approximately 0.96--0.97 for ${\abs{\eta} < 1.3}$ and 0.96--0.99 for ${1.3 < 
\abs{\eta} < 2.5}$, when requiring jet ${\pt>200\GeV}$.

\begin{figure*}[htb]
  \centering
  \includegraphics[width=0.45\textwidth]{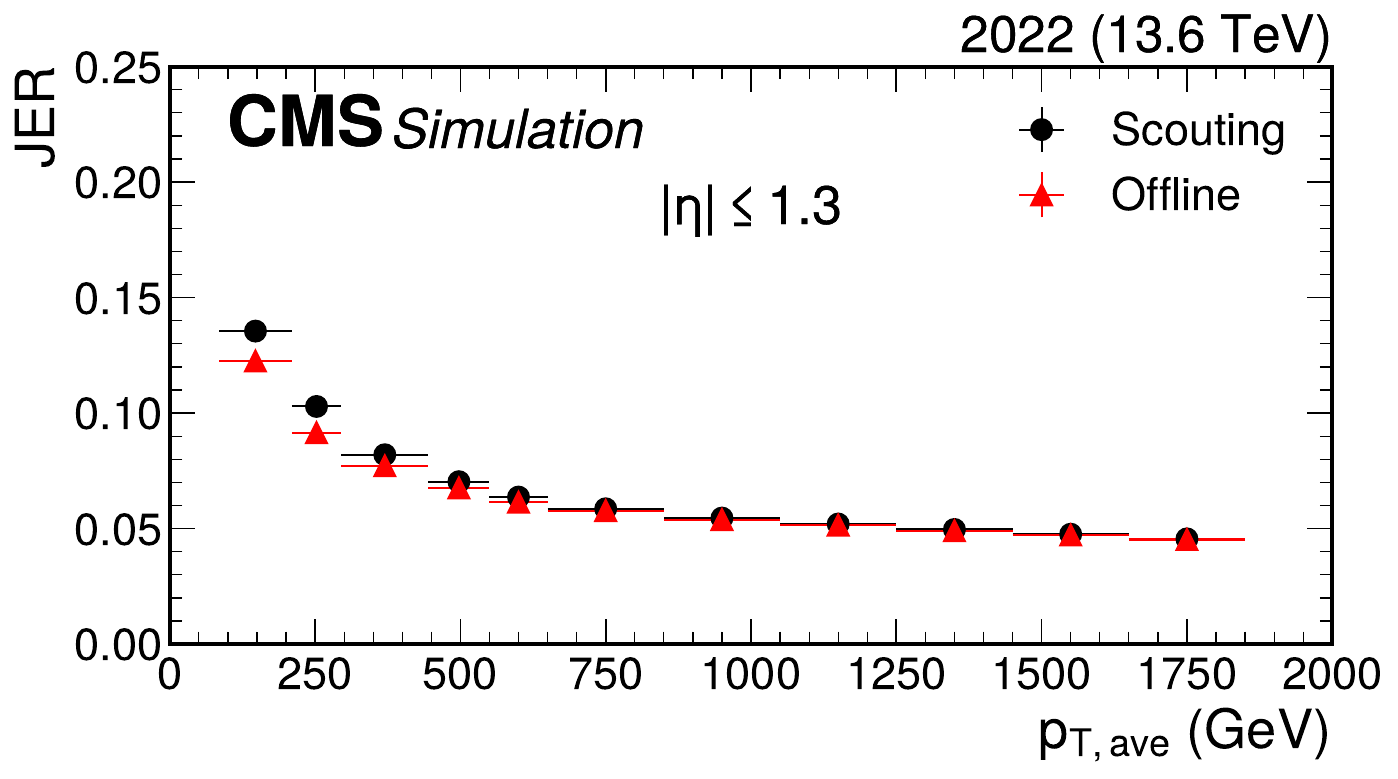}
  \includegraphics[width=0.45\textwidth]{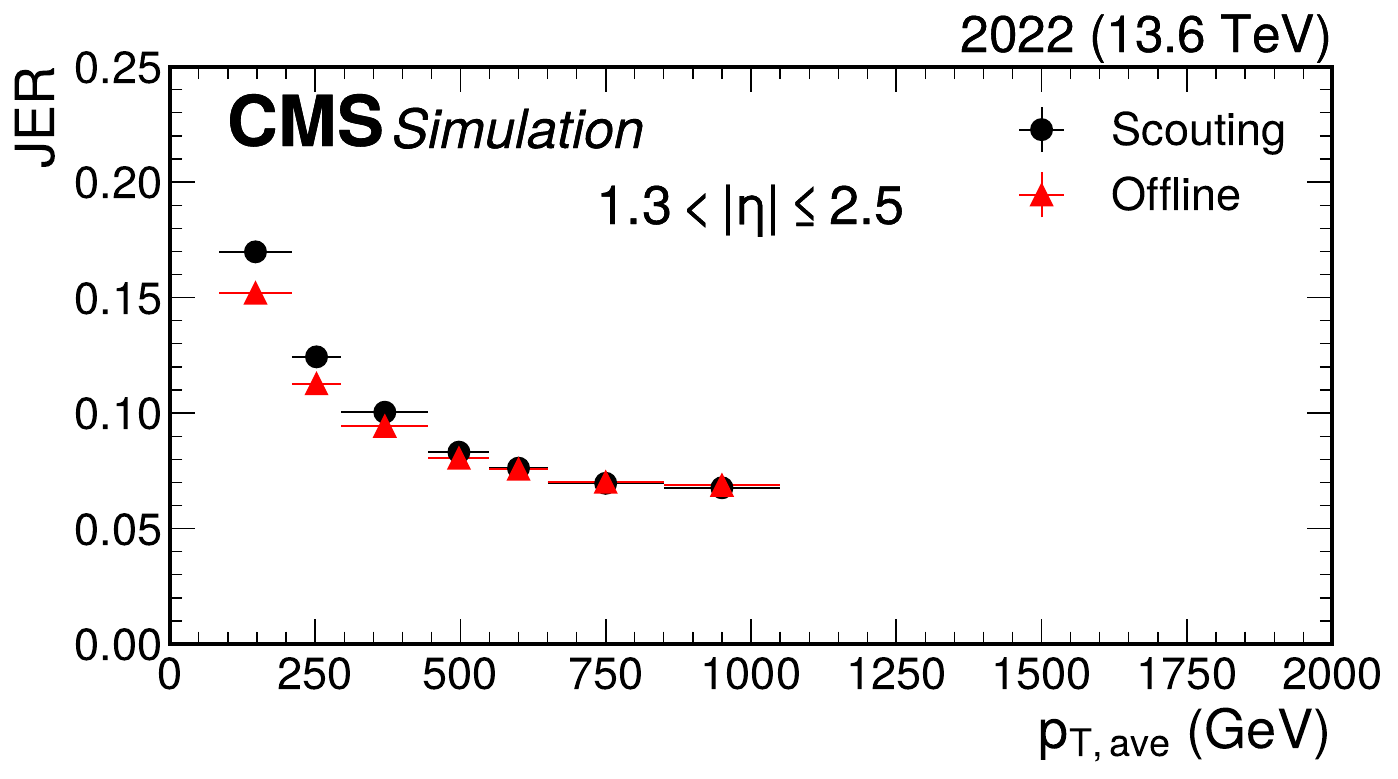}
  \includegraphics[width=0.45\textwidth]{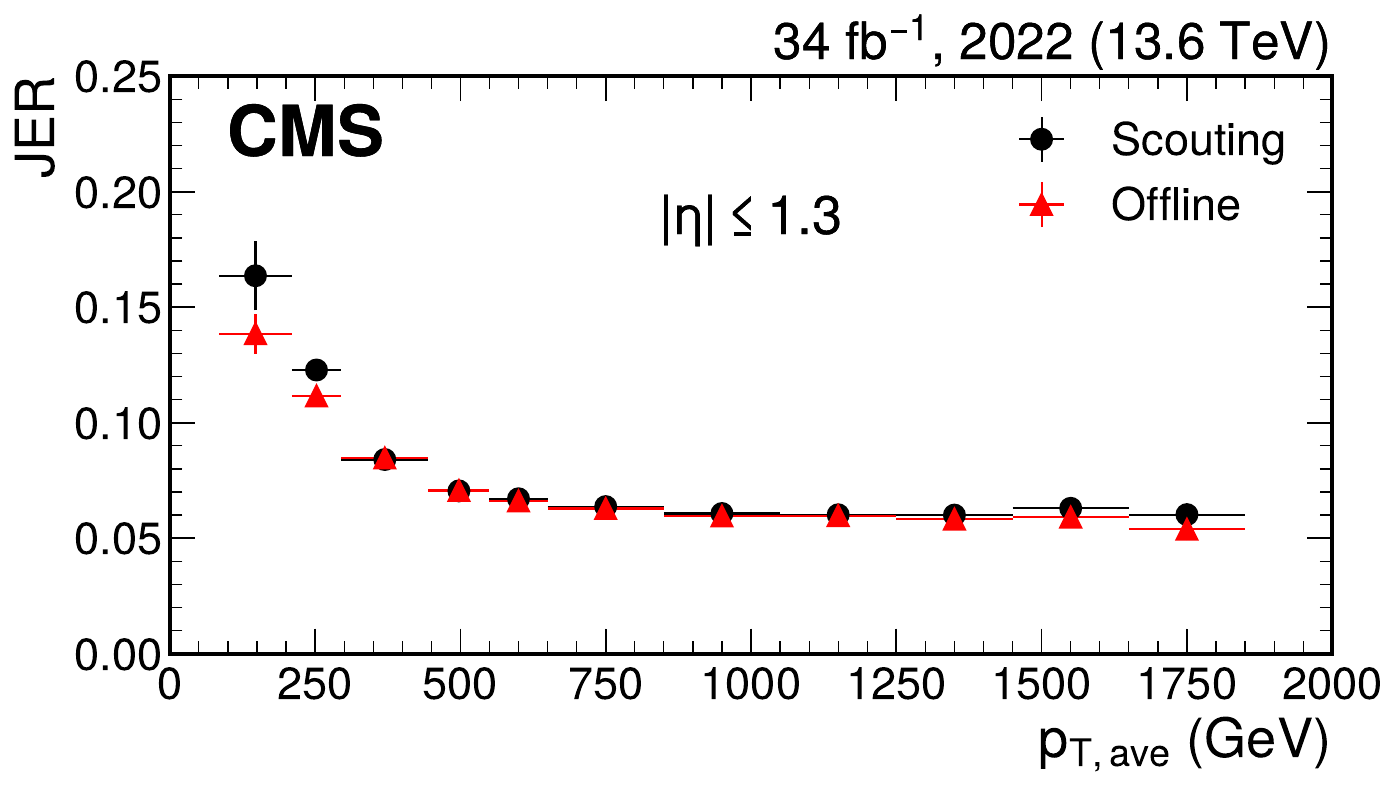}
  \includegraphics[width=0.45\textwidth]{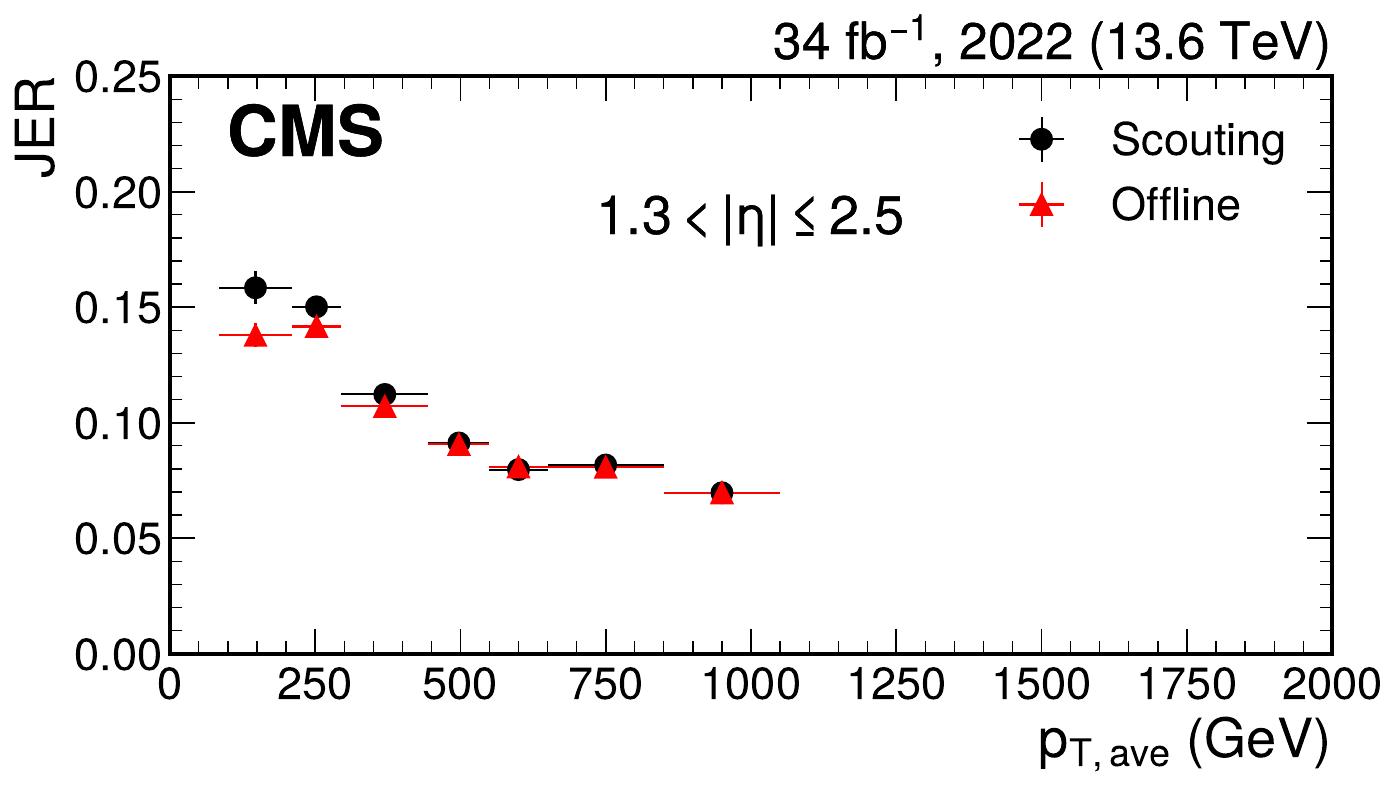}
  \caption{The JER as a function of average \pt. The JER is computed from simulated (upper) and recorded (lower) events, by requiring the two leading jets to have ${\abs{\eta} < 1.3}$ (left) and ${1.3 < \abs{\eta} < 2.5}$ (right). The red and black data points denote 2022 collision data reconstructed by the scouting and offline algorithms, respectively.}
  \label{plot:run 3 jer}
\end{figure*}

\begin{figure*}[!htb]
  \centering
  \includegraphics[width=0.45\textwidth]{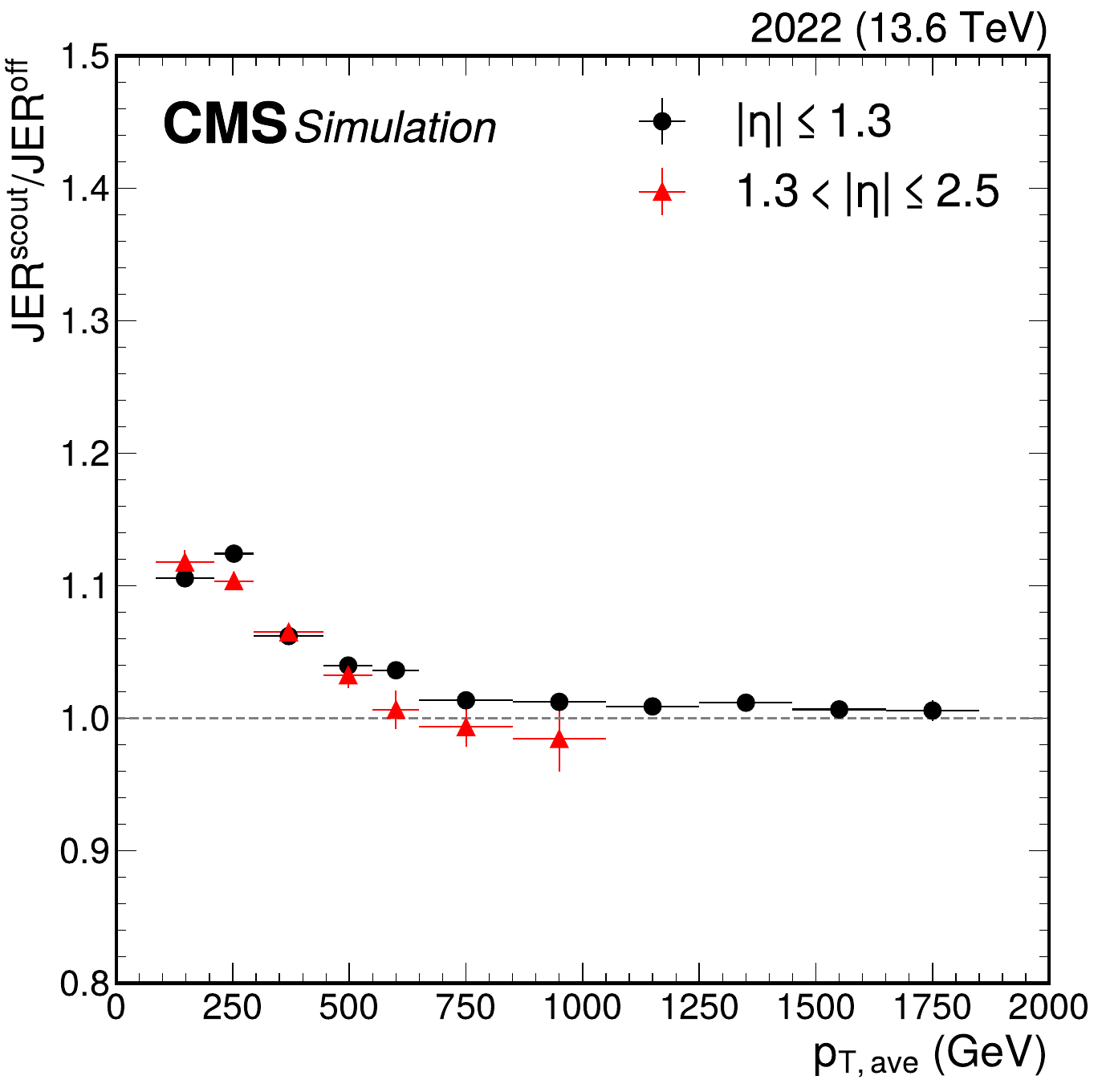}
  \includegraphics[width=0.45\textwidth]{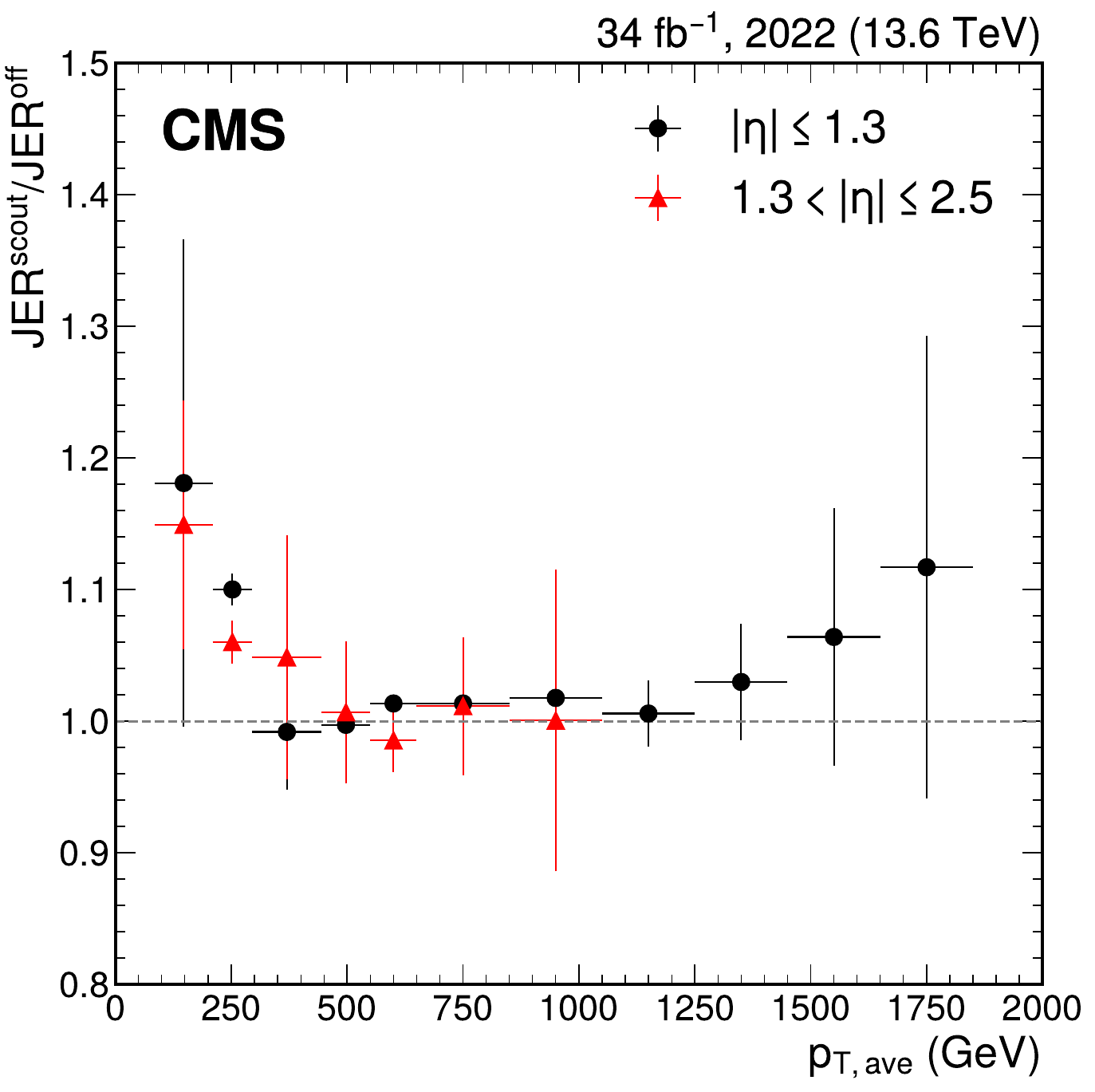}
  \caption{The ratio of the JER derived from scouting events to the JER derived from offline events as a function of average \pt. The ratio is computed from simulated (left) and recorded (right) events, by requiring the two leading jets to have ${\abs{\eta} < 1.3}$ (red points) and ${1.3 < \abs{\eta} < 2.5}$ (black points).}
  \label{plot:run 3 jer ratio}
\end{figure*}

The result of the JER measurement is presented in Figs.~\ref{plot:run 3 jer}~and~\ref{plot:run 3 jer ratio}, as function of the average \pt ($p_{T,\text{ave}} = (p_{T\text{, 1st jet}} + p_{T\text{, 2nd jet}}) / 2$) of the two highest-\pt jets in the event. The JER is stable from  an average \pt value above 500\GeV, measuring approximately 5\% (6\%) in the barrel (endcap) region for both offline and scouting jets. For jet \pt below 500\GeV, scouting jets feature ${\approx}10\%$ worse resolution compared to offline jets.

The jet performance discussed so far and the physics results presented earlier in Section~\ref{sec:Run2ScoutingPhysicsResults} demonstrate that exploiting scouting jets is an effective strategy for resonance searches in hadronic final states. In addition, we now show that the scouting strategy provides a large data sample that is sensitive to a final-state topology featuring Higgs boson decays to bottom quark-antiquark pairs (${\PH \to \PQb \PAQb}$). Recent searches by CMS for Higgs boson production in bottom~\cite{CMS:2020zge,HIG-21-020} or charm~\cite{CMS:2022fxs} quark decay channels improved signal sensitivity by targeting boosted final-state topologies that require reconstruction of the Higgs boson decay products within a single large-radius jet. The boosted object can then be identified with jet tagging techniques, \eg, by exploiting neural networks \cite{PNet:Qu_2020, Qu:2022mxj}. As a preliminary study, we investigate the production of boosted Higgs bosons produced via gluon-gluon fusion (ggF) decaying to bottom quark-antiquark pairs.

\begin{figure*}[!htb]
    \centering
    \includegraphics[width=0.45\textwidth]{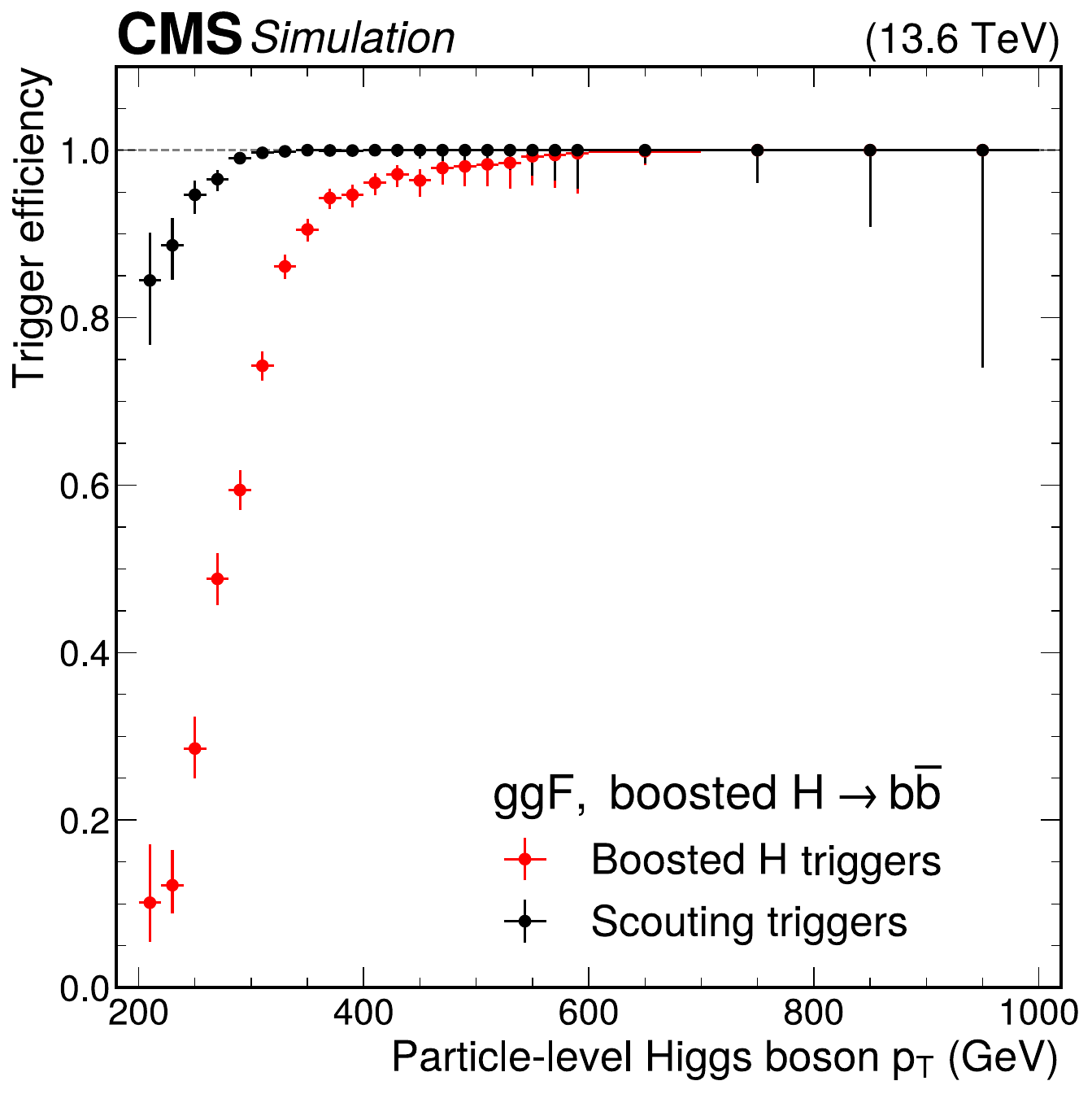}
    \hspace{0.03\textwidth}
    \includegraphics[width=0.45\textwidth]{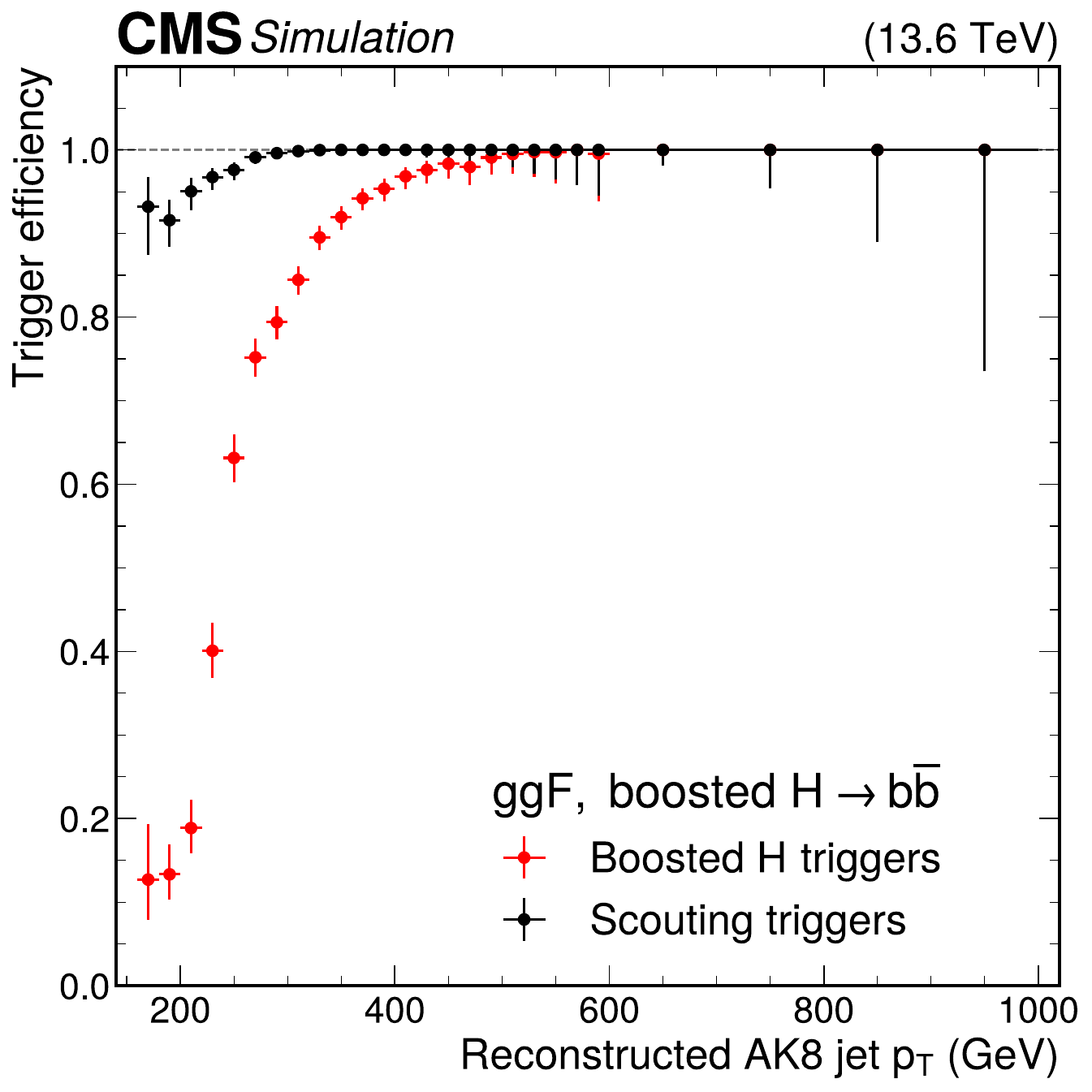}
    \caption{Trigger efficiency for ggF boosted $\PH \to \PQb \PAQb$ events as a function of the highest particle-level Higgs boson \pt (left) and highest offline-reconstructed AK8 jet \pt (right), as determined from simulation. The black and red points correspond to the scouting and the standard trigger selection, respectively.
    }
    \label{plot:Hbb_trigger_efficiency}
\end{figure*}

\begin{figure*}[!htb]
    \centering
    \includegraphics[width=0.45\textwidth]{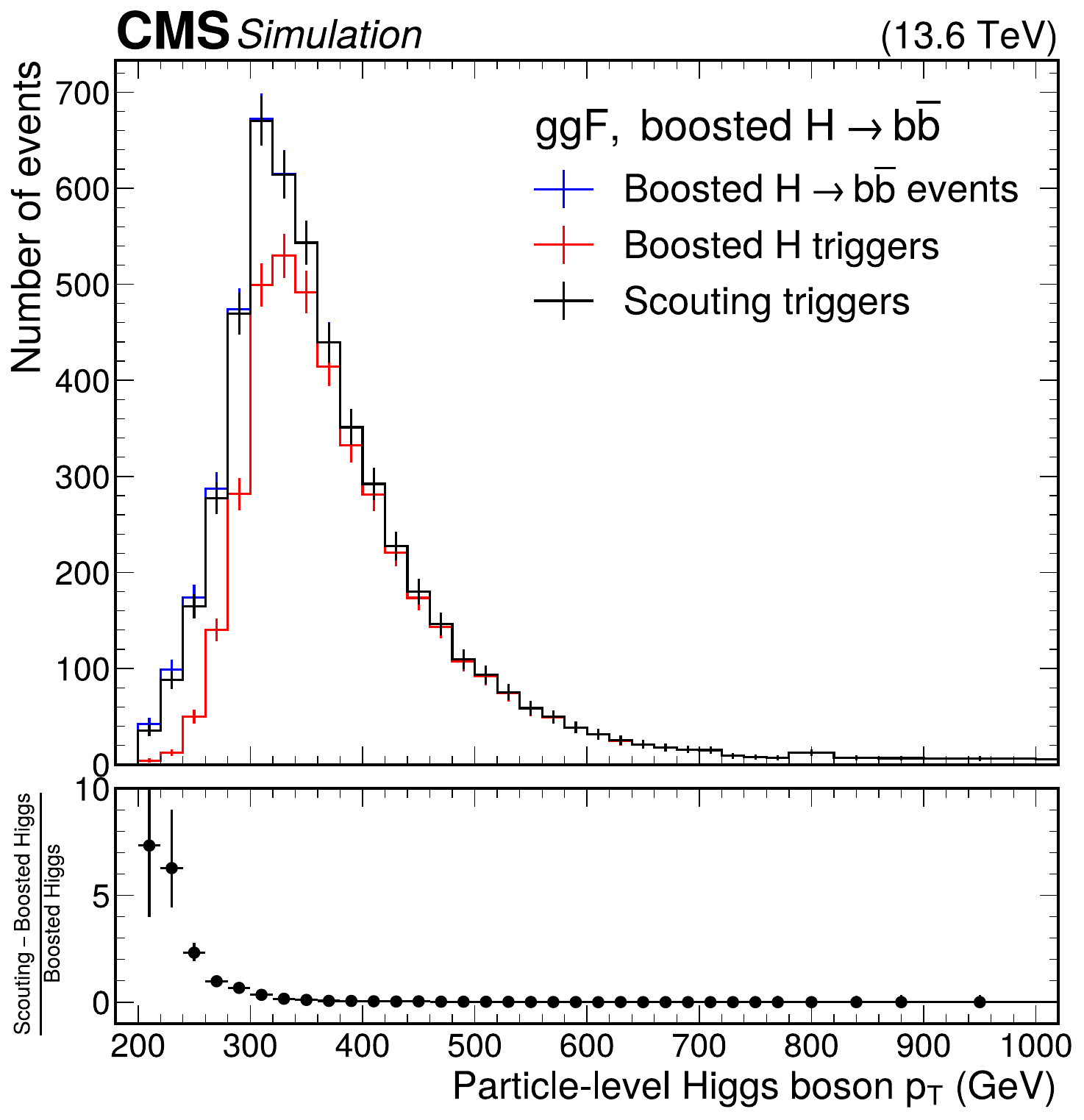}
    \hspace{0.03\textwidth}
    \includegraphics[width=0.45\textwidth]{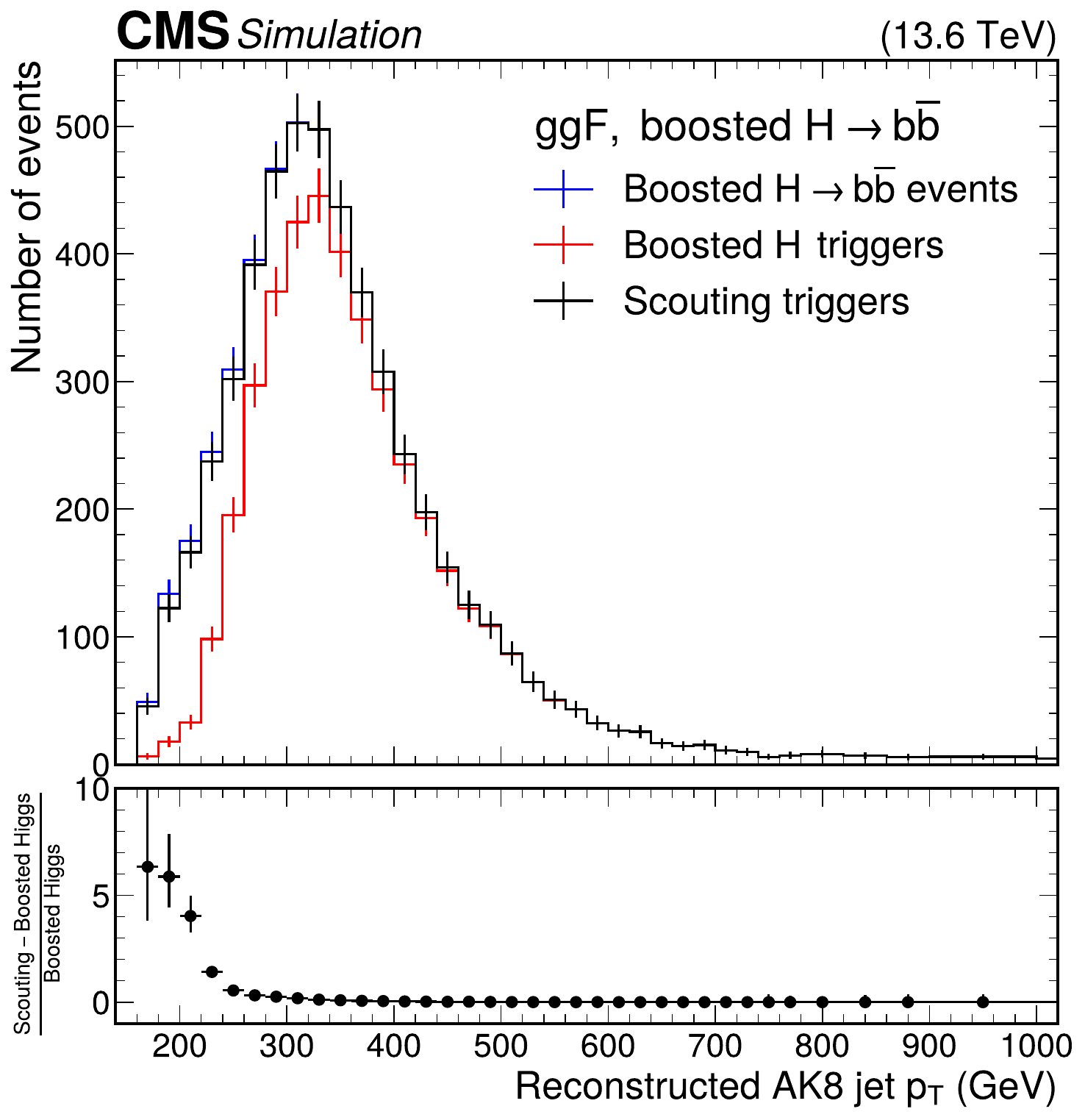}
    \caption{Number of ggF boosted $\PH \to \PQb \PAQb$ events as a function of the highest particle-level Higgs boson \pt (left) and highest offline-reconstructed AK8 jet \pt (right). The large-radius jet with highest \pt in each event is required to have a maximum angular distance ${\deltar < 0.8}$ from the two final-state \PQb quarks (blue). The events are then required to pass either the standard (red) or scouting (black) trigger selection. The number of events is computed from simulation with projected ${\Lint= 100\fbinv}$.
    }
    \label{plot:Hbb_eventyield}
\end{figure*}

The study is performed using a simulated sample of boosted Higgs boson events. Events are required to pass either a logical ``OR'' expression of jet-based scouting triggers or dedicated triggers targeting boosted Higgs boson topologies, deployed during Run~3 as part of the standard stream. Figures~\ref{plot:Hbb_trigger_efficiency} and~\ref{plot:Hbb_eventyield} show the performance of each trigger selection in terms of the trigger efficiency and the number of boosted Higgs boson events collected, respectively. The identification of such events is based on the requirement that the particle-level Higgs boson, together with its decay products, the bottom quark and antiquark, have a maximum angular distance ${\deltar < 0.8}$ from the reconstructed AK8 jet with the highest \pt. The performance is shown as a function of both particle-level Higgs boson \pt and reconstructed AK8 jet \pt. The latter corresponds to the reconstructed jets originating from the boosted Higgs boson decays. The study indicates that data scouting increases the overall event selection efficiency for boosted ${\PH \to \PQb \PAQb}$ decays by approximately 20\% compared to the standard triggers, particularly at low \pt. Since these results are preliminary and based only on simulated samples, this value is best interpreted as an upper bound and can be  affected by several factors, such as the background increase in scouting data or scouting jet-tagging efficiency, which require further investigation.

\subsection{Muons}
\label{subsec:run3muons}

\begin{figure*}[!htb]
\centering
\includegraphics[width=0.8\textwidth]{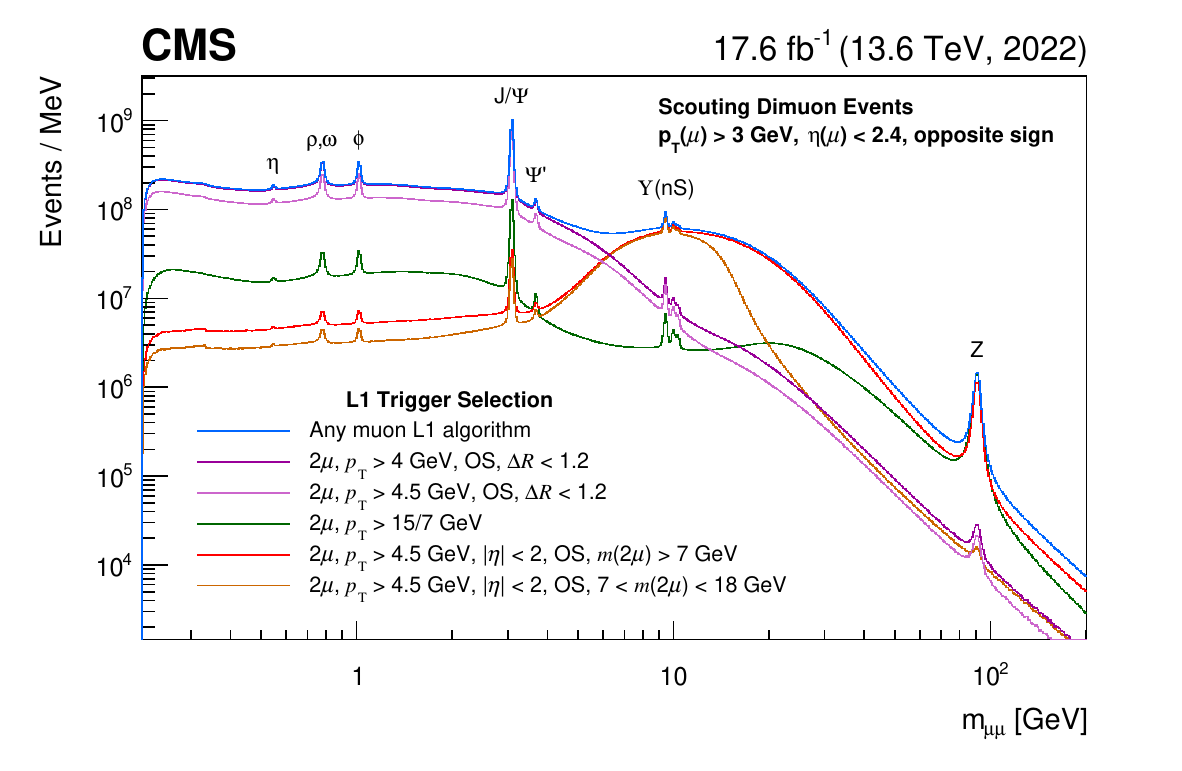}
\caption{Invariant mass distribution of opposite-sign muon pairs obtained with the scouting triggers, collected during 2022 with all Run~3 dimuon algorithms (blue curve), and with each individual algorithm (remaining colors).}
\label{fig:muRun3vsRun2_mass}
\end{figure*}

The dimuon scouting strategy, first developed in Run~2, offers numerous opportunities, as detailed in  Section~\ref{sec:Run2ScoutingPhysicsResults}. In Run~3, the set of L1 algorithms used as input to the scouting stream in 2022 is the same as in Run~2, except for the temporary removal of the dimuon trigger with the loosest transverse momentum requirement, which is relevant for events in the very low dimuon mass window. This trigger was restored for the 2023 data-taking period. The dimuon mass spectrum obtained from opposite-sign muon pairs selected by requiring at least one of the 2022 L1 triggers to be satisfied is shown in Fig.~\ref{fig:muRun3vsRun2_mass}. Considering data collected in 2022, corresponding to 17.6\fbinv, all well-known dimuon resonances from meson or \PZ boson decays are visible. The breakdown of the individual L1 seed contributions is also shown. 

\begin{figure*}[!htb]
\centering
\includegraphics[width=0.7\textwidth]{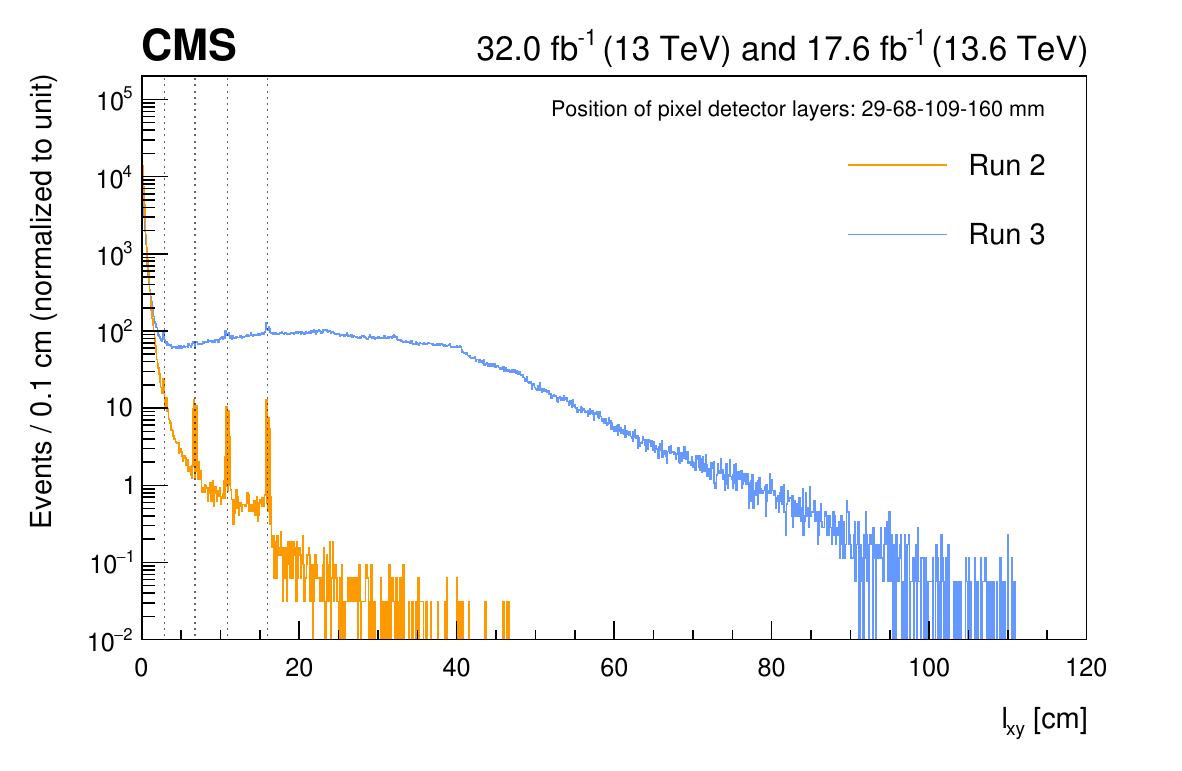}
\caption{Comparison between the \lxy distribution for Run~2 (orange) and Run~3 (blue) events in data that contain dimuon pairs with a common displaced vertex and a minimal selection on the vertex quality. The dashed vertical lines, placed at radii of 29, 68, 109, and 160\mm, correspond to the positions of the pixel layers where photons undergo conversion processes in the material, causing the observed peaks in the \lxy distribution.}
\label{fig:fig:muRun3vsRun2_lxy}
\end{figure*}

As mentioned in Section~\ref{subsec:run3scouting}, the removal of the requirement on the minimum number of hits in the pixel layers is a novelty with respect to Run~2. This update enables the collection of a larger number of events with high values of the dimuon transverse displacement, thus enhancing the ability to search for LLPs decaying to muons. The change occurred concurrently with the implementation of an updated version of the muon track finder algorithm in the barrel region at L1, which mainly improved the displaced-muon
triggering performance~\cite{CMS:2023gfb}.
Figure~\ref{fig:fig:muRun3vsRun2_lxy} shows the distribution of the dimuon vertex transverse displacement \lxy for events that contain at least one pair of OS muons associated with a selected secondary vertex.
Events are collected with dimuon displacements up to ${\approx}100\cm$, corresponding to the end of the sensitive region of the tracker. At the positions of the pixel layers, with radii of 29, 68, 109, and 160\mm, photons undergoing conversion processes in the material lead to peaks in the \lxy distribution. These peaks are less pronounced in the Run~3 distribution because of the removal of the pixel-hit requirement, which leads to higher efficiency, but also lower purity -- if no additional analysis-specific quality criteria are required, as is the case here.

\begin{figure*}[!htb]
\centering
\includegraphics[width=0.7\textwidth]{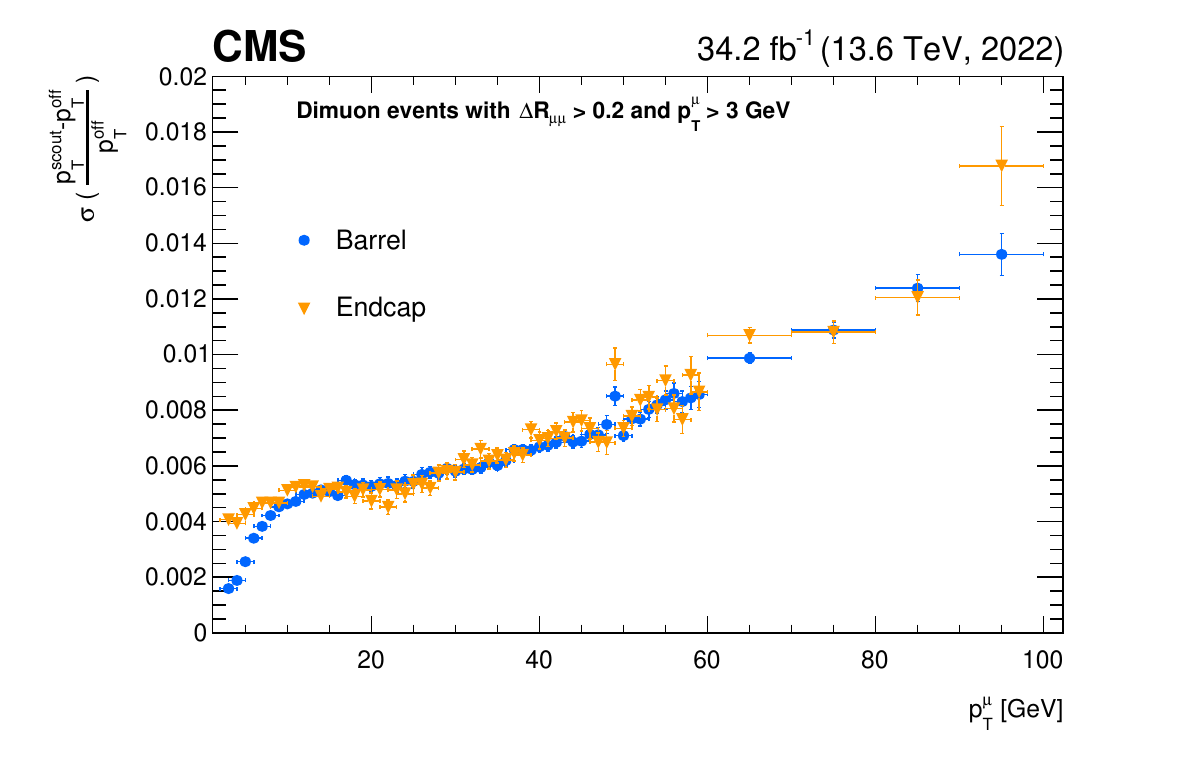}
\caption{Resolution on the transverse momentum of scouting muons compared to offline muons using data collected in 2022. Differences in muon momentum scale between the scouting and offline reconstruction algorithms are studied in bins of 1\GeV (10\GeV) for muon \pt smaller (larger) than 60\GeV. Values for the barrel (blue circles) and endcap (orange triangles) sections are shown separately.}
\label{fig:muRun3_res}
\end{figure*}

The performance achieved by the online and offline reconstruction methods are compared using 2022 data collected with the scouting monitoring triggers.
The resolution of the transverse momenta of muons reconstructed with the scouting algorithm is studied separately for muons reconstructed in the barrel and endcap regions. While the \pt range of interest for scouting muons is below 50\GeV, the study is performed for various \pt intervals between 3 and 100\GeV.
Figure~\ref{fig:muRun3_res} shows the \pt resolution of scouting muons with respect to offline muons, computed as the standard deviation ($\sigma$) of the Gaussian fit to the following quantity: 
\begin{equation*}
\frac{\ptscout-\ptoff}{\ptscout}.
\label{eq:ptres}
\end{equation*}

Leading and subleading \pt muons in events with exactly two muons are required to have ${\deltar > 0.2}$ and to be geometrically matched to the corresponding offline muons. Differences in muon momentum resolution between the scouting and offline reconstruction algorithms are found to be less than 1\% for muons with ${\pt<60\GeV}$ and up to 1.5\% for higher \pt values, in both barrel and endcap regions. 

\begin{figure*}[!htb]
\centering
\includegraphics[width=0.65\textwidth]{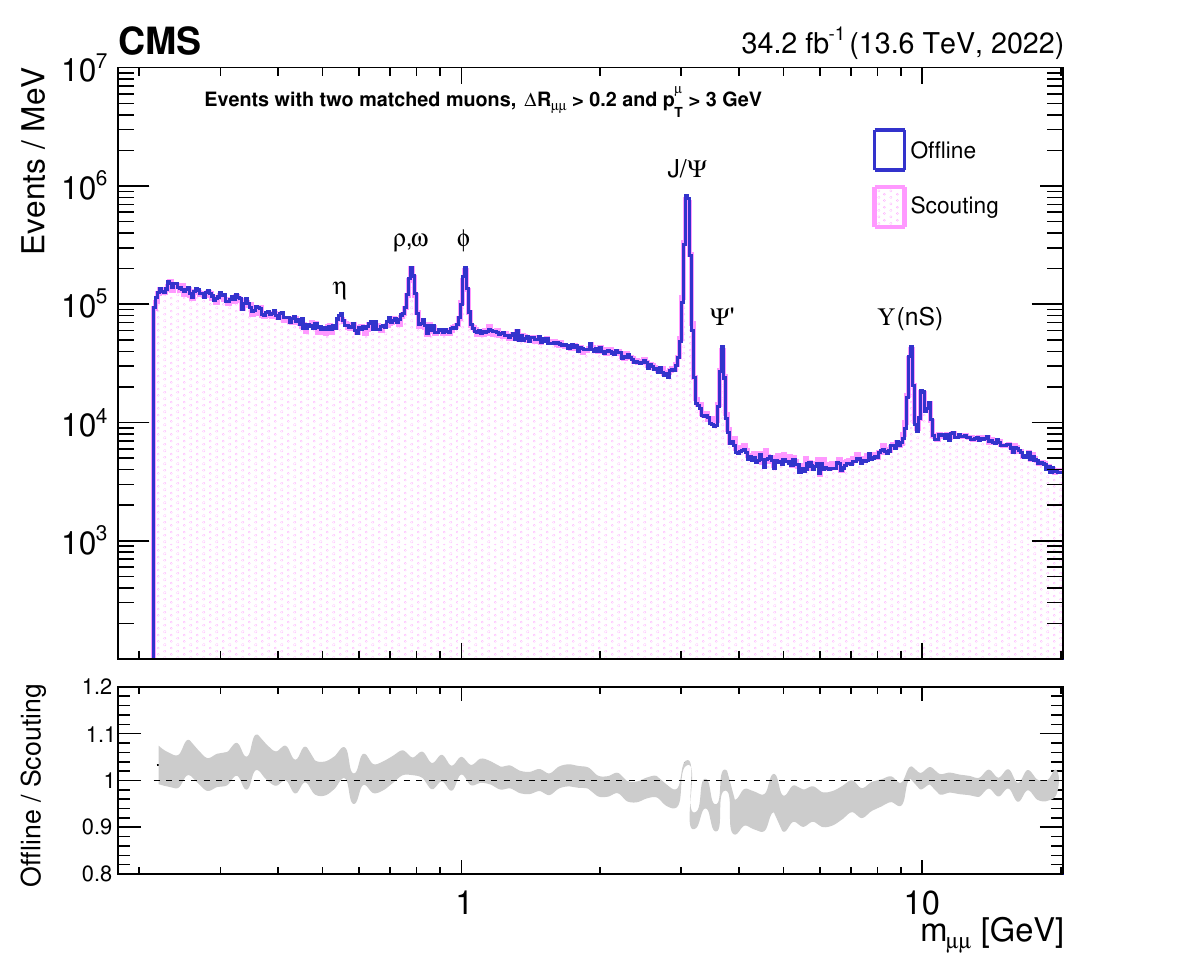}
\caption{Comparison of the dimuon spectra obtained with scouting (pink filled histogram) and offline (blue solid line) muons during the 2022 data-taking period. The ratio between the two distributions with a wider binning is also shown in the bottom panel as a gray band.}
\label{fig:muRun3_mass}
\end{figure*}

Figure~\ref{fig:muRun3_mass} presents a comparison of the dimuon spectra obtained with the scouting and offline reconstruction, showing excellent agreement. The former is reconstructed with pairs of online muons associated with a common vertex and matched to the corresponding offline muons within ${\deltar < 0.1}$, while the latter is composed of pairs of offline muons from selected events with exactly two muons. All dimuon resonances in the very low mass range below 11\GeV are reconstructed with excellent resolution compared to the offline algorithm. Differences between the mass resolution obtained with scouting and standard muons are observed to be less than 1.0--1.5\%, both in the barrel and endcap regions.

\subsection{Electrons and photons}
\label{subsec:run3electronsAndPhotons}

The endeavor to reduce trigger thresholds for physics studies with electrons and photons represents a major and challenging novelty of the Run~3 scouting strategy. Background processes from soft hadronic interactions in \pp collisions, electromagnetic activity within jets, and other low-energy deposits from pileup pose technical difficulties when lowering the trigger thresholds on the amount of energy in the ECAL. This complexity is further exacerbated by the time required to execute the GSF tracking algorithm for electrons. The inclusion of electrons and photons in the scouting stream was made possible by applying a background rejection strategy that focuses on the shower shape of the energy deposits in the ECAL, and by the reduced event content achieved.

The L1 and HLT requirements for the electron and photon scouting triggers are reported in Table~\ref{tab:run3_L1_Thr}. The efficiency of the scouting triggers to select events with a single electron or photon is calculated with an unbiased data set collected using a jet-based reference trigger. 
For the trigger efficiency measurements, both offline and scouting electrons and photons are used. 
An offline electron (photon) that passes a set of medium (loose) identification criteria is required to be within $\deltar < 0.06$ of a scouting electron (photon). The medium and loose identification criteria are defined to achieve an isolated \egamma selection efficiency of 80\% and 90\%, respectively~\cite{CMS:2020uim}. The trigger efficiency is computed as the ratio of the number of events that pass both the scouting single-\egamma trigger and the reference trigger, compared to the number of events that only pass the reference trigger.

\begin{figure*}[!htb]
  \centering
  \includegraphics[width=0.49\textwidth]{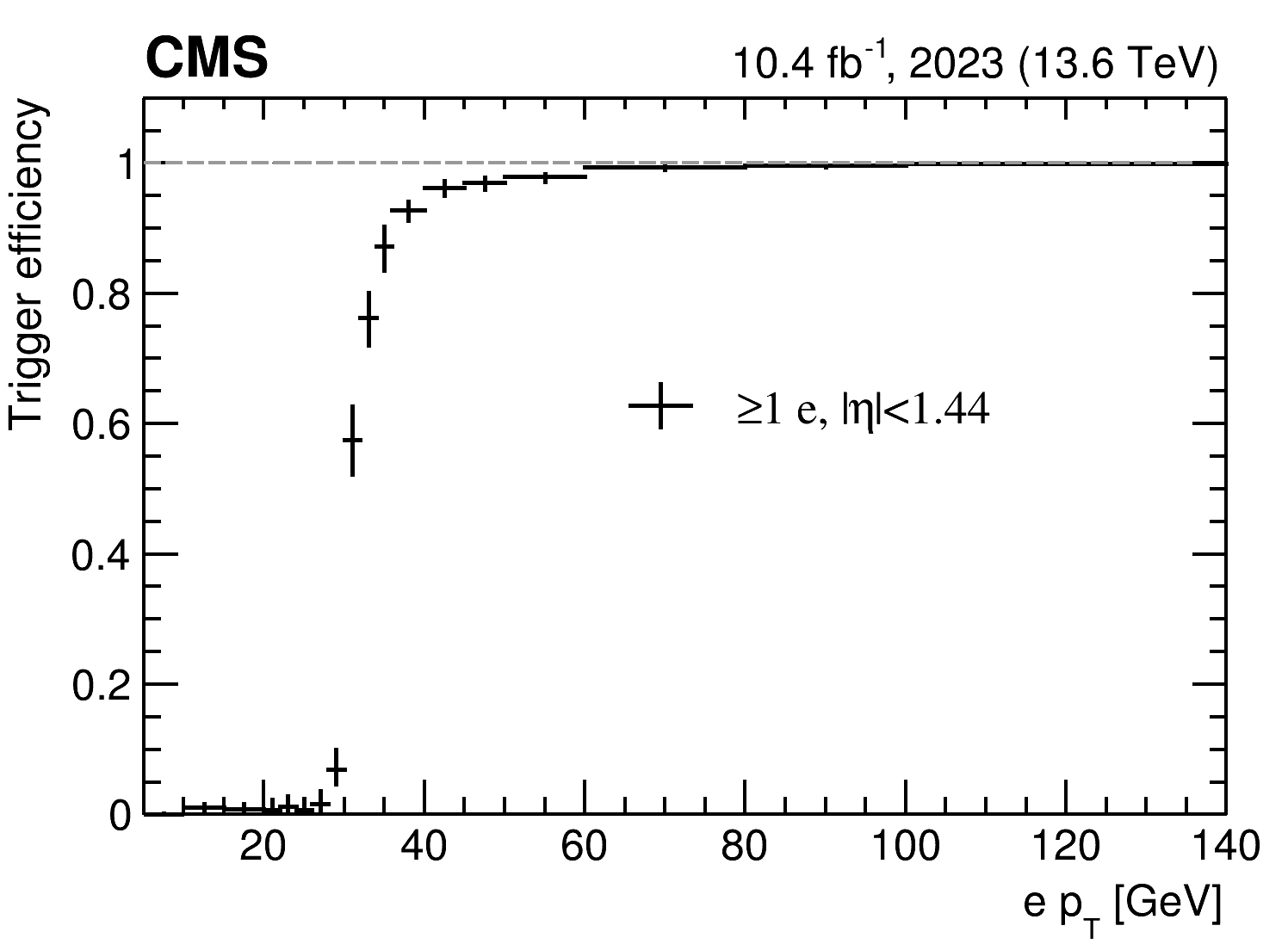}
  \includegraphics[width=0.49\textwidth]{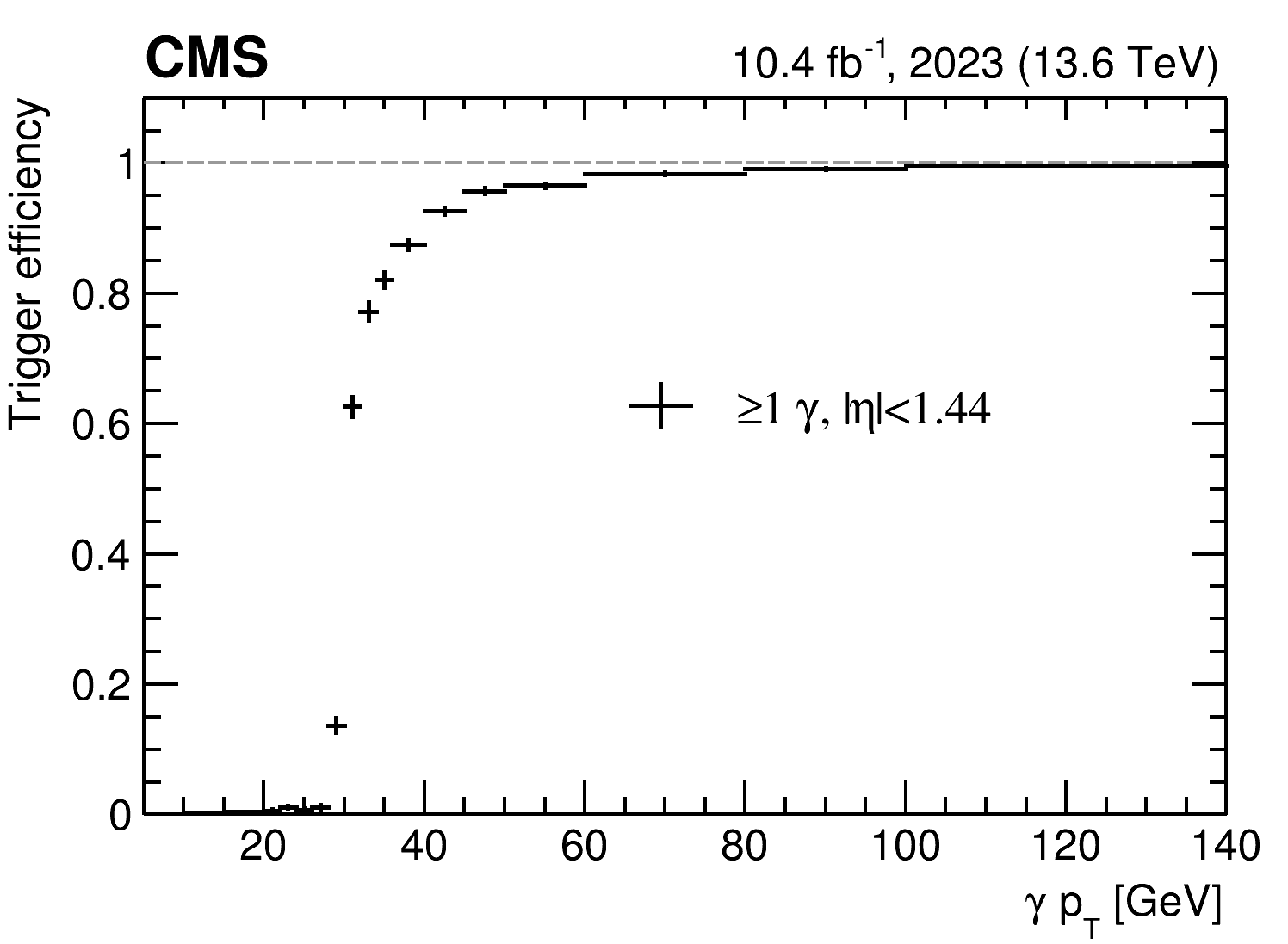}
  \caption{Trigger efficiencies for the scouting trigger paths seeded by L1 algorithms targeting either single-electron events (left) or single-photon events (right), as a function of the respective object \pt reconstructed offline. To be considered for scouting, the leading electron or photon must have ${\pt>30\GeV}$. Results are only shown for electrons or photons detected in the barrel region. 
  }
  \label{fig:run3_el_pt_trig_eff}
\end{figure*}

Figure~\ref{fig:run3_el_pt_trig_eff} shows the trigger efficiencies for electrons and photons in events triggered by the single-\egamma trigger as a function of the offline object \pt. The trigger efficiency 
increases sharply for ${\pt \approx 30\GeV}$
and reaches a plateau with ${>}90\%$ efficiency for ${\pt > 45\GeV}$. The low-energy reach for photon-triggered events in the scouting collection is therefore much improved compared to the 200\GeV (nonisolated) and 110\GeV (barrel only, isolated) thresholds in the standard trigger paths~\cite{CMS:2021zxu}. The trigger threshold for single-electron events in the scouting collection is at the minimum L1 threshold for triggering ECAL energy deposits. A minimal identification criteria is applied at the trigger level for both single-electron and single-photon paths. The typical offline criteria employed for physics analyses are expected to be much tighter than these selections. The scouting strategy thus maximizes the trigger efficiency for events with single electrons and photons in the CMS detector at the lowest energies.

The ability to perform physics studies with a combination of physics objects is new in Run~3 scouting. The reconstruction of scouting electrons and photons in paths seeded by L1 muons, jets and \HT becomes efficient at lower \pt thresholds compared to the thresholds required by trigger algorithms that exclusively target these objects, which must ensure that the trigger rates remain affordable. The scouting reconstruction efficiency of electrons and photons is identical to the one from the online HLT reconstruction.

\begin{figure*}[!htb]
  \centering
  \includegraphics[width=0.8\textwidth]{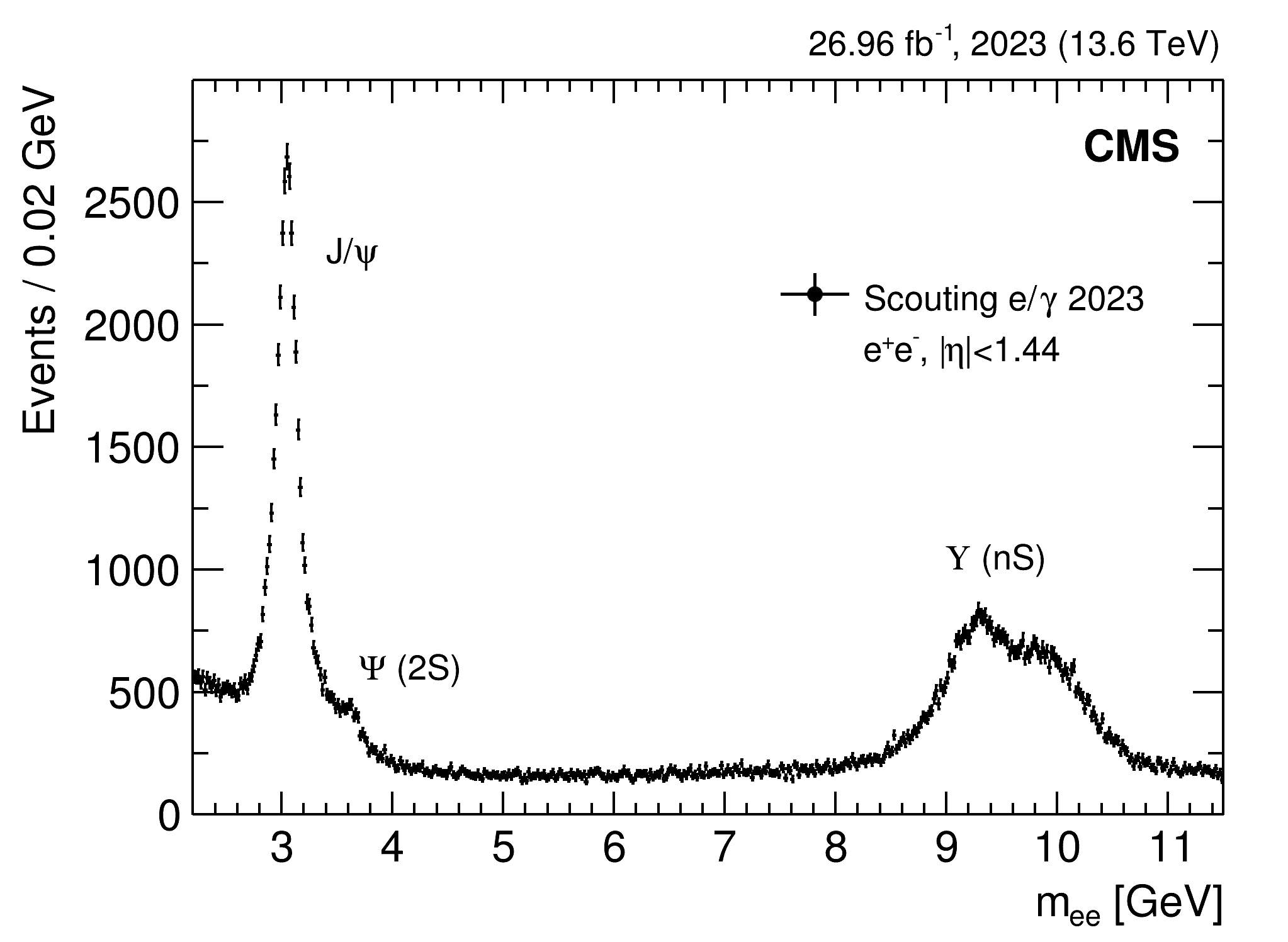}
  \caption{Dielectron mass distribution observed with Run~3 scouting data collected during the 2023 data-taking period. The \PJGy and two of the \PGU meson peaks are visible.
  }
  \label{fig:run3_el_invm}
\end{figure*}

The high quality of scouting electron objects is demonstrated by the ability to resolve decays of light mesons  (${\mEE < 12\GeV}$) to an electron-positron pair.
Figure~\ref{fig:run3_el_invm} shows the scouting dielectron mass spectrum where the \PJGy, \Pgy, and two of the resolved \PGU meson peaks (1S and 2S) are visible. Events from the single- and double-\egamma scouting trigger paths collected during the 2023 data-taking period were combined for Fig.~\ref{fig:run3_el_invm} including selections and corrections as follows. An electron-positron pair, each with ${\pt > 12\GeV}$, is required to pass an identification selection developed for mesons decaying to such pairs. To maximize the resonance signal over the background, both electrons are required to pass a tight identification selection. The reconstructed energy of the electron was corrected based on its position and shower shape in the ECAL using corrections derived from simulation. As a result, the peaks are shifted by ${<}2\%$ with respect to the actual mass of the corresponding SM mesons. The reconstructed dielectron mass resolution for the \PJGy peak is approximately 3\%. Further calibration of the scouting electrons could make it possible to improve the resolution, and will depend on the physics analysis under consideration.

The enhanced scouting program of CMS will play a pivotal role in carrying out low-mass searches and precision measurements during Run~3. The addition of pixel tracks will notably enhance the capability to conduct efficient searches and pioneer novel analysis strategies. This evolution will not only extend the results achieved during Run~2, but also broaden the data scouting physics program of CMS.

\section{Data parking in Run~1 and Run~2}
\label{ch:parking_run2}
The data parking strategy addresses key computational challenges that impede prompt event reconstruction of all the data collected by the experiment. In Run~1, numerous physics processes were considered in the original parking approach, which is discussed in Section~\ref{sec:run1altParking}. In Run~2, the focus of the parking strategy shifted to the \PB physics program, apart from specialized backup parking data sets collected in 2016 and 2017 that remain unprocessed. 
Section~\ref{sec:bparking_run2} describes in detail the physics motivation and trigger strategy of the \bparking campaign in Run~2. Finally, Section~\ref{sec:physics-2018} presents an overview of the physics results obtained with this unique \bparking data set.

\subsection{Data parking in Run 1}
\label{sec:run1altParking}

An overview of the data parking triggers in 2012, which stored events at a total rate of about 300--350\unit{Hz} with $\Linst \approx \sci{4}{33}\invcms$, is shown in Table~\ref{tab:run1_parking}. The initial parking strategy focused on dijet and double-\PGt triggers for VBF topologies and Higgs boson measurements, multijet and single-photon plus \ptmiss triggers to target compressed SUSY and DM models, and various dimuon triggers for \PB physics measurements.

\begin{table*}[!htb]
    \centering
    \topcaption{Summary of the main physics targets of the Run 1 parking strategy and the corresponding parking triggers in 2012. The average HLT rates reserved for Higgs boson measurements, \PB physics measurements, and new-physics searches are also quoted for ${\Linst \approx \sci{4}{33}\invcms}$.} 
    \renewcommand{\arraystretch}{1.3}
    \begin{tabular}{lcc}
    Physics motivation & Parking triggers & Average HLT rate [Hz] \\
    \hline
    VBF topologies & Dijet (large $\Delta \eta_{jj}$) &  \multirow{2}{*}{
            $\left\}\begin{array}{l} 150 \end{array}\right.$}  \\
    Higgs boson measurements & Double \tauh &  \\
    \hline
    B physics measurements & Double $\mu$ & 95 \\
    \hline
    Multijet searches & Four-jet & \multirow{4}{*}{
            $\left\}\begin{array}{l} \\ 105 \\ \\ \end{array}\right.$} \\
    DM and dark photons & \ptmiss + \PGg &\\
    DM production & Razor variables &\\
    SUSY hadronic searches & \HT+ $\alpha_{\text{T}}$  &  \\
    \end{tabular}
    \label{tab:run1_parking}
\end{table*}

A brief overview of the CMS results published with the Run~1 parking triggers is provided in the following paragraphs.

\subsubsection{Higgs boson measurements}

The distinctive pattern of VBF production, characterized by two jets with considerable angular separation in pseudorapidity and a high dijet invariant mass, was exploited for dedicated parking triggers that collected data in 2012, corresponding to \Lint of about 18.3\fbinv. This inclusive trigger required two jets with \pt greater than 35 and 30\GeV, respectively, ${\abs{\Delta\eta_{jj}}>3.5}$, and ${\mjj>700\GeV}$.
Two analyses were performed assuming a VBF production mode: a search for invisible Higgs boson decays~\cite{CMS:2016dhk}, and a search for the SM Higgs boson decaying to a bottom quark-antiquark pair~\cite{CMS:2015ebl_run1parking_Hbb}. In both cases, the analysis sensitivity was significantly enhanced by the use of triggers recorded in the parked data stream.
This trigger strategy has been resumed and refined in Run~3, as discussed in Section~\ref{subsec:run3parking_vbf}. 

Similarly, new trigger paths with lower thresholds on the object transverse momentum were exploited to target both hadronic and leptonic decays of the tau lepton. In particular, the requirement of two isolated \tauh objects, each with ${\pt>35\GeV}$ and ${\abs{\eta}<2.1}$ enhanced the search for ${\PH \to \PGt\PGt}$ decays in both SM~\cite{CMS:2014wdm_run1parking_Htautau} and BSM~\cite{CMS:2014ccx_run1parking_MSSMHtautau} scenarios.

\subsubsection{Dimuon final states for \texorpdfstring{\PB}{B} physics}

Final states with two muons provide a clean and distinctive experimental signature to identify interesting and rare processes involving for example \PQb hadron, charmonium, or bottomonium decays. Inclusive low-\pt dimuon triggers without any mass constraint are ideal to cover a wide breadth of \PB physics analyses with maximum acceptance and simplicity, and were thus employed in CMS starting from 2011. In 2012, the instantaneous luminosity delivered by the LHC doubled relative to 2011, reaching ${\sci{8}{33}\invcms}$ and saturating the prompt event processing capacity of CMS. Dimuon triggers were nevertheless retained in 2012 but the reconstruction of collected events was postponed to 2013, during the LS1 of the LHC.
This parked dimuon data set collected in 2012 allowed CMS to make important contributions to the accurate measurements of various \PQb quark hadron lifetimes, including the rare and heavy \PBc meson and the \PBs meson in two final states corresponding to different mass eigenstate admixtures~\cite{CMS:2017ygm}. Moreover, CMS contributed to the measurement of angular parameters of the decays ${\PBz \to \PKstz \PGmp \PGmm}$~\cite{CMS:2015bcy,CMS:2017rzx}, ${\PBp \to \PKp \PGmp \PGmm}$~\cite{CMS:2018qih}, and ${\PBp \to \PKstp \PGmp \PGmm}$~\cite{CMS:2020oqb}, which are sensitive to new physics contributions in processes described by ``penguin'' diagrams and relevant to the ongoing puzzle surrounding flavor anomalies~\cite{doi:10.1146/annurev-nucl-102020-090209}.
In Run~2, inclusive dimuon triggers 
could not be maintained anymore because both the instantaneous luminosity and the center-of-mass energy nearly doubled relative to 2012. To keep the rate under control, these triggers were replaced by more complex and restrictive triggers, which were used to collect data in the prompt data sets. Additional requirements beyond the presence of two muons were applied, such as additional tracks to form three- or four-body vertices, dimuon mass requirements, higher dimuon \pt thresholds, and displacement conditions. Starting from 2022, thanks to a different resource allocation and to the enhanced capacity of the CMS DAQ and computing systems in Run 3, inclusive dimuon triggers were reinstated, as discussed in Section~\ref{sec:dimuon-2022}.

\subsubsection{Searches for BSM physics}

New trigger paths were introduced in the Run~1 data parking stream to recover sensitivity to new physics models in regions of phase space not covered by the standard trigger paths. The collection of parked data sets targeted three main physics cases: multijet searches, monophoton searches, and SUSY hadronic searches. 

A four-jet trigger with loose thresholds on the transverse momentum of the jets, lowered to 45--50\GeV, was designed to target multijet searches and look for the top quark superpartner (top squark, or stop) predicted by natural SUSY models~\cite{cmsstop8}. A subset of the 2012 data corresponding to 12.4\fbinv was parked and used to extend the search in the mass region below 300\GeV, taking advantage of the lower jet \pt threshold.

A new parking trigger designed to extend the physics reach of monophoton searches in the low photon \pt and low-\ptmiss phase space was also introduced. It required a low-\pt photon with thresholds of 22 and 30\GeV at the L1 and at the HLT, respectively, and \ptmiss of at least 35\GeV, reducing the thresholds on these reconstructed objects by a factor of 2--3 compared to the standard triggers. Parking data collected at 8\TeV and corresponding to ${\Lint = 7.3\fbinv}$ were used to set limits on the exotic decays of the SM Higgs boson, and results were interpreted in the context of dark photon and dark matter pair production models~\cite{CMS:2015ifd_run1parking_HundetectedGamma}.

For dark matter searches, new algorithms were designed to collect events with a large momentum imbalance, requiring at least two jets and no isolated leptons. Dedicated ``razor'' variables~\cite{Rogan:2010kb,CMS:2011xie} were computed from the momenta of the two leading jets and the \ptmiss in the event, in order to quantify the transverse momentum balance of the jet momenta. Trigger paths with loose requirements on these kinematic variables were introduced in the data parking stream. Parked data corresponding to ${\Lint = 18.8\fbinv}$ were collected at 8\TeV and enabled the exploration of events with moderate jet \pt, thereby improving the sensitivity to direct dark matter production~\cite{CMS:2016gox_run1parking_razor_EX0-14-004}.

Dedicated trigger algorithms that relied on the dimensionless variable $\alpha_{\text{T}}$ were used for SUSY searches in final states with jets and \ptmiss~\cite{CMS:2016rjk_run1parking_SUS-14-006_alphaT}. This kinematic variable is computed from the system of the two leading jets in the event. It is defined as the ratio between the transverse energy of the less energetic jet and the transverse mass of the dijet system, and it is used to discriminate between events with genuine \ptmiss associated with unobserved particles (\eg, neutralinos) and spurious values of \ptmiss arising from jet energy mismeasurements (\eg, QCD multijet background). The data sample, corresponding to ${\Lint = 18.5\fbinv}$, was used to search for evidence of SUSY models involving the pair production of top squarks. 
Parking data was recorded with a lower \HT threshold, extending the acceptance to a wide array of compressed-SUSY models, where the top squark and the lightest neutralino (a DM candidate) are nearly degenerate in mass.

\subsection{Data parking for \texorpdfstring{\PB}{B} physics in Run 2}
\label{sec:bparking_run2}

This section details the main data parking strategy in Run~2, which focused on \PB physics. The physics motivation, experimental challenges, trigger strategy, and the performance of the parked triggers are all discussed in the following subsections. The physics results obtained with this approach are presented in Section~\ref{sec:physics-2018}.

\subsubsection{Physics motivation}

At the present time, several measurements of observables related to rare \PQb hadron decays present some tension with respect to their predicted
values from the SM. Collectively, these measurements are known as the ``\PB flavor anomalies'' and they are being interpreted by many in the physics community as potential evidence for 
BSM physics~\cite{doi:10.1146/annurev-nucl-102020-090209,Hurth:2023jwr}. These anomalies have been observed throughout the last decade in both charged-current \btoclnu and neutral-current \btosll transitions by the BaBar~\cite{BaBar:2001yhh}, Belle~\cite{Belle:2000cnh}, and LHCb~\cite{LHCb:2008vvz} Collaborations.

The anomalous measurements can be divided into two categories of physics observables. First, there are those relating to (differential) branching fractions and the parametrization of four-body angular distributions for decays via the \btosmm transition. Second, there are observables constructed from ratios of branching fractions for semileptonic decays with final states that differ only by the lepton flavor. Several ratios \RX can be measured, where X represents the final-state hadron produced in the semileptonic decay.

The \PB flavor observables are particularly powerful probes of BSM physics because of the availability of both precise theoretical predictions and clean experimental signatures for processes involving semileptonic (and fully leptonic) decays of \PQb hadrons. For instance, the \RX observables are sensitive to the violation of an accidental symmetry within the SM, known as lepton flavor universality (LFU), whereby the interactions between the gauge bosons and charged leptons are identical for all three lepton generations (beyond kinematical effects due to their differing masses). Confirmation of LFU violation would be a striking proof of the existence of BSM physics. 
In recent years, several key observables have received significant attention, some examples of which are given below.

The branching fraction ${\BF(\bstomm)}$ is an excellent probe to test the flavor sector of the SM, given its precise theoretical prediction and clean experimental signature. Furthermore, possible modifications of ${\BF(\bstomm)}$ relative to the SM expectations can be related to the same new physics operators responsible for LFU violation.
The ATLAS~\cite{ATLAS_2008}, CMS~\cite{CMS:2008xjf}, and LHCb Collaborations have reported several independent measurements~\cite{CMS:2022mgd, LHCb:2021awg, CMS:2019bbr,  ATLAS:2018cur}, as well as combined measurements~\cite{CMS:2014xfa}, in recent years. These measurements constitute one of the cleanest inputs to global fits aimed at providing a coherent global interpretation of the flavor anomalies~\cite{doi:10.1146/annurev-nucl-102020-090209}.

In an effective field theory framework, the angular distributions of three- and four-body decays arising from \btosll transitions offer sensitivity to new-physics operators. Multiple measurements of the \PfivePrime observable~\cite{Descotes-Genon:2012isb}, constructed from the Wilson coefficients associated with these operators and optimized to mitigate the impact of QCD uncertainties, have been conducted using the \BKstmm process. Most of these measurements have indicated tensions with the SM since 2013~\cite{LHCb:2013ghj, LHCb:2015svh, LHCb:2020lmf, CMS:2017rzx}. The theoretical predictions are affected by the limited knowledge of long-distance charm loop contributions, which might enhance the apparent discrepancy with the SM. Discussion of the recent progress in this area can be found in Refs.~\cite{Ciuchini:2021smi,Gubernari:2023puw,Gubernari:2024ews}.

The ratios of branching fractions $\RD = \BF(\btodtaunu) / \BF(\btodlnu)$ and $\RDst =\BF(\btodsttaunu) / \BF(\btodstlnu)$ $(\Pell = \Pe, \PGm)$ involve tree-level \btoclnu transitions. Two further observables are $\RK = \BF(\BKmm) / \BF(\BKee)$ and $\RKst = \BF(\BKstmm) / \BF(\BKstee)$, which involve loop-level \btosll transitions. Numerous measurements for \RDstar~\cite{BaBar:2012obs, BaBar:2013mob, Belle:2015qfa, LHCb:2015gmp} and \RKstar~\cite{LHCb:2014vgu, LHCb:2017avl, LHCb:2019hip, BELLE:2019xld, Belle:2019oag} have been reported since 2012, culminating in a reported evidence of LFU violation for \RK in 2022~\cite{LHCb:2021trn}. 
Most recently, the LHCb Collaboration has provided combined measurements for \RD and \RDst~\cite{LHCb:2023zxo}, and for \RK and \RKst~\cite{LHCb:2022qnv, LHCb:2022vje}, which are now consistent with the SM at the level of 1.9 and 0.2$\sigma$, respectively, and the latter result supersedes the one reported in Ref.~\cite{LHCb:2021trn}. Regardless of this recent dilution of a pattern of anomalous behavior, there remains the potential for LFU violating processes and there is still substantial interest from the physics community for new results pertaining to the \PB flavor anomalies.

The CMS Collaboration recorded a unique data set of \pp collisions at ${\sqrt{s} = 13\TeV}$ in 2018 with the primary aim of extending its program of LFU tests. Dedicated trigger and data storage strategies were developed to record a large sample of events containing 10~billion unbiased \PQb hadron decays. The reconstruction of physics events from the raw data sample was delayed until computing resources were available in 2019. The trigger and data storage strategies (known henceforth as ``\PB parking''), the defining characteristics of the resulting data set, and some of the key physics results are described in the following sections.

\subsubsection{Experimental challenges}
\label{sec:challenges-2018}

Prior to 2018, measurements targeting the \PB flavor anomalies within the \PB physics program of CMS were restricted to observables involving dimuon final states, a consequence of the available dimuon trigger algorithms. Examples of measurements from CMS include ${\BF(\bstomm)}$ \cite{CMS:2022mgd} and several angular observables~\cite{CMS:2017rzx, CMS:2018qih, CMS:2020oqb}. These successes can be contrasted with the absence of measurements of observables such as \RD and \RDst, which rely on the reconstruction of single-muon final states resulting from the \btoclnu transitions. Prior to 2018, the single-muon trigger algorithms were typically geared towards high-\pt physics processes, such as \PW boson production. The typical kinematical (${\pt > 20\GeV}$) and topological (isolation from neighboring particles) requirements on the muon suppressed the acceptance to \btomux decays. Furthermore, prior to 2018, no trigger algorithms provided adequate fiducial acceptance to final states containing electron pairs that arise from rare \btosee transitions; typically, the single- and dielectron trigger algorithms imposed \pt thresholds of 30 and 20\GeV, respectively, as well as isolation requirements. These constraints have thus far prohibited the measurement of observables such as \RK and \RKst.

A new trigger strategy was devised and implemented in time for the LHC \pp collision run of 2018. The trigger strategy relies on the accumulation of a very large sample of \PQb quark-antiquark (\bbbar) pairs using a ``tag-side'' trigger logic that identifies the semileptonic decay of one of the \PQb hadrons to a final state containing at least one displaced muon above an evolving \pt threshold in the range 7--12\GeV. Decays via the transition \btoctomux have a total branching fraction of 18\%~\cite{Workman:2022ynf} and thus approximately one in three \bbbar events results in a final state containing at least one muon. The other ``probe-side'' \PQb hadron is able to decay to all possible final states (including any flavor of lepton) with minimal kinematic bias from the tag-side trigger requirements. Thus, the study of \PQb hadron decays that lead to final-state muons can focus on the tag-side muon candidate identified by the trigger system, while the study of processes involving other lepton flavors or nonleptonic final states can rely on the probe-side decays. Rare processes with branching fractions as small as $\mathcal{O}(10^{-7})$ are accessible if the sample of \bbbar pairs is sufficiently large, \eg, $\mathcal{O}(10^{10})$. 

Early feasibility studies, based on simulated data, demonstrated that a suitable data sample could be accumulated if high trigger rates and purities could be sustained throughout the 2018 data-taking period. The number of \bbbar pairs $N_{\bbbar}$ produced at the LHC and subsequently recorded by the trigger logic was estimated with the following expression: ${N_{\bbbar} = t_{\text{LHC}}\mathcal{R}P_{\bbbar}}$, where $t_{\text{LHC}}$ is the LHC operational running time (seconds), $\mathcal{R}$ is the rate in Hz at which the trigger logic returns a positive decision, and $P_{\bbbar}$ is the purity of the resulting data stream, defined as the ratio of the numbers of genuine \bbbar pairs and \pp collision events recorded by the trigger system. The value of $t_{\text{LHC}}$ is $\sci{6}{6}\unit{s}$ when assuming 140\unit{days} of LHC operations over a 6-month period and a beam duty cycle of 0.5. Hence, by maintaining an average trigger rate $\mathcal{R} \approx 2\unit{kHz}$ throughout each LHC fill, and assuming a purity $P \approx 80\%$, it is possible to accumulate a data sample comprising \sci{1.2}{10} events that contain $10^{10}$ \bbbar pairs. Data parking is thus required to handle the high trigger rates, and the strategy was based on experience accumulated during Run~1, as detailed in Section~\ref{sec:run1altParking}.

Equivalently, $N_{\bbbar}$ can be estimated from the expression ${N_{\bbbar} = \Lint \sigma_{\bbbar} \BF^\prime\ \varepsilon}$, where the data recorded by the muon-based triggers correspond to ${\Lint = 41.6\fbinv}$ (described later in Section~\ref{sec:characteristics}), the inclusive \bbbar cross section is ${\sigma_{\bbbar} = \sci{4.7}{11}\unit{fb}}$ at ${\sqrt{s} = 13\TeV}$~\cite{Cacciari:1998it, Cacciari:2001td,Cacciari:2012ny, Cacciari:2015fta}, the term ${\BF^\prime = 1-(1-\mathcal{B)}^2}$ accounts for the fact that either \PQb hadron can decay into a muon and the branching fraction for the \btoctomux decay is ${\BF = 18\%}$~\cite{Workman:2022ynf}, and the efficiency of the muon-based triggers to record an event containing at least one \btomux decay is ${\varepsilon \approx \sci{2}{-3}}$. These values yield ${N_{\bbbar} \approx 10^{10}}$.

A precise measurement of the \RDst observable is feasible by reconstructing the \btodstmunu and \btodsttautomunu decays using the large sample of $10^{10}$ tag-side muons: the sample is expected to contain $\mathcal{O}(10^6)$ and $\mathcal{O}(10^4)$ candidates for these decays, respectively. Similar arguments for precision can be made for other observables involving muon-based signatures, such as \RD and the numerator of the \RK observable, \BF(\BKmm).

The limiting factor in the precision of a measurement of \RK is related to the number of reconstructed \BKee decays that can be identified unambiguously above background contributions. The number of \BKee decays found within the fiducial acceptance can be estimated from the expression ${N_{\BKee} \approx N_{\bbbar}f_{\PBp}\BF\mathcal{A}}$, where: ${N_{\bbbar} = 10^{10}}$, the fragmentation fraction is 
${f_{\PBp} = 0.4}$~\cite{Workman:2022ynf}; the branching fraction is ${\BF(\BKee) = \sci{4.5}{-7}}$~\cite{Workman:2022ynf}; and the fiducial acceptance, defined by the fraction of probe-side \PBp decays with all daughter particles satisfying the requirements ${\pt > 0.5\GeV}$ and ${\abs{\eta} < 2.5}$, is ${\mathcal{A} \approx 55\%}$. Thus, approximately 1000 probe-side \BKee decays are expected within the fiducial volume. Subsequent reconstruction and selection requirements will further reduce the sample of identifiable \BKee decays.

\subsubsection{Trigger strategy}
\label{sec:trigger-strategy-2018}

The primary aim of the trigger strategy described here is to maximize the number of recorded \bbbar events by maximizing the data stream purity and operating the trigger system close to its design limits. Crucially, this mode of operation must not compromise the availability of online resources for the core CMS physics program.

The new trigger strategy adopted a two-step optimization of the L1 and HLT trigger algorithms given the following key design constraints for  2018: the total L1 trigger rate was restricted to 90\unit{kHz} to maintain acceptable dead time from the subdetector readout systems; and the bandwidth of the \bparking data stream could not substantially exceed an average of 2\gbs because of the finite capacity of buffers at the CMS experimental site.

\subsubsection{The L1 optimization}
\label{sec:Bparking_L1_optimisation}

During an LHC fill, the \Linst slowly decreases with time.  As a consequence, the number of \pp interactions that occur within the same LHC bunch crossing, the L1 and HLT trigger rates, and the per-event HLT computational load are all observed to decrease with time. Hence, the availability of idle resources increases during the ongoing LHC fills, which can be leveraged by the trigger strategy described here.

\begin{figure*}[!htb]
  \centering
  \includegraphics[width=0.46\textwidth]{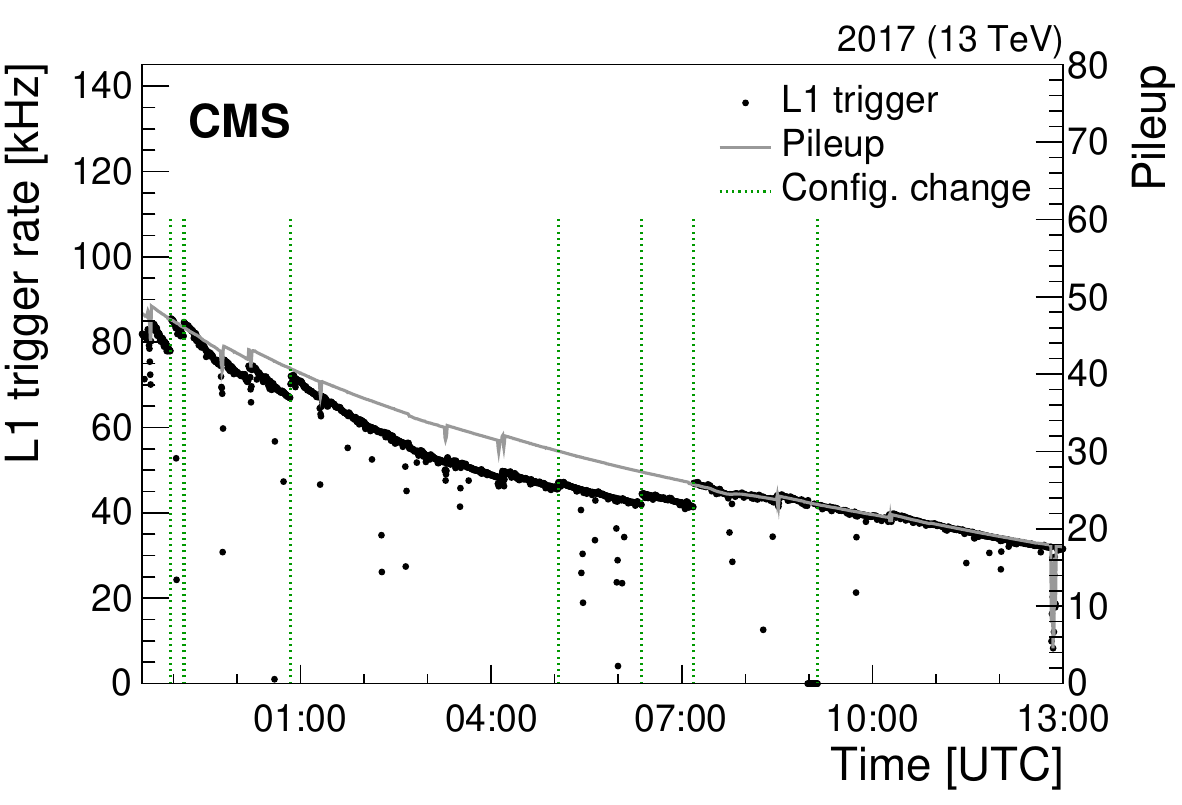}
  \includegraphics[width=0.46\textwidth]{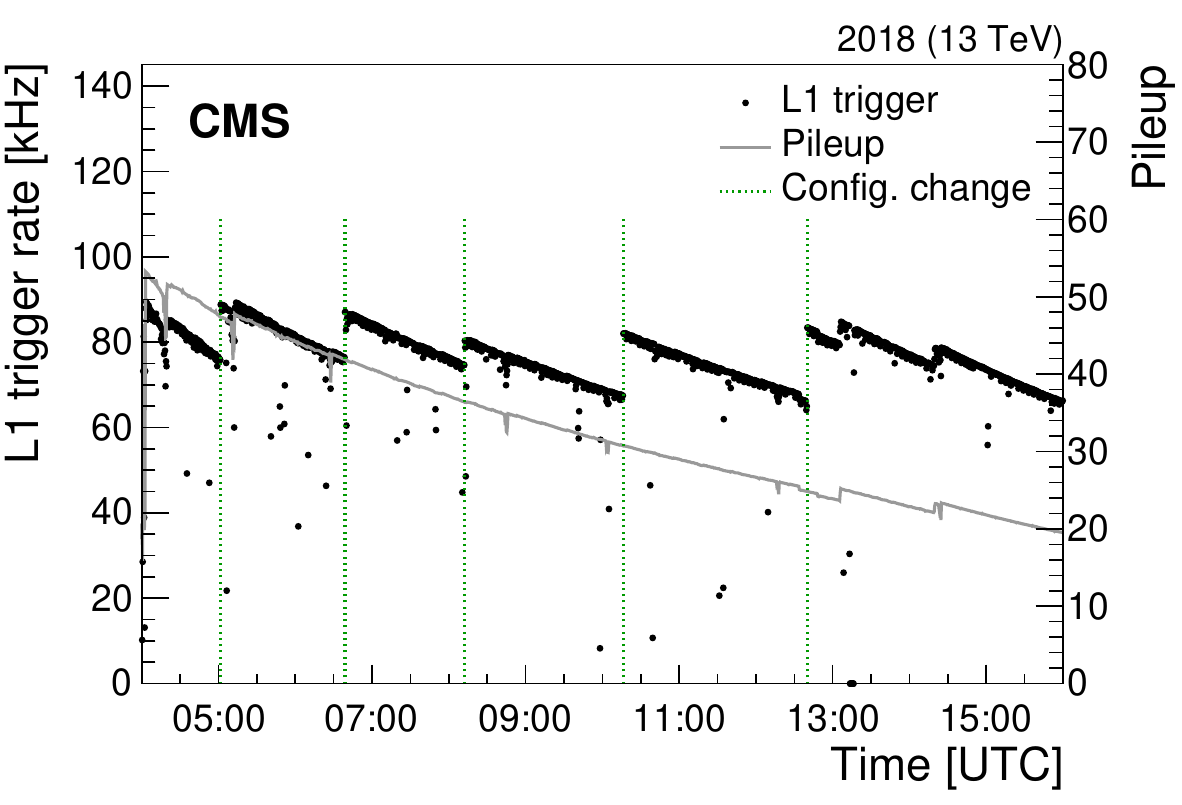}
  \caption{The L1 trigger rate and the amount of pileup as a function of time, shown for representative LHC fills during 2017 (\cmsLeft) and 2018 (\cmsRight). Occasional lower rates are observed due to transient effects, such as the throttling of the trigger system in response to subdetector dead time~\cite{CMS:2016ngn}. Changes in the trigger configuration are indicated by vertical green dashed lines.}
  \label{fig:CMS-L1-trigger-rates-2017-2018}
\end{figure*}

The \cmsLeft panel of Fig.~\ref{fig:CMS-L1-trigger-rates-2017-2018} shows the evolution of the total L1 trigger rate and pileup as a function of time during a typical LHC fill in the 2017 \pp collisions run, prior to the implementation of the trigger strategy discussed here. Over a period of 14.5\unit{hours}, the pileup value decreases from 48 to 18 and, correspondingly, the total L1 trigger rate also decreases.
At the beginning of an LHC fill, the L1 system typically operates at a total trigger rate of 90\unit{kHz}; towards the end of an LHC fill, there are up to several tens of kHz of spare-rate capacity available.

The new strategy for the L1 system repurposes existing algorithms to identify low-\pt muon candidates with high efficiency and purity. Simple kinematical requirements are used to identify interesting muon candidates. First, each muon candidate is required to be found centrally in the detector by satisfying ${\abs{\eta} < 1.5}$, where the L1 muon identification and momentum-resolution performance is generally optimal; this requirement simultaneously improves the trigger purity and enhances the fiducial acceptance for the probe-side \PQb hadron decays. Second, a variable \pt threshold is applied within the range of 12\GeV at ${\Linst = \sci{1.7}{34}\invcms}$ down to 7\GeV at ${\Linst = \sci{0.9}{34}\invcms}$. The threshold is progressively loosened within this range (via changes in the trigger configuration) as \Linst and the pileup decreases. Importantly, the threshold is tuned to ensure that the L1 system operates close to its design limit, \ie, at ${\approx}90\unit{kHz}$ throughout the LHC fill, but not beyond so as to keep dead time from the subdetector readout systems to below ${\approx}1\%$. Thus, there is negligible impact on the accumulated \Lint and the wider CMS physics program. While the evolution of the threshold improves the acceptance to \bbbar candidates, it also degrades the purity of the data stream from the L1 system. The average purity (in terms of the fraction of selected events containing a \bbbar candidate, based on studies of simulated data) is ${\approx}0.3$. The remainder of the events contain muons from the direct production of charm mesons and their semileptonic decays, muons from kaon or pion decays, and misidentified muons.

The \cmsRight panel of Fig.~\ref{fig:CMS-L1-trigger-rates-2017-2018} shows the L1 trigger rate as a function of time during a typical LHC fill in the 2018 \pp collisions run. While the average pileup during an LHC fill decreases similarly in 2017 and 2018, the trigger rates do not. The large instantaneous increases in the L1 trigger rate are coincident with changes in the trigger configuration. The total L1 trigger rate peaks at values close to 90\unit{kHz} throughout the LHC fill; this is because of an increasing rate contribution, as high as 40\unit{kHz}, from the L1 single-muon algorithm as the \pt threshold is reduced. At the start of each LHC fill, the trigger algorithms that serve the core physics program operate with a total L1 rate close to the 90\unit{kHz} ceiling; only once \Linst has dropped below $\sci{1.7}{34}\invcms$ are the dedicated triggers enabled, when sufficient online resources are available.

\subsubsection{The HLT optimization}

The purity of the data stream from the L1 system is substantially improved by the use of tailored muon algorithms at the HLT. The algorithms provide superior performance relative to the L1 logic, in terms of muon identification and momentum scale and resolution, because of the ability of the HLT software to reconstruct muons using tracking information from the silicon pixel and strip trackers.

The \bparking data throughput, given by the product of the HLT trigger rate and the triggered event size, was limited in 2018 to an average of 2\gbs for timescales longer than 24\unit{hours} because of the limited buffer capacity, as described in Section~\ref{sec:data-parking-2018}. The triggered event size has a linear dependence on \Linst and thus higher HLT trigger rates are accessible later during an LHC fill as both \Linst and the event size, with reduced pileup, decrease.

\begin{figure*}[!htb]
  \centering
  \includegraphics[width=0.47\textwidth]{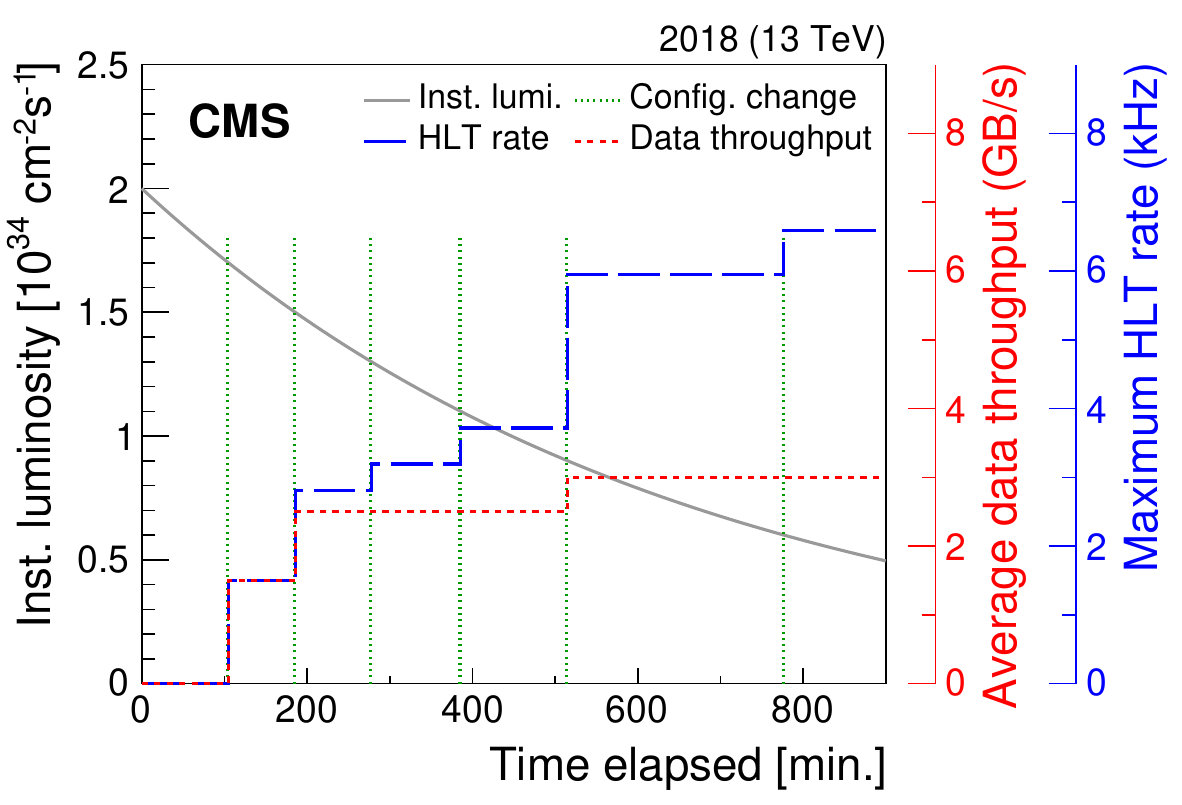}
  \includegraphics[width=0.47\textwidth]{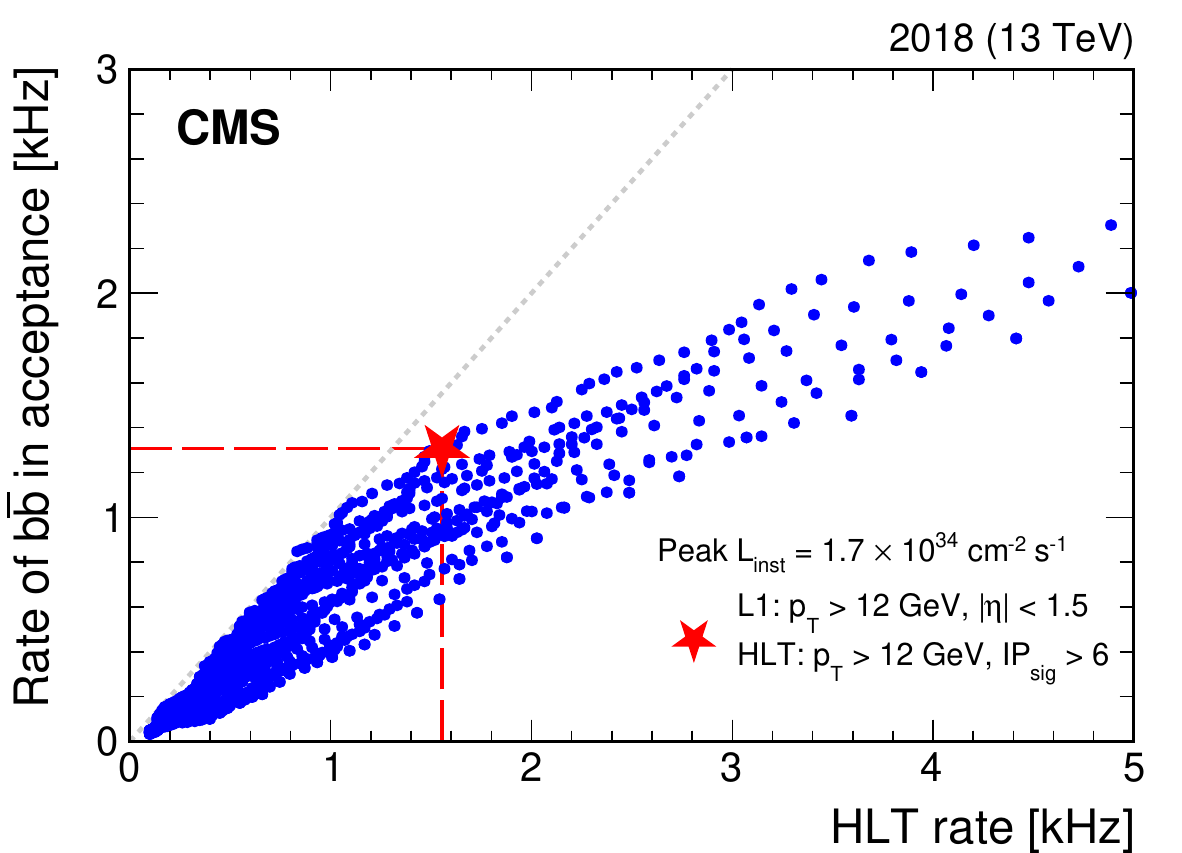}
  \caption{Left: an example scenario in which the \bparking data throughput per L1 trigger setting is adjusted to maintain an average of approximately 2\gbs throughout an LHC fill. The dotted red and dashed blue lines trace the \bparking data throughput and maximum allowed HLT rate, respectively, determined for each trigger configuration. Changes in the trigger configuration are indicated by vertical green dashed lines. The trigger logic is adjusted to operate close to the permitted HLT rate. Right: rate of \bbbar events in acceptance versus HLT rate for a parameter scan over \pt and \ipsig thresholds imposed in the HLT logic for  the \Linst and L1 requirements indicated in the legend; each point (blue circle) represents a unique pair of thresholds and the red star indicates the optimal pairing of ${\pt > 12\GeV}$ and ${\ipsig > 6}$ at a peak ${\Linst = \sci{1.7}{34}\invcms}$ and an HLT rate close to the maximum allowed value of ${\approx}1.5\unit{kHz}$. 
    \label{fig:parking-throughput-2018}}
\end{figure*}

Various scenarios involving different assumptions on the LHC performance and load-balancing of the DAQ system, in terms of varying data throughput during an LHC fill, were investigated. The \cmsLeft panel of Fig.~\ref{fig:parking-throughput-2018} shows an example scenario in which the data throughput is allowed to evolve during an LHC fill while ensuring that the average does not exceed 2\gbs. The left panel also indicates the maximum HLT trigger rate permitted for each L1 trigger configuration, which changes as a function of \Linst. The thresholds of the HLT trigger algorithms are therefore adjusted to operate close to these maximum values. This scheme, which allows for HLT trigger rates in the multi-\unit{kHz} range, is optimized for long LHC fills, of duration 15\unit{hours} or more.

The HLT algorithms are tuned to maximize the number of recorded \bbbar events and satisfy the trigger rate constraints. Two of the most discriminating variables are the muon \pt and the transverse impact parameter of the muon expressed in terms of its measurement significance, \ipsig, relative to the \pp luminous region. By requiring a nonzero \ipsig value, the relatively long lifetime of the \PQb hadron and the characteristic displacement (relative to the primary interaction point) of the muon from \btomux decays are leveraged to reduce prompt muon production from background processes such as \PD mesons and charmonium decay.

For each L1 trigger configuration, which changes as a function of \Linst, a parameter scan is performed across all feasible combinations of \pt and \ipsig thresholds imposed by the HLT algorithm. The L1 configuration is determined by the procedure described above. 
The right panel of Fig.~\ref{fig:parking-throughput-2018} shows the rate of \bbbar candidates found within the experimental acceptance and the corresponding HLT trigger rate obtained for each unique (\pt, \ipsig) combination at the peak ${\Linst = \sci{1.7}{34}\invcms}$. The optimal combination is one that lies along the upper boundary of the point set shown in the right panel of Fig.~\ref{fig:parking-throughput-2018} and is as close as possible but does not exceed the maximum allowed HLT trigger rate for the given \Linst value. For the example shown, the procedure indicates that the HLT thresholds ${\pt > 12\GeV}$ and ${\ipsig > 6}$ accumulate \bbbar candidates at a rate of 1.3\unit{kHz} at a trigger rate of 1.5\unit{kHz}, which corresponds to a trigger purity of 85\%. The procedure is repeated for each L1 trigger configuration, which varies with \Linst during an LHC fill.

\begin{figure*}[!htb]
  \centering
  \includegraphics[width=0.6\textwidth]{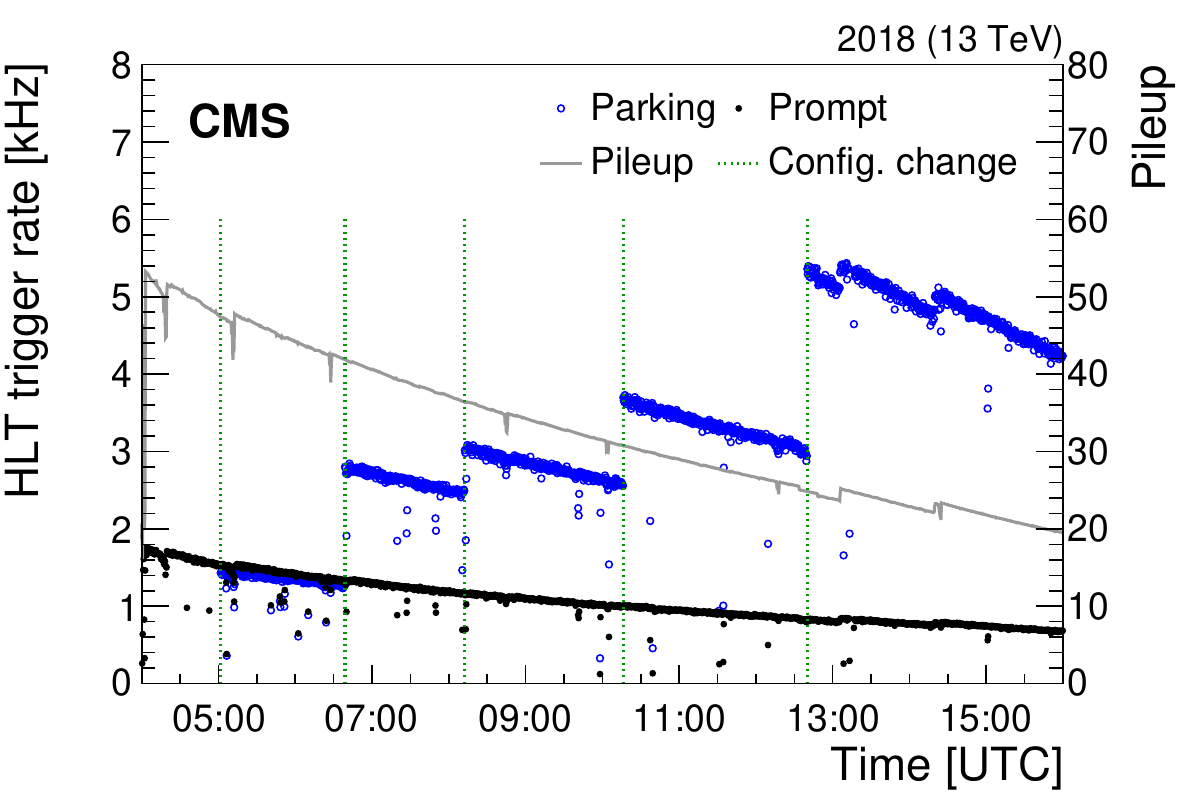}
  \caption{HLT trigger rates and the number of pileup events shown as a function of time during a representative LHC fill in 2018. The rates for the promptly reconstructed core physics (black solid markers) and \bparking (blue open markers) data streams are shown separately. Occasional lower rates are observed due to transient effects, such as the throttling of the trigger system in response to subdetector dead time~\cite{CMS:2016ngn}. Changes in the trigger  configuration are indicated by vertical green dashed lines. 
  }
  \label{fig:CMS-HLT-trigger-rates-2018}
\end{figure*}

As a result, the lower-bound thresholds on both \pt and \ipsig are relaxed, within the ranges 7--12\GeV and 3--6, respectively, as \Linst decreases during an LHC fill.
Table~\ref{tab:CMS-trigger-rates-2018} summarizes the evolution of the trigger thresholds used to record \pp collision data during the LHC fill shown in Figs.~\ref{fig:CMS-L1-trigger-rates-2017-2018} and~\ref{fig:CMS-HLT-trigger-rates-2018}, as well as the resulting peak trigger rates and data stream purities. The settings are deployed sequentially via changes in the trigger configuration at different values of peak \Linst; each new setting corresponds to the enabling of new trigger logic with lower thresholds until the end of the LHC fill. No dedicated trigger logic is enabled for values of \Linst above $\sci{1.7}{34}\invcms$. Minor adjustments were made to these settings during 2018 in response to the evolving LHC performance. This intra-fill evolution of trigger settings maximizes the estimated number of \bbbar events recorded in the \bparking data stream. Furthermore, the recording of \bbbar events at higher rates towards the end of an LHC fill ensures that the pileup, and thus the additional activity, in these events is low.

\begin{table*}[!htb]
  \centering 
  \caption{Single-muon trigger settings used during a typical LHC fill. The kinematical thresholds are changed when \Linst (and consequently pileup) fall below the listed values. Also listed are the lower-bound thresholds on the tag-side muon \pt (L1 and HLT) and \ipsig (HLT only), the corresponding L1 and HLT peak trigger rates, and the HLT data stream purity estimated from simulated events. No dedicated trigger is enabled at the start of each LHC fill when \Linst is typically above $\sci{1.7}{34}\invcms$.}
  \renewcommand{\arraystretch}{1.3}
  \begin{tabular}{cccccccc}
    \Linst & Pileup  & L1 \PGm \pt  & HLT \PGm \pt
    & HLT \PGm & Peak L1 & Peak HLT & Purity \\
    $[10^{34}\invcms]$ & & $[\GeVns]$ & $[\GeVns]$ &
    \ipsig & rate [kHz] & rate [kHz] & [\%] \\
    \hline
    2.0 & 54.0 & \NA & \NA & \NA & \NA & \NA  & \NA \\
    1.7 & 45.9 & 12 & 12 & 6  & 20 & 1.5 & $92 \pm 5$ \\
    1.5 & 42.8 & 10 & 9  & 6  & 30 & 2.8 & $87 \pm 4$ \\
    1.3 & 35.1 & 9  & 9  & 5  & 32 & 3.0 & $86 \pm 4$ \\
    1.1 & 29.7 & 8  & 8  & 5  & 43 & 3.7 & $83 \pm 4$ \\
    0.9 & 24.3 & 7  & 7  & 4  & 53 & 5.4 & $59 \pm 3$ \\
  \end{tabular}
  \label{tab:CMS-trigger-rates-2018}
\end{table*} 

Figure~\ref{fig:CMS-HLT-trigger-rates-2018} shows the HLT trigger rates for both the promptly reconstructed data stream that serves the core CMS physics program, with a monotonically decreasing behavior, and the \bparking data stream provided by the single-muon trigger algorithms. In the latter case, rates as high as 5.5\unit{kHz} are obtained late in the LHC fill because of the relaxed kinematical and topological thresholds.

\subsubsection{Trigger purity}
\label{sec:trigger-purity-2018}

The purity of the data stream is estimated from simulated events to be in the range 60--90\%, depending on the trigger thresholds, as indicated in Table~\ref{tab:CMS-trigger-rates-2018}, with an average of ${\approx}80\%$. The estimates each have an associated uncertainty of 5\%, arising from sources such as the kinematical modeling of the \PB meson decays.

The average purity is also determined from an analysis of the data sample itself, by estimating the total number of \bbbar events contained in the sample by reconstructing \PDstp candidates from the production mode $\PBz \to \PDstp \PGmm \PAGn$ and the subsequent decay chain $\PDstp \to \PDz\PGp^+_{\text{soft}} \to (\PKm\PGpp)\PGp^+_{\text{soft}}$, where $\PGp_{\text{soft}}$ indicates a low-momentum pion. The decay mode via the \PDstp state is chosen for the purity measurement because of its large branching fraction and the fully reconstructable \PDstp decay chain.

\begin{figure*}[!htb]
  \centering
  \includegraphics[width=0.6\textwidth]{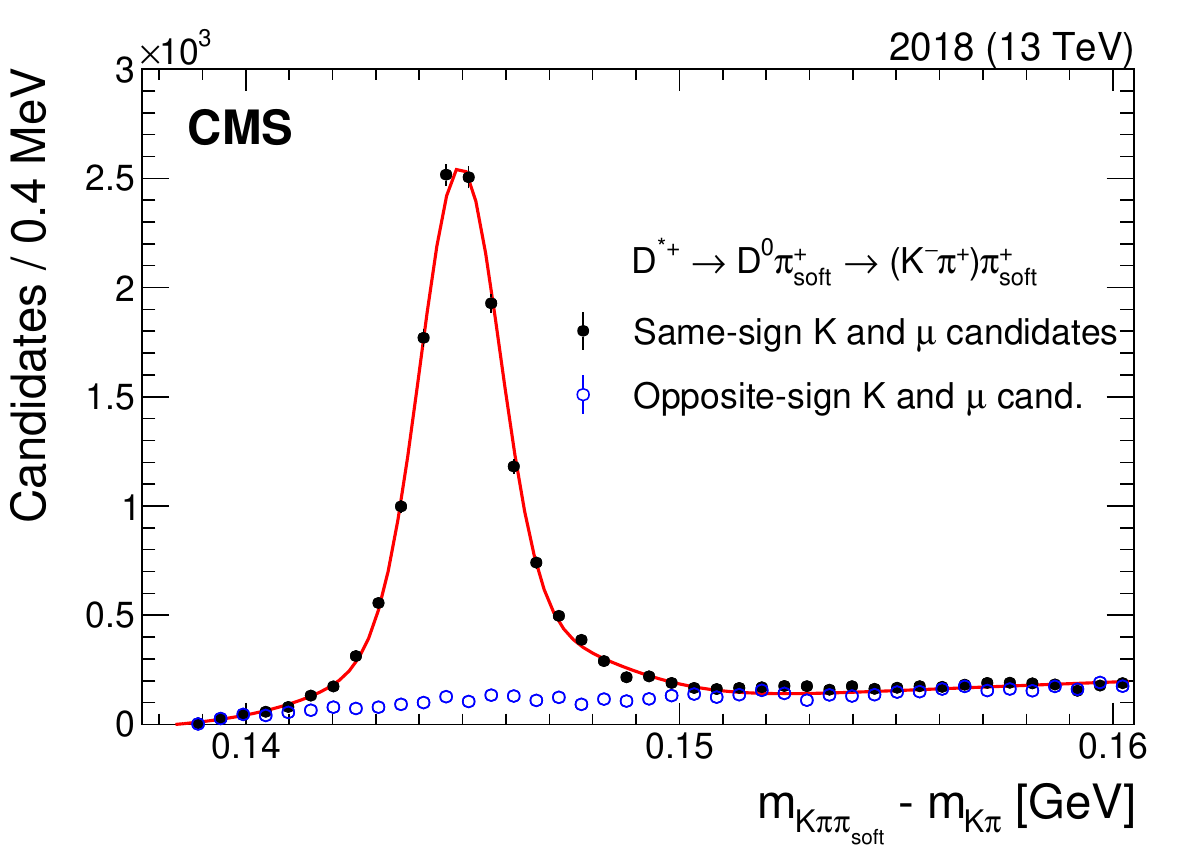}
  \caption{Mass difference between reconstructed \PDstp and \PDz candidates from the production mode $\PBz \to \PDstp \PGmm \PAGn$ and the subsequent decay chain $\PDstp \to \PDz\PGpp_{\text{soft}} \to (\PKm\PGpp)\PGpp_{\text{soft}}$. Events containing kaon and muon candidates with same-sign (opposite-sign) charges are indicated by solid (open) markers.}
  \label{fig:HLT-single-mu-trigger-purity}   
\end{figure*}

The method relies on extracting the number of \PDstp candidates by exploiting the mass difference between the reconstructed \PDstp and \PDz candidates, as shown in Fig.~\ref{fig:HLT-single-mu-trigger-purity}. The value of $\Delta m = m(\PKm\PGpp\PGp^+_{\text{soft}}) - m(\PKm\PGpp)$ is expected to peak at the mass difference between the \PDstp and \PDz mesons if the kaon and muon candidates have same-sign charges, whereas a smooth combinatorial background shape is expected for opposite-sign charges.

In order to identify the $\PBz \to \PDstp\PGmm\PAGn$ decay, the muon responsible for the positive trigger decision and neighboring charged-particle tracks are considered. A candidate $\PDz \to \PKm\PGpp$ decay is identified by considering pairs of oppositely charged tracks that form a vertex. The leading (subleading) track is required to satisfy $\pt > 5 (3)\GeV$ and quality criteria, and the vertex is subject to both displacement and quality criteria. The particle track with the same (opposite) electrical charge as the muon is assigned the mass of the kaon (pion). Only \PDz candidates with a reconstructed mass within a ${\pm}30\MeV$ (corresponding to a ${\pm}3\sigma$) window of 1.86\GeV~\cite{Workman:2022ynf} are selected for further consideration. Finally, a candidate $\PDstp \to \PDz\PGpp_{\text{soft}}$ decay is identified by combining the selected \PDz candidate with a ``soft'' pion candidate (${\pt > 300\MeV}$).

The number of $\PBz \to \PDstp \PGmm \PAGn$ decays is obtained by performing a binned maximum likelihood fit to the mass difference distribution obtained from data using Gaussian and second-order polynomial functions for the same-sign and opposite-sign charge hypotheses, respectively. The yield obtained from the fit is then corrected to determine the number of \bbbar candidates $N_{\bbbar}$ contained in the data sample by accounting for differences in the fiducial acceptances and reconstruction efficiencies determined from simulation, and branching fractions for the \btomux decay and the $\PBz \to \PDstp \PGmm \PAGn$ decay chain. The procedure yields ${N_{\bbbar} \approx \sci{9}{9}}$ with an associated uncertainty of 5\%. Given that the number of recorded LHC events is $\sci{11.9}{9}$, the purity obtained from data is ${P_{\bbbar} = 0.75 \pm 0.04}$, which agrees with the estimate obtained from simulation.

\subsubsection{Data parking and processing}
\label{sec:data-parking-2018}

The \bparking data are not transferred immediately to the permanent data processing center, but, instead, are temporarily stored in local buffers at the CMS experimental site and later transferred unprocessed to permanent tape storage. The buffers are capable of handling a total data throughput of 2\gbs, limited by the maximum data transfer rate achievable from the buffer to the tape resources. The effective limit on data throughput is higher because of the LHC inter-fill downtime; for instance, a total of 3\gbs when averaged over the timescale of a week. Hence, a throughput of 2\gbs can be sustained for the \bparking data stream, in addition to an allocation of 1\gbs for the core CMS physics streams. 
Tape storage resources, originally allocated to a data parking stream of 500\unit{Hz} throughout Run~2 and sufficient to allow for primary and backup copies of the data, were reallocated in full to the \bparking campaign in 2018. Only a single copy of the \bparking data is kept on tape.

The data parking strategy exploits the opportunistic use of computational resources for the delayed processing (\ie, global event reconstruction) of very large event samples that would otherwise not be possible during the LHC running periods. This includes the short end-of-year and long end-of-run shutdowns of the LHC complex, when the resource load from the core CMS physics program is reduced. Three processing campaigns of the \bparking data set have been undertaken so far. An early ``pilot'' reconstruction campaign was performed in 2018 on a small fraction (${\approx}5\%$) of the data set to check the performance of the trigger strategy, via purity measurements (described in Section~\ref{sec:trigger-strategy-2018}), and to validate new reconstruction algorithms, such as the one described in this section. The first full processing of the parked data was performed between May and December 2019, during the LHC LS2. Finally, the full data set was reprocessed in 2022 with the ultimate ``legacy'' reconstruction software and calibrations that provide the optimal physics performance for future data analysis.

\subsubsection{\texorpdfstring{Low-\pt}{Low-pT} electrons}
\label{sec:lowpt-electrons}

\begin{figure*}[!htb]
  \centering
  \includegraphics[width=0.47\textwidth]{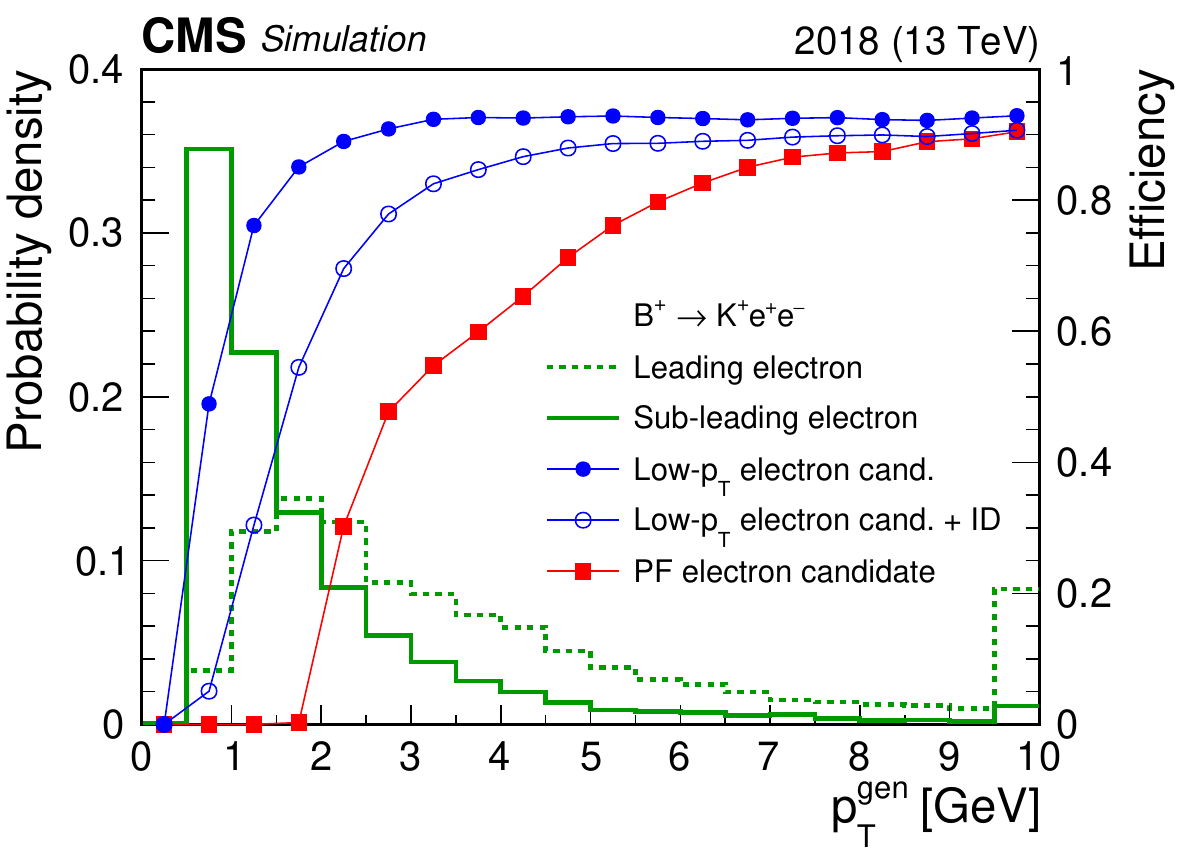}
  \includegraphics[width=0.47\textwidth]{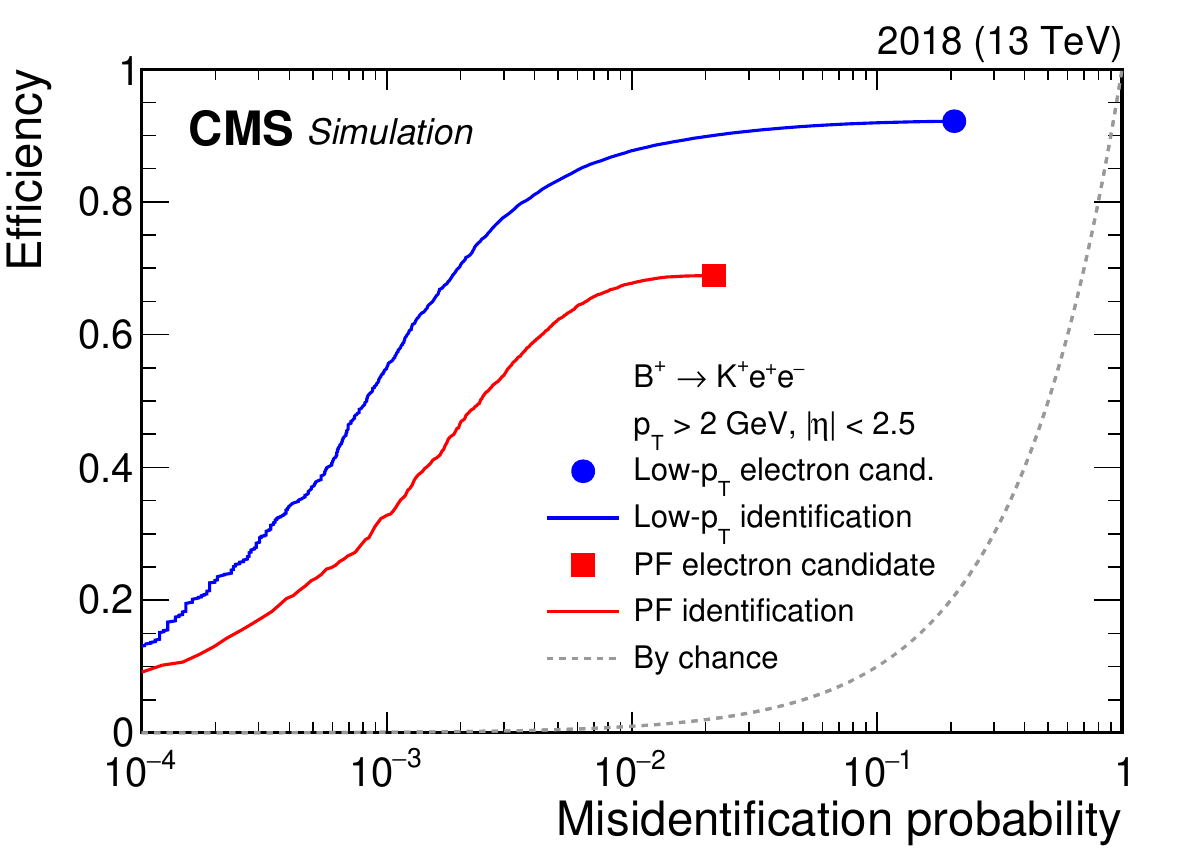}
  \caption{The \cmsLeft panel shows the \ptgen spectra of the leading and subleading electrons (dashed and solid green histograms) from \BKee decays, and the efficiency to identify genuine electrons as a function of \ptgen for PF (solid red squares) and low-\pt electron candidates (solid blue circles). Efficiencies for the low-\pt electron candidates that satisfy an ID score threshold, tuned to give the same misidentification probability as for PF electron candidates, are also shown (open markers). The \cmsRight panel shows the performance of the PF (solid red square) and low-\pt (solid blue circle) reconstruction algorithms and their corresponding ID algorithms (curves). The efficiencies and misidentification probabilities are determined relative to charged-particle tracks obtained from simulation, for both \BKee decays and background processes, and satisfying ${\pt > 2\GeV}$ and ${\abs{\eta} < 2.5}$.}
  \label{fig:soft-electrons-2018}
\end{figure*}

The \bparking data set provides access to a large sample of unbiased \PQb hadron decays. The particles produced in these decays typically have very low \pt values. The \cmsLeft panel of Fig.~\ref{fig:soft-electrons-2018} shows the \ptgen distributions for the leading- and subleading-\ptgen electrons from the \BKee decay, where \ptgen is the generator-level \pt quantity obtained from simulation. The kinematical requirements of ${\pt > 7\GeV}$ and ${\abs{\eta} < 1.5}$ are imposed on the tag-side muon. The most probable \ptgen values are below 2\GeV. The standard PF-based electron reconstruction algorithm~\cite{Sirunyan:2017ulk, CMS:2020uim} relies primarily on information from both the tracking and ECAL subdetector systems and it is optimized to identify electrons with high efficiency for \ptgen values above 10\GeV. The performance of the PF electron algorithm worsens significantly below 10\GeV, with efficiencies falling to zero at 2\GeV and below, as shown in the \cmsLeft panel of Fig.~\ref{fig:soft-electrons-2018}.

Given the limited low-\pt performance of the existing electron algorithm, a custom multistage electron reconstruction algorithm has been developed to improve the reconstruction and identification of genuine electrons with \pt values below 10\GeV. Given the superior physics performance of the tracker system relative to the calorimeter systems at very low \pt, in terms of momentum or energy scale and resolution, the new algorithm uses a tracker-seeded approach that first considers charged-particle tracks
and attempts to match each one with a compatible pattern of calorimetric energy deposits. Primarily, the low-\pt electron algorithm targets electrons from \PQb hadron decays, but it exhibits comparable performance for electrons, originating promptly or otherwise, from a broad range of physics processes.

Electrons undergo photon bremsstrahlung when traversing detector material. Thus, to accommodate these energy losses, the accurate determination of the track parameters for electron candidates relies on a technique that employs a GSF~\cite{Adam_2005}, as described in Section~\ref{subsubsec:eg_reco_desc}. The GSF approach delivers the optimal electron momentum scale and resolution, albeit at a high computational cost. Consequently, it is preceded by a more computationally efficient seeding algorithm.

The first stage of the reconstruction chain comprises seeding logic that exploits two BDT algorithms. The BDTs consider a range of kinematical and topological variables constructed from tracker- and calorimetry-based measurements that include: the shape of electromagnetic showers, the spatial compatibility between the track trajectory and energy deposits within the calorimetry systems, the compatibility of momentum and energy measurements, and the differences in the momentum measurements determined at the innermost and outermost layers of the tracker systems. The set of variables considered is analogous to that used by the ID algorithms reported in Ref.~\cite{CMS:2020uim}. A simplified version of the GSF track fit, with a reduced set of parameters, is used, which is then compared with the nominal tracking fit algorithm based on the Kalman filter~\cite{CMS:2014pgm}. The seeding logic is carefully tuned to balance signal-to-background discrimination performance against computational load. One BDT uses a kinematically agnostic approach that exploits only the aforementioned variables. The other BDT provides a kinematically aware model that also uses the \pt, $\eta$, and the transverse impact parameter significance \dxysig of the electron candidate to discriminate between genuine signal electrons from \BKee decays and misidentified electrons from background physics processes.

In order for the reconstruction to proceed to the next stage, the seeding logic requires the score produced by each BDT to satisfy an independent threshold value. The full GSF-based track fit is subsequently performed on each electron seed to determine the optimal track parameters. The resulting trajectory is used to identify a spatially compatible energy cluster in the ECAL that is assumed to be the electromagnetic shower from the incident electron. Additional clusters of energy that are spatially compatible with the expected position of bremsstrahlung photons within the ECAL are associated with the original cluster to form a super cluster. The logic is tuned to ensure that the electron seeds are promoted to electron candidates with high efficiency (${>}95\%$). The associated cost of high-rate particle misidentification is mitigated in the next stage.

The final stage of the low-\pt algorithm aims to differentiate between genuine electrons and misidentified particles with the highest possible performance. A further BDT model takes information from all the preceding stages, namely the seeding, GSF-based tracking, and super-clustering algorithms. An expanded set of input variables is used relative to the seeding BDTs, such as improved track parameter estimates from the full GSF-based tracking stage, and variables that test for consistency between the supercluster substructure and a bremsstrahlung energy-loss pattern. The BDTs are trained with a simulated sample of \BKee events using the \textsc{XGBoost} package~\cite{Chen:2016:XST:2939672.2939785}.

Electron candidates are considered to be interesting for further analysis if they satisfy a threshold applied to the BDT discrimination score, known as an ID working point. Low-\pt electron candidates are reconstructed and identified for the kinematical regime ${\pt > 0.5\GeV}$ and ${\abs{\eta} < 2.5}$, whereas the PF electron algorithm is restricted to ${\pt > 2\GeV}$. The \cmsRight panel of Fig.~\ref{fig:soft-electrons-2018} shows the performance of the two algorithms for electron candidates originating from both \BKee decays and randomly selected charged-particle tracks that satisfy $\pt > 2\GeV$. The PF electron candidates are reconstructed with an efficiency and misidentification probability of 69\% and 2\%, respectively. The low-\pt reconstruction algorithm provides electron candidates with a higher efficiency of 92\%, but also a substantially higher misidentification probability of 21\%.

The ID performance for both the low-\pt and PF electron candidates is quantified in terms of the efficiency and misidentification probability as a function of the ID score threshold, as shown in the \cmsRight panel of Fig.~\ref{fig:soft-electrons-2018}. The PF-based BDT model is based on the standard approach defined in Ref.~\cite{CMS:2020uim}, but it is retrained specifically for an extended kinematical regime down to 2\GeV. The performance of the low-\pt and PF IDs can be compared at working points that yield the same misidentification probability: for instance, efficiencies of 56\% and 33\% are obtained for low-\pt and PF electrons, respectively, for a misidentification probability of 0.1\%. For the regime ${0.5 < \pt < 2\GeV}$, the low-\pt ID yields an efficiency of 30\% for a misidentification probability of 0.1\%. Finally, a tag-and-probe technique~\cite{CMS:2014pkt} is utilized with a sample of \BKJpee decays in data to check the accuracy of the simulation modeling of the input variables to the BDTs and their output scores. The ID score distributions for both the low-\pt and PF electrons in data are consistent with those in simulation, within statistical uncertainties.

\subsubsection{Characterization of the data set}
\label{sec:characteristics}

\begin{table*}[!htb]
  \centering 
  \caption{Trigger configurations, defined by unique combinations of L1 \pt, HLT \pt, and \ipsig thresholds, used to record events containing \btomux decays. The \Lint value, the mean number of pileup interactions, and the number of events recorded by each combination are aggregated over the periods for which each combination provided the lowest enabled L1 \pt threshold.}
  \def\arraystretch{1.2}
  \begin{tabular}{cccccc}
    L1 \pt     & HLT \pt    & HLT    & \Lint      & Mean & Events             \\
    $[\GeVns]$ & $[\GeVns]$ & \ipsig & $[\fbinv]$ & PU   & $[{\times}10^{9}]$ \\
    \hline
    12   & 12  & 6   & 8.1    & 37.7 & 0.72 \\
    10   & 9   & 6   & 8.4    & 32.9 & 1.67 \\
    10   & 9   & 5   & 1.6    & 33.9 & 0.37 \\
    9    & 9   & 6   & 1.6    & 28.2 & 0.34 \\
    9    & 9   & 5   & 5.2    & 28.3 & 1.30 \\
    9    & 8   & 5   & 1.6    & 29.2 & 0.52 \\
    8    & 9   & 6   & 1.8    & 24.2 & 0.40 \\
    8    & 9   & 5   & 3.8    & 23.9 & 1.00 \\
    8    & 8   & 5   & 1.7    & 24.2 & 0.60 \\
    8    & 7   & 4   & 1.5    & 24.5 & 0.84 \\
    7    & 8   & 3   & 0.8    & 19.1 & 0.45 \\
    7    & 7   & 4   & 5.5    & 18.6 & 3.56 \\
    \multicolumn{2}{c}{Other combinations} & & 0.3 & \NA & 0.12 \\
    \hline
    Total & & & 41.9 & 22.7 & 11.9 \\
  \end{tabular}
  \label{tab:trigger-configuration-2018}
\end{table*}

The \bparking data set comprises \sci{1.2}{10}~events and contains $\sci{1}{10}$ unbiased \PQb hadron decays. The size of the single-copy unprocessed data sample is 7.6\unit{PB}. The reconstruction-level MINIAOD~\cite{Petrucciani_2015} data sample has a reduced footprint of 950\unit{TB}. Custom analysis-specific data samples, based on a common analysis-level data format (known as NANOAOD~\cite{Peruzzi_2020}), typically have a footprint of around 10\unit{TB}.

\begin{figure*}[!htb]
  \centering
  \includegraphics[width=0.67\textwidth]{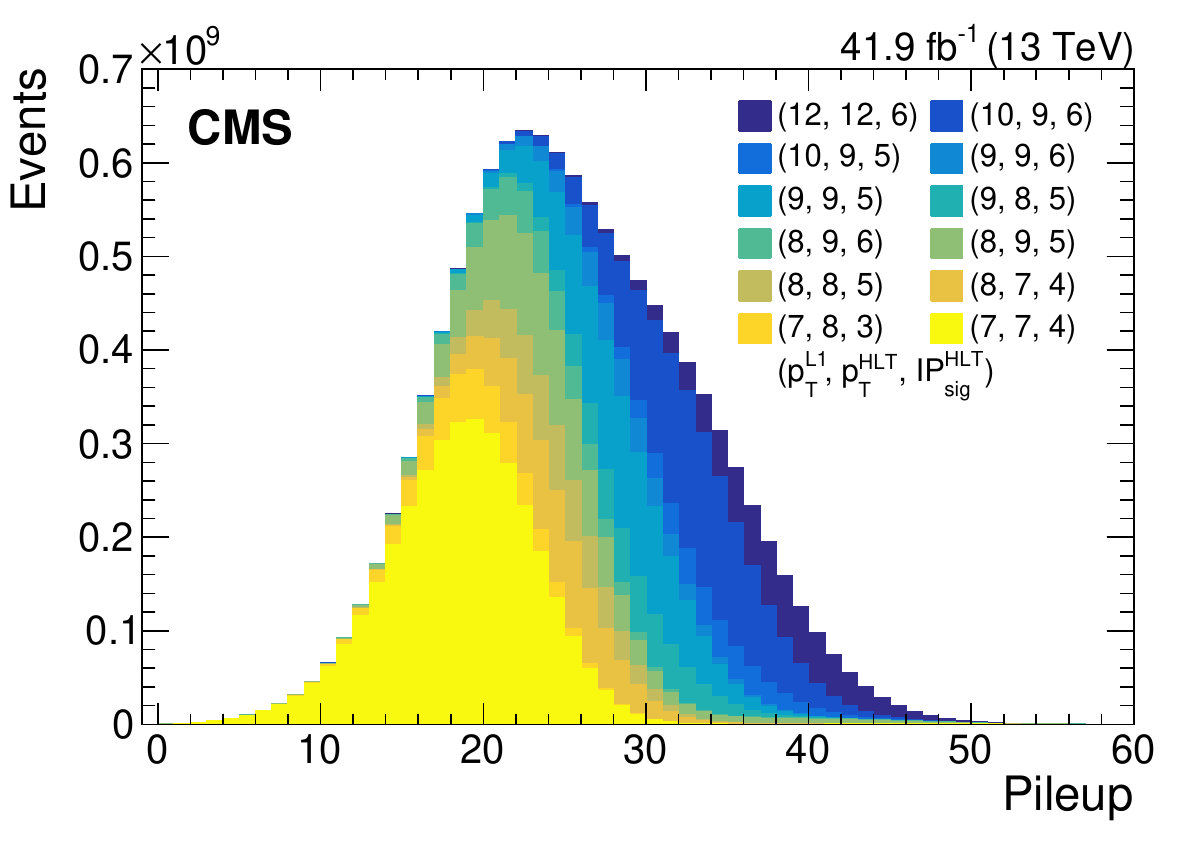}
  \caption{The pileup distribution obtained from the \bparking data set. Contributions from each trigger combination are shown, with the histogram areas normalized to the number of events recorded by each trigger.}
   \label{fig:pileup-2018}
\end{figure*}

Table~\ref{tab:trigger-configuration-2018} summarizes all the unique combinations of thresholds used in the L1 and HLT algorithms. In total, the suite of algorithms recorded 41.9\fbinv. The \Lint value, the mean number of pileup interactions, and the number of events recorded by each combination are provided. The values are determined from the periods for which each combination provided the lowest enabled L1 \pt threshold. The highest HLT rates are observed later in a fill and so a larger fraction of the data are recorded at lower pileup values than for the standard physics data streams. Figure~\ref{fig:pileup-2018} shows the pileup distribution for the \bparking data set, along with the contributions from each of the individual trigger combinations.

\begin{figure*}[!htb]
  \centering
  \includegraphics[width=0.67\textwidth]{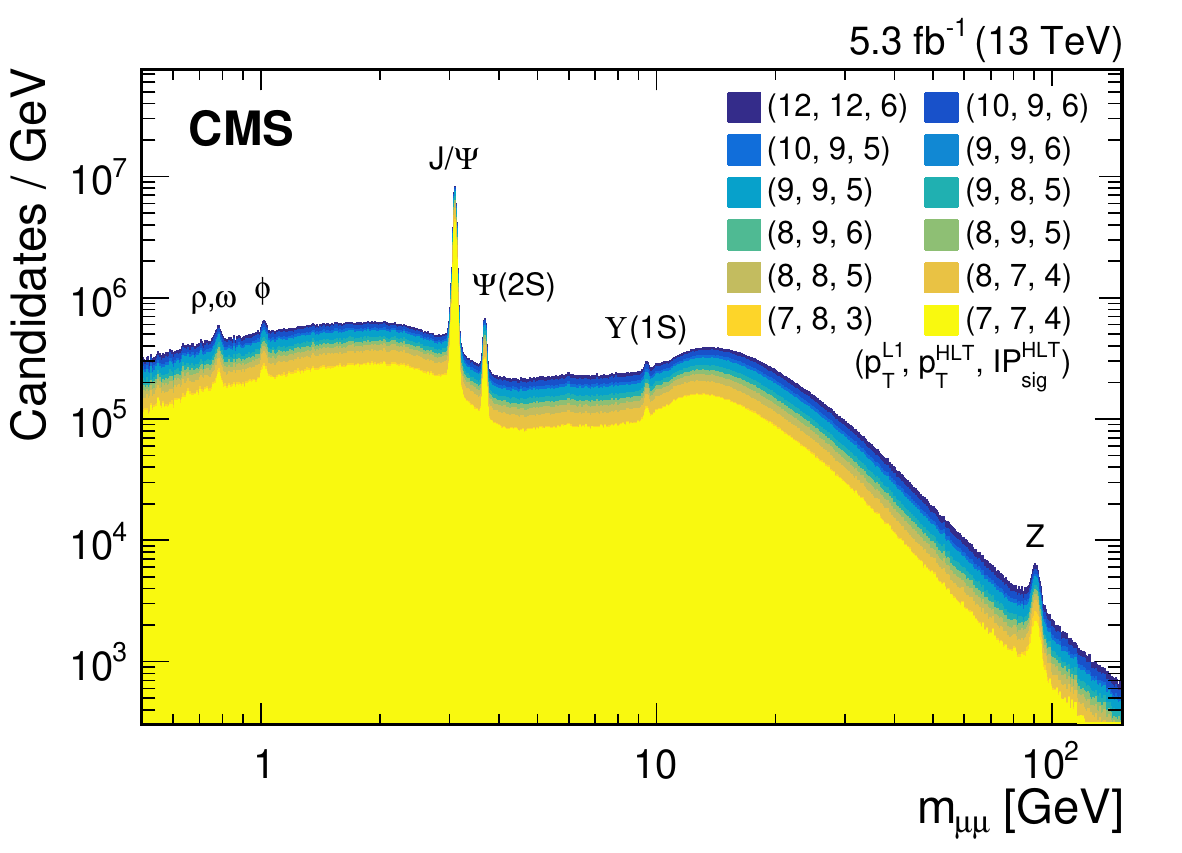}
  \caption{The invariant mass distribution for pairs of oppositely charged muons originating from a common vertex, obtained from a subset of the \bparking data.}
  \label{fig:dimuon-inv-mass-2018}
\end{figure*}

Figure~\ref{fig:dimuon-inv-mass-2018} shows the invariant mass distributions for pairs of oppositely charged muons originating from a common vertex, as obtained from the \bparking data set. Both muons are required to satisfy a minimal set of kinematical and ID criteria, and one of the muons must be matched to the candidate responsible for the positive trigger decision. Peaks in the data resulting from the $\PGr/\PGo$, \PGf, \PJGy, \Pgy, \PgUa, and \PZ resonances are visible above the continuous background.

\subsection{Physics results with the \texorpdfstring{\PB}{B} parking data set}
\label{sec:physics-2018}

Various searches for LFU violation are being considered or are currently underway, \eg, measuring \RDst using \btodstlnu decays and the tag-side muon; searching for lepton flavor violation in (tag-side) $\PBzs \to \PGmm \Pep$ decays; and studying charge-parity violating processes in fully reconstructed hadronic final states using probe-side \PQb hadron decays, such as $\PDz \to \PKS \PKS$ and $\PBzs \to \PGf(\to\PKp\PKm)\PGf(\to\PKp\PKm)$.

Beyond \PB physics, the data set provides a rich opportunity for the discovery of a broad range of BSM scenarios and will serve novel physics analyses for several years. It provides access to BSM models with low-mass states and/or very rare decays, a parameter space complementary to the one offered by data sets that serve the high-\pt searches typical at the LHC, and thus substantially extends the reach of the CMS physics program.

In addition to the ${\approx}10^{10}$ \bbbar events, the \bparking data set contains ${\approx}\sci{3}{11}$ \pp pileup interactions, which may be of interest to searches for BSM processes that yield ``untriggerable'' signatures (\ie, for which there is no feasible trigger algorithm). Further, experimentally difficult signatures from the decay of low-mass BSM particles, which may be sensitive to combinatorial backgrounds from a high-pileup environment, can exploit the relatively low pileup observed for the \bparking data set, particularly for data collected later in the LHC fills by the trigger algorithms with low \pt thresholds.

The following subsections highlight two key physics results obtained with the \bparking data set: a measurement of \RK and of the differential branching fraction for the \BKmm process, and a BSM search for heavy neutral leptons in \PQb hadron decays.

\subsubsection{Measurement of the \texorpdfstring{\RK}{RK} observable}
\label{sec:rk_run2}

Given the negligible masses of both electrons and muons in comparison to the \PB meson mass, the ratio of SM branching fractions, $\BF(\BKee)$ and $\BF(\BKmm)$, is very close to unity. The presence of BSM physics could induce appreciable modifications to the branching fractions of the different lepton generations, such as via a leptoquark with flavor-dependent couplings.
The observation of a nonunity ratio would provide compelling evidence for LFU violation through a BSM mechanism.

To enhance the sensitivity to LFU violation in \btosll decays, it is advantageous to employ an observable that minimizes any associated theoretical uncertainties. The ratio of branching fractions $\BF(\BKee) / \BF(\BKmm)$, determined within a specific range of the dilepton mass squared $\qsqmin < \qsq < \qsqmax$, is robust against potentially substantial long-range corrections~\cite{Ciuchini:2021smi} and is known to be close to unity with a precision of approximately 1\%~\cite{Hiller:2003js, Bordone:2016gaq, Isidori:2020acz, Isidori:2022bzw}. Additionally, experimental systematic uncertainties (such as those related to lepton acceptances, momentum scale, and identification) can be reduced by measuring a double ratio, normalized to the corresponding \BKJpll decay channels:
\begin{equation*}
  \label{eq:RKJ}
  \RK(\qsq)[\qsqmin,\qsqmax]=\left.\frac{\BF(\BKmm)[\qsqmin,\qsqmax]}{\BF(\BKJpmm)}
    \middle/ \frac{\BF(\BKee)[\qsqmin,\qsqmax]}{\BF(\BKJpee)} \right.
\end{equation*}
where the ratio $\BF(\BKJpmm) / \BF(\BKJpee)$ has been experimentally determined to be nearly unity with a precision of 0.7\%~\cite{Workman:2022ynf}. 

The \RK observable has been measured by the CMS Collaboration in the \qsq range spanning from 1.1 to $6.0\GeV^2$~\cite{CMS:2023klk}. Within the same analysis, a differential branching fraction measurement is performed for the \BKmm channel, divided into 15 bins of \qsq ranging from 0.98 to $22.9\GeV^2$ and excluding the resonance regions of \PJGy and \Pgy.

To determine \RK, events are chosen in which either a \BKmm candidate is identified on the tag side of the event (\ie, one of the muons is responsible for the positive trigger decision) or a \BKee candidate is identified from the sample of unbiased decays of the other \PQb hadrons in the event. 

The \PBp candidates are constructed by pairing two same-flavor leptons with opposite charges, whose invariant mass falls below 5\GeV, along with a positively charged track to which the kaon mass
is assigned (in the absence of particle identification for pions and kaons). Rigorous quality criteria are imposed for each candidate, whether it be a muon, electron, or track, to minimize the occurrence of misreconstructed objects.

The tracks of the three particles constituting the \PBp candidate, including that of the muon responsible for the positive trigger decision, are required to have the same point of origin. The tracks are then used in a vertex fitting procedure, which relies on their measured momentum vectors along with associated uncertainties. This procedure enhances the accuracy of mass measurements for both the \PBp candidate and the lepton pair.

The kinematic fit algorithm~\cite{kinFit}, which is based on a least mean square minimization approach, constrains the tracks associated with each \PBp candidate, ensuring they originate from a common vertex. The particle trajectories are then refitted taking into account the common vertex as an additional constraint, and their momenta are recalculated.

In the electron channel, to enhance the sample purity, at least one of the electrons is required to be reconstructed by the PF algorithm. The second electron can be reconstructed using either the PF algorithm or the dedicated low-\pt (LP) algorithm described in Section~\ref{sec:lowpt-electrons}.

The \PBp candidate is subjected to a minimal set of kinematic, topological, and quality criteria. A significant fraction of semileptonic decays of heavy-flavor hadrons, containing a $\PDz \to \PKm\PGpp$ decay, remains within the sample, primarily due to the misidentification of the \PKm meson as a lepton. To address this issue, a dedicated veto is implemented to remove such decays.

The final selection in each channel relies on a BDT, which combines several variables into a classifier constructed using \textsc{XGBoost}. In the electron channel, independent BDTs are trained for the two exclusive PF-PF and PF-LP event categories, using the same input variables. Given the notably higher background levels in the electron channels relative to the muon channel, the corresponding BDTs utilize a greater number of input variables. The BDTs use a supervised training, which relies on simulated \BKll decays in the low-\qsq bin for signal, and data sidebands to the low-\qsq bin for background. The optimal working point for each BDT is selected to maximize the expected significance of the \BKll signal within the low-\qsq region. Various checks are performed to ensure that the BDT performance is unbiased and robust for all three channels (muon, PF-PF, and PF-LP) using multiple data control regions.

\begin{figure*}[!htb]
  \centering
  \includegraphics[width=0.49\textwidth]{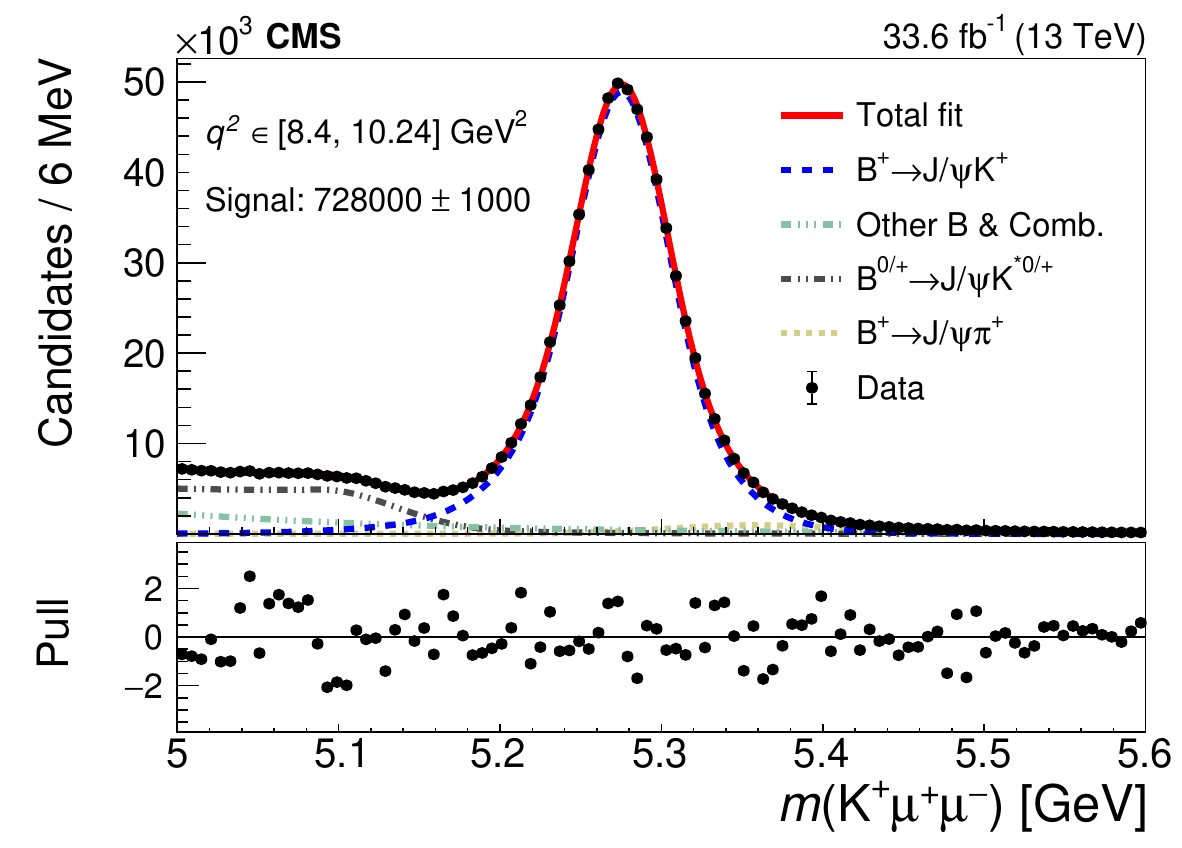}
  \includegraphics[width=0.49\textwidth]{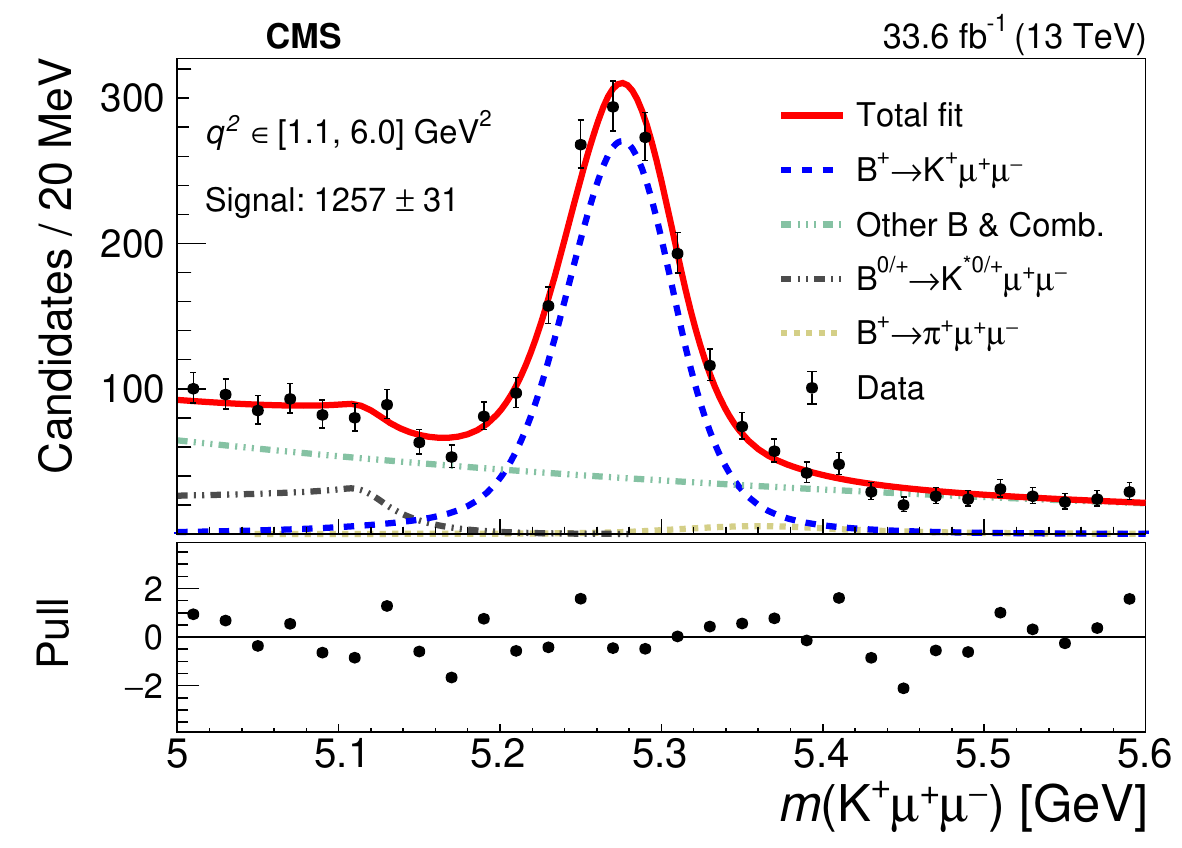}
  \caption{Results of an unbinned likelihood fit to the invariant mass distributions for the \BKJpmm (\cmsLeft) and the \BKmm (\cmsRight) channels. Various functions are used to parametrize the contributions from the signal and various background processes. Taken from Ref.~\cite{CMS:2023klk}.
  }
  \label{fig:mass_fit_muons}
\end{figure*}

\begin{figure*}[!htb]
  \centering
  \includegraphics[width=0.49\textwidth]{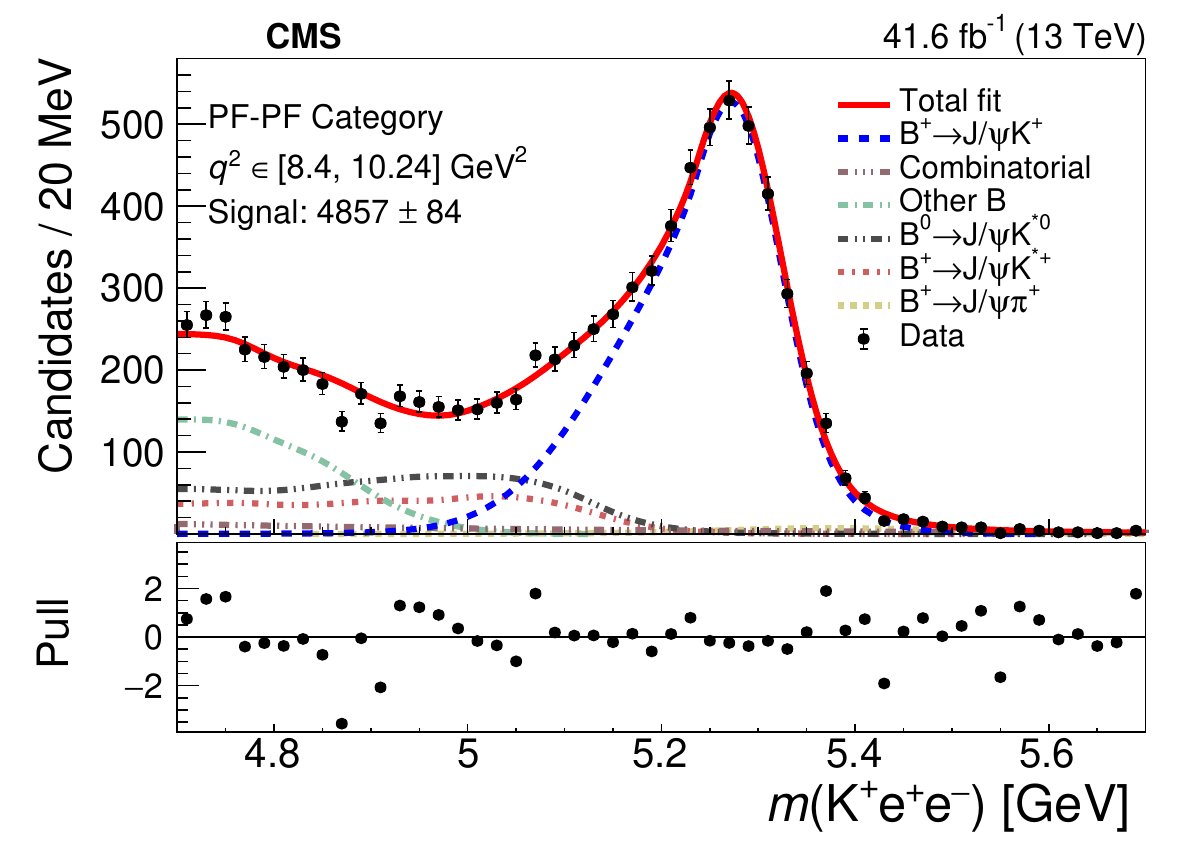}
  \includegraphics[width=0.49\textwidth]{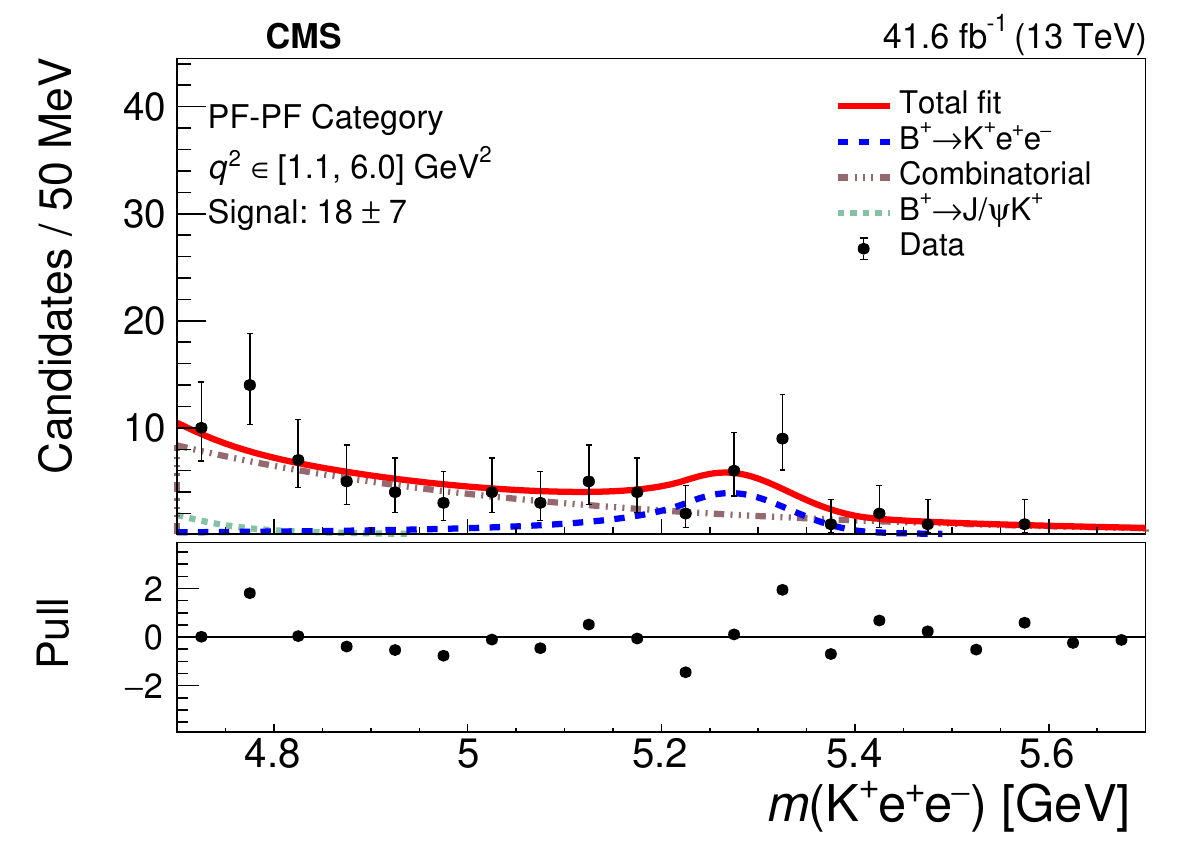} 
  \includegraphics[width=0.49\textwidth]{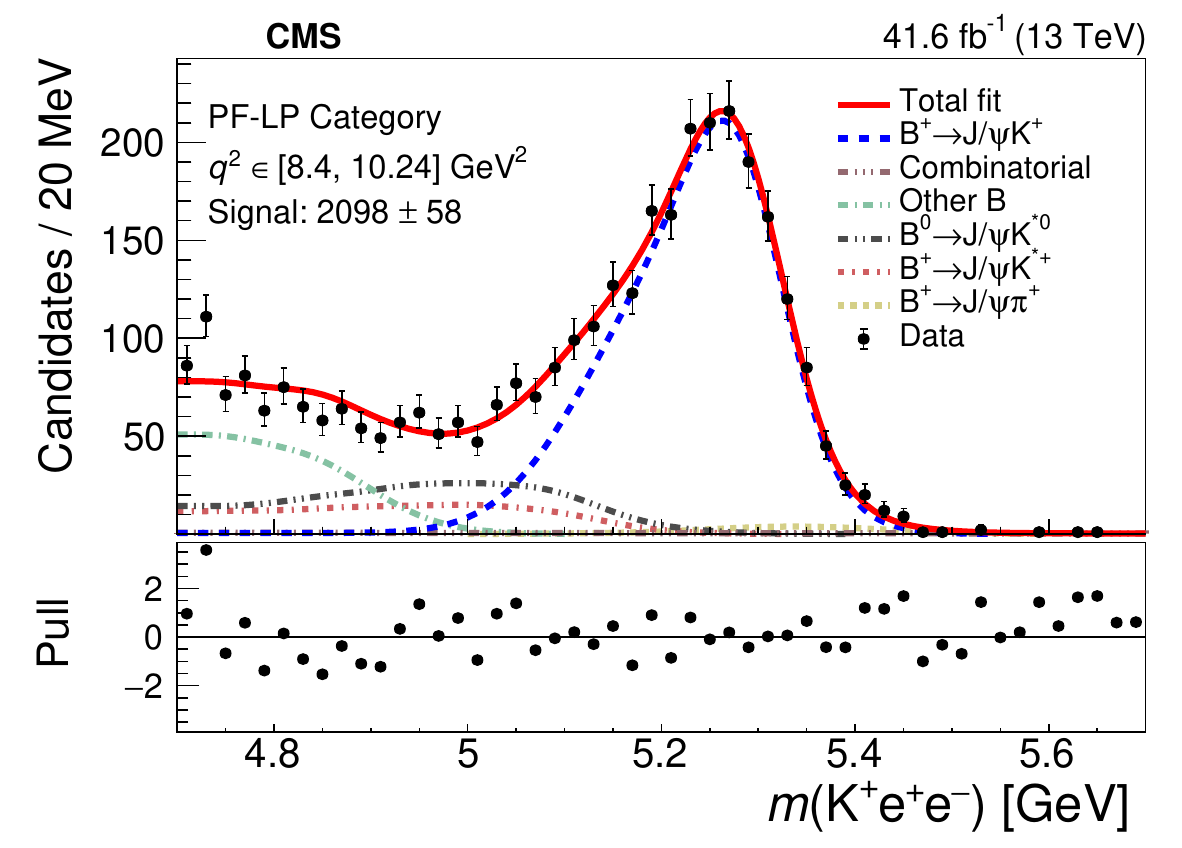}
  \includegraphics[width=0.49\textwidth]{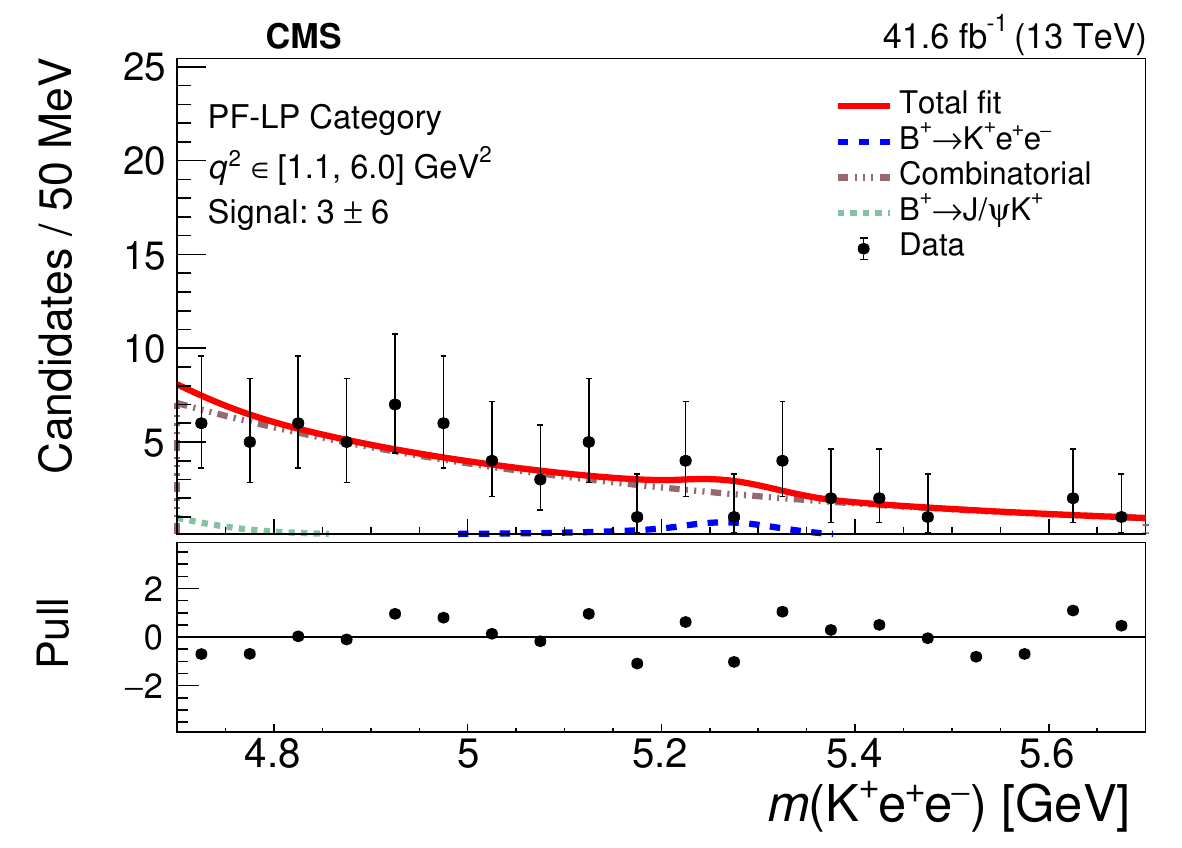}
  \caption{Results of an unbinned likelihood fit to the invariant mass distributions for the \BKJpee (\cmsLeft) and the \BKee (\cmsRight) channels. The upper (lower) panels are for candidates using the PF-PF (PF-LP) category. Various functions are used to parametrize the contributions from the signal and various background processes. Taken from Ref.~\cite{CMS:2023klk}. 
  }
  \label{fig:mass_fit_electrons}
\end{figure*}

After the final selection of \PBp candidates using the BDTs, the mass distributions obtained from the reconstructed \PBp candidates in each channel are fitted using analytical functions, as shown in Figs.~\ref{fig:mass_fit_muons}~and~\ref{fig:mass_fit_electrons}. The fits for the muon, PF-PF, and PF-LP channels are statistically combined to yield the final \RK measurement. Systematic uncertainties are assessed independently for all channels. The uncertainty in the final measurement is statistically limited. The measurement of \RK is 
\begin{equation*}
\RK = 0.78^{+0.46}_{-0.23}\stat^{+0.09}_{-0.05}\syst = 0.78^{+0.47}_{-0.23},
\end{equation*}
which is consistent with the SM prediction within one standard deviation and can be compared with the corresponding LHCb measurement of ${\RK = 0.949^{+0.048}_{-0.047}}$~\cite{LHCb:2022qnv}.

\begin{figure*}[!htb]
  \centering
  \includegraphics[width=0.67\textwidth]{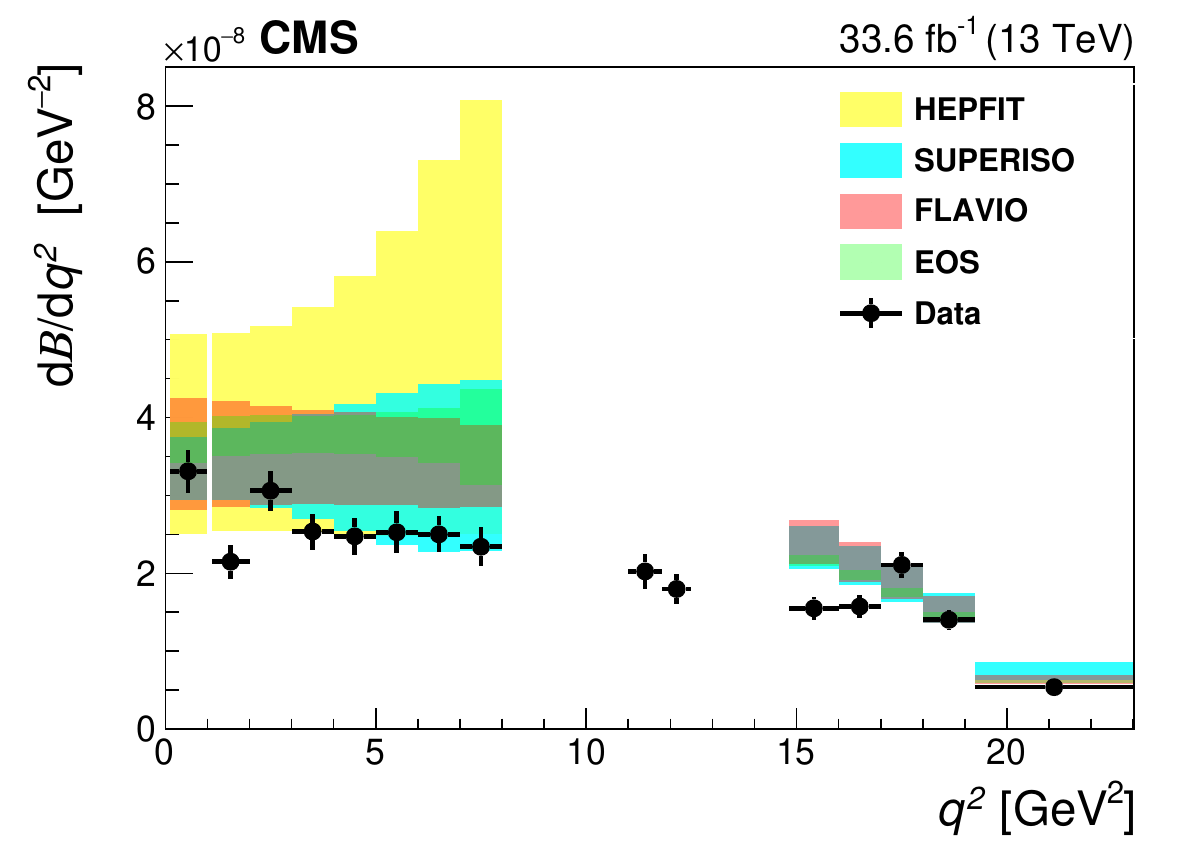}
  \caption{Comparison of the measured differential \BKmm branching fraction with the theoretical predictions obtained using \textsc{hepfit}, \textsc{superiso}, \textsc{flavio}, and \textsc{eos} packages. Reliable predictions are not available between the \PJGy and \Pgy resonance regions. The \textsc{hepfit} predictions are available only for ${\qsq < 8\GeV^2}$. Taken from Ref.~\cite{CMS:2023klk}.}
\label{fig:diff_theory}
\end{figure*}

The sample of \BKmm candidates is sufficiently large to enable both an inclusive and a differential measurement of $\BF(\BKmm)$. Figure~\ref{fig:diff_theory} shows the measured differential branching fraction of the \BKmm decay, as a function of \qsq, along with corresponding SM theoretical predictions as determined by various
packages. The inclusive value for $\BF(\BKmm)$ within the \qsq region $1.1 < \qsq < 6.0\GeV^2$ is $\BF(\BKmm) = (12.42 \pm 0.68) \times 10^{-8}$, which is statistically limited and consistent with the world average~\cite{Workman:2022ynf}.

\subsubsection{Search for long-lived heavy neutrinos in \texorpdfstring{\PB}{B} meson decays}

Heavy neutrinos (\PN) that feebly mix with the SM neutrinos constitute a possible appealing answer to several open questions in the SM, notably the evidence for nonzero neutrino masses in neutrino oscillations~\cite{Bilenky:2016pep}, the large amount of dark matter inferred from astrophysical and cosmological measurements~\cite{bertone_2010}, and the baryon asymmetry in the universe~\cite{PhysRevD.50.774}. The small mixing amplitudes, \Ve, \Vu, and \Vt, between heavy neutrinos and their SM counterparts in the three flavor families imply that heavy neutrinos are long-lived. Indeed, the \PN particle proper lifetime scales as ${\ctauN \propto \mN^{-5}\abs{V_\PN}^{-2}}$, where \mN is the mass of the heavy neutrino and \VV is defined as ${\VV=\VVe+\VVu+\VVt}$. The large signal lifetimes give rise to displaced signatures. Moreover, thanks to mixing, flavor conservation may be violated.

\begin{figure*}[!htb]
\centering
\includegraphics[width=0.75\textwidth]{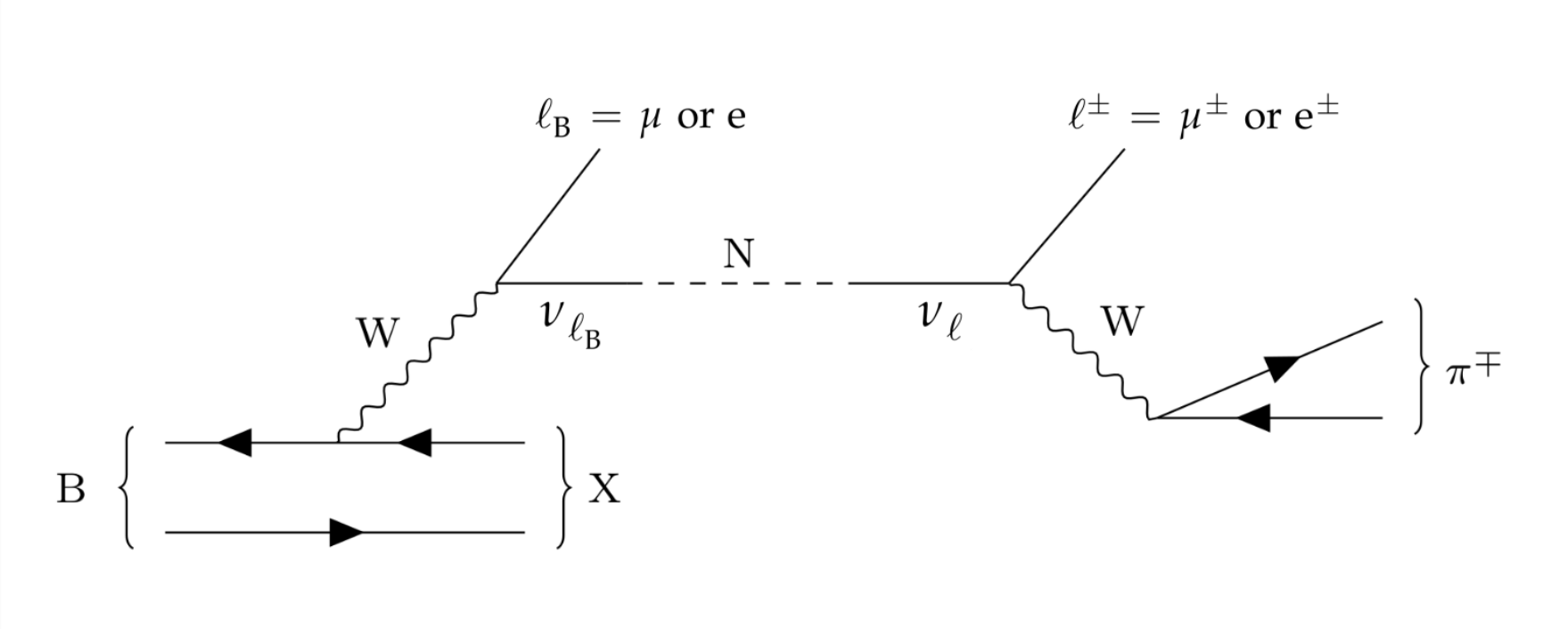} 
\caption{Feynman diagram of a semileptonic decay of a \PB meson into the primary lepton ($\ell_{P}$), a hadronic system (X), and an SM neutrino, which contains a small admixture of a heavy neutrino (\PN). The \PN decays weakly into a charged lepton $\ell^\pm$ and a charged pion \PGpmp, forming a vertex displaced from the \pp interaction point. Taken from Ref.~\cite{CMS-PAS-EXO-22-019}.}
\label{fig:feynMain}
\end{figure*}

Numerous searches have been performed at the most diverse experiments covering a vast range of heavy neutrino masses, from the \keVns to the \TeVns scale, and mixing amplitudes, down to ${\VV \approx 10^{-7}}$. In CMS, virtually all searches for \PN with \mN up to around 100\GeV relied on the process ${\PW\to \ell \PN}$, where events are collected by high-\pt, single isolated lepton triggers~\cite{CMS-PAS-EXO-23-006}. However, electroweak processes with on-shell gauge boson production are not the sole or even the most abundant source of neutrinos, and hence also potentially of heavy neutrinos, at the LHC: leptonic and semileptonic \PB meson decays (shown in Fig.~\ref{fig:feynMain}) are more than a thousand times more copious thanks to the much larger ${\Pp\Pp\to\bbbar}$ cross section. Furthermore, the final-state particles produced in \PB meson decays have lower momenta than those produced in \PW boson decays, because the \PB mesons have a lower mass than the \PW boson. For long-lived signatures, the softer momentum spectrum is an advantage as it leads to a higher fraction of \PN particles that decay within the CMS acceptance. 

The 2018 \PB parking data set (described in Section~\ref{sec:bparking_run2}) enabled CMS to perform a search for long-lived heavy neutrinos in \PB meson decays, using events with one lepton plus one displaced vertex comprising a lepton-pion pair and compatible with the decay of a long-lived heavy neutrino $\PN\to\ell^\pm\PGpmp$~\cite{CMS-PAS-EXO-22-019}. Both muons and electrons are considered as long as at least one muon matched to a \PB parking trigger algorithm is present. This search would have not been possible with standard triggers.

\begin{figure*}[!htb] 
\centering
\includegraphics[width=0.49\textwidth]{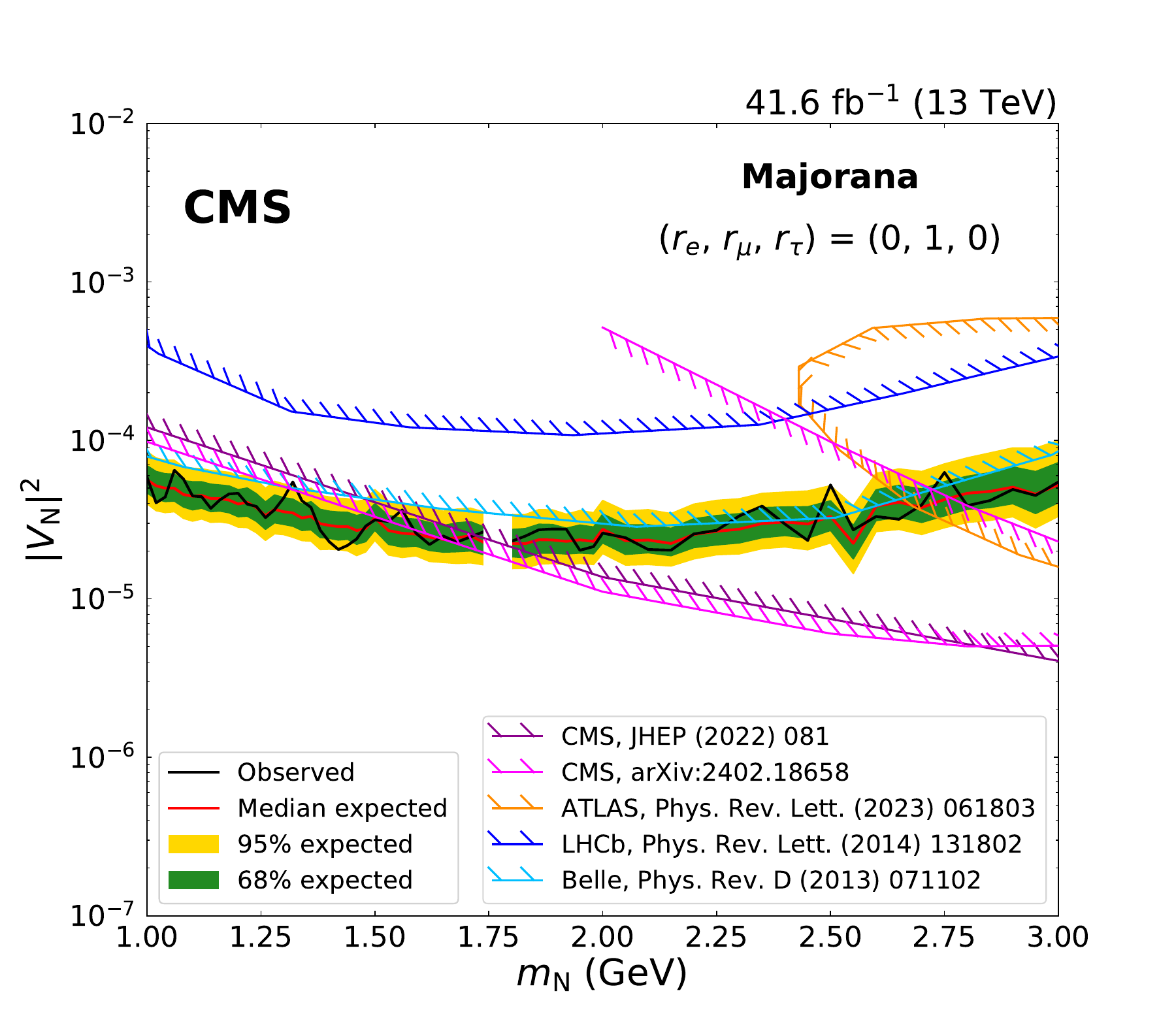}
\includegraphics[width=0.49\textwidth]{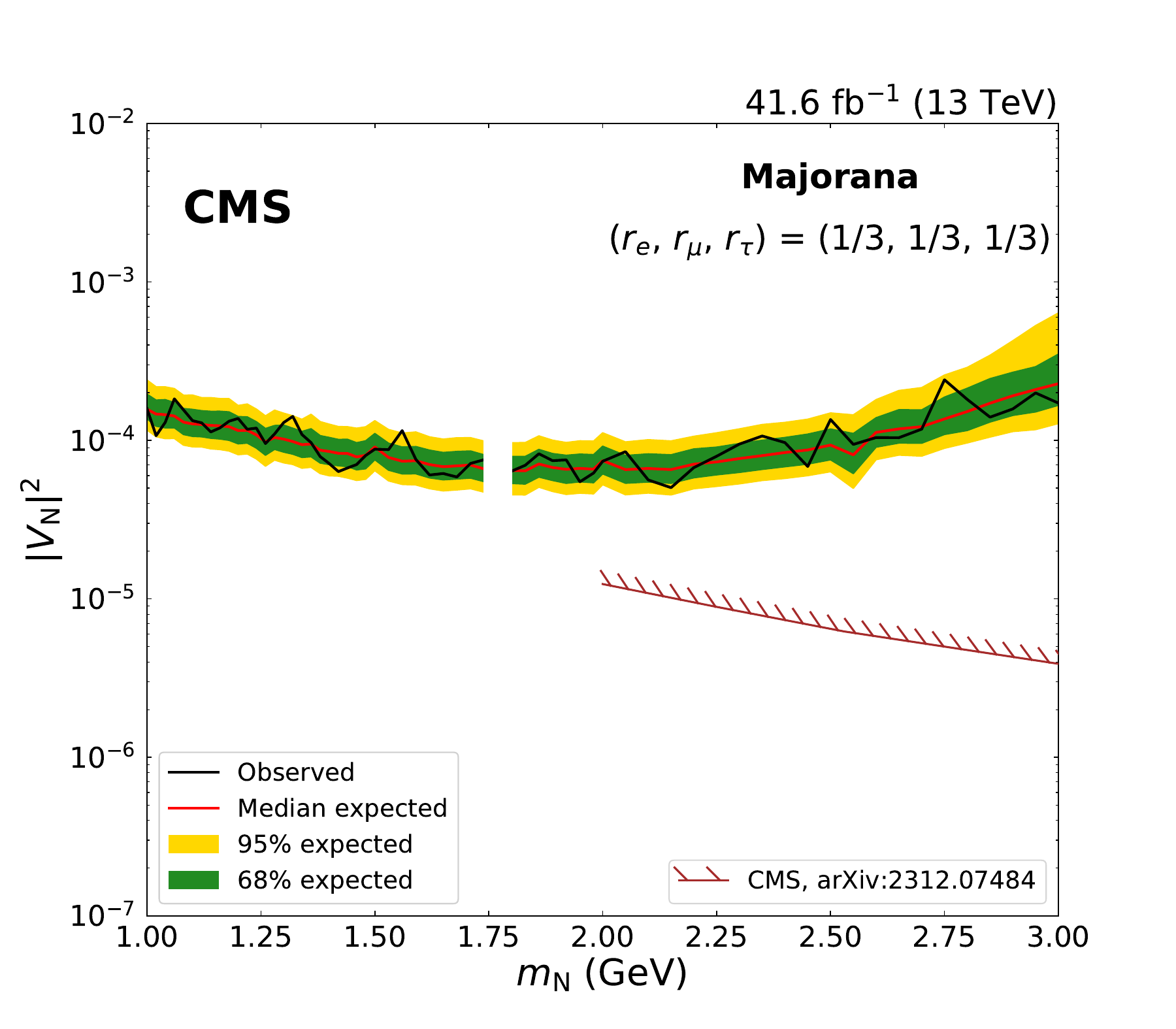}
\caption{Expected and observed 95\%~\CL limits on \VV as a function of \mN in the Majorana scenario, for the coupling hypotheses (\rehnl, \ruhnl, \rthnl) = (0, 1, 0) on the left and (\rehnl, \ruhnl, \rthnl) = (1/3, 1/3, 1/3) on the right. The mass range with no limits shown corresponds to the \PDz meson veto employed by the search. Taken from Ref.~\cite{CMS-PAS-EXO-22-019}.}
\label{figure_limits}
\end{figure*}

\begin{figure*}[!htb] 
\centering
\includegraphics[width=0.47\textwidth]{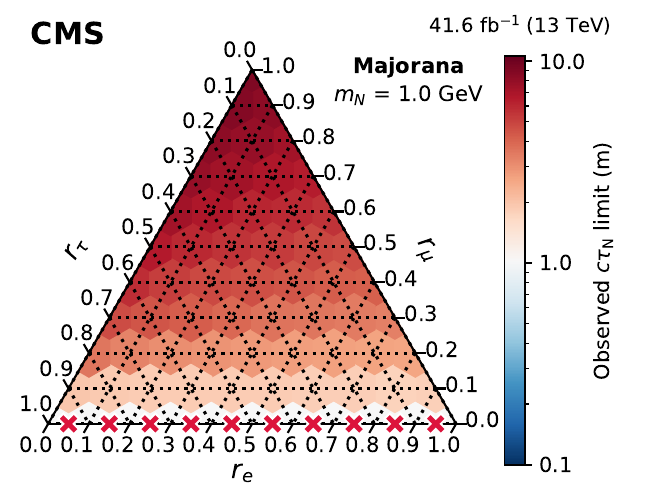}
\includegraphics[width=0.47\textwidth]{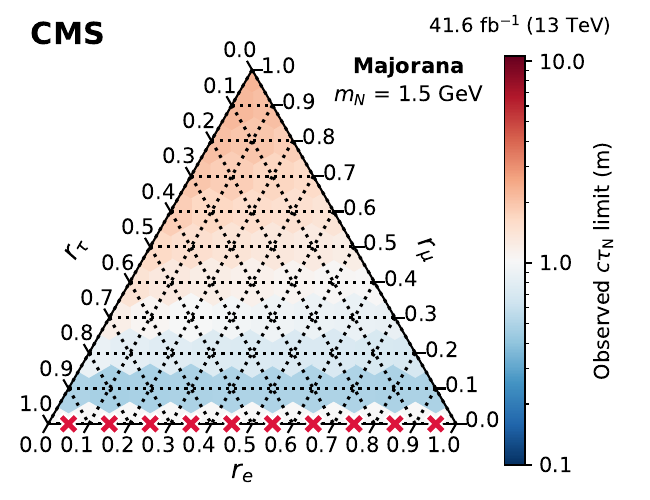}
\includegraphics[width=0.47\textwidth]{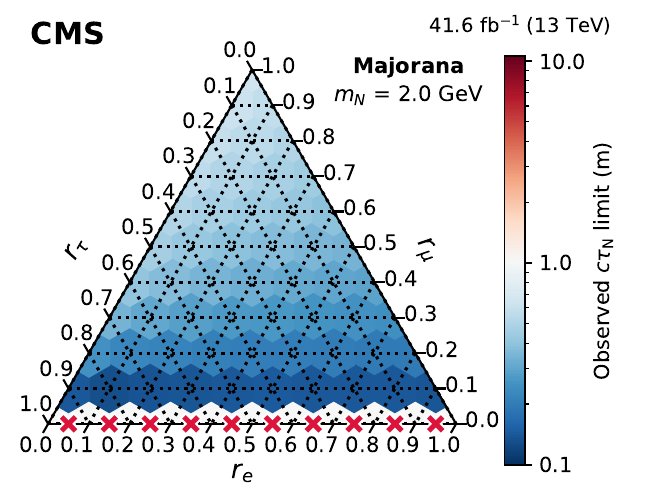}
\caption{Observed limits on $c\tau_{\PN}$ as a function of the coupling ratios (\rehnl, \ruhnl, \rthnl) for fixed \PN masses of 1\GeV (upper left), 1.5\GeV (upper right), and 2\GeV (lower center), in the Majorana scenario. A red cross indicates that no exclusion limit was set for that point. The tick orientation indicates the direction of reading. Taken from Ref.~\cite{CMS-PAS-EXO-22-019}.}
\label{figure_ternary_majorana}
\end{figure*}

This analysis sets the most stringent upper limits, among those obtained at collider experiments, at 95\%~\CL on \VV for heavy neutrinos with masses between 1.0 and 1.7\GeV and does so with unprecedented resolution as the \PN decay is fully reconstructed. Results are interpreted in various scenarios specified by different values of the mixing ratios ${r_\ell\equiv\abs{V_{\ell\PN}}^2/\VV}$, $\ell=(\Pe, \,\PGm, \,\PGt)$, as well as under either the Majorana (shown in Fig.~\ref{figure_limits}) or Dirac hypotheses. For masses $\mN = 1.0, 1.5, 2.0\GeV$, lower limits on \ctauN are provided for sixty-six possible flavor violating scenarios, depicted in Fig.~\ref{figure_ternary_majorana}.

\section{Data parking in Run 3} \label{ch:parking_run3}
The successful execution of the \bparking program during Run~2 garnered significant interest, propelling the evolution of the parking technique into a comprehensive and diverse program. The enhancement of the \bparking strategy in Run~3 is discussed in Section~\ref{sec:bparking_run3}, which introduces novel inclusive triggers to collect data in the dimuon and dielectron final states. In Section~\ref{sec:run3altParking}, we provide an overview of other diverse opportunities arising from the parking strategy that extend beyond the B physics measurements, referred to as ``alternative'' data parking strategies. 

\subsection{Data parking for \texorpdfstring{\PB}{B} physics in Run~3}
\label{sec:bparking_run3}

The LHC Run~3 period started in 2022 and is currently expected to finish at the end of 2025. The LHC delivered ${\Lint = 73.4\fbinv}$ during 2022 and 2023, and the total may exceed 250\fbinv by the end of the data-taking period in 2025. Thus, Run~3 represents an important opportunity to improve our understanding of \PQb hadron production and decay modes. The LHC center-of-mass energy has also increased from 13 to 13.6\TeV between Runs~2~and~3, yielding a modest increase of 4\%~\cite{Cacciari:1998it, Cacciari:2001td, Cacciari:2012ny, Cacciari:2015fta} in the \bbbar inclusive production cross section. 

A new \bparking trigger strategy was designed and deployed in time for the beginning of \pp collisions in 2022. The new strategy focuses on inclusive dilepton trigger algorithms, operating at high rates, that aim to provide gains in physics reach significantly beyond those expected from an increase in \Lint. The primary aim for Run~3 is to substantially expand the \PB physics program of CMS by using new or improved trigger algorithms and a data parking strategy to provide high acceptance for dimuon and dielectron final states from \PQb hadron decays. Dimuon final states are central to the CMS \PB physics program and the majority of analyses leverage this clean experimental signature. An inclusive dimuon trigger strategy, which differs from the exclusive strategy deployed in Run~2, was identified as a key beneficiary of the data parking approach. Further, new triggers recording dielectron final states are specifically aimed at improving the statistical precision of the \RK measurement.

\subsubsection{Trigger strategy}
\label{sec:trigger_strategy_run3}

The \bparking trigger and data parking strategies of 2018 have been adapted to accommodate changes in LHC operations and the evolving needs of the \PB physics program for Run~3. During 2022, the LHC machine parameters shifted to a new mode of operation in which \Linst is delivered to the experiments at a constant value of ${\approx}\sci{2.0}{34}\invcms$ (corresponding to $\text{PU} \approx 57$) for several hours. This luminosity-leveling mode, introduced in Section~\ref{sec:Challenges}, allows experiments to accumulate large data samples as quickly as possible, under conditions suitable for data analysis---\ie, with a manageable amount of pileup---such that the quality of the event reconstruction and physics performance is not compromised. Following a leveling period of up to six hours in 2022, the \Linst value decreases during the remainder of the LHC fill, typically over a time period of 6--12\unit{hours}.

The change in the LHC machine parameters between Run~2 and Run~3 have important consequences for the trigger and data parking strategies described here. Since there are no idle resources available for the first few hours of an LHC fill, due to the aforementioned leveling, a fraction of the total rate budget for the L1 trigger has to be allocated to the dilepton trigger algorithms described here. This is in contrast to operations in 2018, during which the single-muon trigger algorithms were not enabled when the LHC operated with \Linst values above $\sci{1.7}{34}\invcms$. In the case of the new dielectron trigger algorithms, the allocation is increased later in an LHC fill as  \Linst falls and idle resources become increasingly available.

The standard dimuon trigger algorithms deployed in Run~2 to cater for \PB physics analyses typically required a reconstructed dimuon system plus additional constraints specific to a particular physics process to control trigger rates. For instance, in the HLT algorithms, the reconstructed invariant mass was restricted to a set of windows around the \PBs meson and quarkonia masses (\eg, for the process \bstomm), or the presence of an additional particle track was required (\eg, for \BKmm), or else the dimuon vertex position was required to be displaced with respect to the \pp luminous region (\ie, for any $\PQb\to\MM\PX$ process).

The new dimuon trigger algorithms do not impose any of the rate-reducing constraints highlighted above and are instead designed to provide broad coverage for a range of physics processes, by employing minimal kinematical and topological requirements.  In contrast to the single-muon trigger strategy employed in 2018, where the thresholds were adjusted throughout the fill, the new dimuon triggers are always enabled and employ fixed thresholds. The dimuon trigger algorithms exploit improvements in both the L1 and HLT systems~\cite{CMS:2023gfb} implemented during LS2: the former now performs muon-track finding using the Kalman filter technique~\cite{CMS:2019qux} to provide an improved muon \pt estimate, and the latter uses new algorithms implemented in a heterogeneous computing environment, comprising both CPU and GPU cores. The new algorithms both speed up track reconstruction and improve the track \pt resolution~\cite{Bocci:2020pmi}, which in turn can be exploited by the HLT-based muon algorithms. As a consequence, there is a substantial improvement in the experimental acceptance for a number of interesting production and decay modes that yield prompt and nonprompt dimuon final states, such as $\PGh\to\PGmp\PGmm$ and \bstomm. Furthermore, the new algorithms also improve the acceptance for \PQb hadron decay chains that produce additional particles with nonnegligible lifetimes, such as \bdtokshortjpsi.

The dielectron trigger algorithms target the measurement of the \RK observable. The measurement reported in Section~\ref{sec:rk_run2} is limited by the finite number of reconstructed \BKee decays, because of the challenges associated with small branching fractions and low-\pt daughter particles for the probe-side \PB meson decays, as detailed in Section~\ref{sec:challenges-2018}. Here, an alternative strategy is explored by using the trigger algorithms to directly identify the electron pairs produced in \btosee transitions. This approach removes the aforementioned challenges but requires the use of electron reconstruction and identification algorithms, implemented in the L1 system, with limited performance. These L1 algorithms measure ${\Et = E\sin{\theta}}$, where $\theta$ is the polar angle~\cite{CMS:2008xjf}. Low \Et thresholds are required to provide adequate experimental acceptance for \BKee decays and, as a result, trigger rates in the multi-\unit{kHz} range are required.  The ``dynamic threshold'' strategy used in 2018, \ie, progressively reducing kinematical thresholds as \Linst falls during an LHC fill, is deployed again to access the lowest possible \Et thresholds and maximize the acceptance. 

Additional details on the dimuon and dielectron trigger algorithms are provided in Sections \ref{sec:dimuon-2022} and \ref{sec:di-ele-2022}, respectively.

\subsubsection{Data parking strategy}
\label{sec:data_parking_run3}

In 2018, data parking referred to the delayed reconstruction of trigger data streams, on the order of several months (long-term parking). Since 2022, the definition of data parking has shifted somewhat to include prompt reconstruction (\ie, processing typically starting within 48\unit{hours}), contingent upon the availability of computing resources, and short-term parking, where the reconstruction is delayed until resources become available. Crucially, the ability to promptly reconstruct data streams serving the core CMS physics program must not be compromised by the data-parking strategy.

The initial budgeting of resources for data parking in 2022 was predicated on the assumption that 25\% of the dimuon trigger data stream would undergo prompt reconstruction, while the remaining 75\% would be (short-term) parked for delayed reconstruction. Additionally, it was foreseen that a small fraction of the dielectron data stream would also be promptly reconstructed, on an opportunistic basis and subject to resource availability. In practice, during 2022 and 2023, the data streams generated by both the dimuon and dielectron trigger algorithms were promptly reconstructed in full, since the availability of resources exceeded initial expectations.

\subsubsection{Inclusive triggers for dimuon final states}
\label{sec:dimuon-2022}

The dimuon strategy for Run~3 is implemented in two L1 algorithms and three HLT algorithms. All algorithms were enabled at the beginning of the data-taking period in 2022.

Both L1 algorithms require two oppositely charged muons. The ``central-$\eta$'' algorithm also imposes ${\abs{\eta} < 2}$ and ${\pt > 0\GeV}$ on each muon, even if an implicit threshold of $\pt \gtrsim 3\GeV$ is required if the muons are to reach the muon detectors in the central region, and ${\abs{{\Delta}\eta(\PGm_1,\PGm_2)} < 1.6}$. These thresholds were updated to ${\pt > 3\GeV}$ and ${\Delta}R(\PGm_1,\PGm_2) < 1.4$ in 2023. The ``higher \pt'' algorithm requires ${\pt > 4\GeV}$ and ${\deltar(\PGm_1,\PGm_2) < 1.2}$. The peak L1 trigger rate recorded by these two algorithms was 18\unit{kHz} at ${\Linst = \sci{2.0}{34}\invcms}$, which is approximately 20\% of the total L1 trigger rate. This large allocation was possible as a result of an increase in the total L1 trigger rate from ${\approx}90\unit{kHz}$ in 2018 to ${\approx}100\unit{kHz}$ in 2022, stemming from operational improvements.

The first HLT algorithm, known henceforth as the ``inclusive low-mass dimuon trigger'', imposes the following requirements: two opposite-charge muons, one muon satisfying ${\pt > 4\GeV}$ and the other ${\pt > 3\GeV}$, a dimuon vertex fit probability of ${\pvtx(\PGm \PGm) > 0.5\%}$, and a reconstructed dimuon invariant mass satisfying ${2m_{\PGm}^{\text{PDG}} < \mMM < 8.5\GeV}$. Here the superscript ``PDG" refers to the global average of experimentally measured mass values as reported by the Particle Data Group (PDG)~\cite{Workman:2022ynf}. This algorithm now serves as the main trigger for all physics analyses of final states containing a low-mass dimuon system. The HLT algorithm results in a peak HLT rate of 1.6\unit{kHz} at ${\Linst = \sci{2.0}{34}\invcms}$. 

A second HLT algorithm, known as the ``displaced low-mass dimuon trigger'', imposes tighter requirements: two opposite-charge muons with both muons satisfying ${\pt > 4\GeV}$; $\pvtx(\PGm \PGm) > 10\%$; a reconstructed dimuon vertex that is sufficiently displaced from the \pp luminous region such that it satisfies ${\lxy/\lxyerr > 3}$, where \lxy and \lxyerr are, respectively, the measured transverse decay length and its uncertainty; and ${2m_{\PGm}^{\text{PDG}} < \mMM < 8.5\GeV}$. This trigger, with a lower trigger rate of 0.3\unit{kHz} at ${\Linst = \sci{2.0}{34}\invcms}$, acts as a backup to the inclusive trigger. Its event stream is a subset of that of the inclusive trigger. As a further fail-safe, the majority of all standard dimuon algorithms used during Run~2 are also maintained during Run~3, even if they also operate largely in the shadow of the new inclusive trigger.

The third algorithm records events that are likely to contain the production and decay of the \PGUPnS resonances, where ${n = 1, 2, 3}$. The requirements comprise: two oppositely charged muons, ${\pvtx(\PGm \PGm) > 0.5\%}$, a reconstructed dimuon system satisfying ${\pt > 10\GeV}$, ${\abs{y} < 1.4}$, and ${8.5 < \mMM < 11.5\GeV}$.

\begin{figure*}[!htb]
  \centering
  \includegraphics[width=0.87\textwidth]{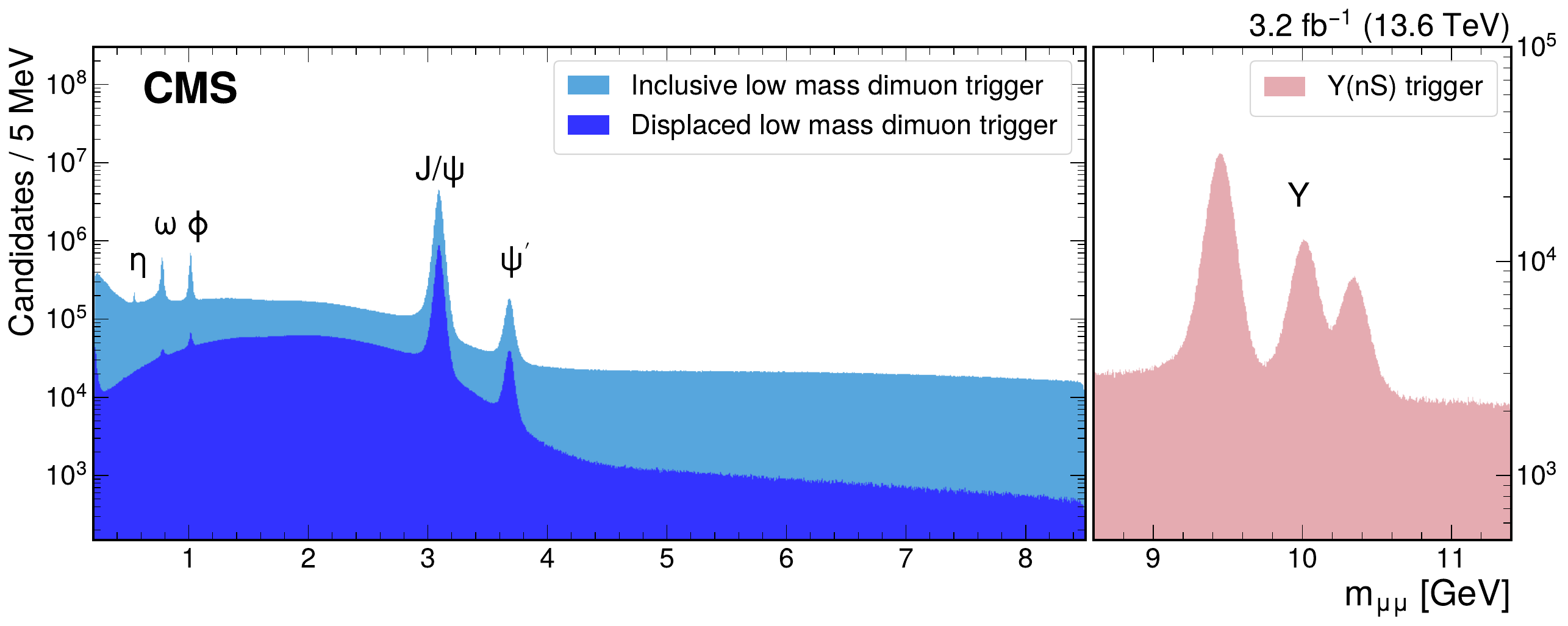}
  \caption{Dimuon mass spectra obtained from data recorded in 2022 during Run~3, corresponding to ${\Lint = 3.2\fbinv}$. In the range ${2m_{\PGm}^{\text{PDG}} < \mMM < 8.5\GeV}$, the light blue distribution represents the subset of dimuon events triggered by the inclusive low-mass trigger algorithm, while the dark blue distribution shows the subset of dimuon events triggered by the displaced low-mass trigger path. In the range ${8.5 < \mMM < 11.5\GeV}$, dimuon events are instead triggered by the HLT paths targeting the \PGUPnS resonances, which are shown by the pink  distribution. }
  \label{ref:DiMuonSpectrum}
\end{figure*}

Figure~\ref{ref:DiMuonSpectrum} shows the invariant mass distribution of dimuon final states for events from the data streams recorded by the trigger algorithms described above. The reconstructed dimuon candidates are required to satisfy loose kinematical, topological, and quality criteria. Several resonances are clearly visible over the continuum background, such as \PGh, \PGo, \PGf, \PJGy, \Pgy, and \PGUPnS ($n = 1, 2, 3$).

\begin{figure*}[!htb]
  \centering
  \includegraphics[width=0.49\textwidth]{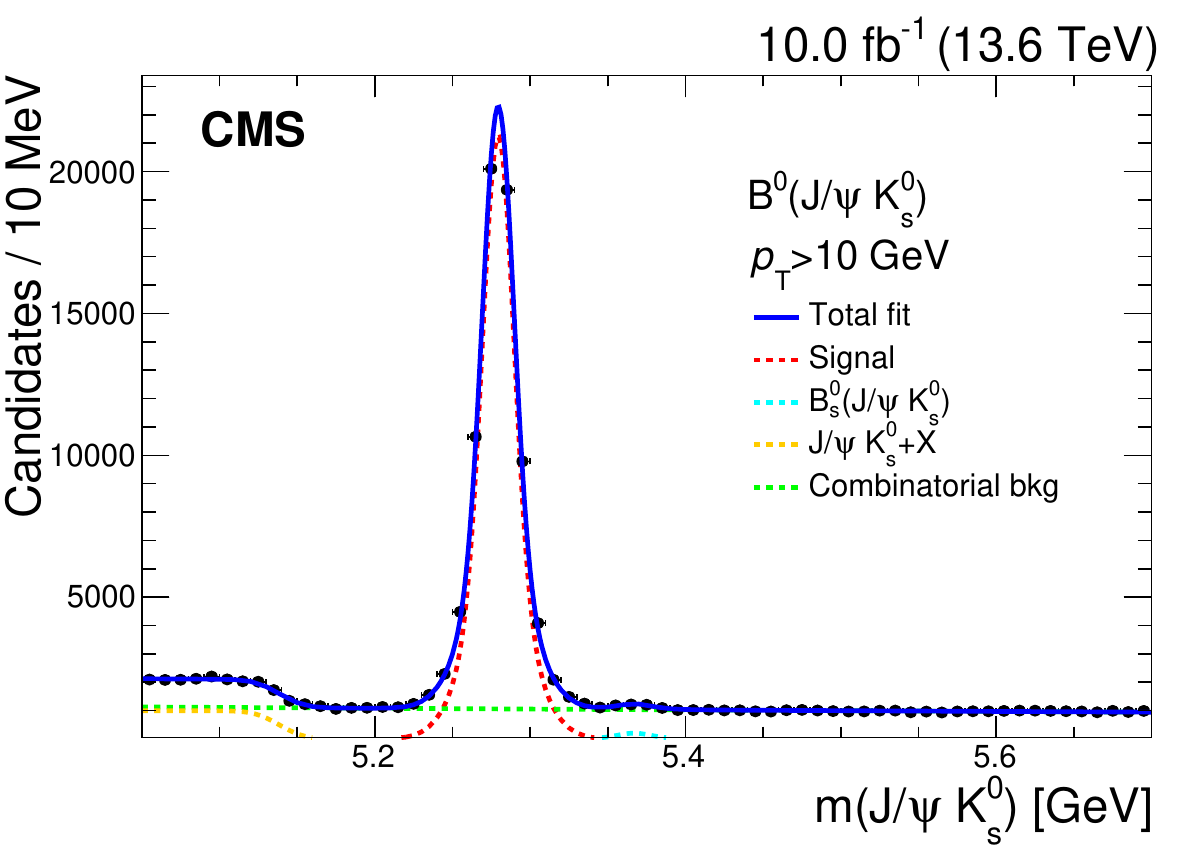}
  \includegraphics[width=0.49\textwidth]{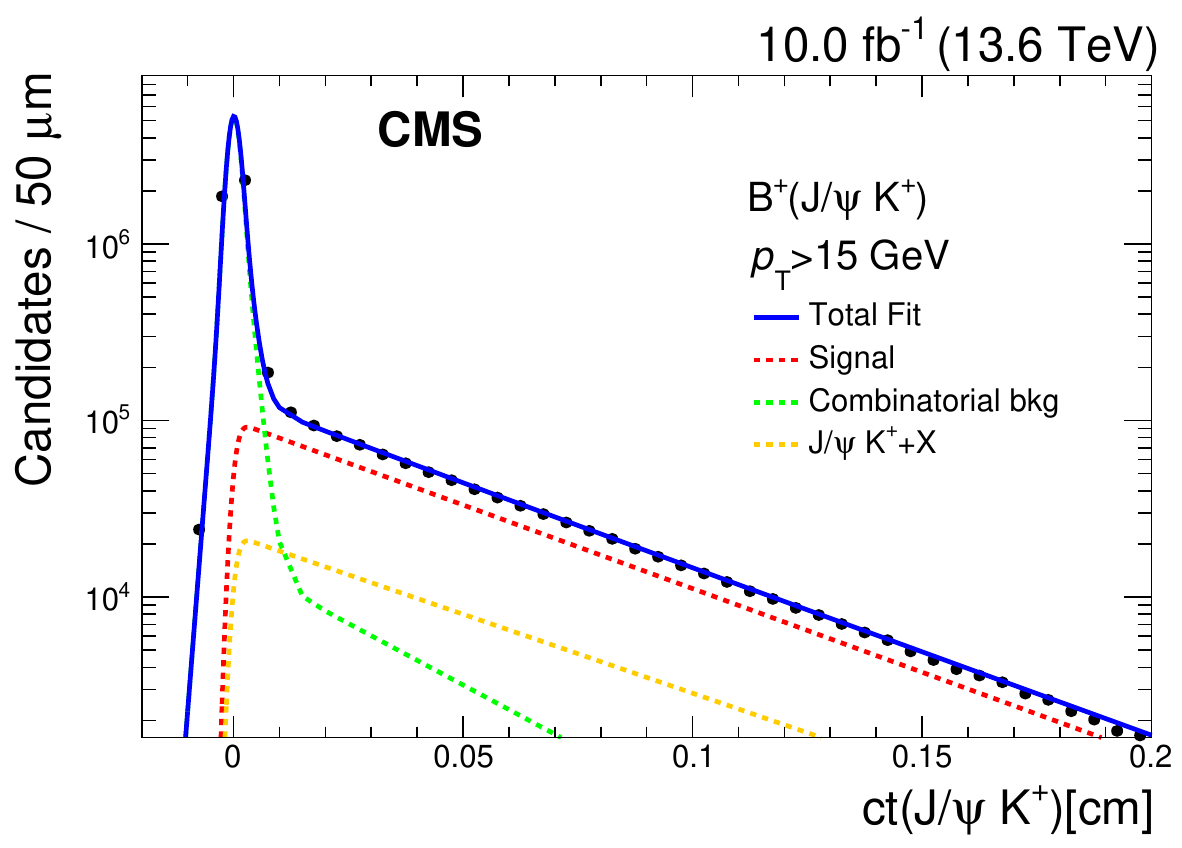}
  \caption{Left: invariant mass distribution for candidate \bdtokshortjpsi decays. Right: proper decay length (ct) distribution obtained from candidate \BKJpmm decays. Both types of candidates are reconstructed from events recorded using the dimuon triggers.}
  \label{fig:dimuon-benefits}
\end{figure*}

Several physics analyses stand to benefit from the new inclusive approach. For example, the \bdtokshortjpsi decay is used to measure the $\sin(2\beta)$ angle in the CKM unitary triangle~\cite{LHCb:2015brj}. The new inclusive trigger provides a factor~of~12 increase in the signal yield per \fbinv with respect to the standard dimuon triggers available in Run~2. The \cmsLeft panel of Fig.~\ref{fig:dimuon-benefits} shows the mass distribution of \bdtokshortjpsi candidates determined from events recorded by the dimuon triggers and selected according to the following requirements: ${\abs{m_{\PGpp\PGpm} - m^{\text{PDG}}_{\PKzS}} < 10\MeV}$, the reconstructed $\PJGy\to\PGmp\PGmm$ candidate satisfies ${\abs{\mMM - m^{\text{PDG}}_{\PJGy}} < 150\MeV}$ and ${\pt > 8\GeV}$, the reconstructed \PBz candidate satisfies ${\pt > 10\GeV}$, and the reconstructed \PJGy and \PKzS candidates form a vertex with a fit probability greater than 10\%. A one-dimensional unbinned extended maximum likelihood fit is performed to the mass distribution: two Gaussian functions with a common mean are used to model the signal distribution, an exponential function is used for the combinatorial background, and an error function is used to describe the $\PJGy\PKzS+\PX$ background.

Another example of a substantial gain is in the decay $\PGLb \to \PJGy (\to \PGmp\PGmm)\PGL$, which would permit a new angular analysis of this channel in CMS. In addition, significantly improved acceptance to low-mass resonances such as \PGo and \PGf allow for CPV measurements in decays such as $\PBzs \to \PGf(\to\MM)\PGf(\to\PKp\PKm)$.

The \cmsRight panel of Fig.~\ref{fig:dimuon-benefits} shows the proper decay length (ct) distribution as obtained from a control sample of \BKJpmm decays in the events recorded by the dimuon triggers. The following selection criteria are also applied: the reconstructed kaon, \PJGy, and \PBp candidates are required to satisfy \pt thresholds of 2, 8, and 15\GeV, respectively, and the \PJGy and kaon candidates are required to form a vertex with a fit probability greater than 15\%. The \PJGy candidate is also required to satisfy ${\abs{\mMM - m^{\textrm{PDG}}_{\PJGy}} < 150\MeV}$. The ct distributions for both the signal and $\PJGy\PKp + \PX$ background processes are modeled with decaying exponential functions of the form ($e^{-\mathrm{ct}/\lambda}$); the combinatorial background uses a decaying exponential function with both negative and positive terms. All exponential functions are convolved with a Gaussian function to reflect the finite time resolution. The absence of any displacement requirements in the inclusive trigger is expected to improve the accuracy of the lifetime measurements of the comparatively shorter-lived \PBpc meson. 

\begin{figure*}[!htb]
  \centering
  \includegraphics[width=0.49\textwidth]{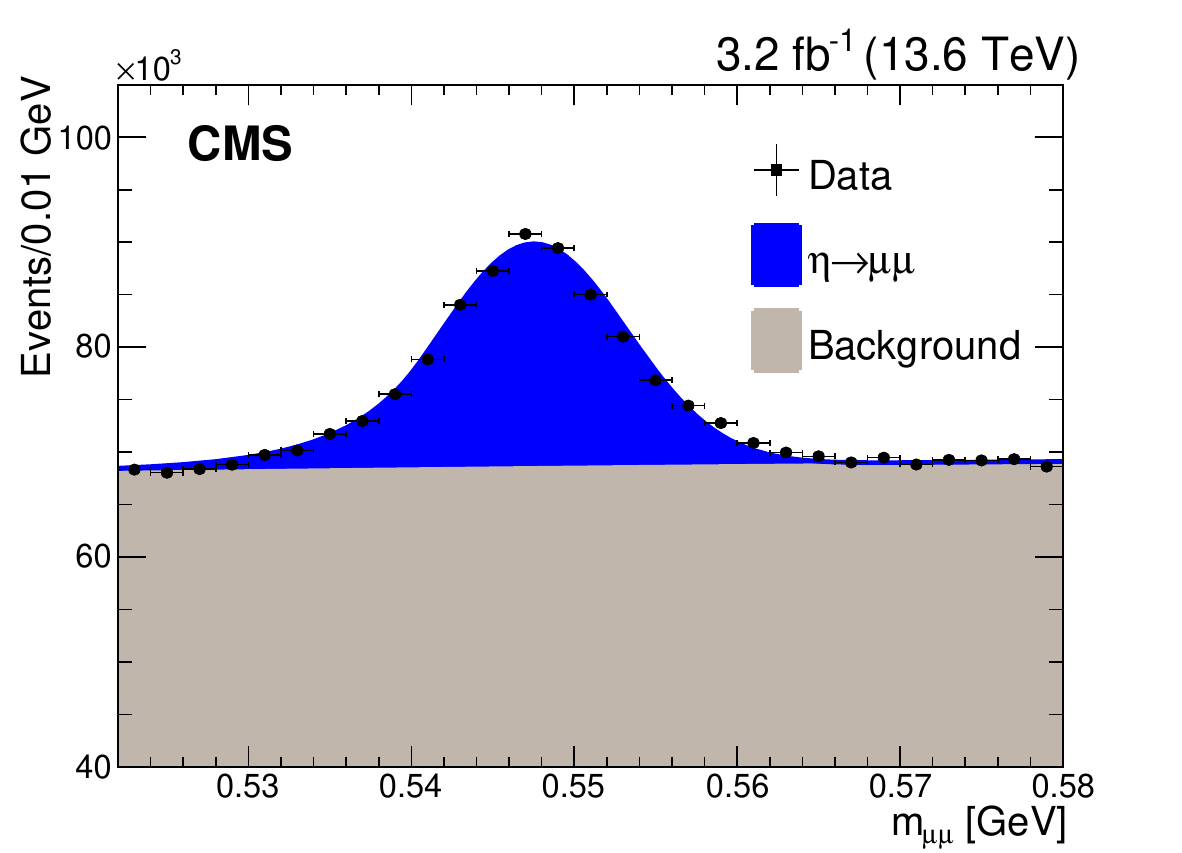}
  \includegraphics[width=0.49\textwidth]{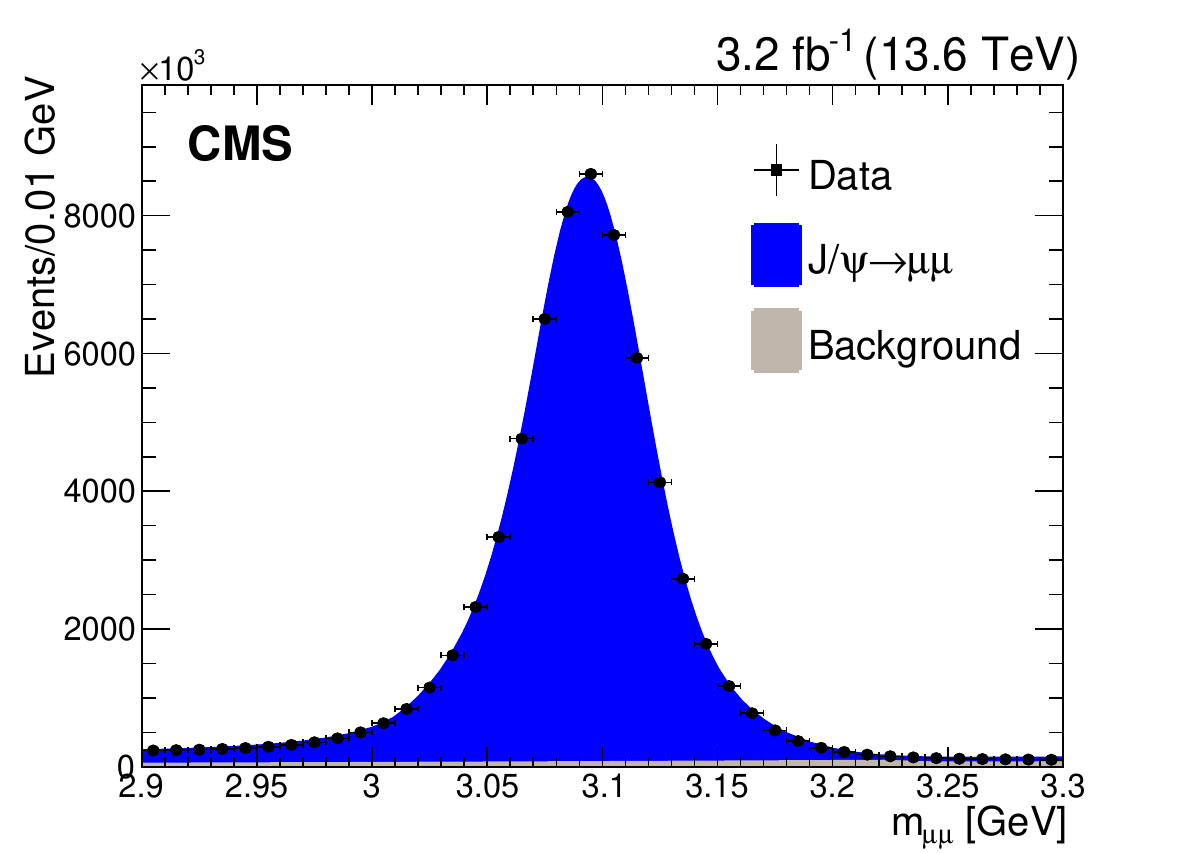}
  \caption{Dimuon invariant mass distributions in the \PGh (\cmsLeft) and \PJGy (\cmsRight) mass regions, as obtained from data recorded by the inclusive low-mass dimuon trigger algorithm.}
  \label{ref:DiMuonSpectrumFitPart1}
\end{figure*}

Figure~\ref{ref:DiMuonSpectrumFitPart1} shows regions of the invariant mass distribution, as recorded by the inclusive low-mass trigger algorithm, that highlight the \PGh and \PJGy resonances. Fits are performed to the data in these regions, as well as to other regions that cover the \PGf, \Pgy, and \PGUPnS ($n = 1, 2, 3$) resonances. These fits are used to extract the measured dimuon mass \mMM and resolution $\sigma(\mMM)$ parameters for each resonance. The measured dimuon mass scale $s_{\MM}$ is defined in terms of the PDG mass value~\cite{Workman:2022ynf} for the resonance under consideration: ${s_{\MM} = \abs{\mMM - m^{\text{PDG}}_{\MM}}/m^{\text{PDG}}_{\MM}}$. For the \PJGy, \Pgy and \PGUPnS resonances, the intrinsic width of the resonance is negligible with respect to the experimental resolution. These resonances are modeled by the CB and Gaussian functions that share the same \mMM parameter. The latter function is used to correctly model effects due to the finite experimental resolution.
The continuous background is described by an exponential function. For some resonances, such as the \PGf, the CB and Gaussian functions are convolved with a Breit--Wigner function that models the natural lineshape for the resonance. The dimuon mass scale is accurate to the per-mil level, and the relative dimuon mass resolution $\sigma(\mMM)/\mMM$ is measured to be in the range 0.8--1.6\% for the different resonances.

\subsubsection{Inclusive triggers for dielectron final states}
\label{sec:di-ele-2022}

The dielectron trigger algorithms require the presence of two electrons that satisfy loose kinematical and ID quality requirements. The L1 algorithm identifies electron/photon (\egamma) candidates from patterns of measured energy deposits in the ECAL system. Both \egamma candidates must satisfy $\abs{\eta} < 1.2$ and a variable threshold applied to the \Et measurement of each object. The highest \Et threshold applied is 11\GeV, which results in an L1 trigger rate of ${\approx}2\unit{kHz}$ for ${\Linst = \sci{2.0}{34}\invcms}$. Copies of the algorithm exist with lower thresholds, available in 0.5\GeV steps down to 5\GeV, which are enabled later during an LHC fill when the \Linst and pileup are smaller and idle resources become available. 
The evolution of thresholds is tuned such that the L1 system operates close to its design limit of 100\unit{kHz} throughout each fill.  The L1 trigger algorithms also impose an upper bound on the separation distance \deltar between the two \egamma candidates; the upper bound evolves between values from 0.6 to 0.9 with the decreasing \Et threshold. This requirement helps to control the trigger rate, while maintaining an efficiency of 95\% for physics processes that produce two final-state electrons with four-momenta that yield an invariant mass below 6\GeV.

Positive trigger decisions from the L1 algorithms act as seeds for companion algorithms implemented in the HLT, which are able to exploit superior physics performance for reconstructed particle candidates by combining information from the tracker, ECAL, and HCAL subdetectors. Each version of the L1 algorithm, with its unique \Et threshold, is paired with a version of the HLT logic that imposes a corresponding threshold on the measured \Et values determined by the HLT algorithms. The HLT \Et thresholds, which range from 4 to 6.5\GeV, are lower than the L1 \Et thresholds so as to maintain high efficiency while controlling purity. Several quality-related criteria are also imposed by the HLT algorithms to help identify genuine electrons from interesting physics processes, while rejecting other particles misidentified as electrons. The ID criteria follow closely those used by the standard HLT electron reconstruction algorithms~\cite{CMS:2020uim}, with minor changes that improve the per-electron efficiency from 75 to 90\%, relative to the L1 \egamma object. An upper bound of 6\GeV is imposed on the invariant mass of the dielectron system, which maps closely to the \deltar requirements imposed at L1. Overall, the HLT requirements are tuned with an emphasis on high signal efficiency, while substantially improving the purity of the L1 data stream.

Table~\ref{tab:CMS-trigger-rates-2022-new} lists the unique combinations of \Et and \deltar thresholds used in the L1 and HLT algorithms to record events containing dielectron final states. Settings with lower \Et thresholds are enabled as the \Linst decreases during an LHC fill. At least one setting is enabled from the start of every LHC fill, \ie throughout the period of luminosity leveling. In the L1 system, the highest \Et threshold is enabled for \Linst values of $\sci{2.0}{34}\invcms$ and above, which corresponds to as many as 57 pileup interactions. For each setting, the \Lint and the mean number of pileup interactions are aggregated over the periods for which the setting provides the lowest enabled L1 \Et threshold. The peak L1 and HLT trigger rates are listed for each setting. The efficiency for recording events containing \BKee decays changes significantly across the different settings, differing by a factor of ${\approx}20$ for the settings with the lowest and highest L1 \Et thresholds.

\begin{table*}[!htb]
  \centering
  \caption{Trigger configurations, defined by unique combinations of L1 and HLT \Et thresholds (applied to each electron candidate) and L1 \deltar, used to record events containing dielectron final states. The thresholds on the L1 and HLT \Et (L1 \deltar) values are lower (upper) bounds. The \Lint value and the mean number of pileup interactions recorded by each trigger combination are aggregated over periods for which each combination provided the lowest enabled L1 \Et threshold. Representative peak L1 and HLT trigger rates are given for each setting.}
  \renewcommand{\arraystretch}{1.3}
  \begin{tabular}{ccccccc}
    L1 \Et     & L1 \deltar & HLT \Et    & \Lint      & Mean & Peak L1    & Peak HLT   \\
    $[\GeVns]$ &            & $[\GeVns]$ & $[\fbinv]$ & PU   & rate [kHz] & rate [kHz] \\
    \hline
    11.0 & 0.6 & 6.5 & 1.6 & 45.6 & 2.2 & 0.1 \\
    10.5 & 0.6 & 6.5 & 1.1 & 42.2 & 3.0 & 0.3 \\
    9.0  & 0.7 & 6.0 & 8.8 & 47.4 & 9.3 & 0.6  \\
    8.5  & 0.7 & 5.5 & 3.3 & 46.2 & 13  & 0.9  \\
    8.0  & 0.7 & 5.0 & 6.9 & 39.1 & 16  & 1.2  \\
    7.5  & 0.7 & 5.0 & 1.6 & 40.3 & 23  & 1.4  \\
    7.0  & 0.8 & 5.0 & 2.7 & 36.3 & 27  & 1.3  \\
    6.5  & 0.8 & 4.5 & 3.6 & 31.2 & 35  & 1.3  \\
    6.0  & 0.8 & 4.0 & 2.5 & 27.4 & 46  & 1.4  \\
    5.5  & 0.8 & 4.0 & 0.7 & 23.6 & 54  & 1.0  \\
    \multicolumn{3}{l}{Other combinations} & 1.0 & \NA & \NA & \NA \\
    \hline
    \multicolumn{3}{l}{Total} & 33.9 & 34.8 & \NA & \NA \\
  \end{tabular}
  \label{tab:CMS-trigger-rates-2022-new}
\end{table*}

\begin{figure*}[!htb]
  \centering
  \includegraphics[width=0.49\textwidth]{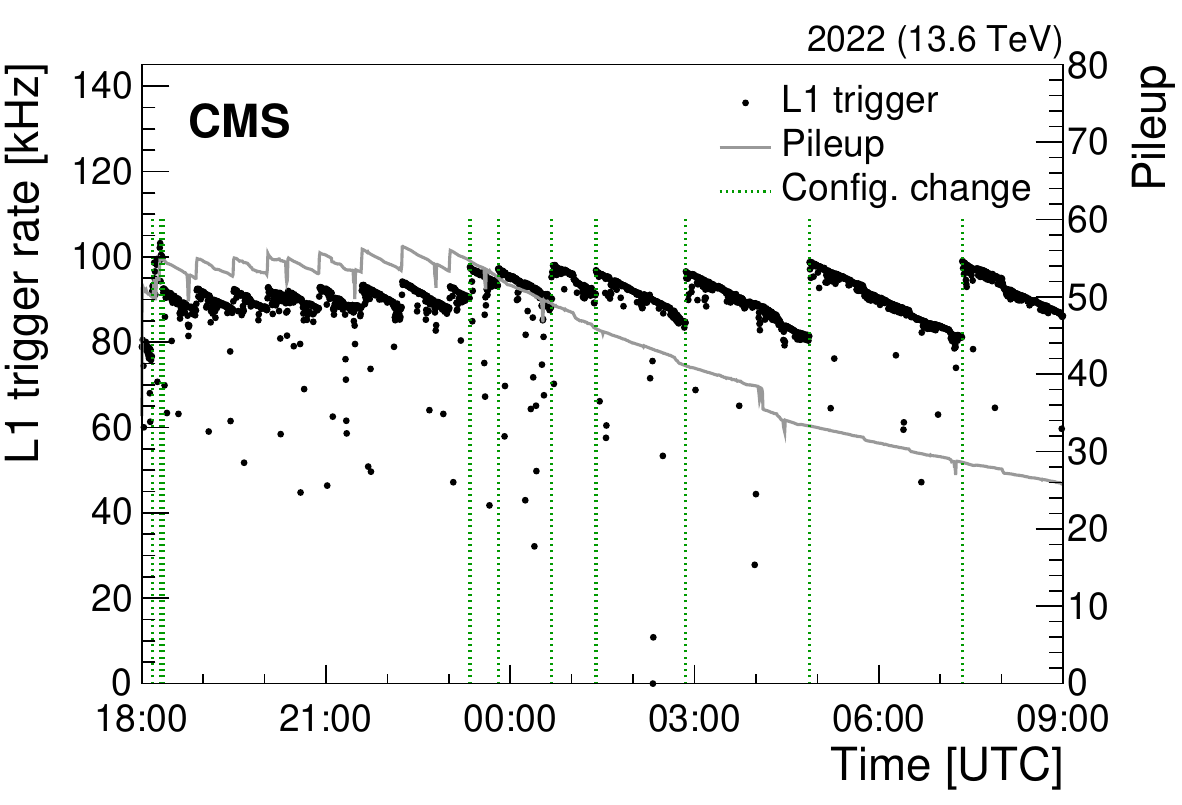}
  \includegraphics[width=0.49\textwidth]{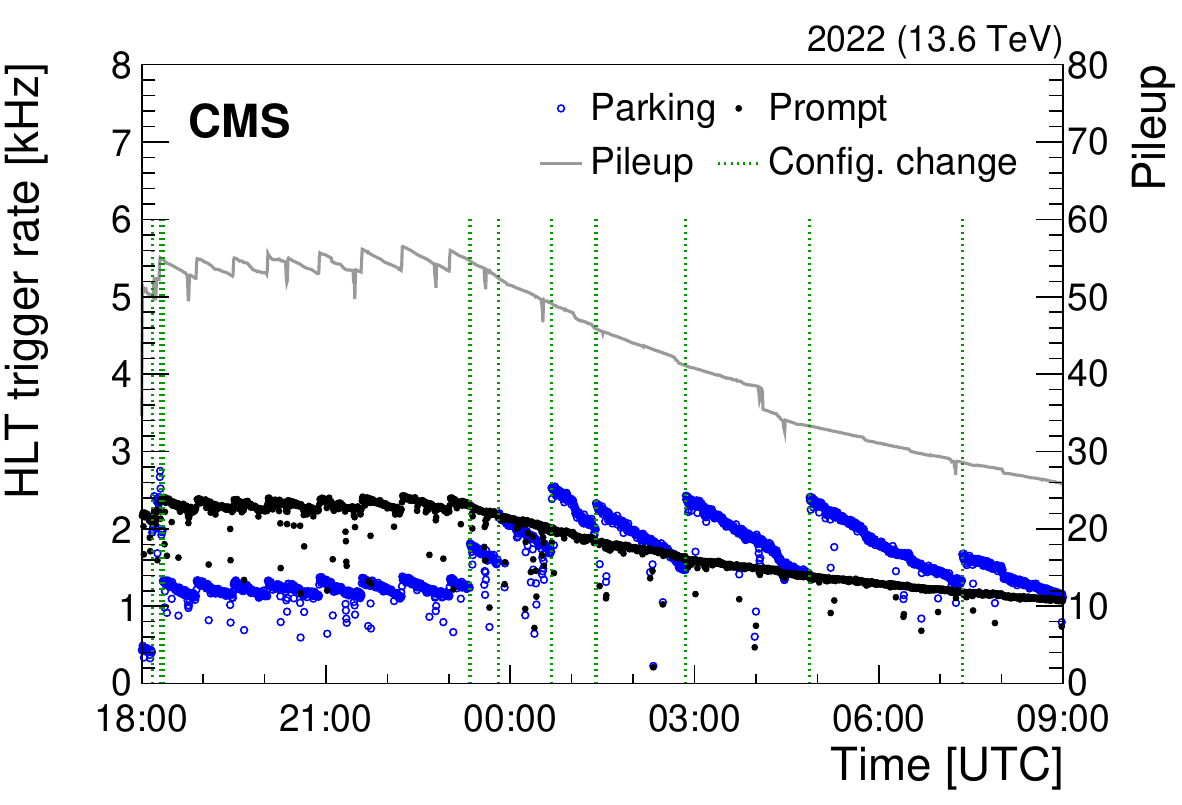}
  \caption{Total L1 (\cmsLeft) and HLT (\cmsRight) trigger rates, and the number of pileup interactions, shown as a function of time for a representative LHC fill during 2022. The rates for the promptly reconstructed core physics (black solid markers) and \bparking (blue open markers) data streams are shown separately in the \cmsRight panel. Occasional lower rates are observed due to transient effects, such as the throttling of the trigger system in response to subdetector dead time~\cite{CMS:2016ngn}. Changes in the trigger configuration are indicated by vertical green dashed lines.}
  \label{fig:diele-rates-2022}
\end{figure*}

The \cmsLeft and \cmsRight panels of Fig.~\ref{fig:diele-rates-2022} illustrate how the L1 and HLT trigger rates evolve during a typical LHC fill during 2022. Similar to 2018, instantaneous increases in rate are observed when changes in the trigger configuration occur, and the L1 system operates with a trigger rate close to 100\unit{kHz} throughout the LHC fill. The L1-based algorithms for the dielectron and dimuon triggers (the latter are described in Section~\ref{sec:dimuon-2022}) contribute up to 54 and 18\unit{kHz} to the total rate shown in the \cmsLeft panel of Fig.~\ref{fig:diele-rates-2022}.

The \cmsRight panel of Fig.~\ref{fig:diele-rates-2022} shows the total HLT trigger rates for both the promptly reconstructed data stream, which serves the core CMS physics program, and the \bparking stream. The latter stream includes contributions from both the dielectron and dimuon algorithms. The HLT algorithms for the dielectron triggers typically operate at a rate of ${\approx}100\unit{Hz}$ during the luminosity leveling period and as high as ${\approx}1.3\unit{kHz}$ later in an LHC fill. The dimuon algorithms operate at a total rate of up to ${\approx}1.2\unit{kHz}$, which monotonically falls with \Linst during an LHC fill.

\begin{figure*}[!htb]
  \centering
  \includegraphics[width=0.65\textwidth]{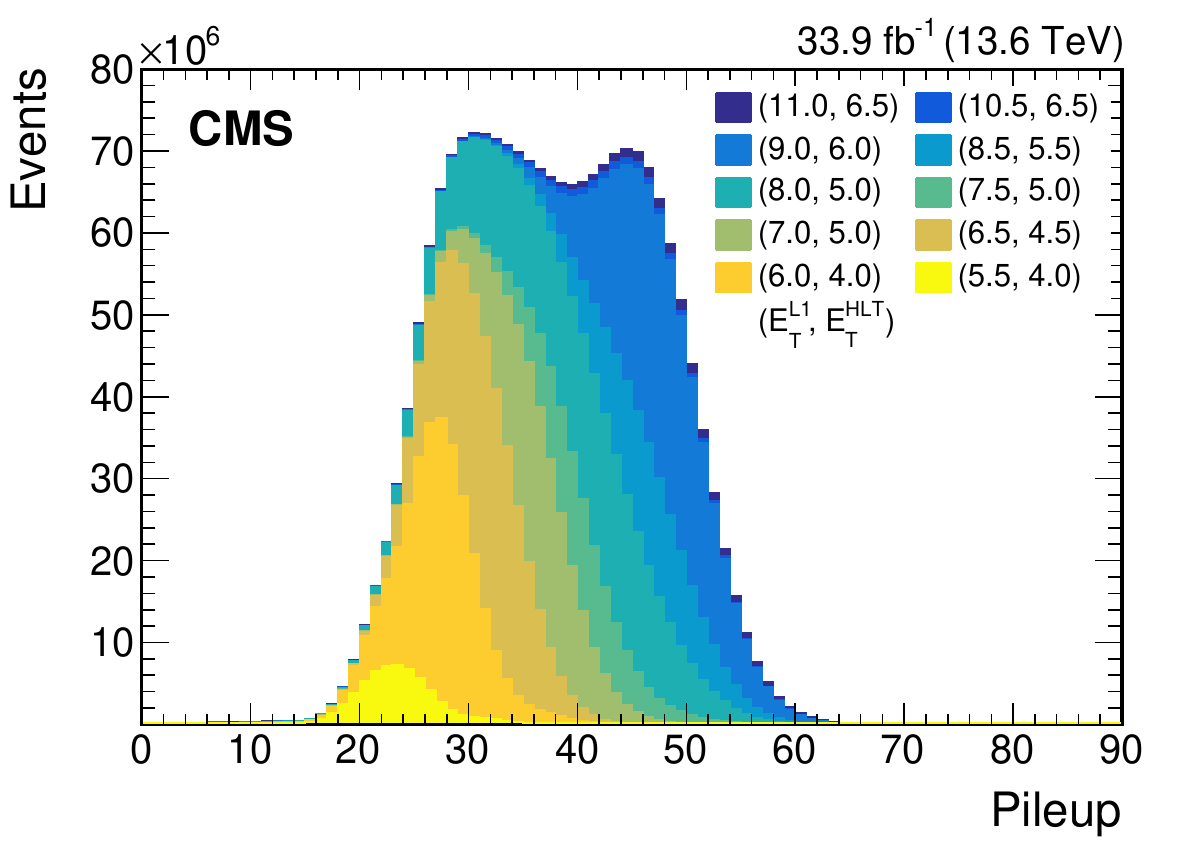}
  \caption{Pileup distribution measured in the dielectron data set. Contributions from each trigger combination are shown, with the histogram areas normalized to the number of events recorded by each trigger. \label{fig:diele-pileup-2022} }
\end{figure*}

\begin{figure*}[!htb]
  \centering
  \includegraphics[width=0.65\textwidth]{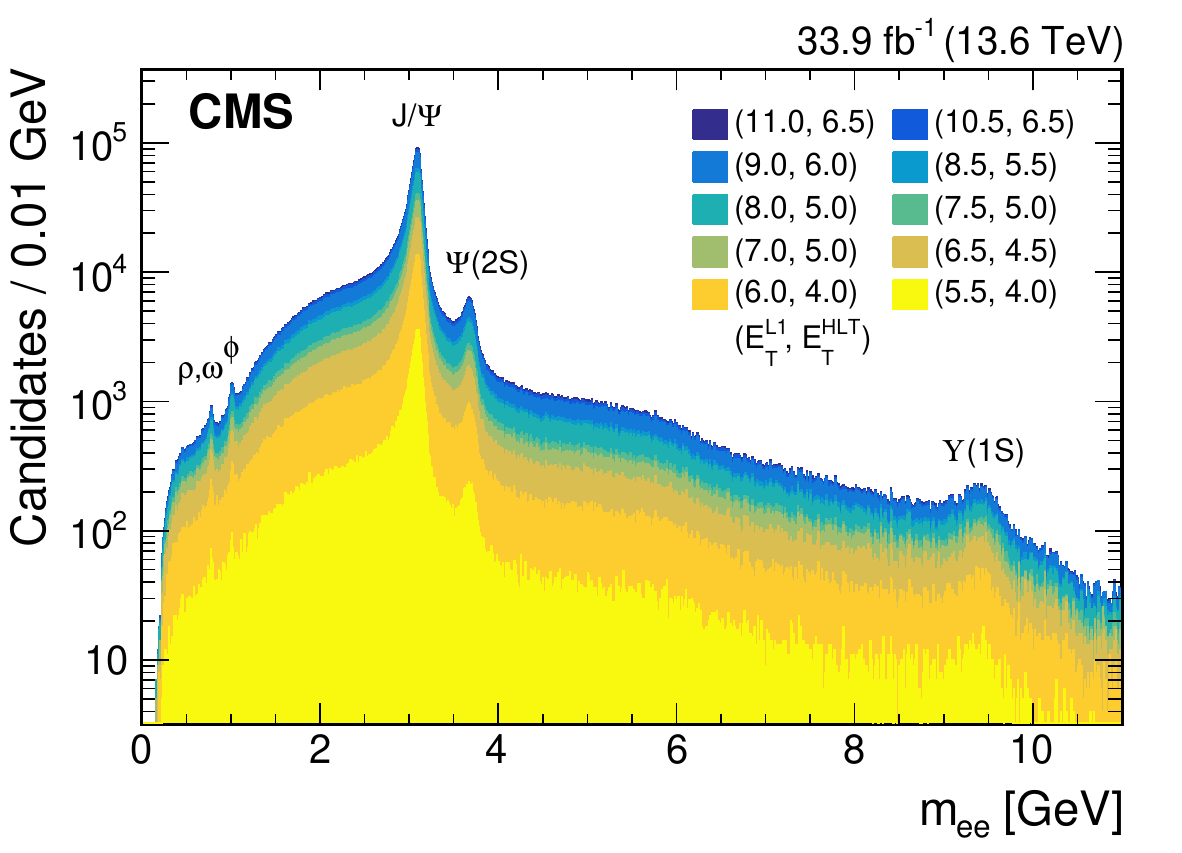}
  \caption{The invariant mass distribution for pairs of oppositely charged electrons originating from a common vertex, reconstructed from the dielectron data set. \label{fig:diele-inv-mass-2022} }
\end{figure*}

Figure~\ref{fig:diele-pileup-2022} shows the pileup distribution for the dielectron data set, along with the contributions from each of the individual trigger combinations. Figure~\ref{fig:diele-inv-mass-2022} shows the invariant mass distributions for pairs of oppositely charged electrons originating from a common vertex, as obtained from the dielectron data set. Both electrons are required to satisfy a minimal set of kinematic and ID criteria, and both electrons must be matched to the electron candidates responsible for the positive trigger decision. Peaks in the data resulting from the $\PGr/\PGo$, \PGf, \PJGy, \Pgy, and \PgUa resonances are clearly visible.

\begin{figure*}[!htb]
  \centering
  \includegraphics[width=0.65\textwidth]{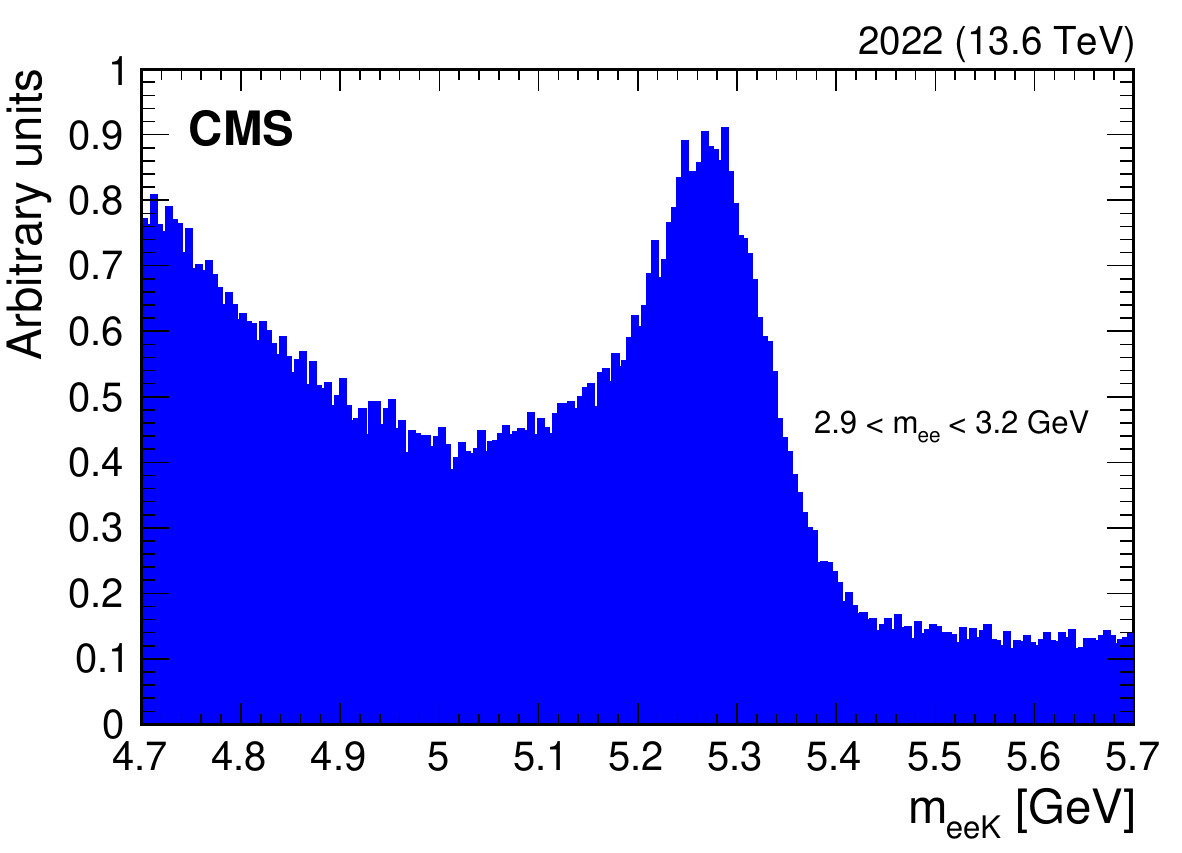}
  \caption{The invariant mass distribution for candidate \BKJpee decays, reconstructed from the dielectron data set. The histogram is normalized to unit area.\label{fig:btokjpsi-2022} }
\end{figure*}

Figure~\ref{fig:btokjpsi-2022} shows the invariant mass distributions for candidate \BKJpee decays, as obtained from the dielectron data set. A pair of oppositely charged electrons originating from a common vertex, along with a charged-particle track assumed to be the kaon, are required to satisfy a minimal set of kinematic and ID criteria, and both electrons must be matched to the electron candidates responsible for the positive trigger decision. The invariant mass of the dielectron system is required to satisfy ${2.9 < \mEE < 3.2\GeV}$ so as to be consistent with originating from the \PJGy meson decay.

\subsubsection{Physics potential}

The new dilepton triggers will accumulate large data samples in Run~3, with substantially improved acceptances for many physics processes relative to Run~2. The dimuon triggers will provide substantial gains, beyond those expected from an increase in \Lint alone, for a range of interesting physics processes. In particular, processes that involve the associated production of \MM pairs and additional particles will benefit. Examples include \BKmm, \bctojpsilnu, \bdtokshortjpsi, and \bstojpsiphi decays. The new dielectron triggers aim primarily to allow a precise measurement of \RK using the \BKee decay. Investigations are ongoing for the potential to perform an angular analysis or differential branching fraction measurements for decays such as \BKee and \BKstee. Furthermore, both the dimuon and dielectron triggers provide ample scope for novel BSM searches. Both triggers also use L1 algorithms that are being adopted by the data scouting trigger streams to further improve acceptance to low-mass states.

\subsection{Alternative data parking strategies in Run 3}
\label{sec:run3altParking}

After a break in Run~2 when the parking strategy focused predominantly on \PB physics analyses, the original idea of a more diversified parking strategy, meant to complement the existing standard triggers, was revived in Run~3. The main goal is to overcome the limited HLT bandwidth and to collect a sufficient number of events for specific physics goals, such as studying final states produced via VBF, exploring possible anomalous $\PH\PH$ production, and improving the sensitivity to searches for exotic LLPs. These three novel parking approaches are detailed in the following sections.
    
\subsubsection{VBF parking}
\label{subsec:run3parking_vbf}

Higgs boson production via the VBF channel is of paramount importance for the experimental study of the Higgs boson at the LHC. It is the second most common production mechanism, contributing about 10\% of the total \PH production cross section. The sensitivity to several \PH decay modes, such as ${\PH \to \PGtp\PGtm}$~\cite{CMS-PAPERS-HIG-16-043}, ${\PH \to \text{invisible}}$~\cite{CMS-PAPERS-HIG-20-003}, and ${\PH \to \PGmp\PGmm}$~\cite{CMS-PAPERS-HIG-19-006}, is driven by the sensitivity to \PH VBF production. The VBF production is also important for a variety of measurements, \eg, in effective field theory measurements that constrain dimension-6 operators~\cite{paper:SMEFT}, and in  $\PH\PH$ production, where it grants unique access to the ${\PV\PV\PH\PH}$ coupling~\cite{CMS-PAPERS-B2G-22-003}. The VBF triggers constitute an interesting workaround to the low signal efficiencies obtained with triggers that must be sufficiently restrictive to keep rates under control. Instead of restricting the kinematic properties of the central physics objects from the signal of interest, VBF triggers place tighter constraints on the auxiliary jets. These requirements are often sufficient to significantly loosen or even entirely remove the selection on the central objects. 

A dedicated inclusive VBF L1 seed was already introduced in 2017~\cite{CMS-DP-2017-022}. This seed requires at least two jets with $\pt > 110$ and 35\GeV, respectively, and at least one pair of jets satisfying ${\mjj>650\GeV}$ among all pairs of jets with ${\pt>35\GeV}$ in the event. 

In 2023, the VBF trigger strategy was extended with the introduction of a retuned inclusive VBF L1 seed, as well as a set of exclusive seeds, each requiring either one additional muon, \tauh, \egamma object, or \ptmiss, or two additional central jets, contributing an L1 rate of about 10\unit{kHz} with ${\Linst = \sci{2}{34}\invcms}$. These L1 triggers are described in Table~\ref{tab:2023_VBFseeds}.

\begin{table*}[!htb]
    \centering
    \topcaption{Definition and rates of the VBF algorithms at L1. The quoted rates are for ${\Linst = \sci{2}{34}\invcms}$ and do not account for overlaps with other seeds.}
    \renewcommand{\arraystretch}{1.3}
    \cmsTable{
    \begin{tabular}{lccr}
      L1 trigger seed & \begin{tabular}[c]{@{}c@{}}VBF requirements\\ ($\pt^{j1}$, $\pt^{j2}$, \mjj)\end{tabular} & \begin{tabular}[c]{@{}c@{}}Requirements on \\ additional objects\end{tabular} & Rate {[}kHz{]} \\
    \hline
    VBF inclusive        & (90, 30, 800)   & \NA                                    & 5.0    \\
    \hline
    VBF + 2 central jets & (60, 30, 500)   & 2 central jets, $\pt^j>50\GeV$         & 3.0     \\
    VBF + \ptmiss        & (65, 30, 500)   & $\ptmiss>65\GeV$                       & 2.9     \\
    VBF + $\PGm$         & (90, 30, 500)   & 1 muon, $\pt^\PGm>3\GeV$               & 2.3     \\
    VBF + \tauh          & (35, 35, 450)   & 1 isolated \tauh, $\pt^\PGt > 35\GeV$  & 3.1     \\
    VBF + \egamma        & (40, 40, 450)   & 1 isolated \egamma object, $\mathrm{E}_{\egamma}>15\GeV$  & 1.0   \\
    \end{tabular}
    }
    \label{tab:2023_VBFseeds}
\end{table*}

The new L1 VBF triggers were used to define a set of loose HLT paths, described in Table~\ref{tab:2023_VBFpaths}, contributing a total HLT rate of about 1.2\unit{kHz}, directed to a dedicated VBF parking stream. The introduction of these new HLT paths is expected to significantly improve the acceptance to SM-like VBF signals. 
The low thresholds on central objects achieved with the exclusive paths will enable the efficient probing of final-state topologies sensitive to BSM models, such as dark photons~\cite{CMS-PAPERS-EXO-20-005}, signatures with soft unclustered energy patterns (SUEPs)~\cite{paper:SUEPs}, and any currently unexplored experimental signatures that may become of interest in the future.

\begin{table*}[!htb]
    \centering
    \topcaption{Definition and rates of the VBF paths at the HLT. The quoted rates are for ${\Linst = \sci{2}{34}\invcms}$ and do not account for overlaps with other seeds.}
    \renewcommand{\arraystretch}{1.3}
    \begin{tabular}{lccr}
    HLT trigger path  & \begin{tabular}[c]{@{}c@{}}VBF requirements\\ ($\pt^{j1}$, $\pt^{j2}$, \mjj, $\Delta\eta_{\text{jj}}$)\end{tabular} & \begin{tabular}[c]{@{}c@{}}Requirements on \\ additional objects\end{tabular}  & Rate {[}Hz{]}\\
    \hline
    VBF inclusive         & (105, 40, 1000, 3.5)   & \NA                              & 800          \\
    \hline
    VBF + 2 central jets  & (70, 40, 600, 2.5)     & 2 central jets, $\pt^j>60\GeV$   & 380          \\
    VBF + \ptmiss         & (75, 40, 500, 2.5)     & $\ptmiss>85\GeV$                 & 110          \\
    VBF + \PGm            & (90, 40, 600, 2.5)     & 1 muon, $\pt^\PGm>3\GeV$         & 120          \\
    VBF + \tauh           & (45, 45, 500, 2.5)     & 1 isolated \tauh, $\pt^\PGt > 45\GeV$ &  40     \\
    VBF + \PGg            & (45, 45, 500, 2.5)     & 1 photon, $\pt^\PGg > 17\GeV$         & 100      \\
    VBF + \Pe             & (45, 45, 500, 2.5)     & 1 isolated electron,  $\pt^\Pe > 17\GeV$    &   2          \\
    \end{tabular}
\label{tab:2023_VBFpaths}
\end{table*}

\begin{figure*}[!htb]
  \centering
  \includegraphics[width=0.49\textwidth]{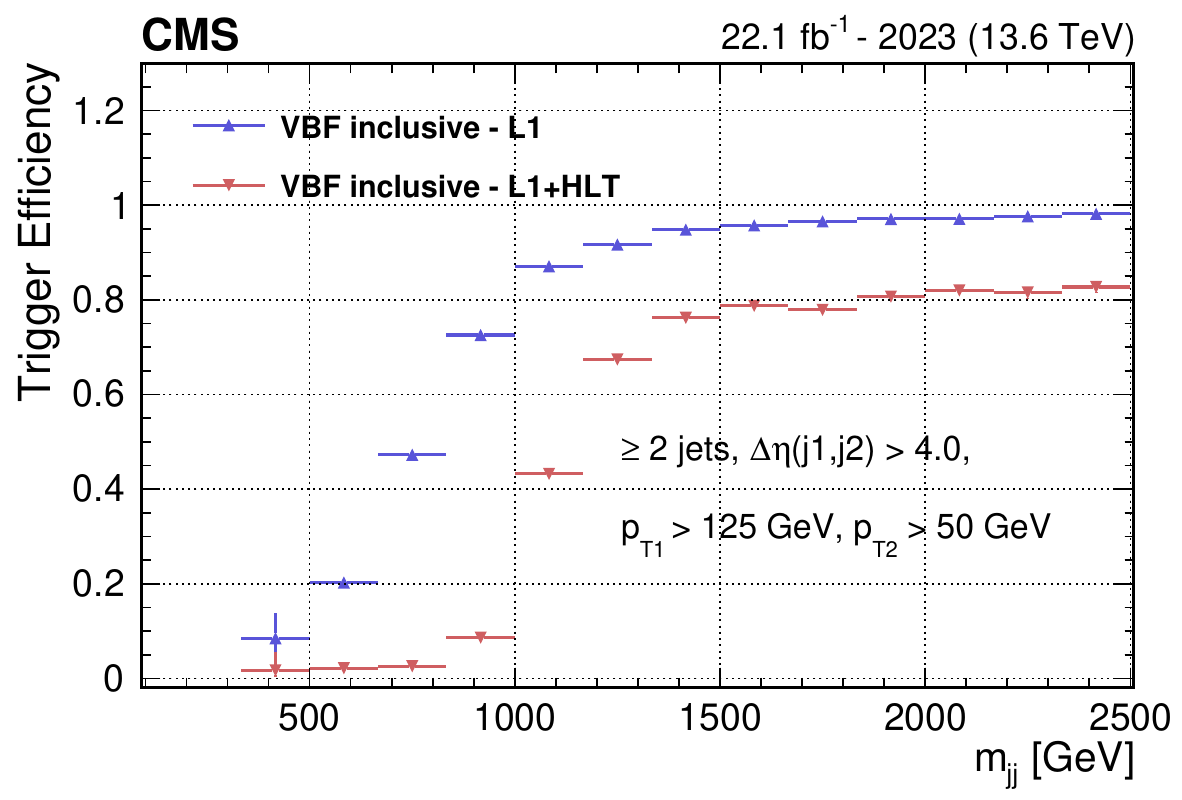}
  \includegraphics[width=0.49\textwidth]{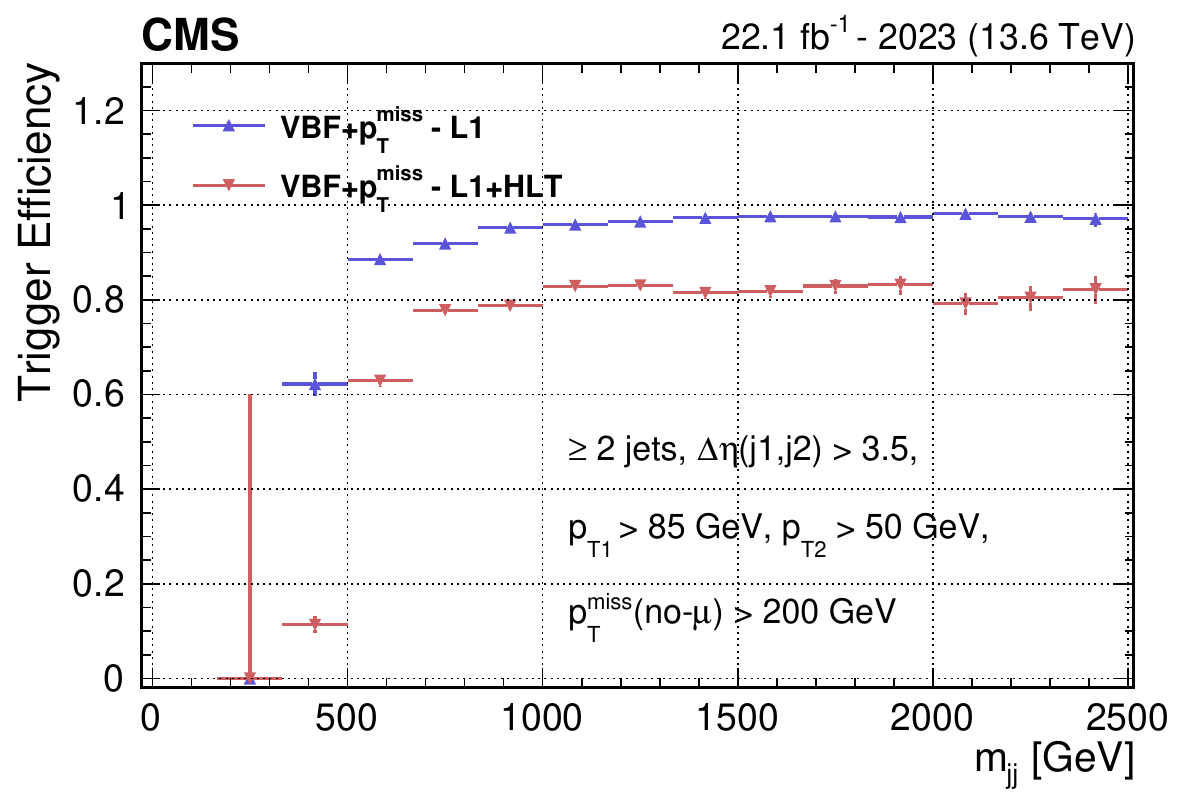}
  \caption{The L1 (blue) and L1+HLT (red) efficiencies as a function of \mjj for the VBF inclusive (left) and VBF+\ptmiss (right) parking triggers. In the right figure, \ptmiss(no-\PGm) refers to the event \ptmiss corrected for muons.}
  \label{fig:VBFparking_L1_L1HLT}
\end{figure*}

The efficiencies of the new VBF triggers were measured in data with a set of events passing a reference single-muon trigger available in the 2023 trigger menu, with muon \pt threshold of 27\GeV. Offline, events are required to contain exactly one isolated and well-identified muon with ${\pt > 30\GeV}$. Additional offline requirements on the kinematic variables of the VBF jets and of any central objects are applied to the events, depending on the trigger under consideration, in order to match the requirements made at the trigger level. Figure~\ref{fig:VBFparking_L1_L1HLT} shows the efficiencies of the inclusive VBF and VBF+\ptmiss triggers, both at the L1 and HLT, as functions of \mjj. The HLT efficiencies plateau in both cases at around 80\% because of suboptimal jet quality requirements in the online implementation. This has been fixed for the 2024 data-taking period.

\begin{figure*}[!htb]
  \centering
  \includegraphics[width=0.47\textwidth]{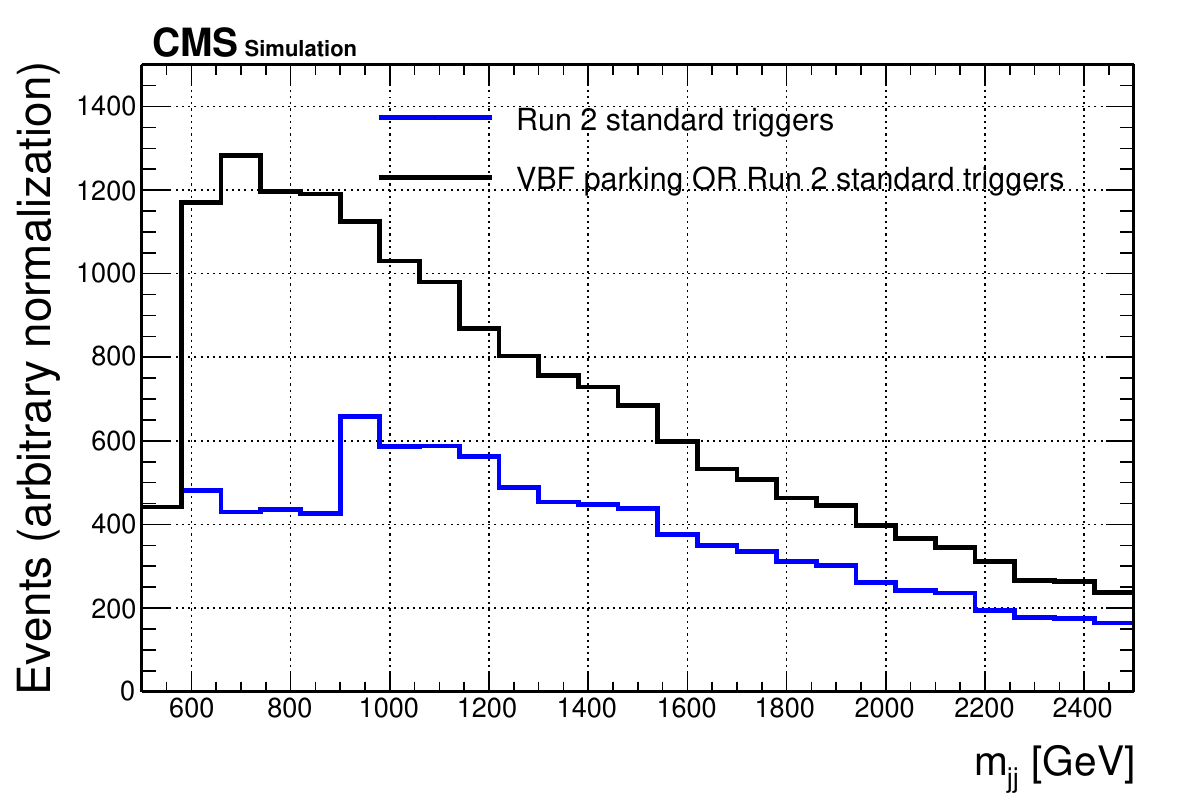}
  \caption{Distributions of \mjj for VBF $\PH \to \text{invisible}$ events passing the triggers used in the Run~2 analysis (blue), compared to events passing either one of the Run~2 triggers, the VBF+\ptmiss parking trigger, or the VBF inclusive parking trigger implemented in Run 3 (black). The Run~2 trigger selection includes the VBF+\ptmiss trigger algorithm introduced in Run~2, plus the standard trigger requiring $\ptmiss > 120\GeV$ and $\mht> 120\GeV$. In all cases, loose offline selections are applied to match the trigger-level requirements.}
  \label{fig:VBFparking_vsPrompt}
\end{figure*}

Figure~\ref{fig:VBFparking_vsPrompt} compares the \mjj distribution of the VBF $\PH \to \text{invisible}$ events that pass only the Run~2 standard trigger path versus events that pass either the standard triggers or the VBF triggers (inclusive or VBF+\ptmiss).
The new algorithms show a significant acceptance improvement, ranging from 20\% at high \mjj to more than 300\% at lower \mjj. The impact of the new VBF triggers on the overall acceptance to other benchmark signals is illustrated in Table~\ref{tab:2023_VBFacceptances}.

\begin{table*}[!htpb]
    \centering
    \topcaption{Gains in signal acceptance from the new individual VBF parking paths for a selection of benchmark signals with respect to the relevant Run~2 triggers used to collect data in the standard data set.}
    \renewcommand{\arraystretch}{1.3}
    \begin{tabular}{lccr}

 Benchmark signal & Standard triggers & Parking triggers & Acceptance gain\\
    \hline
    VBF $\PH \to \text{SUEPs}$ & \HT & VBF inclusive & +2500\% \\
    \hline
    Fully hadronic VBS & Multijet, \HT & VBF + 2 central jets & +30\% \\
    VBF $\PH \to \text{invisible}$ & \ptmiss, tight VBF+\ptmiss & VBF + \ptmiss & +60\% \\
    VBF $\PH \to\PGt\PGt(\to\tauh\tauh)$ & Di-\tauh, VBF+di-\tauh & VBF + \tauh & +50\% \\
    VBF $\PH \to\PGt\PGt(\to\PGm\tauh)$ & Single-\PGm, $\PGm+\tauh$ & VBF + \PGm & +30\% \\
    VBF $\PH \to\PGt\PGt(\to \Pe\tauh)$ & Single-\Pe, $\Pe+\tauh$ & VBF + \Pe & +40\% \\
    VBF $\PH \to \PGr\PGg$ & $\PGg + 2$~collimated tracks & VBF + \PGg & +25\% \\
\end{tabular}
\label{tab:2023_VBFacceptances}
\end{table*}

\subsubsection{The \texorpdfstring{$\PH\PH$}{HH} parking}
\label{subsec:run3parking_hh}

The Higgs boson self-coupling is a key parameter of the Higgs potential that remains unmeasured. It determines the strength of the double ($\PH\PH$) and triple ($\PH\PH\PH$) Higgs production at the LHC, which are sensitive probes of the electroweak symmetry breaking mechanism. The dominant decay mode of the Higgs boson is ${\PH \to \bbbar}$, leading to final states with multiple \PQb jets in both $\PH\PH$ and $\PH\PH\PH$ searches. The most promising channels for $\PH\PH$ observation are ${\PH\PH \to 4\PQb}$, ${\PH\PH \to 2\PQb 2\PGt}$ and ${\PH\PH \to 2\PQb2\PGg}$. The photon triggers achieve high efficiency because of the calorimeter performance. The \PQb jets and tau lepton triggers are subject to a larger contamination from jets faking their experimental signature in the detector. This effect is mitigated by applying an additional selection on the \PQb tagging and \PGt lepton identification algorithms which results in lower efficiencies at the HLT. 

In Run~2, the dedicated HLT path targeting the ${\PH\PH \to 4\PQb}$ signal recorded events with multiple small-radius jets, of which at least three were identified as \PQb jets with the online \textsc{DeepCSV}~\cite{BTagCSVandDeepCSV:Sirunyan_2018} algorithm. The HLT path was seeded by L1 seeds that required \HT values of at least 280\GeV (2016) and 360\GeV (2018). At the HLT, the trigger required values of \HT above 340\GeV. The four most energetic small-radius jets were required to have \pt above 70, 50, 45, and 40\GeV, and at least three of these jets must pass a requirement on the \textsc{DeepCSV} \PQb jet discriminant.

In early 2022, a computationally lighter version of the \textsc{ParticleNet} \PQb tagging algorithm was deployed online \cite{CMS-DP-2023-021}. The $\PH\PH \to 4\PQb$ trigger was updated accordingly to benefit from the improved \PQb tagging performance with respect to previous online algorithms. The switch to \textsc{ParticleNet} allowed for more relaxed \pt thresholds for the jets (70, 50, 40, and 35\GeV) and a looser set of \PQb tagging selection criteria, involving only the two most ``\PQb jet like'' jets. 
At the L1, the \HT requirement remained 360\GeV. To increase the acceptance to the $\PH\PH \to 4\PQb$ signal, in the low \HT region, the L1 \HT requirement was relaxed to 280\GeV in early 2023.
Moreover, by exploiting the larger rate available via the data parking method, the trigger was updated to record events with at least four jets with \pt above 30\GeV and a looser threshold on the \PQb tagging score, using the \textsc{ParticleNet} online \PQb tagging algorithm. In late 2023, this trigger achieved a rate of 180\unit{Hz} with ${\Linst = \sci{2}{34}\invcms}$.

\begin{figure*}[!htb]
  \centering
  \includegraphics[width=0.49\textwidth]{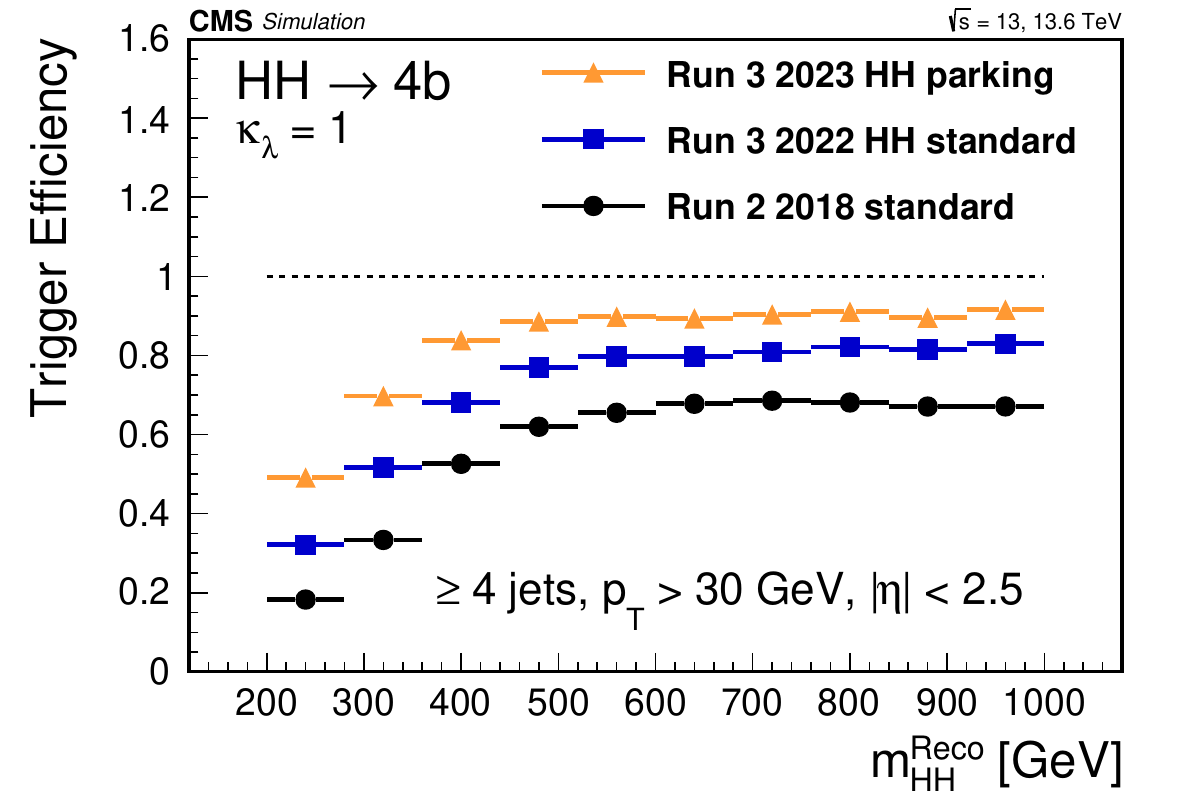}
  \includegraphics[width=0.49\textwidth]{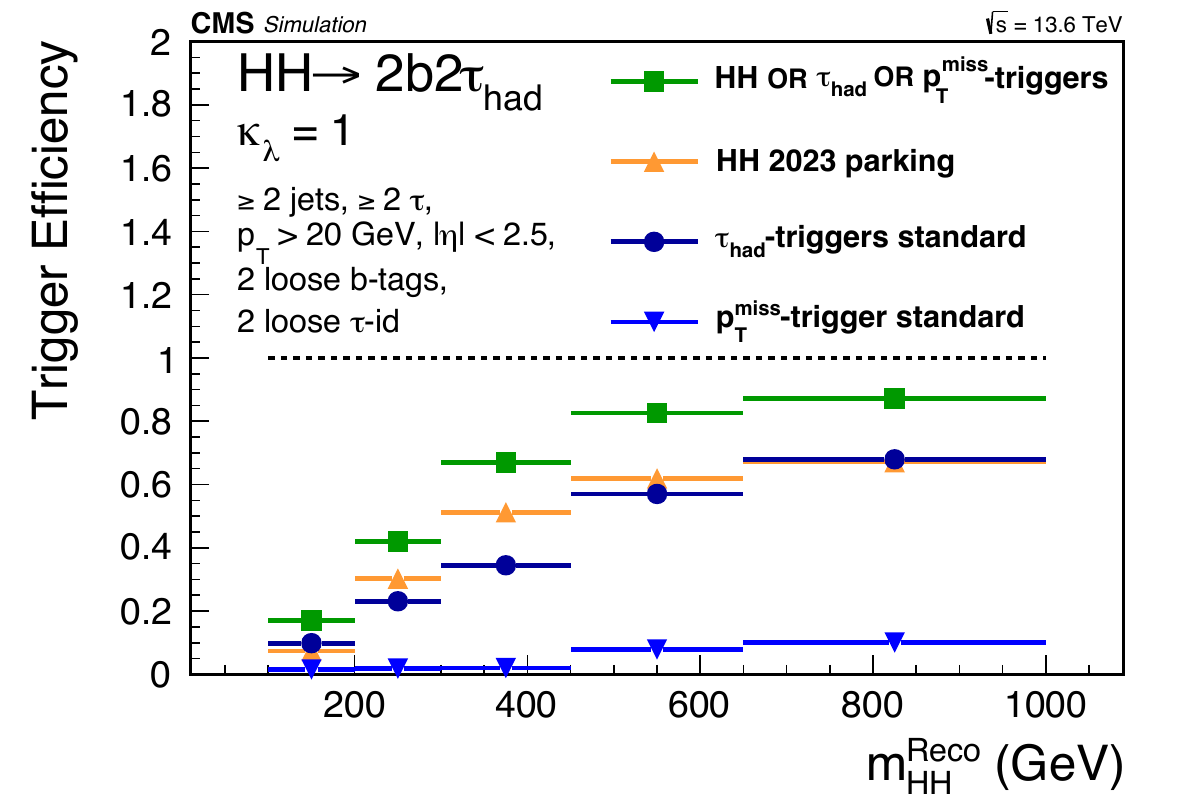}
  \caption{Trigger efficiency for selecting signal $\PH\PH$ events, plotted as a function of the reconstructed invariant mass of the two Higgs bosons, as measured in simulated ${\PH\PH \to 4\PQb}$ (left) and ${\PH\PH \to 2\PQb2\PGt}$ (right) samples corresponding to nominal Run~3 conditions.}
  \label{fig:hh-parking-trigger-efficiency}
\end{figure*}

The trigger efficiency is measured in simulated ggF ${\PH\PH \to 4\PQb}$ and ${\PH\PH \to 2\PQb2\tauh}$ events and defined as the number of events accepted by the signal triggers and an offline baseline selection, relative to the number of events satisfying the offline baseline selection alone. The offline selection requires the presence of at least four small-radius jets with \pt greater than 30\GeV and ${\abs{\eta}<2.5}$. The four most ``\PQb jet like'' jets are used to reconstruct the $\PH\PH$ pair. The trigger efficiency as a function of the reconstructed invariant mass of the two Higgs bosons is shown in Fig.~\ref{fig:hh-parking-trigger-efficiency}. 

For the ${\PH\PH \to 4\PQb}$ events (left panel of Fig.~\ref{fig:hh-parking-trigger-efficiency}), the 2023 $\PH\PH$ trigger achieved an overall signal efficiency of 82\%, corresponding to an improvement of about 60\% (20\%) with respect to the Run~2 (2022) trigger. Since the 2022 and 2023 di-Higgs triggers are designed to record events with at least two \PQb jets in the final state, they are also suitable to record ${\PH\PH \to 2\PQb2\PGt}$ signal events (right panel of Fig.~\ref{fig:hh-parking-trigger-efficiency}). For the given offline selection, the triggers requiring at least two \tauh candidates~\cite{CMS-DP-2023-024} achieve a signal efficiency of 34\%, while the $\PH\PH$ parking trigger results in an efficiency of 43\%. By requiring events to satisfy either one of the two sets of triggers, the efficiency reaches 58\%, demonstrating the complementarity of both sets in selecting signal events. 

\subsubsection{The LLP parking}
\label{subsec:run3parking_llp}

In 2023, new triggers were added to the parking stream to extend the physics reach in the search for exotic LLPs~\cite{LLPRun3TriggerDPNote}. Particles with long lifetimes are often predicted in BSM theories and thus constitute interesting probes of new physics. Most conventional searches at the LHC target promptly decaying particles, and there are still vast regions of parameter space in the context of LLPs that remain unexplored. Thus, searches for new LLPs have a great potential for discovery. 

The LLPs have distinct experimental signatures. They can decay far away from the \pp interaction point, leaving decay products that are displaced from the PV. Specific examples of LLP signatures include displaced and delayed leptons~\cite{EXO-18-003,EXO-21-006,CMS:2024qxz,EXO-20-014}, photons~\cite{CMS:2019zxa}, and jets~\cite{CMS:2020iwv,EXO-19-013,EXO-19-001,CMS-PAS-EXO-21-014,CMS:2024trg}; disappearing tracks~\cite{SUS-21-006,CMS:2020atg}; and nonstandard tracks produced by monopoles or heavy stable charged particles~\cite{Khachatryan:2016sfv}. Conventional trigger paths, object reconstruction algorithms, and background estimation strategies are usually inadequate for LLP searches because they are designed for promptly decaying particles, and custom techniques are needed for the online selection of interesting events and the offline analysis of the data.

The LLPs decaying to hadrons produce jets that contain tracks originating from a SV, spatially displaced from the PV (displaced jets). In addition, since massive LLPs often have nonrelativistic velocities, the signal energy deposits in the calorimeters are expected to arrive late in time compared to those of relativistic background particles produced at the PV (delayed jets). In 2023, two sets of LLP triggers were added to the parking stream, targeting both displaced jet and delayed jet signatures.

A suite of displaced dijet paths, already available in the standard Run~2 data stream~\cite{CMS:2020iwv}, underwent major improvements for the Run~3 data taking. The new displaced dijet paths in Run~3 require events passing two sets of selection criteria: either  $\HT > 430\GeV$ plus two jets with zero prompt tracks, or $\HT > 240\GeV$ plus a soft muon with $\pt>6\GeV$, plus two jets with zero prompt tracks and at least one displaced track. In all cases, the \HT is computed from calorimeter quantities. These new triggers used in Run~3 provide better efficiency in selecting low-mass LLPs, especially those that undergo heavy-flavor decays. In addition to the displaced dijet paths in the prompt reconstruction stream, several paths with lower \HT thresholds (down to 360\GeV) were added to the parking stream. The lower thresholds provide an increase in signal acceptance of 40--100\% relative to prompt triggers for Higgs bosons that decay to long-lived scalars, with masses between 20 and 50\GeV and proper lifetimes ${c\tau =  1\text{--}1000\mm}$.

In addition to the displaced-jet triggers, new HLT paths were added in Run~3 that make use of ECAL timing. Two different types of delayed-jet triggers use the ECAL timing at the HLT: paths that are seeded by \HT, and paths that are seeded by L1 \PGt objects. Depending on the L1 seed, different requirements are placed at the HLT, \eg, one or two delayed jets, number of matched tracks in each jet, and the timing delay. In addition, for both types of seeds, some paths were added to the parking stream with a reduced timing delay. The latter increases the efficiency by a factor ranging from 30 to 800\% for BSM delayed-jet signals, such as those produced by long-lived scalars that decay to four \PQb jets or four \PGt leptons. 

The combined trigger rate for the LLP parking triggers is 150\unit{Hz} with ${\Linst = \sci{2}{34}\invcms}$. The parked LLP triggers in Run~3 will play a crucial role in extending the sensitivity reach of displaced-jet and delayed-jets searches.

\section{Summary}
\label{sec:summary}

The extreme collision rate of the LHC, coupled with the data size needed to characterize complex interactions, poses a fundamental problem for collider experiments. Trigger, data acquisition, and downstream computing systems have finite resources that the experiments carefully allocate to the various parts of their physics programs. Searches for new physics at the energy frontier and measurements of electroweak-scale particles are one of the centerpieces of the CMS physics program, and therefore a significant portion of data acquisition resources is dedicated to triggers focusing on heavy-mass and high-\pt particles. There are, however, compelling reasons that new physics might manifest itself in other ways, for example in the existence of new light particles or, indirectly, as unexpected anomalies in precision measurements of benchmark standard model (SM) processes.

This review has highlighted two innovative techniques employed by CMS to extend the physics reach of the experiment beyond the one achieved with the original design of the detector and of the data processing pipeline: data parking and data scouting. After their inception in Run~1, both data scouting and data parking were expanded in Run~2, and became established and widely employed techniques in Run~3.

Data scouting records high-rate streams of data at the cost of a reduced event content utilizing physics objects such as jets, muons, and electrons reconstructed at the trigger level. The momentum and energy resolutions of these objects approach those achieved by the full offline event reconstruction, thereby facilitating searches for new physics in previously unexplored regions of phase space at the LHC. Notably, data scouting has enabled groundbreaking searches employing jet and muon objects that have already been published, while ongoing efforts involving electron and tau lepton objects further promise to extend the potential of this innovative technique.

The incorporation of jet objects obtained with data scouting has proven instrumental in pushing the boundaries of resonance searches in both dijet and multijet channels. These objects, which closely mirror nominal jets in energy resolution, have extended the reach of resonance searches to previously unexplored low-mass regions, overcoming limitations faced by standard analyses. Notably, the search for dijet resonances has been extended from the conventional lower limit of 1.5\TeV to 600\GeV. Moreover, multijet searches employing data scouting exhibit remarkable sensitivity down to masses as low as 70\GeV for both two- and three-parton decays, leveraging resolved and merged jets. The cross section sensitivity achieved in multijet analyses with data scouting surpasses previous searches by an order of magnitude, enabling probes of new physics sectors with electroweak couplings, such as Higgsinos, in fully hadronic final states.
Furthermore, muon scouting has allowed the triggering of very low \pt dimuon events, including both prompt- and displaced-decay scenarios. These events have significantly extended the searches for new particles decaying to muon pairs, reaching the $2m_\PGm$ kinematic limit. In the pursuit of long-lived dimuon resonances, the results are competitive with those from LHCb and \PB factories within certain mass and lifetime ranges. Additionally, decays to four muons have been successfully studied, culminating in the groundbreaking first-time observation of the rare decay of the \PGh meson into four muons, $\PGh \to \PGmp \PGmm \PGmp \PGmm$, with a measured branching fraction on the order of $10^{-9}$. The result is in agreement with theoretical predictions and improves on the precision of previous upper-limit measurements by more than 5 orders of magnitude. 

In the data-parking technique, data are recorded by high-rate triggers and temporarily stored (\ie, parked) in raw format until computing resources become available for full event reconstruction, for example during the end of the year or even during long shutdown periods. This contrasts with the typical reconstruction workflow, where data reconstruction begins within 48\unit{hours} after it is recorded. Throughout Run~1, different parking strategies were explored. Data collected with parking triggers during 2012 served the publication of impactful physics analyses including Higgs boson measurements, with a focus on vector boson fusion production, as well as precise measurements of various \PQb quark hadron lifetimes in final states involving pairs of muons. Searches for beyond SM physics ranged from multijet searches to investigations into dark matter and supersymmetry in hadronic final states. 

The 2018 data-parking analyses emphasized \PB physics. CMS collected a large data set comprising $\mathcal{O}(10^{10})$ \bbbar events using a broad collection of single, displaced muon triggers.
One distinctive feature of the parking data set is that different triggers with progressively lower thresholds were activated at different times during an LHC fill to maintain approximately constant (and high) L1 trigger rates, as the instantaneous luminosity decreased over the duration of the fill. With this method, we maximized the number of \bbbar events recorded. At the start of Run~3, the parking strategy for \PB physics was expanded, notably including dedicated low-\pt triggers to collect dimuon and dielectron events.  
The single- and double-muon, and double-electron parked data sets enable CMS to perform a variety of \PB physics analyses of rare SM processes, as well as precision tests of lepton flavor universality.
In addition, the \bparking data enables several innovative searches beyond heavy flavor physics, such as for long-lived heavy neutrinos. 

Ongoing efforts in Run~3 have also seen the enhancement of alternative parking strategies inspired by the experience acquired in Runs~1~and~2. Dedicated parking triggers targeting vector boson fusion and double Higgs boson production have been designed to augment sensitivity to tests of the Higgs sector. Moreover, the implementation of distinct parking triggers dedicated to the exploration of long-lived particles decaying into jets, leptons, and photons offers the opportunity to substantially expand the CMS physics capabilities, extending its role at the forefront of high priority searches in the field.

These two techniques, scouting and parking, serve complementary purposes: data scouting is designed to accommodate searches akin to those conducted with standard triggers and data sets while overcoming the restrictions of stringent trigger thresholds, while data parking is particularly beneficial for analyses focused on optimal precision and accuracy but that also require a higher rate of data collection.
Searches for new physics with low-\pt jets in the final state require scouting data streams. The parking data sets provide sensitivity to processes involving low-\pt single-muon and dilepton final states. Both scouting and parking data can be utilized to design comprehensive search and measurement strategies for low-\pt dimuon final states.

The comprehensive investigations conducted via scouting and parking analyses have significantly expanded the parameter space boundaries of CMS sensitivity, pushing beyond the anticipated limits of hadron colliders. The improvements achieved in trigger thresholds and data collection bandwidth, coupled with the implementation of pioneering methods for the reconstruction of essential entities such as jets, electrons, muons, tau leptons decaying hadronically, and photons, stand poised to elevate not only the quantity but also the caliber of data amassed with these sophisticated techniques. This breakthrough promises to deepen our understanding and unlock novel insights into the underlying physics, marking a significant advancement in the capabilities of the CMS experiment.

The importance of the data scouting and data parking approaches goes beyond enriching the current CMS physics program. The classic data acquisition model is not guaranteed to remain sustainable with the luminosity expected to be delivered by the LHC at the end of Run~3, and, crucially, during the high-luminosity LHC era. Moreover, computing and data acquisition resources will be designed to cope with peak luminosity demands, leaving significant spare computing power and data bandwidth at non-peak times. It is therefore essential to devise ways to develop real-time physics selection and analysis within the trigger and data acquisition systems themselves, and to optimize the utilization of idle computing resources during periods when the LHC is not operating at its maximum capacity. The CMS experience with data scouting and data parking in the last decade will prove decisive to tackling these challenges for particle physics experiments in the coming years.

\begin{acknowledgments}
\hyphenation{Bundes-ministerium Forschungs-gemeinschaft Forschungs-zentren Rachada-pisek} We congratulate our colleagues in the CERN accelerator departments for the excellent performance of the LHC and thank the technical and administrative staffs at CERN and at other CMS institutes for their contributions to the success of the CMS effort. In addition, we gratefully acknowledge the computing centers and personnel of the Worldwide LHC Computing Grid and other centers for delivering so effectively the computing infrastructure essential to our analyses. Finally, we acknowledge the enduring support for the construction and operation of the LHC, the CMS detector, and the supporting computing infrastructure provided by the following funding agencies: the Armenian Science Committee, project no. 22rl-037; the Austrian Federal Ministry of Education, Science and Research and the Austrian Science Fund; the Belgian Fonds de la Recherche Scientifique, and Fonds voor Wetenschappelijk Onderzoek; the Brazilian Funding Agencies (CNPq, CAPES, FAPERJ, FAPERGS, and FAPESP); the Bulgarian Ministry of Education and Science, and the Bulgarian National Science Fund; CERN; the Chinese Academy of Sciences, Ministry of Science and Technology, the National Natural Science Foundation of China, and Fundamental Research Funds for the Central Universities; the Ministerio de Ciencia Tecnolog\'ia e Innovaci\'on (MINCIENCIAS), Colombia; the Croatian Ministry of Science, Education and Sport, and the Croatian Science Foundation; the Research and Innovation Foundation, Cyprus; the Secretariat for Higher Education, Science, Technology and Innovation, Ecuador; the Estonian Research Council via PRG780, PRG803, RVTT3 and the Ministry of Education and Research TK202; the Academy of Finland, Finnish Ministry of Education and Culture, and Helsinki Institute of Physics; the Institut National de Physique Nucl\'eaire et de Physique des Particules~/~CNRS, and Commissariat \`a l'\'Energie Atomique et aux \'Energies Alternatives~/~CEA, France; the Shota Rustaveli National Science Foundation, Georgia; the Bundesministerium f\"ur Bildung und Forschung, the Deutsche Forschungsgemeinschaft (DFG), under Germany's Excellence Strategy -- EXC 2121 ``Quantum Universe" -- 390833306, and under project number 400140256 - GRK2497, and Helmholtz-Gemeinschaft Deutscher Forschungszentren, Germany; the General Secretariat for Research and Innovation and the Hellenic Foundation for Research and Innovation (HFRI), Project Number 2288, Greece; the National Research, Development and Innovation Office (NKFIH), Hungary; the Department of Atomic Energy and the Department of Science and Technology, India; the Institute for Studies in Theoretical Physics and Mathematics, Iran; the Science Foundation, Ireland; the Istituto Nazionale di Fisica Nucleare, Italy; the Ministry of Science, ICT and Future Planning, and National Research Foundation (NRF), Republic of Korea; the Ministry of Education and Science of the Republic of Latvia; the Research Council of Lithuania, agreement No.\ VS-19 (LMTLT); the Ministry of Education, and University of Malaya (Malaysia); the Ministry of Science of Montenegro; the Mexican Funding Agencies (BUAP, CINVESTAV, CONACYT, LNS, SEP, and UASLP-FAI); the Ministry of Business, Innovation and Employment, New Zealand; the Pakistan Atomic Energy Commission; the Ministry of Education and Science and the National Science Center, Poland; the Funda\c{c}\~ao para a Ci\^encia e a Tecnologia, grants CERN/FIS-PAR/0025/2019 and CERN/FIS-INS/0032/2019, Portugal; the Ministry of Education, Science and Technological Development of Serbia; MCIN/AEI/10.13039/501100011033, ERDF ``a way of making Europe", Programa Estatal de Fomento de la Investigaci{\'o}n Cient{\'i}fica y T{\'e}cnica de Excelencia Mar\'{\i}a de Maeztu, grant MDM-2017-0765, projects PID2020-113705RB, PID2020-113304RB, PID2020-116262RB and PID2020-113341RB-I00, and Plan de Ciencia, Tecnolog{\'i}a e Innovaci{\'o}n de Asturias, Spain; the Ministry of Science, Technology and Research, Sri Lanka; the Swiss Funding Agencies (ETH Board, ETH Zurich, PSI, SNF, UniZH, Canton Zurich, and SER); the Ministry of Science and Technology, Taipei; the Ministry of Higher Education, Science, Research and Innovation, and the National Science and Technology Development Agency of Thailand; the Scientific and Technical Research Council of Turkey, and Turkish Energy, Nuclear and Mineral Research Agency; the National Academy of Sciences of Ukraine; the Science and Technology Facilities Council, UK; the US Department of Energy, and the US National Science Foundation.

Individuals have received support from the Marie-Curie programme and the European Research Council and Horizon 2020 Grant, contract Nos.\ 675440, 724704, 752730, 758316, 765710, 824093, 101115353,101002207, and COST Action CA16108 (European Union) the Leventis Foundation; the Alfred P.\ Sloan Foundation; the Alexander von Humboldt Foundation; the Belgian Federal Science Policy Office; the Fonds pour la Formation \`a la Recherche dans l'Industrie et dans l'Agriculture (FRIA-Belgium); the Agentschap voor Innovatie door Wetenschap en Technologie (IWT-Belgium); the F.R.S.-FNRS and FWO (Belgium) under the ``Excellence of Science -- EOS" -- be.h project n.\ 30820817; the Beijing Municipal Science \& Technology Commission, No. Z191100007219010; the Ministry of Education, Youth and Sports (MEYS) of the Czech Republic; the Shota Rustaveli National Science Foundation, grant FR-22-985 (Georgia); the Hungarian Academy of Sciences, the New National Excellence Program - \'UNKP, the NKFIH research grants K 131991, K 133046, K 138136, K 143460, K 143477, K 146913, K 146914, K 147048, 2020-2.2.1-ED-2021-00181, and TKP2021-NKTA-64 (Hungary); the Council of Scientific and Industrial Research, India; ICSC -- National Research Center for High Performance Computing, Big Data and Quantum Computing, funded by the EU NexGeneration program, Italy; the Latvian Council of Science; the Ministry of Education and Science, project no. 2022/WK/14, and the National Science Center, contracts Opus 2021/41/B/ST2/01369 and 2021/43/B/ST2/01552 (Poland); the Funda\c{c}\~ao para a Ci\^encia e a Tecnologia, grant FCT CEECIND/01334/2018; the National Priorities Research Program by Qatar National Research Fund; the Programa Estatal de Fomento de la Investigaci{\'o}n Cient{\'i}fica y T{\'e}cnica de Excelencia Mar\'{\i}a de Maeztu, grant MDM-2017-0765 and projects PID2020-113705RB, PID2020-113304RB, PID2020-116262RB and PID2020-113341RB-I00, and Programa Severo Ochoa del Principado de Asturias (Spain); the Chulalongkorn Academic into Its 2nd Century Project Advancement Project, and the National Science, Research and Innovation Fund via the Program Management Unit for Human Resources \& Institutional Development, Research and Innovation, grant B37G660013 (Thailand); the Kavli Foundation; the Nvidia Corporation; the SuperMicro Corporation; the Welch Foundation, contract C-1845; and the Weston Havens Foundation (USA).  
\end{acknowledgments}

\bibliography{auto_generated}

\providecommand{\href}[2]{#2}\begingroup\raggedright\begin{thebibliography}{100}%
\makeatletter
\providecommand{\hrefCMSnoop }[0]{\@secondoftwo}%
\makeatother
\providecommand{\doi}{\texttt{doi:}\begingroup \urlstyle{tt}\Url}

\bibitem{CMS-PAPERS-HIG-19-004}
\hrefCMSnoop {}{{CMS Collaboration}, ``{A measurement of the Higgs boson mass
  in the diphoton decay channel}'',} \textit{ Phys. Lett. B} \textbf{ 805}
  (2020) 135425,
  \href{http://dx.doi.org/10.1016/j.physletb.2020.135425}{\doi{10.1016/j.physletb.2020.135425}},
  \href{http://www.arXiv.org/abs/2002.06398}{\texttt{arXiv:2002.06398}}.

\bibitem{CMS-PAPERS-HIG-19-001}
\hrefCMSnoop {}{{CMS Collaboration}, ``{Measurements of production cross
  sections of the Higgs boson in the four-lepton final state in
  proton\textendash{}proton collisions at $\sqrt{s} = 13\,\text {Te}\text {V}
  $}'',} \textit{ Eur. Phys. J. C} \textbf{ 81} (2021) 488,
  \href{http://dx.doi.org/10.1140/epjc/s10052-021-09200-x}{\doi{10.1140/epjc/s10052-021-09200-x}},
  \href{http://www.arXiv.org/abs/2103.04956}{\texttt{arXiv:2103.04956}}.

\bibitem{CMS-PAPERS-HIG-19-002}
\hrefCMSnoop {}{{CMS Collaboration}, ``{Measurement of the inclusive and
  differential Higgs boson production cross sections in the leptonic WW decay
  mode at $\sqrt{s} =$ 13 TeV}'',} \textit{ JHEP} \textbf{ 03} (2021) 003,
  \href{http://dx.doi.org/10.1007/JHEP03(2021)003}{\doi{10.1007/JHEP03(2021)003}},
  \href{http://www.arXiv.org/abs/2007.01984}{\texttt{arXiv:2007.01984}}.

\bibitem{CMS:2022dwd}
\hrefCMSnoop {}{{CMS Collaboration}, ``A portrait of the {Higgs} boson by the
  {CMS} experiment ten years after the discovery.'',} \textit{ Nature} \textbf{
  607} (2022) 60,
  \href{http://dx.doi.org/10.1038/s41586-022-04892-x}{\doi{10.1038/s41586-022-04892-x}},
  \href{http://www.arXiv.org/abs/2207.00043}{\texttt{arXiv:2207.00043}}.

\bibitem{Agrawal:2021dbo}
\hrefCMSnoop {}{P.~Agrawal { et~al.}, ``Feebly-{Interacting} {Particles}:
  {FIPs} 2020 workshop report'',} \textit{ Eur. Phys. J. C} \textbf{ 81} (2021)
  1015,
  \href{http://dx.doi.org/10.1140/epjc/s10052-021-09703-7}{\doi{10.1140/epjc/s10052-021-09703-7}},
  \href{http://www.arXiv.org/abs/2102.12143}{\texttt{arXiv:2102.12143}}.

\bibitem{Antel:2023hkf}
\hrefCMSnoop {}{C.~Antel { et~al.}, ``Feebly {Interacting} {Particles}: {FIPs}
  2022 workshop report'',} \textit{ Eur. Phys. J. C} \textbf{ 83} (2023) 1122,
  \href{http://dx.doi.org/10.1140/epjc/s10052-023-12168-5}{\doi{10.1140/epjc/s10052-023-12168-5}},
  \href{http://www.arXiv.org/abs/2305.01715}{\texttt{arXiv:2305.01715}}.

\bibitem{Bruning:782076}
O.~S. Bruning { et~al.}, ``{LHC Design Report Vol.1: The LHC Main Ring}''.
\newblock CERN Yellow Reports: Monographs. CERN, Geneva, 2004.
\newblock
  \href{http://dx.doi.org/10.5170/CERN-2004-003-V-1}{\doi{10.5170/CERN-2004-003-V-1}}.

\bibitem{CMS:2020cmk}
\hrefCMSnoop {}{{CMS Collaboration}, ``{Performance of the CMS Level-1 trigger
  in proton-proton collisions at $\sqrt{s} = 13$\,TeV}'',} \textit{ JINST}
  \textbf{ 15} (2020) P10017,
  \href{http://dx.doi.org/10.1088/1748-0221/15/10/P10017}{\doi{10.1088/1748-0221/15/10/P10017}},
  \href{http://www.arXiv.org/abs/2006.10165}{\texttt{arXiv:2006.10165}}.

\bibitem{CMS:2016ngn}
\hrefCMSnoop {}{{CMS Collaboration}, ``{The CMS trigger system}'',} \textit{
  JINST} \textbf{ 12} (2017) P01020,
  \href{http://dx.doi.org/10.1088/1748-0221/12/01/P01020}{\doi{10.1088/1748-0221/12/01/P01020}},
\href{http://www.arXiv.org/abs/1609.02366}{\texttt{arXiv:1609.02366}}.

\bibitem{CMS:2006myw}
\href {https://cds.cern.ch/record/922757}{{CMS Collaboration}, ``{CMS Physics}:
  {Technical Design Report Volume 1: Detector Performance and Software}'',}
  Technical Report CERN-LHCC-2006-001, CMS-TDR-8-1, 2006.

\bibitem{CMS:2000mvk}
\href {https://cds.cern.ch/record/706847}{{CMS Collaboration}, ``{CMS. The
  TriDAS project. Technical design report, vol. 1: The trigger systems}'',}
  Technical Report CERN-LHCC-2000-038, 2000.

\bibitem{Sphicas:2002gg}
\href {https://cds.cern.ch/record/578006}{{CMS Collaboration}, ``{CMS: The
  TriDAS project. Technical design report, Vol. 2: Data acquisition and
  high-level trigger}'',} Technical Report CERN-LHCC-2002-026, 2002.

\bibitem{Cerminara:2015hov}
\hrefCMSnoop {}{G.~Cerminara and B.~van Besien, ``{Automated workflows for
  critical time-dependent calibrations at the CMS experiment}'',} \textit{ J.
  Phys. Conf. Ser.} \textbf{ 664} (2015) 072009,
  \href{http://dx.doi.org/10.1088/1742-6596/664/7/072009}{\doi{10.1088/1742-6596/664/7/072009}}.

\bibitem{CMS:2012ScoutingParking}
\href {https://cds.cern.ch/record/1480607}{{CMS Collaboration}, ``Data parking
  and data scouting at the {CMS Experiment}'',} CMS Detector Performance Note
  CMS-DP-2012-022, 2012.

\bibitem{EXO-14-005}
\hrefCMSnoop {}{{CMS Collaboration}, ``Search for narrow resonances in dijet
  final states at $\sqrt{s}= 8$ {TeV} with the novel {CMS} technique of data
  scouting'',} \textit{ Phys. Rev. Lett.} \textbf{ 117} (2016) 031802,
  \href{http://dx.doi.org/10.1103/PhysRevLett.117.031802}{\doi{10.1103/PhysRevLett.117.031802}},
  \href{http://www.arXiv.org/abs/1604.08907}{\texttt{arXiv:1604.08907}}.

\bibitem{Benson:2015yzo}
\hrefCMSnoop {}{S.~Benson, V.~V. Gligorov, M.~A. Vesterinen, and M.~Williams,
  ``{The LHCb Turbo Stream}'',} \textit{ J. Phys. Conf. Ser.} \textbf{ 664}
  (2015) 082004,
  \href{http://dx.doi.org/10.1088/1742-6596/664/8/082004}{\doi{10.1088/1742-6596/664/8/082004}}.

\bibitem{ATLAS:2018qto}
\hrefCMSnoop {}{{ATLAS Collaboration}, ``Search for low-mass dijet resonances
  using trigger-level jets with the {ATLAS} detector in $pp$ collisions at
  $\sqrt{s}=13$ {TeV}'',} \textit{ Phys. Rev. Lett.} \textbf{ 121} (2018)
  081801,
  \href{http://dx.doi.org/10.1103/PhysRevLett.121.081801}{\doi{10.1103/PhysRevLett.121.081801}},
  \href{http://www.arXiv.org/abs/1804.03496}{\texttt{arXiv:1804.03496}}.

\bibitem{Aad:1735492_AtlasDelayedStream}
\hrefCMSnoop {}{{ATLAS Collaboration}, ``Search for new phenomena in the dijet
  mass distribution using $\pp$ collision data at $\sqrt{s}=8$ {TeV} with the
  {ATLAS} detector'',} \textit{ Phys. Rev. D} \textbf{ 91} (2015) 052007,
  \href{http://dx.doi.org/10.1103/PhysRevD.91.052007}{\doi{10.1103/PhysRevD.91.052007}},
  \href{http://www.arXiv.org/abs/1407.1376}{\texttt{arXiv:1407.1376}}.

\bibitem{Antonioli:2013ppp}
\href {https://cds.cern.ch/record/1603472}{{ALICE Collaboration}, ``Upgrade of
  the {ALICE} readout $\&$ trigger system'',} Technical Design Report
  CERN-LHCC-2013-019, ALICE-TDR-015, 2013.

\bibitem{Kvapil:2021tuj}
{Kvapil, Jakub}\hrefCMSnoop {}{ { et~al.}, ``{ALICE} central trigger system for
  {LHC} {Run 3}'',} \textit{ EPJ Web Conf.} \textbf{ 251} (2021) 04022,
  \href{http://dx.doi.org/10.1051/epjconf/202125104022}{\doi{10.1051/epjconf/202125104022}},
  \href{http://www.arXiv.org/abs/2106.08353}{\texttt{arXiv:2106.08353}}.

\bibitem{Sirunyan:2017ulk}
\hrefCMSnoop {}{{CMS Collaboration}, ``Particle-flow reconstruction and global
  event description with the {CMS} detector'',} \textit{ JINST} \textbf{ 12}
  (2017) P10003,
  \href{http://dx.doi.org/10.1088/1748-0221/12/10/P10003}{\doi{10.1088/1748-0221/12/10/P10003}},
  \href{http://www.arXiv.org/abs/1706.04965}{\texttt{arXiv:1706.04965}}.

\bibitem{Petrucciani_2015}
\hrefCMSnoop {}{G.~Petrucciani, A.~Rizzi, and C.~Vuosalo, ``{Mini-AOD}: A new
  analysis data format for {CMS}'',} \textit{ J. Phys. Conf. Ser.} \textbf{
  664} (2015) 072052,
  \href{http://dx.doi.org/10.1088/1742-6596/664/7/072052}{\doi{10.1088/1742-6596/664/7/072052}},
  \href{http://www.arXiv.org/abs/1702.04685}{\texttt{arXiv:1702.04685}}.

\bibitem{Peruzzi_2020}
\hrefCMSnoop {}{M.~Peruzzi, G.~Petrucciani, and A.~Rizzi, ``The {NanoAOD} event
  data format in {CMS}'',} \textit{ J. Phys. Conf. Ser.} \textbf{ 1525} (2020)
  012038,
  \href{http://dx.doi.org/10.1088/1742-6596/1525/1/012038}{\doi{10.1088/1742-6596/1525/1/012038}}.

\bibitem{doi:10.1146/annurev-nucl-102020-090209}
\hrefCMSnoop {}{D.~London and J.~Matias, ``B flavor anomalies: 2021 theoretical
  status report'',} \textit{ Annu. Rev. Nucl. Part. Sci.} \textbf{ 72} (2022)
  37,
  \href{http://dx.doi.org/10.1146/annurev-nucl-102020-090209}{\doi{10.1146/annurev-nucl-102020-090209}},
  \href{http://www.arXiv.org/abs/2110.13270}{\texttt{arXiv:2110.13270}}.

\bibitem{Hiller:2003js}
\hrefCMSnoop {}{G.~Hiller and F.~Kruger, ``{More model-independent analysis of
  $b \to s$ processes}'',} \textit{ Phys. Rev. D} \textbf{ 69} (2004) 074020,
  \href{http://dx.doi.org/10.1103/PhysRevD.69.074020}{\doi{10.1103/PhysRevD.69.074020}},
  \href{http://www.arXiv.org/abs/hep-ph/0310219}{\texttt{arXiv:hep-ph/0310219}}.

\bibitem{Bordone:2016gaq}
\hrefCMSnoop {}{M.~Bordone, G.~Isidori, and A.~Pattori, ``On the standard model
  predictions for {$R_K$} and {$R_{K^*}$}'',} \textit{ Eur. Phys. J. C}
  \textbf{ 76} (2016) 440,
  \href{http://dx.doi.org/10.1140/epjc/s10052-016-4274-7}{\doi{10.1140/epjc/s10052-016-4274-7}},
  \href{http://www.arXiv.org/abs/1605.07633}{\texttt{arXiv:1605.07633}}.

\bibitem{Isidori:2020acz}
\hrefCMSnoop {}{G.~Isidori, S.~Nabeebaccus, and R.~Zwicky, ``{QED corrections
  in $ \overline{B}\to \overline{K}{\mathrm{\ell}}^{+}{\mathrm{\ell}}^{-} $ at
  the double-differential level}'',} \textit{ JHEP} \textbf{ 12} (2020) 104,
  \href{http://dx.doi.org/10.1007/JHEP12(2020)104}{\doi{10.1007/JHEP12(2020)104}},
  \href{http://www.arXiv.org/abs/2009.00929}{\texttt{arXiv:2009.00929}}.

\bibitem{Isidori:2022bzw}
\hrefCMSnoop {}{G.~Isidori, D.~Lancierini, S.~Nabeebaccus, and R.~Zwicky,
  ``{QED in $ \overline{B} \to \overline{K}
  $\ensuremath{\ell}$^{+}$\ensuremath{\ell}$^{-}$ LFU ratios: theory versus
  experiment, a Monte Carlo study}'',} \textit{ JHEP} \textbf{ 10} (2022) 146,
  \href{http://dx.doi.org/10.1007/JHEP10(2022)146}{\doi{10.1007/JHEP10(2022)146}},
  \href{http://www.arXiv.org/abs/2205.08635}{\texttt{arXiv:2205.08635}}.

\bibitem{Phase1Pixel}
\hrefCMSnoop {}{{CMS Tracker} Collaboration, ``The {CMS} {Phase-1} pixel
  detector upgrade'',} \textit{ JINST} \textbf{ 16} (2021) P02027,
  \href{http://dx.doi.org/10.1088/1748-0221/16/02/P02027}{\doi{10.1088/1748-0221/16/02/P02027}},
  \href{http://www.arXiv.org/abs/2012.14304}{\texttt{arXiv:2012.14304}}.

\bibitem{CMS:2008xjf}
\hrefCMSnoop {}{{CMS Collaboration}, ``The {CMS} experiment at the {CERN}
  {LHC}'',} \textit{ JINST} \textbf{ 3} (2008) S08004,
  \href{http://dx.doi.org/10.1088/1748-0221/3/08/S08004}{\doi{10.1088/1748-0221/3/08/S08004}}.

\bibitem{CMS:2023gfb}
\hrefCMSnoop {}{{CMS Collaboration}, ``Development of the {CMS} detector for
  the {CERN} {LHC} {Run 3}'',} \textit{ JINST} \textbf{ 19} (2024) P05064,
  \href{http://dx.doi.org/10.1088/1748-0221/19/05/P05064}{\doi{10.1088/1748-0221/19/05/P05064}},
  \href{http://www.arXiv.org/abs/2309.05466}{\texttt{arXiv:2309.05466}}.

\bibitem{PhaseTwoUpgradeDAQ2021}
\hrefCMSnoop {}{G.~Badaro { et~al.}, ``The {Phase-2} upgrade of the {CMS} data
  acquisition'',} \textit{ EPJ Web Conf.} \textbf{ 251} (2021) 04023,
  \href{http://dx.doi.org/10.1051/epjconf/202125104023}{\doi{10.1051/epjconf/202125104023}}.

\bibitem{CMS-TDR-022}
\href {https://cds.cern.ch/record/2759072}{{CMS Collaboration}, ``The {Phase-2}
  upgrade of the {CMS} data acquisition and high level trigger'',} Technical
  Report CERN-LHCC-2021-007, CMS-TDR-022, 2021.

\bibitem{CMS:2014pgm}
\hrefCMSnoop {}{{CMS Collaboration}, ``{Description and performance of track
  and primary-vertex reconstruction with the CMS tracker}'',} \textit{ JINST}
  \textbf{ 9} (2014) P10009,
  \href{http://dx.doi.org/10.1088/1748-0221/9/10/P10009}{\doi{10.1088/1748-0221/9/10/P10009}},
  \href{http://www.arXiv.org/abs/1405.6569}{\texttt{arXiv:1405.6569}}.

\bibitem{CMS-TDR-15-02}
\href {http://cds.cern.ch/record/2020886}{{CMS Collaboration}, ``Technical
  proposal for the {Phase-II} upgrade of the {Compact Muon Solenoid}'',} CMS
  Technical Proposal CERN-LHCC-2015-010, CMS-TDR-15-02, 2015.

\bibitem{Bocci:2020pmi}
A.~Bocci\hrefCMSnoop {}{ { et~al.}, ``Heterogeneous reconstruction of tracks
  and primary vertices with the {CMS} pixel tracker'',} \textit{ Front. Big
  Data} \textbf{ 3} (2020) 601728,
  \href{http://dx.doi.org/10.3389/fdata.2020.601728}{\doi{10.3389/fdata.2020.601728}},
  \href{http://www.arXiv.org/abs/2008.13461}{\texttt{arXiv:2008.13461}}.

\bibitem{ECAL:CMS:2020xlg}
\hrefCMSnoop {}{{CMS Collaboration}, ``Reconstruction of signal amplitudes in
  the {CMS} electromagnetic calorimeter in the presence of overlapping
  proton-proton interactions'',} \textit{ JINST} \textbf{ 15} (2020) P10002,
  \href{http://dx.doi.org/10.1088/1748-0221/15/10/P10002}{\doi{10.1088/1748-0221/15/10/P10002}},
  \href{http://www.arXiv.org/abs/2006.14359}{\texttt{arXiv:2006.14359}}.

\bibitem{CMS-EGM-18-002}
\hrefCMSnoop {}{{CMS Collaboration}, ``{Performance of the CMS electromagnetic
  calorimeter in pp collisions at $\sqrt{s}=13$ TeV}'',} \textit{ JINST}
  \textbf{ 19} (2024) P09004,
  \href{http://dx.doi.org/10.1088/1748-0221/19/09/P09004}{\doi{10.1088/1748-0221/19/09/P09004}},
  \href{http://www.arXiv.org/abs/2403.15518}{\texttt{arXiv:2403.15518}}.

\bibitem{CMS:2020uim}
\hrefCMSnoop {}{{CMS Collaboration}, ``Electron and photon reconstruction and
  identification with the {CMS} experiment at the {CERN LHC}'',} \textit{
  JINST} \textbf{ 16} (2021) P05014,
  \href{http://dx.doi.org/10.1088/1748-0221/16/05/P05014}{\doi{10.1088/1748-0221/16/05/P05014}},
  \href{http://www.arXiv.org/abs/2012.06888}{\texttt{arXiv:2012.06888}}.

\bibitem{HCAL:CMS:2023lqq}
\hrefCMSnoop {}{{CMS Collaboration}, ``Performance of the local reconstruction
  algorithms for the {CMS} hadron calorimeter with {Run 2} data'',} \textit{
  JINST} \textbf{ 18} (2023) P11017,
  \href{http://dx.doi.org/10.1088/1748-0221/18/11/P11017}{\doi{10.1088/1748-0221/18/11/P11017}},
  \href{http://www.arXiv.org/abs/2306.10355}{\texttt{arXiv:2306.10355}}.

\bibitem{CMS:2018rym}
\hrefCMSnoop {}{{CMS Collaboration}, ``Performance of the {CMS} muon detector
  and muon reconstruction with proton-proton collisions at $\sqrt{s}=$ 13
  {TeV}'',} \textit{ JINST} \textbf{ 13} (2018) P06015,
  \href{http://dx.doi.org/10.1088/1748-0221/13/06/P06015}{\doi{10.1088/1748-0221/13/06/P06015}},
  \href{http://www.arXiv.org/abs/1804.04528}{\texttt{arXiv:1804.04528}}.

\bibitem{Cacciari:2008gp}
\hrefCMSnoop {}{M.~Cacciari, G.~P. Salam, and G.~Soyez, ``The anti-\kt jet
  clustering algorithm'',} \textit{ JHEP} \textbf{ 04} (2008) 063,
  \href{http://dx.doi.org/10.1088/1126-6708/2008/04/063}{\doi{10.1088/1126-6708/2008/04/063}},
  \href{http://www.arXiv.org/abs/0802.1189}{\texttt{arXiv:0802.1189}}.

\bibitem{Cacciari:2011ma}
\hrefCMSnoop {}{M.~Cacciari, G.~P. Salam, and G.~Soyez, ``{FastJet} user
  manual'',} \textit{ Eur. Phys. J. C} \textbf{ 72} (2012) 1896,
  \href{http://dx.doi.org/10.1140/epjc/s10052-012-1896-2}{\doi{10.1140/epjc/s10052-012-1896-2}},
\href{http://www.arXiv.org/abs/1111.6097}{\texttt{arXiv:1111.6097}}.

\bibitem{PUPPI:Sirunyan_2020foa}
\hrefCMSnoop {}{{CMS Collaboration}, ``{Pileup mitigation at CMS in 13 TeV
  data}'',} \textit{ JINST} \textbf{ 15} (2020) P09018,
  \href{http://dx.doi.org/10.1088/1748-0221/15/09/p09018}{\doi{10.1088/1748-0221/15/09/p09018}},
  \href{http://www.arXiv.org/abs/2003.00503}{\texttt{arXiv:2003.00503}}.

\bibitem{PUPPI:Bertolini_2014bba}
\hrefCMSnoop {}{D.~Bertolini, P.~Harris, M.~Low, and N.~Tran, ``Pileup per
  particle identification'',} \textit{ JHEP} \textbf{ 10} (2014) 059,
  \href{http://dx.doi.org/10.1007/JHEP10(2014)059}{\doi{10.1007/JHEP10(2014)059}},
\href{http://www.arXiv.org/abs/1407.6013}{\texttt{arXiv:1407.6013}}.

\bibitem{JEC:CMS_2016lmd}
\hrefCMSnoop {}{{CMS Collaboration}, ``{Jet energy scale and resolution in the
  CMS experiment in pp collisions at 8 TeV}'',} \textit{ JINST} \textbf{ 12}
  (2017) P02014,
  \href{http://dx.doi.org/10.1088/1748-0221/12/02/P02014}{\doi{10.1088/1748-0221/12/02/P02014}},
\href{http://www.arXiv.org/abs/1607.03663}{\texttt{arXiv:1607.03663}}.

\bibitem{qgl_old}
\href {https://cds.cern.ch/record/2256875}{{CMS Collaboration}, ``Jet
  algorithms performance in 13 {TeV} data'',} CMS Physics Analysis Summary
  CMS-PAS-JME-16-003, 2017.

\bibitem{CMS:2014rsx}
\hrefCMSnoop {}{{CMS Collaboration}, ``Identification techniques for highly
  boosted {W} bosons that decay into hadrons'',} \textit{ JHEP} \textbf{ 12}
  (2014) 017,
  \href{http://dx.doi.org/10.1007/JHEP12(2014)017}{\doi{10.1007/JHEP12(2014)017}},
  \href{http://www.arXiv.org/abs/1410.4227}{\texttt{arXiv:1410.4227}}.

\bibitem{Krohn:2009th}
\hrefCMSnoop {}{D.~Krohn, J.~Thaler, and L.-T. Wang, ``Jet trimming'',}
  \textit{ JHEP} \textbf{ 02} (2010) 084,
  \href{http://dx.doi.org/10.1007/JHEP02(2010)084}{\doi{10.1007/JHEP02(2010)084}},
  \href{http://www.arXiv.org/abs/0912.1342}{\texttt{arXiv:0912.1342}}.

\bibitem{CMS:2020poo}
\hrefCMSnoop {}{{CMS Collaboration}, ``{Identification of heavy, energetic,
  hadronically decaying particles using machine-learning techniques}'',}
  \textit{ JINST} \textbf{ 15} (2020) P06005,
  \href{http://dx.doi.org/10.1088/1748-0221/15/06/P06005}{\doi{10.1088/1748-0221/15/06/P06005}},
  \href{http://www.arXiv.org/abs/2004.08262}{\texttt{arXiv:2004.08262}}.

\bibitem{Thaler:2010tr}
\hrefCMSnoop {}{J.~Thaler and K.~Van~Tilburg, ``{Identifying boosted objects
  with $N$-subjettiness}'',} \textit{ JHEP} \textbf{ 03} (2011) 015,
  \href{http://dx.doi.org/10.1007/JHEP03(2011)015}{\doi{10.1007/JHEP03(2011)015}},
  \href{http://www.arXiv.org/abs/1011.2268}{\texttt{arXiv:1011.2268}}.

\bibitem{Moult:2016cvt}
\hrefCMSnoop {}{I.~Moult, L.~Necib, and J.~Thaler, ``{New angles on energy
  correlation functions}'',} \textit{ JHEP} \textbf{ 12} (2016) 153,
  \href{http://dx.doi.org/10.1007/JHEP12(2016)153}{\doi{10.1007/JHEP12(2016)153}},
  \href{http://www.arXiv.org/abs/1609.07483}{\texttt{arXiv:1609.07483}}.

\bibitem{BTagCSV:Chatrchyan:2012jua}
\hrefCMSnoop {}{{CMS Collaboration}, ``{Identification of b-quark jets with the
  CMS experiment}'',} \textit{ JINST} \textbf{ 8} (2013) P04013,
  \href{http://dx.doi.org/10.1088/1748-0221/8/04/P04013}{\doi{10.1088/1748-0221/8/04/P04013}},
  \href{http://www.arXiv.org/abs/1211.4462}{\texttt{arXiv:1211.4462}}.

\bibitem{BTagCSVandDeepCSV:Sirunyan_2018}
\hrefCMSnoop {}{{CMS Collaboration}, ``Identification of heavy-flavour jets
  with the {CMS} detector in pp collisions at 13 {TeV}'',} \textit{ JINST}
  \textbf{ 13} (2018) P05011,
  \href{http://dx.doi.org/10.1088/1748-0221/13/05/p05011}{\doi{10.1088/1748-0221/13/05/p05011}},
  \href{http://www.arXiv.org/abs/1712.07158}{\texttt{arXiv:1712.07158}}.

\bibitem{BTagDeepJet:Bols_2020}
E.~Bols\hrefCMSnoop {}{ { et~al.}, ``Jet flavour classification using
  {DeepJet}'',} \textit{ JINST} \textbf{ 15} (2020) P12012,
  \href{http://dx.doi.org/10.1088/1748-0221/15/12/P12012}{\doi{10.1088/1748-0221/15/12/P12012}},
  \href{http://www.arXiv.org/abs/2008.10519}{\texttt{arXiv:2008.10519}}.

\bibitem{BTagDeepJet:CMS-DP-2018-058}
\href {http://cds.cern.ch/record/2646773}{{CMS Collaboration}, ``{Performance
  of the DeepJet b tagging algorithm using 41.9/fb of data from proton-proton
  collisions at 13 TeV with Phase 1 CMS detector}'',} CMS Detector Performance
  Note CMS-DP-2018-058, 2018.

\bibitem{PNet:Qu_2020}
\hrefCMSnoop {}{H.~Qu and L.~Gouskos, ``Jet tagging via particle clouds'',}
  \textit{ Phys. Rev. D} \textbf{ 101} (2020) 056019,
  \href{http://dx.doi.org/10.1103/physrevd.101.056019}{\doi{10.1103/physrevd.101.056019}},
  \href{http://www.arXiv.org/abs/1902.08570}{\texttt{arXiv:1902.08570}}.

\bibitem{Fruhwirth:178627}
\hrefCMSnoop {}{R.~Fr{\"u}hwirth, ``{Application of Kalman filtering to track
  and vertex fitting}'',} \textit{ Nucl. Instrum. Methods Phys. Res. A}
  \textbf{ 262} (1987) 444,
  \href{http://dx.doi.org/10.1016/0168-9002(87)90887-4}{\doi{10.1016/0168-9002(87)90887-4}}.

\bibitem{CMS:2021yvr}
\hrefCMSnoop {}{{CMS Collaboration}, ``{Performance of the CMS muon trigger
  system in proton-proton collisions at $\sqrt{s} =$ 13 TeV}'',} \textit{
  JINST} \textbf{ 16} (2021) P07001,
  \href{http://dx.doi.org/10.1088/1748-0221/16/07/P07001}{\doi{10.1088/1748-0221/16/07/P07001}},
  \href{http://www.arXiv.org/abs/2102.04790}{\texttt{arXiv:2102.04790}}.

\bibitem{Adam_2005}
\hrefCMSnoop {}{W.~Adam, R.~Fr{\"u}hwirth, A.~Strandlie, and T.~Todorov,
  ``Reconstruction of electrons with the {Gaussian}-sum filter in the {CMS}
  tracker at the {LHC}'',} \textit{ J. Phys. G: Nucl. Part. Phys.} \textbf{ 31}
  (2005) N9,
  \href{http://dx.doi.org/10.1088/0954-3899/31/9/n01}{\doi{10.1088/0954-3899/31/9/n01}},
  \href{http://www.arXiv.org/abs/physics/0306087}{\texttt{arXiv:physics/0306087}}.

\bibitem{MET:2019ctu}
\hrefCMSnoop {}{{CMS Collaboration}, ``Performance of missing transverse
  momentum reconstruction in proton-proton collisions at $\sqrt{s} = 13$\,{TeV}
  using the {CMS} detector'',} \textit{ JINST} \textbf{ 14} (2019) P07004,
  \href{http://dx.doi.org/10.1088/1748-0221/14/07/P07004}{\doi{10.1088/1748-0221/14/07/P07004}},
\href{http://www.arXiv.org/abs/1903.06078}{\texttt{arXiv:1903.06078}}.

\bibitem{Sirunyan:2018pgf}
\hrefCMSnoop {}{{CMS Collaboration}, ``{Performance of reconstruction and
  identification of $\tau$ leptons decaying to hadrons and $\nu_\tau$ in pp
  collisions at $\sqrt{s}=$ 13 TeV}'',} \textit{ JINST} \textbf{ 13} (2018)
  P10005,
  \href{http://dx.doi.org/10.1088/1748-0221/13/10/P10005}{\doi{10.1088/1748-0221/13/10/P10005}},
\href{http://www.arXiv.org/abs/1809.02816}{\texttt{arXiv:1809.02816}}.

\bibitem{CMS:2022prd}
\hrefCMSnoop {}{{CMS Collaboration}, ``{Identification of hadronic tau lepton
  decays using a deep neural network}'',} \textit{ JINST} \textbf{ 17} (2022)
  P07023,
  \href{http://dx.doi.org/10.1088/1748-0221/17/07/P07023}{\doi{10.1088/1748-0221/17/07/P07023}},
  \href{http://www.arXiv.org/abs/2201.08458}{\texttt{arXiv:2201.08458}}.

\bibitem{Cacciari:2007fd}
\hrefCMSnoop {}{M.~Cacciari and G.~P. Salam, ``Pileup subtraction using jet
  areas'',} \textit{ Phys. Lett. B} \textbf{ 659} (2008) 119,
  \href{http://dx.doi.org/10.1016/j.physletb.2007.09.077}{\doi{10.1016/j.physletb.2007.09.077}},
  \href{http://www.arXiv.org/abs/0707.1378}{\texttt{arXiv:0707.1378}}.

\bibitem{EXO-16-056}
\hrefCMSnoop {}{{CMS Collaboration}, ``Search for narrow and broad dijet
  resonances in proton-proton collisions at $\sqrt{s} = 13$ {TeV} and
  constraints on dark matter mediators and other new particles'',} \textit{
  JHEP} \textbf{ 08} (2018) 130,
  \href{http://dx.doi.org/10.1007/jhep08(2018)130}{\doi{10.1007/jhep08(2018)130}},
  \href{http://www.arXiv.org/abs/1806.00843}{\texttt{arXiv:1806.00843}}.

\bibitem{NIPS2017_f22e4747}
M.~Zaheer\href
  {https://proceedings.neurips.cc/paper/2017/file/f22e4747da1aa27e363d86d40ff442fe-Paper.pdf}{
  { et~al.}, ``Deep sets'',} in \textit{ Advances in Neural Information
  Processing Systems}, I.~Guyon { et~al.}, eds., volume~30, p.~3392.
\newblock Curran Associates, Inc., 2017.

\bibitem{EXO-21-004}
\hrefCMSnoop {}{{CMS Collaboration}, ``{Searches for pair-produced multijet
  resonances using data scouting in proton-proton collisions at $\sqrt{s}$ = 13
  TeV}'',} (2024).
  \href{http://www.arXiv.org/abs/2404.02992}{\texttt{arXiv:2404.02992}}.
  {Accepted for publication by PRL}.

\bibitem{EXO-19-018}
\hrefCMSnoop {}{{CMS Collaboration}, ``Search for a narrow resonance lighter
  than 200 {GeV} decaying to a pair of muons in proton-proton collisions at
  $\sqrt{s}=13$ {TeV}'',} \textit{ Phys. Rev. Lett.} \textbf{ 124} (2020)
  131802,
  \href{http://dx.doi.org/10.1103/PhysRevLett.124.131802}{\doi{10.1103/PhysRevLett.124.131802}},
  \href{http://www.arXiv.org/abs/1912.04776}{\texttt{arXiv:1912.04776}}.

\bibitem{EXO-20-014}
\hrefCMSnoop {}{{CMS Collaboration}, ``Search for long-lived particles decaying
  into muon pairs in proton-proton collisions at $ \sqrt{s} = 13$ {TeV}
  collected with a dedicated high-rate data stream'',} \textit{ JHEP} \textbf{
  04} (2022) 062,
  \href{http://dx.doi.org/10.1007/JHEP04(2022)062}{\doi{10.1007/JHEP04(2022)062}},
  \href{http://www.arXiv.org/abs/2112.13769}{\texttt{arXiv:2112.13769}}.

\bibitem{BDT_ROE2005577}
B.~P. Roe\hrefCMSnoop {}{ { et~al.}, ``Boosted decision trees as an alternative
  to artificial neural networks for particle identification'',} \textit{ Nucl.
  Instrum. Methods Phys. Res. A} \textbf{ 543} (2005) 577,
  \href{http://dx.doi.org/10.1016/j.nima.2004.12.018}{\doi{10.1016/j.nima.2004.12.018}},
  \href{http://www.arXiv.org/abs/physics/0408124}{\texttt{arXiv:physics/0408124}}.

\bibitem{EXO-21-005}
\hrefCMSnoop {}{{CMS Collaboration}, ``{Search for direct production of
  GeV-scale resonances decaying to a pair of muons in proton-proton collisions
  at $\sqrt{s}$ = 13 TeV}'',} \textit{ JHEP} \textbf{ 12} (2023) 070,
  \href{http://dx.doi.org/10.1007/JHEP12(2023)070}{\doi{10.1007/JHEP12(2023)070}},
  \href{http://www.arXiv.org/abs/2309.16003}{\texttt{arXiv:2309.16003}}.

\bibitem{crystalball-1}
\href {http://www.slac.stanford.edu/pubs/slacreports/slac-r-236.html}{M.~J.
  Oreglia, ``A study of the reactions $\psi^\prime \to \gamma \gamma \psi$''}.
\newblock PhD thesis, Stanford University, 1980.
\newblock {SLAC} Report {SLAC-R-236}, see Appendix {D}.

\bibitem{crystalball-2}
\href
  {http://slac.stanford.edu/pubs/slacreports/reports02/slac-r-255.pdf}{J.~E.
  Gaiser, ``{Charmonium spectroscopy from radiative decays of the $J/\psi$ and
  $\psi^\prime$}''}.
\newblock PhD thesis, Stanford University, 1982.
\newblock {SLAC} Report {SLAC-R-255}.

\bibitem{Harris:2011bh}
\hrefCMSnoop {}{R.~M. Harris and K.~Kousouris, ``Searches for dijet resonances
  at hadron colliders'',} \textit{ Int. J. Mod. Phys. A} \textbf{ 26} (2011)
  5005,
  \href{http://dx.doi.org/10.1142/S0217751X11054905}{\doi{10.1142/S0217751X11054905}},
  \href{http://www.arXiv.org/abs/1110.5302}{\texttt{arXiv:1110.5302}}.

\bibitem{Dobrescu:2013cmh}
\hrefCMSnoop {}{B.~A. Dobrescu and F.~Yu, ``{Coupling-Mass Mapping of Dijet
  Peak Searches}'',} \textit{ Phys. Rev. D} \textbf{ 88} (2013) 035021,
  \href{http://dx.doi.org/10.1103/PhysRevD.88.035021}{\doi{10.1103/PhysRevD.88.035021}},
  \href{http://www.arXiv.org/abs/1306.2629}{\texttt{arXiv:1306.2629}}.
  [Erratum: Phys.Rev. D 90 (2014) 079901].

\bibitem{EXO-16-032}
\hrefCMSnoop {}{{CMS Collaboration}, ``Search for dijet resonances in
  proton-proton collisions at $\sqrt{s} = 13$ {TeV} and constraints on dark
  matter and other models'',} \textit{ Phys. Lett. B} \textbf{ 769} (2017) 520,
  \href{http://dx.doi.org/10.1016/j.physletb.2017.02.012}{\doi{10.1016/j.physletb.2017.02.012}},
  \href{http://www.arXiv.org/abs/1611.03568}{\texttt{arXiv:1611.03568}}.
  [Erratum: \DOI{10.1016/j.physletb.2017.09.029}].

\bibitem{EXO-19-004}
\hrefCMSnoop {}{{CMS Collaboration}, ``Search for dijet resonances using events
  with three jets in proton-proton collisions at $\sqrt{s} = 13$ {TeV}'',}
  \textit{ Phys. Lett. B} \textbf{ 805} (2020) 135448,
  \href{http://dx.doi.org/10.1016/j.physletb.2020.135448}{\doi{10.1016/j.physletb.2020.135448}},
  \href{http://www.arXiv.org/abs/1911.03761}{\texttt{arXiv:1911.03761}}.

\bibitem{EXO-17-001}
\hrefCMSnoop {}{{CMS Collaboration}, ``Search for low mass vector resonances
  decaying into quark-antiquark pairs in proton-proton collisions at $\sqrt{s}
  = 13$ {TeV}'',} \textit{ JHEP} \textbf{ 01} (2018) 097,
  \href{http://dx.doi.org/10.1007/JHEP01(2018)097}{\doi{10.1007/JHEP01(2018)097}},
  \href{http://www.arXiv.org/abs/1710.00159}{\texttt{arXiv:1710.00159}}.

\bibitem{EXO-17-027}
\hrefCMSnoop {}{{CMS Collaboration}, ``Search for low-mass quark-antiquark
  resonances produced in association with a photon at $\sqrt {s} = 13$
  {TeV}'',} \textit{ Phys. Rev. Lett.} \textbf{ 123} (2019) 231803,
  \href{http://dx.doi.org/10.1103/PhysRevLett.123.231803}{\doi{10.1103/PhysRevLett.123.231803}},
  \href{http://www.arXiv.org/abs/1905.10331}{\texttt{arXiv:1905.10331}}.

\bibitem{Barbier:2004ez}
R.~Barbier\hrefCMSnoop {}{ { et~al.}, ``R-parity violating supersymmetry'',}
  \textit{ Phys. Rept.} \textbf{ 420} (2005) 1,
  \href{http://dx.doi.org/10.1016/j.physrep.2005.08.006}{\doi{10.1016/j.physrep.2005.08.006}},
  \href{http://www.arXiv.org/abs/hep-ph/0406039}{\texttt{arXiv:hep-ph/0406039}}.

\bibitem{DDT}
J.~Dolen\hrefCMSnoop {}{ { et~al.}, ``Thinking outside the {ROCs}: Designing
  decorrelated taggers ({DDT}) for jet substructure'',} \textit{ JHEP} \textbf{
  05} (2016) 156,
  \href{http://dx.doi.org/10.1007/jhep05(2016)156}{\doi{10.1007/jhep05(2016)156}},
  \href{http://www.arXiv.org/abs/1603.00027}{\texttt{arXiv:1603.00027}}.

\bibitem{cdfmultijets}
\hrefCMSnoop {}{{CDF} Collaboration, ``First search for multijet resonances in
  $\sqrt{s} = 1.96$ {TeV} $ p\bar{p}$ collisions'',} \textit{ Phys. Rev. Lett.}
  \textbf{ 107} (2011) 042001,
  \href{http://dx.doi.org/10.1103/PhysRevLett.107.042001}{\doi{10.1103/PhysRevLett.107.042001}},
\href{http://www.arXiv.org/abs/1105.2815}{\texttt{arXiv:1105.2815}}.

\bibitem{atlas8}
\hrefCMSnoop {}{{ATLAS Collaboration}, ``Search for massive supersymmetric
  particles decaying to many jets using the {ATLAS} detector in $\pp$
  collisions at $\sqrt{s} = 8$ {TeV}'',} \textit{ Phys. Rev. D} \textbf{ 91}
  (2015) 112016,
  \href{http://dx.doi.org/10.1103/PhysRevD.91.112016}{\doi{10.1103/PhysRevD.91.112016}},
  \href{http://www.arXiv.org/abs/1502.05686}{\texttt{arXiv:1502.05686}}.
[Erratum: \DOI{10.1103/PhysRevD.93.039901}].

\bibitem{atlas13}
\hrefCMSnoop {}{{ATLAS Collaboration}, ``Search for {R}-parity-violating
  supersymmetric particles in multi-jet final states produced in p-p collisions
  at $\sqrt{s} =13$ {TeV} using the {ATLAS} detector at the {LHC}'',} \textit{
  Phys. Lett. B} \textbf{ 785} (2018) 136,
  \href{http://dx.doi.org/10.1016/j.physletb.2018.08.021}{\doi{10.1016/j.physletb.2018.08.021}},
\href{http://www.arXiv.org/abs/1804.03568}{\texttt{arXiv:1804.03568}}.

\bibitem{gluino2011}
\hrefCMSnoop {}{{CMS Collaboration}, ``Search for three-jet resonances in pp
  collisions at $\sqrt{s}=7$ {TeV}'',} \textit{ Phys. Lett. B} \textbf{ 718}
  (2012) 329,
  \href{http://dx.doi.org/10.1016/j.physletb.2012.10.048}{\doi{10.1016/j.physletb.2012.10.048}},
  \href{http://www.arXiv.org/abs/1208.2931}{\texttt{arXiv:1208.2931}}.

\bibitem{gluino2012}
\hrefCMSnoop {}{{CMS Collaboration}, ``Searches for light- and heavy-flavour
  three-jet resonances in pp collisions at $\sqrt{s} = 8$ {TeV}'',} \textit{
  Phys. Lett. B} \textbf{ 730} (2014) 193,
  \href{http://dx.doi.org/10.1016/j.physletb.2014.01.049}{\doi{10.1016/j.physletb.2014.01.049}},
\href{http://www.arXiv.org/abs/1311.1799}{\texttt{arXiv:1311.1799}}.

\bibitem{gluino2017}
\hrefCMSnoop {}{{CMS Collaboration}, ``Search for pair-produced three-jet
  resonances in proton-proton collisions at $\sqrt s = 13$ {TeV}'',} \textit{
  Phys. Rev. D} \textbf{ 99} (2019) 012010,
  \href{http://dx.doi.org/10.1103/PhysRevD.99.012010}{\doi{10.1103/PhysRevD.99.012010}},
  \href{http://www.arXiv.org/abs/1810.10092}{\texttt{arXiv:1810.10092}}.

\bibitem{cmsstop13}
\hrefCMSnoop {}{{CMS Collaboration}, ``Search for pair-produced resonances
  decaying to quark pairs in proton-proton collisions at $\sqrt{s}= 13$
  {TeV}'',} \textit{ Phys. Rev. D} \textbf{ 98} (2018) 112014,
  \href{http://dx.doi.org/10.1103/PhysRevD.98.112014}{\doi{10.1103/PhysRevD.98.112014}},
  \href{http://www.arXiv.org/abs/1808.03124}{\texttt{arXiv:1808.03124}}.

\bibitem{CMS:2022usq}
\hrefCMSnoop {}{{CMS Collaboration}, ``Search for resonant and nonresonant
  production of pairs of dijet resonances in proton-proton collisions at
  $\sqrt{s} = 13$ {TeV}'',} \textit{ JHEP} \textbf{ 07} (2023) 161,
  \href{http://dx.doi.org/10.1007/JHEP07(2023)161}{\doi{10.1007/JHEP07(2023)161}},
  \href{http://www.arXiv.org/abs/2206.09997}{\texttt{arXiv:2206.09997}}.

\bibitem{EXO-19-018_LHCb_PhysRevLett.124.041801}
\hrefCMSnoop {}{{LHCb Collaboration}, ``Search for
  ${A}^{\ensuremath{'}}\ensuremath{\rightarrow}{\ensuremath{\mu}}^{+}{\ensuremath{\mu}}^{\ensuremath{-}}$
  decays'',} \textit{ Phys. Rev. Lett.} \textbf{ 124} (2020) 041801,
  \href{http://dx.doi.org/10.1103/PhysRevLett.124.041801}{\doi{10.1103/PhysRevLett.124.041801}},
  \href{http://www.arXiv.org/abs/1910.06926}{\texttt{arXiv:1910.06926}}.

\bibitem{EXO-19-018_EW_Curtin:2014cca}
\hrefCMSnoop {}{D.~Curtin, R.~Essig, S.~Gori, and J.~Shelton, ``Illuminating
  dark photons with high-energy colliders'',} \textit{ JHEP} \textbf{ 02}
  (2015) 157,
  \href{http://dx.doi.org/10.1007/JHEP02(2015)157}{\doi{10.1007/JHEP02(2015)157}},
  \href{http://www.arXiv.org/abs/1412.0018}{\texttt{arXiv:1412.0018}}.

\bibitem{EXO-21-005_LHCb:2020ysn}
\hrefCMSnoop {}{{LHCb Collaboration}, ``{Searches for low-mass dimuon
  resonances}'',} \textit{ JHEP} \textbf{ 10} (2020) 156,
  \href{http://dx.doi.org/10.1007/JHEP10(2020)156}{\doi{10.1007/JHEP10(2020)156}},
  \href{http://www.arXiv.org/abs/2007.03923}{\texttt{arXiv:2007.03923}}.

\bibitem{BaBar:2012wey}
\hrefCMSnoop {}{{BaBar} Collaboration, ``Search for di-muon decays of a
  low-mass {Higgs} boson in radiative decays of the {$\Upsilon$(1S)}'',}
  \textit{ Phys. Rev. D} \textbf{ 87} (2013) 031102,
  \href{http://dx.doi.org/10.1103/PhysRevD.87.031102}{\doi{10.1103/PhysRevD.87.031102}},
  \href{http://www.arXiv.org/abs/1210.0287}{\texttt{arXiv:1210.0287}}.
  [Erratum: \DOI{10.1103/PhysRevD.87.059903}].

\bibitem{Bezrukov:2013fca}
\hrefCMSnoop {}{F.~Bezrukov and D.~Gorbunov, ``{Light inflaton after LHC8 and
  WMAP9 results}'',} \textit{ JHEP} \textbf{ 07} (2013) 140,
  \href{http://dx.doi.org/10.1007/JHEP07(2013)140}{\doi{10.1007/JHEP07(2013)140}},
  \href{http://www.arXiv.org/abs/1303.4395}{\texttt{arXiv:1303.4395}}.

\bibitem{LHCb:2015nkv}
\hrefCMSnoop {}{{LHCb Collaboration}, ``{Search for hidden-sector bosons in
  $B^0 \!\to K^{*0}\mu^+\mu^-$ decays}'',} \textit{ Phys. Rev. Lett.} \textbf{
  115} (2015) 161802,
  \href{http://dx.doi.org/10.1103/PhysRevLett.115.161802}{\doi{10.1103/PhysRevLett.115.161802}},
  \href{http://www.arXiv.org/abs/1508.04094}{\texttt{arXiv:1508.04094}}.

\bibitem{LHCb:2016awg}
\hrefCMSnoop {}{{LHCb Collaboration}, ``{Search for long-lived scalar particles
  in $B^+ \to K^+ \chi (\mu^+\mu^-)$ decays}'',} \textit{ Phys. Rev. D}
  \textbf{ 95} (2017) 071101,
  \href{http://dx.doi.org/10.1103/PhysRevD.95.071101}{\doi{10.1103/PhysRevD.95.071101}},
  \href{http://www.arXiv.org/abs/1612.07818}{\texttt{arXiv:1612.07818}}.

\bibitem{BPH-22-003}
\hrefCMSnoop {}{{CMS Collaboration}, ``Observation of the rare decay of the
  $\eta$ meson to four muons'',} \textit{ Phys. Rev. Lett.} \textbf{ 131}
  (2023) 091903,
  \href{http://dx.doi.org/10.1103/PhysRevLett.131.091903}{\doi{10.1103/PhysRevLett.131.091903}},
  \href{http://www.arXiv.org/abs/2305.04904}{\texttt{arXiv:2305.04904}}.

\bibitem{escribanoDatadrivenApproachPi2018}
\hrefCMSnoop {}{R.~Escribano and S.~{Gonz{\`a}lez-Sol{\'i}s}, ``A data-driven
  approach to {$\pi^0$}, {$\eta$} and {$\eta^\prime$} single and double
  {Dalitz} decays'',} \textit{ Chinese Phys. C} \textbf{ 42} (2018) 023109,
  \href{http://dx.doi.org/10.1088/1674-1137/42/2/023109}{\doi{10.1088/1674-1137/42/2/023109}},
  \href{http://www.arXiv.org/abs/1511.04916}{\texttt{arXiv:1511.04916}}.

\bibitem{Workman:2022ynf}
\hrefCMSnoop {}{{Particle Data Group}, ``{Review of particle physics}'',}
  \textit{ Prog. Theor. Exp. Phys} \textbf{ 2022} (2022) 083C01,
  \href{http://dx.doi.org/10.1093/ptep/ptac097}{\doi{10.1093/ptep/ptac097}}.

\bibitem{CMS-DP-2021-005}
\href {https://cds.cern.ch/record/2765489}{{CMS Collaboration}, ``{PF} jet
  performances at high level trigger using {Patatrack} pixel tracks'',} CMS
  Detector Performance Note CMS-DP-2021-005, 2021.

\bibitem{CMS:2020zge}
\hrefCMSnoop {}{{CMS Collaboration}, ``{Inclusive search for highly boosted
  Higgs bosons decaying to bottom quark-antiquark pairs in proton-proton
  collisions at $\sqrt{s} =$ 13 TeV}'',} \textit{ JHEP} \textbf{ 12} (2020)
  085,
  \href{http://dx.doi.org/10.1007/JHEP12(2020)085}{\doi{10.1007/JHEP12(2020)085}},
  \href{http://www.arXiv.org/abs/2006.13251}{\texttt{arXiv:2006.13251}}.

\bibitem{HIG-21-020}
\hrefCMSnoop {}{{CMS Collaboration}, ``{Measurement of boosted Higgs bosons
  produced via vector boson fusion or gluon fusion in the H $\to
  $$\mathrm{b\bar{b}}$ decay mode using LHC proton-proton collision data at
  $\sqrt{s}$ = 13 TeV}'',} {(2024)}.
  \href{http://www.arXiv.org/abs/2407.08012}{\texttt{arXiv:2407.08012}}.
  {Submitted to JHEP}.

\bibitem{CMS:2022fxs}
\hrefCMSnoop {}{{CMS Collaboration}, ``Search for {Higgs} boson and observation
  of {Z} boson through their decay into a charm quark-antiquark pair in boosted
  topologies in proton-proton collisions at $\sqrt{s}=13$ {TeV}'',} \textit{
  Phys. Rev. Lett.} \textbf{ 131} (2023) 041801,
  \href{http://dx.doi.org/10.1103/PhysRevLett.131.041801}{\doi{10.1103/PhysRevLett.131.041801}},
  \href{http://www.arXiv.org/abs/2211.14181}{\texttt{arXiv:2211.14181}}.

\bibitem{Qu:2022mxj}
\hrefCMSnoop {}{H.~Qu, C.~Li, and S.~Qian, ``{Particle Transformer for Jet
  Tagging}'',} 2022.
  \href{http://www.arXiv.org/abs/2202.03772}{\texttt{arXiv:2202.03772}}.

\bibitem{CMS:2021zxu}
\hrefCMSnoop {}{{CMS Collaboration}, ``Search for {W$\gamma$} resonances in
  proton-proton collisions at $\sqrt{s} = 13$ {TeV} using hadronic decays of
  {Lorentz}-boosted {W} bosons'',} \textit{ Phys. Lett. B} \textbf{ 826} (2022)
  136888,
  \href{http://dx.doi.org/10.1016/j.physletb.2022.136888}{\doi{10.1016/j.physletb.2022.136888}},
  \href{http://www.arXiv.org/abs/2106.10509}{\texttt{arXiv:2106.10509}}.

\bibitem{CMS:2016dhk}
\hrefCMSnoop {}{{CMS Collaboration}, ``{Searches for invisible decays of the
  Higgs boson in pp collisions at $\sqrt{s}$ = 7, 8, and 13 TeV}'',} \textit{
  JHEP} \textbf{ 02} (2017) 135,
  \href{http://dx.doi.org/10.1007/JHEP02(2017)135}{\doi{10.1007/JHEP02(2017)135}},
  \href{http://www.arXiv.org/abs/1610.09218}{\texttt{arXiv:1610.09218}}.

\bibitem{CMS:2015ebl_run1parking_Hbb}
\hrefCMSnoop {}{{CMS Collaboration}, ``{Search for the standard model Higgs
  boson produced through vector boson fusion and decaying to $b
  \overline{b}$}'',} \textit{ Phys. Rev. D} \textbf{ 92} (2015) 032008,
  \href{http://dx.doi.org/10.1103/PhysRevD.92.032008}{\doi{10.1103/PhysRevD.92.032008}},
  \href{http://www.arXiv.org/abs/1506.01010}{\texttt{arXiv:1506.01010}}.

\bibitem{CMS:2014wdm_run1parking_Htautau}
\hrefCMSnoop {}{{CMS Collaboration}, ``{Evidence for the 125 GeV Higgs boson
  decaying to a pair of $\tau$ leptons}'',} \textit{ JHEP} \textbf{ 05} (2014)
  104,
  \href{http://dx.doi.org/10.1007/JHEP05(2014)104}{\doi{10.1007/JHEP05(2014)104}},
  \href{http://www.arXiv.org/abs/1401.5041}{\texttt{arXiv:1401.5041}}.

\bibitem{CMS:2014ccx_run1parking_MSSMHtautau}
\hrefCMSnoop {}{{CMS Collaboration}, ``{Search for neutral MSSM Higgs bosons
  decaying to a pair of tau leptons in pp collisions}'',} \textit{ JHEP}
  \textbf{ 10} (2014) 160,
  \href{http://dx.doi.org/10.1007/JHEP10(2014)160}{\doi{10.1007/JHEP10(2014)160}},
  \href{http://www.arXiv.org/abs/1408.3316}{\texttt{arXiv:1408.3316}}.

\bibitem{CMS:2017ygm}
\hrefCMSnoop {}{{CMS Collaboration}, ``Measurement of b hadron lifetimes in pp
  collisions at $\sqrt{s} = 8$~{TeV}'',} \textit{ Eur. Phys. J. C} \textbf{ 78}
  (2018) 457,
  \href{http://dx.doi.org/10.1140/epjc/s10052-018-5929-3}{\doi{10.1140/epjc/s10052-018-5929-3}},
  \href{http://www.arXiv.org/abs/1710.08949}{\texttt{arXiv:1710.08949}}.
  [Erratum: \DOI{10.1140/epjc/s10052-018-6014-7}].

\bibitem{CMS:2015bcy}
\hrefCMSnoop {}{{CMS Collaboration}, ``Angular analysis of the decay {$B^0 \to
  K^{*0} \mu^+ \mu^-$} from pp collisions at $\sqrt{s} = 8$ {TeV}'',} \textit{
  Phys. Lett. B} \textbf{ 753} (2016) 424,
  \href{http://dx.doi.org/10.1016/j.physletb.2015.12.020}{\doi{10.1016/j.physletb.2015.12.020}},
  \href{http://www.arXiv.org/abs/1507.08126}{\texttt{arXiv:1507.08126}}.

\bibitem{CMS:2017rzx}
\hrefCMSnoop {}{{CMS Collaboration}, ``{Measurement of angular parameters from
  the decay $\mathrm{B}^0 \to \mathrm{K}^{*0} \mu^+ \mu^-$ in proton-proton
  collisions at $\sqrt{s} = $ 8 TeV}'',} \textit{ Phys. Lett. B} \textbf{ 781}
  (2018) 517,
  \href{http://dx.doi.org/10.1016/j.physletb.2018.04.030}{\doi{10.1016/j.physletb.2018.04.030}},
  \href{http://www.arXiv.org/abs/1710.02846}{\texttt{arXiv:1710.02846}}.

\bibitem{CMS:2018qih}
\hrefCMSnoop {}{{CMS Collaboration}, ``{Angular analysis of the decay B$^+\to$
  K$^+\mu^+\mu^-$ in proton-proton collisions at $\sqrt{s} =$ 8 TeV}'',}
  \textit{ Phys. Rev. D} \textbf{ 98} (2018) 112011,
  \href{http://dx.doi.org/10.1103/PhysRevD.98.112011}{\doi{10.1103/PhysRevD.98.112011}},
  \href{http://www.arXiv.org/abs/1806.00636}{\texttt{arXiv:1806.00636}}.

\bibitem{CMS:2020oqb}
\hrefCMSnoop {}{{CMS Collaboration}, ``{Angular analysis of the decay B$^+$
  $\to$ K$^*$(892)$^+\mu^+\mu^-$ in proton-proton collisions at $\sqrt{s} =$ 8
  TeV}'',} \textit{ JHEP} \textbf{ 04} (2021) 124,
  \href{http://dx.doi.org/10.1007/JHEP04(2021)124}{\doi{10.1007/JHEP04(2021)124}},
  \href{http://www.arXiv.org/abs/2010.13968}{\texttt{arXiv:2010.13968}}.

\bibitem{cmsstop8}
\hrefCMSnoop {}{{CMS Collaboration}, ``Search for pair-produced resonances
  decaying to jet pairs in proton-proton collisions at $\sqrt{s} = 8$ {TeV}'',}
  \textit{ Phys. Lett. B} \textbf{ 747} (2015) 98,
  \href{http://dx.doi.org/10.1016/j.physletb.2015.04.045}{\doi{10.1016/j.physletb.2015.04.045}},
  \href{http://www.arXiv.org/abs/1412.7706}{\texttt{arXiv:1412.7706}}.

\bibitem{CMS:2015ifd_run1parking_HundetectedGamma}
\hrefCMSnoop {}{{CMS Collaboration}, ``{Search for exotic decays of a Higgs
  boson into undetectable particles and one or more photons}'',} \textit{ Phys.
  Lett. B} \textbf{ 753} (2016) 363,
  \href{http://dx.doi.org/10.1016/j.physletb.2015.12.017}{\doi{10.1016/j.physletb.2015.12.017}},
  \href{http://www.arXiv.org/abs/1507.00359}{\texttt{arXiv:1507.00359}}.

\bibitem{Rogan:2010kb}
\hrefCMSnoop {}{C.~Rogan, ``{Kinematical variables towards new dynamics at the
  LHC}'',} 2010.
  \href{http://www.arXiv.org/abs/1006.2727}{\texttt{arXiv:1006.2727}}.

\bibitem{CMS:2011xie}
\hrefCMSnoop {}{{CMS Collaboration}, ``{Inclusive search for squarks and
  gluinos in $pp$ collisions at $\sqrt{s}=7$ TeV}'',} \textit{ Phys. Rev. D}
  \textbf{ 85} (2012) 012004,
  \href{http://dx.doi.org/10.1103/PhysRevD.85.012004}{\doi{10.1103/PhysRevD.85.012004}},
  \href{http://www.arXiv.org/abs/1107.1279}{\texttt{arXiv:1107.1279}}.

\bibitem{CMS:2016gox_run1parking_razor_EX0-14-004}
\hrefCMSnoop {}{{CMS Collaboration}, ``{Search for dark matter particles in
  proton-proton collisions at $ \sqrt{s}=8 $ TeV using the razor variables}'',}
  \textit{ JHEP} \textbf{ 12} (2016) 088,
  \href{http://dx.doi.org/10.1007/JHEP12(2016)088}{\doi{10.1007/JHEP12(2016)088}},
  \href{http://www.arXiv.org/abs/1603.08914}{\texttt{arXiv:1603.08914}}.

\bibitem{CMS:2016rjk_run1parking_SUS-14-006_alphaT}
\hrefCMSnoop {}{{CMS Collaboration}, ``{Search for top squark pair production
  in compressed-mass-spectrum scenarios in proton-proton collisions at
  $\sqrt{s}$ = 8 TeV using the $\alpha_T$ variable}'',} \textit{ Phys. Lett. B}
  \textbf{ 767} (2017) 403,
  \href{http://dx.doi.org/10.1016/j.physletb.2017.02.007}{\doi{10.1016/j.physletb.2017.02.007}},
  \href{http://www.arXiv.org/abs/1605.08993}{\texttt{arXiv:1605.08993}}.

\bibitem{Hurth:2023jwr}
\hrefCMSnoop {}{T.~Hurth, F.~Mahmoudi, and S.~Neshatpour, ``{$B$ anomalies in
  the post $R_{K^{(*)}}$ era}'',} \textit{ Phys. Rev. D} \textbf{ 108} (2023)
  115037,
  \href{http://dx.doi.org/10.1103/PhysRevD.108.115037}{\doi{10.1103/PhysRevD.108.115037}},
  \href{http://www.arXiv.org/abs/2310.05585}{\texttt{arXiv:2310.05585}}.

\bibitem{BaBar:2001yhh}
\hrefCMSnoop {}{{BaBar} Collaboration, ``{The BaBar detector}'',} \textit{
  Nucl. Instrum. Methods Phys. Res. A} \textbf{ 479} (2002) 1,
  \href{http://dx.doi.org/10.1016/S0168-9002(01)02012-5}{\doi{10.1016/S0168-9002(01)02012-5}},
  \href{http://www.arXiv.org/abs/hep-ex/0105044}{\texttt{arXiv:hep-ex/0105044}}.

\bibitem{Belle:2000cnh}
\hrefCMSnoop {}{{Belle} Collaboration, ``{The Belle Detector}'',} \textit{
  Nucl. Instrum. Methods Phys. Res. A} \textbf{ 479} (2002) 117,
  \href{http://dx.doi.org/10.1016/S0168-9002(01)02013-7}{\doi{10.1016/S0168-9002(01)02013-7}}.

\bibitem{LHCb:2008vvz}
\hrefCMSnoop {}{{LHCb Collaboration}, ``{The LHCb Detector at the LHC}'',}
  \textit{ JINST} \textbf{ 3} (2008) S08005,
  \href{http://dx.doi.org/10.1088/1748-0221/3/08/S08005}{\doi{10.1088/1748-0221/3/08/S08005}}.

\bibitem{ATLAS_2008}
\hrefCMSnoop {}{{ATLAS Collaboration}, ``The {ATLAS} experiment at the {CERN}
  {Large Hadron Collider}'',} \textit{ JINST} \textbf{ 3} (2008) S08003,
  \href{http://dx.doi.org/10.1088/1748-0221/3/08/S08003}{\doi{10.1088/1748-0221/3/08/S08003}}.

\bibitem{CMS:2022mgd}
\hrefCMSnoop {}{{CMS Collaboration}, ``Measurement of the
  {$\mathrm{B}^0_\mathrm{s}\to \mu^+\mu^-$} decay properties and search for the
  {$\mathrm{B}^0\to \mu^+ \mu^-$} decay in proton-proton collisions at
  $\sqrt{s} = 13$ {TeV}'',} \textit{ Phys. Lett. B} \textbf{ 842} (2023)
  137955,
  \href{http://dx.doi.org/10.1016/j.physletb.2023.137955}{\doi{10.1016/j.physletb.2023.137955}},
  \href{http://www.arXiv.org/abs/2212.10311}{\texttt{arXiv:2212.10311}}.

\bibitem{LHCb:2021awg}
\hrefCMSnoop {}{{LHCb Collaboration}, ``{Measurement of the
  $B^0_s\to\mu^+\mu^-$ decay properties and search for the $B^0\to\mu^+\mu^-$
  and $B^0_s\to\mu^+\mu^-\gamma$ decays}'',} \textit{ Phys. Rev. D} \textbf{
  105} (2022) 012010,
  \href{http://dx.doi.org/10.1103/PhysRevD.105.012010}{\doi{10.1103/PhysRevD.105.012010}},
  \href{http://www.arXiv.org/abs/2108.09283}{\texttt{arXiv:2108.09283}}.

\bibitem{CMS:2019bbr}
\hrefCMSnoop {}{{CMS Collaboration}, ``{Measurement of properties of
  B$^0_\mathrm{s}\to\mu^+\mu^-$ decays and search for B$^0\to\mu^+\mu^-$ with
  the CMS experiment}'',} \textit{ JHEP} \textbf{ 04} (2020) 188,
  \href{http://dx.doi.org/10.1007/JHEP04(2020)188}{\doi{10.1007/JHEP04(2020)188}},
  \href{http://www.arXiv.org/abs/1910.12127}{\texttt{arXiv:1910.12127}}.

\bibitem{ATLAS:2018cur}
\hrefCMSnoop {}{{ATLAS Collaboration}, ``{Study of the rare decays of $B^0_s$
  and $B^0$ mesons into muon pairs using data collected during 2015 and 2016
  with the ATLAS detector}'',} \textit{ JHEP} \textbf{ 04} (2019) 098,
  \href{http://dx.doi.org/10.1007/JHEP04(2019)098}{\doi{10.1007/JHEP04(2019)098}},
  \href{http://www.arXiv.org/abs/1812.03017}{\texttt{arXiv:1812.03017}}.

\bibitem{CMS:2014xfa}
\hrefCMSnoop {}{{CMS, LHCb} Collaboration, ``{Observation of the rare
  $B^0_s\to\mu^+\mu^-$ decay from the combined analysis of CMS and LHCb
  data}'',} \textit{ Nature} \textbf{ 522} (2015) 68,
  \href{http://dx.doi.org/10.1038/nature14474}{\doi{10.1038/nature14474}},
  \href{http://www.arXiv.org/abs/1411.4413}{\texttt{arXiv:1411.4413}}.

\bibitem{Descotes-Genon:2012isb}
\hrefCMSnoop {}{S.~Descotes-Genon, J.~Matias, M.~Ramon, and J.~Virto,
  ``{Implications from clean observables for the binned analysis of $B
  \rightarrow K^*\mu^+\mu^-$ at large recoil}'',} \textit{ JHEP} \textbf{ 01}
  (2013) 048,
  \href{http://dx.doi.org/10.1007/JHEP01(2013)048}{\doi{10.1007/JHEP01(2013)048}},
  \href{http://www.arXiv.org/abs/1207.2753}{\texttt{arXiv:1207.2753}}.

\bibitem{LHCb:2013ghj}
\hrefCMSnoop {}{{LHCb Collaboration}, ``Measurement of form-factor-independent
  observables in the decay {$B^{0} \to K^{*0} \mu^+ \mu^-$}'',} \textit{ Phys.
  Rev. Lett.} \textbf{ 111} (2013) 191801,
  \href{http://dx.doi.org/10.1103/PhysRevLett.111.191801}{\doi{10.1103/PhysRevLett.111.191801}},
  \href{http://www.arXiv.org/abs/1308.1707}{\texttt{arXiv:1308.1707}}.

\bibitem{LHCb:2015svh}
\hrefCMSnoop {}{{LHCb Collaboration}, ``{Angular analysis of the $B^{0} \to
  K^{*0} \mu^{+} \mu^{-}$ decay using 3 fb$^{-1}$ of integrated luminosity}'',}
  \textit{ JHEP} \textbf{ 02} (2016) 104,
  \href{http://dx.doi.org/10.1007/JHEP02(2016)104}{\doi{10.1007/JHEP02(2016)104}},
  \href{http://www.arXiv.org/abs/1512.04442}{\texttt{arXiv:1512.04442}}.

\bibitem{LHCb:2020lmf}
\hrefCMSnoop {}{{LHCb Collaboration}, ``Measurement of {$CP$}-averaged
  observables in the {$B^{0}\rightarrow K^{*0}\mu^{+}\mu^{-}$} decay'',}
  \textit{ Phys. Rev. Lett.} \textbf{ 125} (2020) 011802,
  \href{http://dx.doi.org/10.1103/PhysRevLett.125.011802}{\doi{10.1103/PhysRevLett.125.011802}},
  \href{http://www.arXiv.org/abs/2003.04831}{\texttt{arXiv:2003.04831}}.

\bibitem{Ciuchini:2021smi}
M.~Ciuchini\hrefCMSnoop {}{ { et~al.}, ``Charming penguins and lepton
  universality violation in {$b \to s \ell^+ \ell^-$} decays'',} \textit{ Eur.
  Phys. J. C} \textbf{ 83} (2023) 64,
  \href{http://dx.doi.org/10.1140/epjc/s10052-023-11191-w}{\doi{10.1140/epjc/s10052-023-11191-w}},
  \href{http://www.arXiv.org/abs/2110.10126}{\texttt{arXiv:2110.10126}}.

\bibitem{Gubernari:2023puw}
\hrefCMSnoop {}{N.~Gubernari, M.~Reboud, D.~van Dyk, and J.~Virto,
  ``{Dispersive analysis of $B \to K^{(*)}$ and $B_{s} \to \phi$ form
  factors}'',} \textit{ JHEP} \textbf{ 12} (2023) 153,
  \href{http://dx.doi.org/10.1007/JHEP12(2023)153}{\doi{10.1007/JHEP12(2023)153}},
  \href{http://www.arXiv.org/abs/2305.06301}{\texttt{arXiv:2305.06301}}.

\bibitem{Gubernari:2024ews}
\hrefCMSnoop {}{N.~Gubernari, ``{Theoretical predictions for $b\to s \mu^+
  \mu^-$ decays}'',} in \textit{ {58th Rencontres de Moriond on Electroweak
  Interactions and Unified Theories}}.
\newblock 4, 2024.
\newblock
  \href{http://www.arXiv.org/abs/2404.10043}{\texttt{arXiv:2404.10043}}.

\bibitem{BaBar:2012obs}
\hrefCMSnoop {}{{BaBar} Collaboration, ``{Evidence for an excess of $\bar{B}
  \to D^{(*)} \tau^-\bar{\nu}_\tau$ decays}'',} \textit{ Phys. Rev. Lett.}
  \textbf{ 109} (2012) 101802,
  \href{http://dx.doi.org/10.1103/PhysRevLett.109.101802}{\doi{10.1103/PhysRevLett.109.101802}},
  \href{http://www.arXiv.org/abs/1205.5442}{\texttt{arXiv:1205.5442}}.

\bibitem{BaBar:2013mob}
\hrefCMSnoop {}{{BaBar} Collaboration, ``Measurement of an excess of {$\bar{B}
  \to D^{(*)}\tau^- \bar{\nu}_\tau$} decays and implications for charged
  {Higgs} bosons'',} \textit{ Phys. Rev. D} \textbf{ 88} (2013) 072012,
  \href{http://dx.doi.org/10.1103/PhysRevD.88.072012}{\doi{10.1103/PhysRevD.88.072012}},
  \href{http://www.arXiv.org/abs/1303.0571}{\texttt{arXiv:1303.0571}}.

\bibitem{Belle:2015qfa}
\hrefCMSnoop {}{{Belle} Collaboration, ``{Measurement of the branching ratio of
  $\bar{B} \to D^{(\ast)} \tau^- \bar{\nu}_\tau$ relative to $\bar{B} \to
  D^{(\ast)} \ell^- \bar{\nu}_\ell$ decays with hadronic tagging at Belle}'',}
  \textit{ Phys. Rev. D} \textbf{ 92} (2015) 072014,
  \href{http://dx.doi.org/10.1103/PhysRevD.92.072014}{\doi{10.1103/PhysRevD.92.072014}},
  \href{http://www.arXiv.org/abs/1507.03233}{\texttt{arXiv:1507.03233}}.

\bibitem{LHCb:2015gmp}
\hrefCMSnoop {}{{LHCb Collaboration}, ``{Measurement of the ratio of branching
  fractions $\mathcal{B}(\bar{B}^0 \to
  D^{*+}\tau^{-}\bar{\nu}_{\tau})/\mathcal{B}(\bar{B}^0 \to
  D^{*+}\mu^{-}\bar{\nu}_{\mu})$}'',} \textit{ Phys. Rev. Lett.} \textbf{ 115}
  (2015) 111803,
  \href{http://dx.doi.org/10.1103/PhysRevLett.115.111803}{\doi{10.1103/PhysRevLett.115.111803}},
  \href{http://www.arXiv.org/abs/1506.08614}{\texttt{arXiv:1506.08614}}.
  [Erratum: \DOI{10.1103/PhysRevLett.115.159901}].

\bibitem{LHCb:2014vgu}
\hrefCMSnoop {}{{LHCb Collaboration}, ``{Test of lepton universality using
  $B^{+}\rightarrow K^{+}\ell^{+}\ell^{-}$ decays}'',} \textit{ Phys. Rev.
  Lett.} \textbf{ 113} (2014) 151601,
  \href{http://dx.doi.org/10.1103/PhysRevLett.113.151601}{\doi{10.1103/PhysRevLett.113.151601}},
  \href{http://www.arXiv.org/abs/1406.6482}{\texttt{arXiv:1406.6482}}.

\bibitem{LHCb:2017avl}
\hrefCMSnoop {}{{LHCb Collaboration}, ``{Test of lepton universality with
  $B^{0} \rightarrow K^{*0}\ell^{+}\ell^{-}$ decays}'',} \textit{ JHEP}
  \textbf{ 08} (2017) 055,
  \href{http://dx.doi.org/10.1007/JHEP08(2017)055}{\doi{10.1007/JHEP08(2017)055}},
  \href{http://www.arXiv.org/abs/1705.05802}{\texttt{arXiv:1705.05802}}.

\bibitem{LHCb:2019hip}
\hrefCMSnoop {}{{LHCb Collaboration}, ``{Search for lepton-universality
  violation in $B^+\to K^+\ell^+\ell^-$ decays}'',} \textit{ Phys. Rev. Lett.}
  \textbf{ 122} (2019) 191801,
  \href{http://dx.doi.org/10.1103/PhysRevLett.122.191801}{\doi{10.1103/PhysRevLett.122.191801}},
  \href{http://www.arXiv.org/abs/1903.09252}{\texttt{arXiv:1903.09252}}.

\bibitem{BELLE:2019xld}
\hrefCMSnoop {}{{BELLE} Collaboration, ``{Test of lepton flavor universality
  and search for lepton flavor violation in $B \rightarrow K\ell \ell$
  decays}'',} \textit{ JHEP} \textbf{ 03} (2021) 105,
  \href{http://dx.doi.org/10.1007/JHEP03(2021)105}{\doi{10.1007/JHEP03(2021)105}},
  \href{http://www.arXiv.org/abs/1908.01848}{\texttt{arXiv:1908.01848}}.

\bibitem{Belle:2019oag}
\hrefCMSnoop {}{{Belle} Collaboration, ``Test of lepton-flavor universality in
  ${B\to K^\ast\ell^+\ell^-}$ decays at {Belle}'',} \textit{ Phys. Rev. Lett.}
  (2021) 161801,
  \href{http://dx.doi.org/10.1103/PhysRevLett.126.161801}{\doi{10.1103/PhysRevLett.126.161801}},
  \href{http://www.arXiv.org/abs/1904.02440}{\texttt{arXiv:1904.02440}}.

\bibitem{LHCb:2021trn}
\hrefCMSnoop {}{{LHCb Collaboration}, ``{Test of lepton universality in
  beauty-quark decays}'',} \textit{ Nature Phys.} \textbf{ 18} (2022) 277,
  \href{http://dx.doi.org/10.1038/s41567-021-01478-8}{\doi{10.1038/s41567-021-01478-8}},
  \href{http://www.arXiv.org/abs/2103.11769}{\texttt{arXiv:2103.11769}}.
  [Addendum: \DOI{10.1038/s41567-023-02095-3}].

\bibitem{LHCb:2023zxo}
\hrefCMSnoop {}{{LHCb Collaboration}, ``{Measurement of the ratios of branching
  fractions $\mathcal{R}(D^{*})$ and $\mathcal{R}(D^{0})$}'',} \textit{ Phys.
  Rev. Lett.} \textbf{ 131} (2023) 111802,
  \href{http://dx.doi.org/10.1103/PhysRevLett.131.111802}{\doi{10.1103/PhysRevLett.131.111802}},
  \href{http://www.arXiv.org/abs/2302.02886}{\texttt{arXiv:2302.02886}}.

\bibitem{LHCb:2022qnv}
\hrefCMSnoop {}{{LHCb Collaboration}, ``{Test of lepton universality in $b
  \rightarrow s \ell^+ \ell^-$ decays}'',} \textit{ Phys. Rev. Lett.} \textbf{
  131} (2023) 051803,
  \href{http://dx.doi.org/10.1103/PhysRevLett.131.051803}{\doi{10.1103/PhysRevLett.131.051803}},
  \href{http://www.arXiv.org/abs/2212.09152}{\texttt{arXiv:2212.09152}}.

\bibitem{LHCb:2022vje}
\hrefCMSnoop {}{{LHCb Collaboration}, ``{Measurement of lepton universality
  parameters in $B^+\to K^+\ell^+\ell^-$ and $B^0\to K^{*0}\ell^+\ell^-$
  decays}'',} \textit{ Phys. Rev. D} \textbf{ 108} (2023) 032002,
  \href{http://dx.doi.org/10.1103/PhysRevD.108.032002}{\doi{10.1103/PhysRevD.108.032002}},
  \href{http://www.arXiv.org/abs/2212.09153}{\texttt{arXiv:2212.09153}}.

\bibitem{Cacciari:1998it}
\hrefCMSnoop {}{M.~Cacciari, M.~Greco, and P.~Nason, ``{The $p_T$ spectrum in
  heavy-flavour hadroproduction.}'',} \textit{ JHEP} \textbf{ 05} (1998) 007,
  \href{http://dx.doi.org/10.1088/1126-6708/1998/05/007}{\doi{10.1088/1126-6708/1998/05/007}},
  \href{http://www.arXiv.org/abs/hep-ph/9803400}{\texttt{arXiv:hep-ph/9803400}}.

\bibitem{Cacciari:2001td}
\hrefCMSnoop {}{M.~Cacciari, S.~Frixione, and P.~Nason, ``{The $p_T$ spectrum
  in heavy flavor photoproduction}'',} \textit{ JHEP} \textbf{ 03} (2001) 006,
  \href{http://dx.doi.org/10.1088/1126-6708/2001/03/006}{\doi{10.1088/1126-6708/2001/03/006}},
  \href{http://www.arXiv.org/abs/hep-ph/0102134}{\texttt{arXiv:hep-ph/0102134}}.

\bibitem{Cacciari:2012ny}
M.~Cacciari\hrefCMSnoop {}{ { et~al.}, ``{Theoretical predictions for charm and
  bottom production at the LHC}'',} \textit{ JHEP} \textbf{ 10} (2012) 137,
  \href{http://dx.doi.org/10.1007/JHEP10(2012)137}{\doi{10.1007/JHEP10(2012)137}},
  \href{http://www.arXiv.org/abs/1205.6344}{\texttt{arXiv:1205.6344}}.

\bibitem{Cacciari:2015fta}
\hrefCMSnoop {}{M.~Cacciari, M.~L. Mangano, and P.~Nason, ``Gluon {PDF}
  constraints from the ratio of forward heavy-quark production at the {LHC} at
  $\sqrt{s}=7$ and 13 {TeV}'',} \textit{ Eur. Phys. J. C} \textbf{ 75} (2015)
  610,
  \href{http://dx.doi.org/10.1140/epjc/s10052-015-3814-x}{\doi{10.1140/epjc/s10052-015-3814-x}},
  \href{http://www.arXiv.org/abs/1507.06197}{\texttt{arXiv:1507.06197}}.

\bibitem{Chen:2016:XST:2939672.2939785}
\hrefCMSnoop {}{T.~Chen and C.~Guestrin, ``{XGBoost}: A scalable tree boosting
  system'',} in \textit{ Proc. 22nd ACM SIGKDD Int. Conf. on Knowledge
  Discovery and Data Mining}.
\newblock 2016.
\newblock
  \href{http://dx.doi.org/10.1145/2939672.2939785}{\doi{10.1145/2939672.2939785}}.

\bibitem{CMS:2014pkt}
\hrefCMSnoop {}{{CMS Collaboration}, ``Measurement of inclusive {W} and {Z}
  boson production cross sections in pp collisions at $\sqrt{s} = 8$ {TeV}'',}
  \textit{ Phys. Rev. Lett.} \textbf{ 112} (2014) 191802,
  \href{http://dx.doi.org/10.1103/PhysRevLett.112.191802}{\doi{10.1103/PhysRevLett.112.191802}},
  \href{http://www.arXiv.org/abs/1402.0923}{\texttt{arXiv:1402.0923}}.

\bibitem{CMS:2023klk}
\hrefCMSnoop {}{{CMS Collaboration}, ``{Test of lepton flavor universality in
  B$^{\pm}$$\to$ K$^{\pm}\mu^+\mu^-$ and B$^{\pm}$$\to$ K$^{\pm}$e$^+$e$^-$
  decays in proton-proton collisions at $\sqrt{s}$ = 13 TeV}'',} \textit{ Rept.
  Prog. Phys.} \textbf{ 87} (2024) 077802,
  \href{http://dx.doi.org/10.1088/1361-6633/ad4e65}{\doi{10.1088/1361-6633/ad4e65}},
  \href{http://www.arXiv.org/abs/2401.07090}{\texttt{arXiv:2401.07090}}.

\bibitem{kinFit}
\hrefCMSnoop {}{K.~Prokofiev and T.~Speer, ``{A kinematic and a decay chain
  reconstruction library}'',} in \textit{ {14th International Conference on
  Computing in High-Energy and Nuclear Physics}}, p.~411.
\newblock 2005.
\newblock
  \href{http://dx.doi.org/10.5170/CERN-2005-002.411}{\doi{10.5170/CERN-2005-002.411}}.

\bibitem{Bilenky:2016pep}
\hrefCMSnoop {}{S.~Bilenky, ``{Neutrino oscillations: From a historical
  perspective to the present status}'',} \textit{ Nucl. Phys. B} \textbf{ 908}
  (2016) 2,
  \href{http://dx.doi.org/10.1016/j.nuclphysb.2016.01.025}{\doi{10.1016/j.nuclphysb.2016.01.025}},
  \href{http://www.arXiv.org/abs/1602.00170}{\texttt{arXiv:1602.00170}}.

\bibitem{bertone_2010}
G.~Bertone, ed., ``{Particle Dark Matter: Observations, Models and Searches}''.
\newblock Cambridge Univ. Press, Cambridge, 2010.
\newblock
  \href{http://dx.doi.org/10.1017/CBO9780511770739}{\doi{10.1017/CBO9780511770739}},
  ISBN~978-1-107-65392-4.

\bibitem{PhysRevD.50.774}
\hrefCMSnoop {}{G.~R. Farrar and M.~E. Shaposhnikov, ``Baryon asymmetry of the
  universe in the standard model'',} \textit{ Phys. Rev. D} \textbf{ 50} (1994)
  774,
  \href{http://dx.doi.org/10.1103/PhysRevD.50.774}{\doi{10.1103/PhysRevD.50.774}}.

\bibitem{CMS-PAS-EXO-23-006}
\hrefCMSnoop {}{{CMS Collaboration}, ``{Review of searches for vector-like
  quarks, vector-like leptons, and heavy neutral leptons in proton-proton
  collisions at $\sqrt{s}$ = 13 TeV at the CMS experiment}'',} {(2024)}.
  \href{http://www.arXiv.org/abs/2405.17605}{\texttt{arXiv:2405.17605}}.
  {Accepted for publication by Phys. Rept.}

\bibitem{CMS-PAS-EXO-22-019}
\hrefCMSnoop {}{{CMS Collaboration}, ``{Search for long-lived heavy neutrinos
  in the decays of B mesons produced in proton-proton collisions at $ \sqrt{s}
  $ = 13 TeV}'',} \textit{ JHEP} \textbf{ 06} (2024) 183,
  \href{http://dx.doi.org/10.1007/JHEP06(2024)183}{\doi{10.1007/JHEP06(2024)183}},
  \href{http://www.arXiv.org/abs/2403.04584}{\texttt{arXiv:2403.04584}}.

\bibitem{CMS:2019qux}
\hrefCMSnoop {}{{CMS Collaboration}, ``Upgrade of the {CMS} barrel muon track
  finder for {HL-LHC} featuring a {Kalman} filter algorithm and an {ATCA} host
  processor with {Ultrascale+} {FPGAs}'',} \textit{ PoS} \textbf{ 343} (2019)
  139, \href{http://dx.doi.org/10.22323/1.343.0139}{\doi{10.22323/1.343.0139}}.

\bibitem{LHCb:2015brj}
\hrefCMSnoop {}{{LHCb Collaboration}, ``{Measurement of the time-dependent CP
  asymmetries in $B_s^0\rightarrow J/\psi K_{\rm S}^0$}'',} \textit{ JHEP}
  \textbf{ 06} (2015) 131,
  \href{http://dx.doi.org/10.1007/JHEP06(2015)131}{\doi{10.1007/JHEP06(2015)131}},
  \href{http://www.arXiv.org/abs/1503.07055}{\texttt{arXiv:1503.07055}}.

\bibitem{CMS-PAPERS-HIG-16-043}
\hrefCMSnoop {}{{CMS Collaboration}, ``{Observation of the Higgs boson decay to
  a pair of $\tau$ leptons with the CMS detector}'',} \textit{ Phys. Lett. B}
  \textbf{ 779} (2018) 283,
  \href{http://dx.doi.org/10.1016/j.physletb.2018.02.004}{\doi{10.1016/j.physletb.2018.02.004}},
  \href{http://www.arXiv.org/abs/1708.00373}{\texttt{arXiv:1708.00373}}.

\bibitem{CMS-PAPERS-HIG-20-003}
\hrefCMSnoop {}{{CMS Collaboration}, ``Search for invisible decays of the
  {Higgs} boson produced via vector boson fusion in proton-proton collisions at
  $\sqrt{s}=13$ {TeV}'',} \textit{ Phys. Rev. D} \textbf{ 105} (2022) 092007,
  \href{http://dx.doi.org/10.1103/PhysRevD.105.092007}{\doi{10.1103/PhysRevD.105.092007}},
  \href{http://www.arXiv.org/abs/2201.11585}{\texttt{arXiv:2201.11585}}.

\bibitem{CMS-PAPERS-HIG-19-006}
\hrefCMSnoop {}{{CMS Collaboration}, ``{Evidence for Higgs boson decay to a
  pair of muons}'',} \textit{ JHEP} \textbf{ 01} (2021) 148,
  \href{http://dx.doi.org/10.1007/JHEP01(2021)148}{\doi{10.1007/JHEP01(2021)148}},
  \href{http://www.arXiv.org/abs/2009.04363}{\texttt{arXiv:2009.04363}}.

\bibitem{paper:SMEFT}
\hrefCMSnoop {}{{SMEFiT} Collaboration, ``{Combined SMEFT interpretation of
  Higgs, diboson, and top quark data from the LHC}'',} \textit{ JHEP} \textbf{
  11} (2021) 089,
  \href{http://dx.doi.org/10.1007/JHEP11(2021)089}{\doi{10.1007/JHEP11(2021)089}},
  \href{http://www.arXiv.org/abs/2105.00006}{\texttt{arXiv:2105.00006}}.

\bibitem{CMS-PAPERS-B2G-22-003}
\hrefCMSnoop {}{{CMS Collaboration}, ``Search for nonresonant pair production
  of highly energetic {Higgs} bosons decaying to bottom quarks'',} \textit{
  Phys. Rev. Lett.} \textbf{ 131} (2023) 041803,
  \href{http://dx.doi.org/10.1103/PhysRevLett.131.041803}{\doi{10.1103/PhysRevLett.131.041803}},
  \href{http://www.arXiv.org/abs/2205.06667}{\texttt{arXiv:2205.06667}}.

\bibitem{CMS-DP-2017-022}
\href {https://cds.cern.ch/record/2273268}{{CMS Collaboration}, ``Level 1 tau
  trigger performance in 2016 data and {VBF} seeds at {Level 1} trigger'',} CMS
  Detector Performance Note CMS-DP-2017-022, 2017.

\bibitem{CMS-PAPERS-EXO-20-005}
\hrefCMSnoop {}{{CMS Collaboration}, ``{Search for dark photons in Higgs boson
  production via vector boson fusion in proton-proton collisions at $ \sqrt{s}
  $ = 13 TeV}'',} \textit{ JHEP} \textbf{ 03} (2021) 011,
  \href{http://dx.doi.org/10.1007/JHEP03(2021)011}{\doi{10.1007/JHEP03(2021)011}},
  \href{http://www.arXiv.org/abs/2009.14009}{\texttt{arXiv:2009.14009}}.

\bibitem{paper:SUEPs}
J.~Barron\hrefCMSnoop {}{ { et~al.}, ``{Unsupervised hadronic SUEP at the
  LHC}'',} \textit{ JHEP} \textbf{ 12} (2021) 129,
  \href{http://dx.doi.org/10.1007/JHEP12(2021)129}{\doi{10.1007/JHEP12(2021)129}},
  \href{http://www.arXiv.org/abs/2107.12379}{\texttt{arXiv:2107.12379}}.

\bibitem{CMS-DP-2023-021}
\href {https://cds.cern.ch/record/2857440}{{CMS Collaboration}, ``Performance
  of the {ParticleNet} tagger on small and large-radius jets at high level
  trigger in {Run 3}'',} CMS Detector Performance Note CMS-DP-2023-021, 2023.

\bibitem{CMS-DP-2023-024}
\href {https://cds.cern.ch/record/2859462}{{CMS Collaboration}, ``Performance
  of tau lepton reconstruction at high level trigger using 2022 data from the
  {CMS} experiment at {CERN}'',} CMS Detector Performance Note CMS-DP-2023-024,
  2023.

\bibitem{LLPRun3TriggerDPNote}
\href {https://cds.cern.ch/record/2865844}{{CMS Collaboration}, ``Performance
  of long lived particle triggers in {Run 3}'',} CMS Detector Performance Note
  CMS-DP-2023-043, 2023.

\bibitem{EXO-18-003}
\hrefCMSnoop {}{{CMS Collaboration}, ``Search for long-lived particles decaying
  to leptons with large impact parameter in proton-proton collisions at
  $\sqrt{s} = 13$ {TeV}'',} \textit{ Eur. Phys. J. C} \textbf{ 82} (2022) 153,
  \href{http://dx.doi.org/10.1140/epjc/s10052-022-10027-3}{\doi{10.1140/epjc/s10052-022-10027-3}},
  \href{http://www.arXiv.org/abs/2110.04809}{\texttt{arXiv:2110.04809}}.

\bibitem{EXO-21-006}
\hrefCMSnoop {}{{CMS Collaboration}, ``{Search for long-lived particles
  decaying to a pair of muons in proton-proton collisions at $ \sqrt{s} $ = 13
  TeV}'',} \textit{ JHEP} \textbf{ 05} (2023) 228,
  \href{http://dx.doi.org/10.1007/JHEP05(2023)228}{\doi{10.1007/JHEP05(2023)228}},
  \href{http://www.arXiv.org/abs/2205.08582}{\texttt{arXiv:2205.08582}}.

\bibitem{CMS:2024qxz}
\hrefCMSnoop {}{{CMS Collaboration}, ``{Search for long-lived particles
  decaying to final states with a pair of muons in proton-proton collisions at
  $\sqrt{s}$ = 13.6 TeV}'',} \textit{ JHEP} \textbf{ 05} (2024) 047,
  \href{http://dx.doi.org/10.1007/JHEP05(2024)047}{\doi{10.1007/JHEP05(2024)047}},
  \href{http://www.arXiv.org/abs/2402.14491}{\texttt{arXiv:2402.14491}}.

\bibitem{CMS:2019zxa}
\hrefCMSnoop {}{{CMS Collaboration}, ``Search for long-lived particles using
  delayed photons in proton-proton collisions at $\sqrt{s}= 13$ {TeV}'',}
  \textit{ Phys. Rev. D} \textbf{ 100} (2019) 112003,
  \href{http://dx.doi.org/10.1103/PhysRevD.100.112003}{\doi{10.1103/PhysRevD.100.112003}},
  \href{http://www.arXiv.org/abs/1909.06166}{\texttt{arXiv:1909.06166}}.

\bibitem{CMS:2020iwv}
\hrefCMSnoop {}{{CMS Collaboration}, ``{Search for long-lived particles using
  displaced jets in proton-proton collisions at $\sqrt{s} = $ 13 TeV}'',}
  \textit{ Phys. Rev. D} \textbf{ 104} (2021) 012015,
  \href{http://dx.doi.org/10.1103/PhysRevD.104.012015}{\doi{10.1103/PhysRevD.104.012015}},
  \href{http://www.arXiv.org/abs/2012.01581}{\texttt{arXiv:2012.01581}}.

\bibitem{EXO-19-013}
\hrefCMSnoop {}{{CMS Collaboration}, ``{Search for long-lived particles
  decaying to jets with displaced vertices in proton-proton collisions at
  $\sqrt{s}=$ 13 TeV}'',} \textit{ Phys. Rev. D} \textbf{ 104} (2021) 052011,
  \href{http://dx.doi.org/10.1103/PhysRevD.104.052011}{\doi{10.1103/PhysRevD.104.052011}},
  \href{http://www.arXiv.org/abs/2104.13474}{\texttt{arXiv:2104.13474}}.

\bibitem{EXO-19-001}
\hrefCMSnoop {}{{CMS Collaboration}, ``{Search for long-lived particles using
  nonprompt jets and missing transverse momentum with proton-proton collisions
  at $\sqrt{s} =$ 13 TeV}'',} \textit{ Phys. Lett. B} \textbf{ 797} (2019)
  134876,
  \href{http://dx.doi.org/10.1016/j.physletb.2019.134876}{\doi{10.1016/j.physletb.2019.134876}},
  \href{http://www.arXiv.org/abs/1906.06441}{\texttt{arXiv:1906.06441}}.

\bibitem{CMS-PAS-EXO-21-014}
\hrefCMSnoop {}{{CMS Collaboration}, ``{Search for long-lived particles using
  out-of-time trackless jets in proton-proton collisions at $\sqrt{s}$ = 13
  TeV}'',} \textit{ JHEP} \textbf{ 07} (2023) 210,
  \href{http://dx.doi.org/10.1007/JHEP07(2023)210}{\doi{10.1007/JHEP07(2023)210}},
  \href{http://www.arXiv.org/abs/2212.06695}{\texttt{arXiv:2212.06695}}.

\bibitem{CMS:2024trg}
\hrefCMSnoop {}{{CMS Collaboration}, ``{Search for long-lived particles using
  displaced vertices and missing transverse momentum in proton-proton
  collisions at $\sqrt{s}$ = 13 TeV}'',} \textit{ Phys. Rev. D} \textbf{ 109}
  (2024) 112005,
  \href{http://dx.doi.org/10.1103/PhysRevD.109.112005}{\doi{10.1103/PhysRevD.109.112005}},
  \href{http://www.arXiv.org/abs/2402.15804}{\texttt{arXiv:2402.15804}}.

\bibitem{SUS-21-006}
\hrefCMSnoop {}{{CMS Collaboration}, ``Search for supersymmetry in final states
  with disappearing tracks in proton-proton collisions at $\sqrt{s}=13$
  {TeV}'',} \textit{ Phys. Rev. D} \textbf{ 109} (2024) 072007,
  \href{http://dx.doi.org/10.1103/PhysRevD.109.072007}{\doi{10.1103/PhysRevD.109.072007}},
  \href{http://www.arXiv.org/abs/2309.16823}{\texttt{arXiv:2309.16823}}.

\bibitem{CMS:2020atg}
\hrefCMSnoop {}{{CMS Collaboration}, ``Search for disappearing tracks in
  proton-proton collisions at $\sqrt{s} = 13$ {TeV}'',} \textit{ Phys. Lett. B}
  \textbf{ 806} (2020) 135502,
  \href{http://dx.doi.org/10.1016/j.physletb.2020.135502}{\doi{10.1016/j.physletb.2020.135502}},
  \href{http://www.arXiv.org/abs/2004.05153}{\texttt{arXiv:2004.05153}}.

\bibitem{Khachatryan:2016sfv}
\hrefCMSnoop {}{{CMS Collaboration}, ``{Search for long-lived charged particles
  in proton-proton collisions at $\sqrt s=$ 13 TeV}'',} \textit{ Phys. Rev. D}
  \textbf{ 94} (2016) 112004,
  \href{http://dx.doi.org/10.1103/PhysRevD.94.112004}{\doi{10.1103/PhysRevD.94.112004}},
\href{http://www.arXiv.org/abs/1609.08382}{\texttt{arXiv:1609.08382}}.

\end{thebibliography}\endgroup
\appendix

\section{Glossary}
\begin{tabular}{>{\bfseries}ll}
AK & Anti-\kt \\
ALICE & A Large Ion Collider Experiment (experiment) \\
ATLAS & A Toroidal Lhc ApparatuS (experiment) \\
BDT & Boosted decision tree \\
BSM & Beyond standard model \\
CB & Crystal Ball \\
CHS & Charged hadron subtraction (algorithm) \\
CL & Confidence level \\ 
CMS & Compact Muon Solenoid (experiment) \\
CP & Charge-parity (symmetry) \\
CPU & Central processing unit \\ 
CR & Control region \\
DAQ & Data acquisition \\
DDT & Designing decorrelated taggers (procedure) \\
DNN & Deep neural network \\
DT  & Deep tau (identification algorithm) \\
DY & Drell--Yan (process) \\
ECAL & Electromagnetic calorimeter \\
EFT & Effective field theory \\
EB & ECAL barrel section \\
EE & ECAL endcap section \\
EW & Electroweak \\ 
FPGA & Field programmable gate array \\
ggF & Gluon-gluon fusion (production process) \\
GPU & Graphics processing unit \\
GSF & Gaussian sum filter \\
GT & Global trigger \\
HCAL & Hadron calorimeter \\
HB & HCAL barrel section \\
HE & HCAL endcap section \\
HF & HCAL forward section \\
HO & HCAL outer section \\
HH  & Higgs boson pair\\
HL-LHC & High-Luminosity Large Hadron Collider \\
HLT & High-level trigger \\
HPS & Hadrons-plus-strips (tau lepton reconstruction algorithm) \\
IO & Inside-out (method of matching muon tracks between the tracker and the muon detector) \\
IP & Interaction point \\
JER & Jet energy resolution \\
JES & Jet energy scale \\
LHC & Large Hadron Collider \\
LHCb & Large Hadron Collider beauty (experiment) \\
L1 & Level-1 trigger \\
L2 & Level 2 muons (method of reconstructing muon candidates) \\
L3 & Level 3 muons (method of reconstructing muon candidates) \\
\end{tabular}
\newpage
\begin{tabular}{>{\bfseries}ll}
LFU & Lepton flavor universality \\
LLP & Long-lived particle \\
LS1 & Long shutdown 1, during the years 2013--2014 \\
LS2 & Long shutdown 2, during the years 2019--2021 \\
LO & Leading order \\
LP & Low-\pt algorithm (for electron reconstruction) \\
MAHI & Minimization at HCAL, iteratively \\
MC & Monte Carlo (simulation) \\ 
ML & Machine learning \\
MSSM & Minimal supersymmetric standard model \\
MVA & Multi-variate analysis \\
NMSSM & Next-to-minimal supersymmetric standard model \\
NLO & Next-to-leading order \\
NN   & Neural network \\
NWA & Narrow-width approximation \\
OI & Outside-in (method of matching muon tracks between the tracker and the muon detector) \\
PD & Primary data set \\
PDG & Particle Data Group \\
PF  & Particle-flow (method of reconstructing particle candidates) \\
pp & Proton-proton \\ 
PU & Pileup \\
PUPPI & Pileup-per-particle identification (algorithm) \\
PV & Primary vertex \\
QCD & Quantum chromodynamics \\
QGD & Quark-gluon discriminator \\
R   & distance in ($\Delta \eta, \Delta \phi$) space \\
ROC & Receiver operating characteristic \\
RPV & $R$-parity violation (a supersymmetric model) \\
Run 1 & The first run of the LHC, during the years 2010--2012 \\ 
Run 2 & The second run of the LHC, during the years 2015--2018 \\
Run 3 & The third run of the LHC, starting in 2022 \\
SC & Supercluster (in the context of ECAL object reconstruction) \\
SD & Soft-drop (algorithm) \\
SM  & Standard model \\
SR & Signal region \\
SUEP & Soft unclustered energy pattern \\
SUSY & Supersymmetry \\
SV & Secondary vertex \\
TP & Trigger primitive \\
TS & Time sample \\
VBF & Vector boson fusion (production process) \\
VH  & Vector plus Higgs boson (production process or decay channel) \\
2HDM & Two-Higgs-doublet model \\
2HDM+S & Two-Higgs-doublet-plus-additional-singlet model \\
\end{tabular}

\cleardoublepage \section{The CMS Collaboration \label{app:collab}}\begin{sloppypar}\hyphenpenalty=5000\widowpenalty=500\clubpenalty=5000
\cmsinstitute{Yerevan Physics Institute, Yerevan, Armenia}
{\tolerance=6000
A.~Hayrapetyan, A.~Tumasyan\cmsAuthorMark{1}\cmsorcid{0009-0000-0684-6742}
\par}
\cmsinstitute{Institut f\"{u}r Hochenergiephysik, Vienna, Austria}
{\tolerance=6000
W.~Adam\cmsorcid{0000-0001-9099-4341}, J.W.~Andrejkovic, T.~Bergauer\cmsorcid{0000-0002-5786-0293}, S.~Chatterjee\cmsorcid{0000-0003-2660-0349}, K.~Damanakis\cmsorcid{0000-0001-5389-2872}, M.~Dragicevic\cmsorcid{0000-0003-1967-6783}, P.S.~Hussain\cmsorcid{0000-0002-4825-5278}, M.~Jeitler\cmsAuthorMark{2}\cmsorcid{0000-0002-5141-9560}, N.~Krammer\cmsorcid{0000-0002-0548-0985}, A.~Li\cmsorcid{0000-0002-4547-116X}, D.~Liko\cmsorcid{0000-0002-3380-473X}, I.~Mikulec\cmsorcid{0000-0003-0385-2746}, J.~Schieck\cmsAuthorMark{2}\cmsorcid{0000-0002-1058-8093}, R.~Sch\"{o}fbeck\cmsorcid{0000-0002-2332-8784}, D.~Schwarz\cmsorcid{0000-0002-3821-7331}, M.~Sonawane\cmsorcid{0000-0003-0510-7010}, S.~Templ\cmsorcid{0000-0003-3137-5692}, W.~Waltenberger\cmsorcid{0000-0002-6215-7228}, C.-E.~Wulz\cmsAuthorMark{2}\cmsorcid{0000-0001-9226-5812}
% ...
\par}
\vskip\cmsinstskip
\dag:~Deceased\\
$^{1}$Also at Yerevan State University, Yerevan, Armenia\\
$^{2}$Also at TU Wien, Vienna, Austria\\
% ...
\end{sloppypar}
\end{document}